\documentclass[12pt]{article} 

\usepackage{amsmath,amsthm,amssymb,mathrsfs}
\usepackage[english]{babel}
\usepackage{bm}
\usepackage{color}
\usepackage{enumerate}
\usepackage{graphicx}
\usepackage{hyperref}
\usepackage{fullpage}
\usepackage{multirow}
\usepackage{natbib}
\usepackage{rotating}
\usepackage{subcaption}
\usepackage{graphicx}
\usepackage[dvipsnames]{xcolor}
\usepackage[normalem]{ulem}
\usepackage{float}
\usepackage{booktabs}
\usepackage{titlesec}
\titlespacing*{\section}{0pt}{1.5ex plus .5ex minus .2ex}{0.8ex plus .2ex}
\titlespacing*{\subsection}{0pt}{1.2ex plus .4ex minus .2ex}{0.6ex plus .2ex}
\titlespacing*{\subsubsection}{0pt}{1ex plus .3ex minus .2ex}{0.5ex plus .1ex}
\usepackage{url} 
\usepackage[ruled,vlined]{algorithm2e}
\usepackage{algorithmicx,algpseudocode}
\usepackage{bm}

\newtheorem{theorem}{Theorem}
\newtheorem{lemma}{Lemma}
\newtheorem{corollary}{Corollary}
\newtheorem{proposition}{Proposition}

\theoremstyle{definition}

\newtheorem{definition}{Definition}
\newtheorem{example}{Example}
\newtheorem{remark}{Remark}

\definecolor{sy}{RGB}{51,79,255}

\definecolor{yw}{RGB}{20,161,43}

\definecolor{lw}{RGB}{255, 2, 141}

\newif\ifshowchg  \showchgfalse
\newcommand{\chg}[1]{%
  \ifshowchg
    \ifmmode{\color{lw}#1}\else\textcolor{lw}{#1}\fi
  \else #1\fi}
\newenvironment{chgblock}{\ifshowchg\color{lw}\fi}{}

\hypersetup{hypertexnames=false,
    colorlinks=true,
    linkcolor=Maroon,   
    urlcolor=Green,
    citecolor=NavyBlue,
}

\makeatletter
\newcommand*{\addFileDependency}[1]{
  \typeout{(#1)}
  \@addtofilelist{#1}
  \IfFileExists{#1}{}{\typeout{No file #1.}}
}
\makeatother

\def\spacingset#1{\renewcommand{\baselinestretch}%
{#1}\small\normalsize} \spacingset{1}

\newcommand{\argmin}{\operatorname*{\arg\min}}

\newcommand{\tr}{\operatorname{Tr}} 

\newcommand{\E}{\mathbb{E}} 
\newcommand{\OP}{O_{\mathrm P}} 

\newtheorem{assumpB}{Assumption}

\def\boxit#1{\vbox{\hrule\hbox{\vrule\kern6pt
          \vbox{\kern6pt#1\kern6pt}\kern6pt\vrule}\hrule}}

\makeatletter
\let\@fnsymbol\@arabic
\makeatother

\usepackage{longtable}
\SetKwInOut{Parameter}{Parameters}

\newcommand{\bZ}{\bm{Z}}

\newcommand{\bs}{\bm}
\newcommand{\TODO}[1]{\textcolor{red}{[TO VERIFY: #1]}}
\newcommand{\suppanchor}{M}
\expandafter\def\csname theHdefinition\endcsname{\suppanchor.\arabic{definition}}
\expandafter\def\csname theHequation\endcsname{\suppanchor.\arabic{equation}}
\expandafter\def\csname theHsubsection\endcsname{\suppanchor.\arabic{subsection}}
\expandafter\def\csname theHsubsubsection\endcsname{\suppanchor.\arabic{subsubsection}}
\expandafter\def\csname theHtheorem\endcsname{\suppanchor.\arabic{theorem}}
\expandafter\def\csname theHlemma\endcsname{\suppanchor.\arabic{lemma}}
\expandafter\def\csname theHproposition\endcsname{\suppanchor.\arabic{proposition}}
\expandafter\def\csname theHcorollary\endcsname{\suppanchor.\arabic{corollary}}
\expandafter\def\csname theHfigure\endcsname{\suppanchor.\arabic{figure}}
\expandafter\def\csname theHtable\endcsname{\suppanchor.\arabic{table}}
\expandafter\def\csname theHremark\endcsname{\suppanchor.\arabic{remark}}
\expandafter\def\csname theHexample\endcsname{\suppanchor.\arabic{example}}

\begin{document}

\def\spacingset#1{\renewcommand{\baselinestretch}%
{#1}\small\normalsize} \spacingset{1}


\newcommand{\Title}{Causal Learning with Invalid Instruments for High-dimensional Imaging Responses}

\newcommand{\blind}{1}

\if1\blind
{
  \title{\bf \Title}
    \author{
        Yiting Wang\footnote{Department of Statistics, University of Virginia.}
        \and 
        Chunlin Li\footnotemark[1] $^{,}$\footnote{Department of Statistics, Iowa State University.}
        \and 
        Shan Yu\footnotemark[1]
        \and 
        Lily Wang\footnote{Department of Statistics, George Mason University.}
    }
    \date{}
    \maketitle
} \fi

\if0\blind
{
  \bigskip
  \bigskip
  \bigskip
  \title{\bf \Title}
  \date{}
  \maketitle
  \medskip
} \fi

\bigskip
\begin{abstract}
Large-scale neuroimaging studies provide high-dimensional imaging phenotypes for understanding brain-related diseases and their causal relations with biological risk factors. While instrumental variable (IV) methods can address unmeasured confounding in such observational data, existing IV methods face significant limitations when applied to high-dimensional imaging data, particularly in handling large-scale data and leveraging the intrinsic spatial structure of images. {To address these challenges, we propose \textit{High-dimensional Functional-imaging Instrumental Variable Estimation} (Hi-FIVE), a novel framework for estimating spatially varying causal effects when a majority of the candidate instruments may be locally invalid. 
We first establish identifiability conditions under which the spatial continuity of the causal effect surface propagates identification from regions where valid instruments form a plurality into regions where they do not, provided that invalid-IV effects switch on non-smoothly. Motivated by this result, Hi-FIVE combines penalized multivariate splines over triangulations for the causal surface and region-wise penalties for spatially localized invalid-IV effects\chg{, under an anchor constraint that keeps at least one relevant instrument valid on every simplex}. A distributed algorithm parallelizes the computation over simplices for two- and three-dimensional images. We study the asymptotic properties of Hi-FIVE, establish consistency and pointwise asymptotic normality, and construct bootstrap-calibrated simultaneous confidence regions whose finite-sample coverage is evaluated through simulations. Hi-FIVE is illustrated with an analysis of the effect of phosphorylated tau on FDG-PET imaging in the Alzheimer's Disease Neuroimaging Initiative, interpreted under the stated IV assumptions.}
\end{abstract}
\noindent%

\thispagestyle{empty} 

\clearpage
\setcounter{page}{1}   

\newpage
\spacingset{1.65} 

\section{Introduction}

Neuroimaging studies now routinely produce high-dimensional imaging phenotypes, such as structural and functional magnetic resonance imaging (sMRI/fMRI), positron emission tomography, and electroencephalography (EEG) \citep{ombao2016handbookneuroimg, zhu2023statrevneuroimg}. These data offer rich spatially resolved measurements of brain structure and function and have become central to studies of brain development, aging, neurodegeneration, and neurological and psychiatric disorders \citep{wang2024heterogeneity}. In Alzheimer's disease (AD), for example, imaging phenotypes can help characterize spatial patterns of neurodegeneration and connect biological risk factors to downstream clinical impairment. 

These phenotypes, however, pose a distinctive challenge for causal inference. Analyses of such data are predominantly predictive \citep{rahaman2021predimggen, shen2020revimgml} or associational \citep{yu2022mappath}. Such methods are useful for characterizing statistical relationships, but they do not distinguish correlation from causation and therefore may not reveal the biological mechanisms underlying genetic--imaging--clinical pathways \citep{siddiqi2022causalmap, zhu2023statrevneuroimg}. Randomized controlled trials remain the gold standard for causal inference, but they are often infeasible or unethical in neuroscience applications \citep{weichwald2020causalcogneuro}. Consequently, causal analysis of imaging phenotypes must often rely on observational data, where unmeasured confounding can severely bias causal conclusions even after adjustment for measured covariates.

This article develops a causal learning framework for high-dimensional imaging responses under unmeasured confounding. The central idea is that the spatial structure of imaging outcomes serves not only to improve statistical efficiency, as in conventional functional data analysis, but also to strengthen causal identification. In particular, we show that spatial smoothness can help identify spatially varying causal effects in instrumental variable (IV) models even when standard pointwise IV identification conditions fail over substantial portions of an image. This distinction, exploiting spatial structure as a source of causal identification rather than merely a technique for denoising or smoothing, is the key motivation for the proposed approach.

\subsection{Related Work and Challenges}

Our approach builds on IV regression \citep{angrist1996identification}, a widely used framework for addressing unmeasured confounding in causal inference \citep{kang2024identification}. A valid IV must satisfy three core conditions: (V1) relevance, meaning that the IV is associated with the exposure; (V2) exclusion, meaning that the IV affects the outcome only through the exposure; and (V3) exogeneity, meaning that the IV is independent of unmeasured confounders. While relevance can often be empirically assessed, exclusion and exogeneity are typically unverifiable and fragile in applications.

This fragility is especially pronounced in Mendelian randomization (MR), where genetic variants such as single-nucleotide polymorphisms (SNPs) are commonly used as instruments. Genetic variants may violate exclusion or exogeneity through pleiotropy, where a variant affects multiple traits or biological pathways. In neuroimaging MR, this issue is amplified because imaging phenotypes are high-level biological outcomes that aggregate signals from multiple interacting systems. As a result, even genetic variants that are relevant to a scalar biological exposure may influence spatially resolved imaging outcomes through additional biological pathways not mediated by the exposure \citep{taschler2022causmrreview}. Classical IV estimators can be severely biased in the presence of such invalid instruments.

A large literature has developed robust IV and invalid-IV selection methods for scalar outcomes, often relying on the \textit{plurality hypothesis} of valid instruments \citep{kang2016instrumental, windmeijer2019sisviveinf, windmeijer2021confidence, guo2018confidence, kong2022iv, guo2023causal, li2023inference, chen2024discovery, zilinskas2024inferring, lin2023robust, lin2024instrumental}. However, applying these scalar-outcome methods pixel by pixel to imaging data raises two difficulties. First, it requires the plurality condition to hold at every image location, an ``everywhere plurality'' requirement that is often unrealistic for neuroimaging outcomes. Second, it treats each pixel as an isolated causal problem and ignores the spatial organization of the image. Thus, pointwise robust IV methods fail not merely on efficiency grounds but also because they discard information that may be essential for causal identification.

A separate body of work treats imaging outcomes as functional data in image-on-scalar or function-on-scalar regression models \citep{Zhu:Fan:Kong:14, feng2020bayesian, Yu:etal:21, Li:Wang:Wang:21, zeng2022bayesian, zhang2023image, zhou2024npimaging}. These methods borrow strength across neighboring pixels, improve estimation of spatial signals, and provide interpretable effect maps with rigorous inference. Functional data analysis has also been studied in IV settings \citep{florens2015instrumental, benatia2017functional, cai2019inferences, seong2024functional, petrovich2024instrumental}. Existing functional IV methods, however, generally assume that all instruments are valid. They therefore do not address the central difficulty of imaging MR when most candidate instruments may be invalid. More importantly, existing functional approaches use spatial structure primarily for estimation and regularization, not for invalid-IV causal identification.

Scalability presents a further difficulty. Modern imaging datasets often contain thousands to millions of pixels or voxels per subject, with complex anatomical boundaries, spatial dependence, measurement noise, and structured local signals. Existing tensor regression and functional regression methods can be effective for small- to moderate-scale imaging studies \citep{zhou2013tensor, Yu:etal:21}, but conventional functional regression techniques often become computationally prohibitive as image resolution increases. Thus, a useful causal inference framework for imaging outcomes must simultaneously achieve robust identification, spatial regularization, and scalable computation.

\vspace{-0.5cm}
\subsection{Our Contributions}

In this article, we propose \textit{High-dimensional Functional-imaging Instrumental Variable Estimation} (Hi-FIVE), a functional data analysis framework for estimating spatially varying causal effects when a majority of instruments may be invalid.

We first establish causal identifiability conditions for IV regression with imaging responses that relax the ``everywhere-plurality'' requirement of pointwise robust-IV methods. 
The key idea is a separation of smoothness classes: the causal effect surface is assumed to be $C^{r}$, whereas an instrument whose invalid effect switches on inside a region where plurality fails must do so non-smoothly. Under spatial identifiability conditions of Theorem \ref{theorem:identifiability}, identification then propagates from the region where valid instruments form a plurality to the entire image domain. Thus, spatial structure contributes to causal identification rather than merely smoothing: pixelwise estimation followed by post-smoothing cannot recover the causal effect where pointwise plurality fails, whereas the joint spatial formulation can. 
Although our primary motivation is neuroimaging, the identifiability argument applies more broadly to function-valued outcomes.

Second, we develop an estimation procedure that operationalizes this identification strategy. Hi-FIVE models the causal effect as a smooth spatial function and represents it using multivariate penalized splines over triangulations. Triangulations are well suited for imaging data because they can accommodate irregular anatomical domains and complex boundaries \citep{Yu:etal:21, wang2020simultaneous, Li:Wang:Wang:21}. To account for invalid instruments, Hi-FIVE introduces region-wise selection penalties that allow an IV to be valid in some spatial regions and invalid in others. This is important because different brain regions, tissue types, or image segments may be affected by different biological pathways. In addition, \chg{we introduce an anchor constraint that guarantees at least one relevant instrument remains valid on every simplex, mirroring the identification condition (A1) in Theorem~\ref{theorem:identifiability}.} The resulting method estimates a smooth causal effect map while simultaneously identifying locally invalid IV effects. A key distinction from post-smoothing approaches \citep{fan1999statistical} is that Hi-FIVE uses spatial information during the invalid-IV selection and causal estimation steps. In non-causal image-on-scalar regression, one may estimate a coefficient surface pointwise and then smooth the estimates afterward. In causal imaging analysis, this logic breaks down: smoothing biased pointwise IV estimates does not restore identification when plurality fails locally. Hi-FIVE instead incorporates spatial structure directly into the causal identification and estimation problem, which enables consistent estimation under settings where pointwise robust IV methods and their post-smoothed variants may remain biased.

Third, we design Hi-FIVE to be computationally scalable for high-resolution imaging data through a domain decomposition scheme built on distributed learning for functional data regression \citep{yu2025distributed}. By partitioning the imaging domain into triangulated subregions containing relatively small number of pixels or voxels and associated local parameters, invalid-IV selection and effect estimation can be performed locally and in parallel, substantially reducing the computational burden. In particular, because the Bernstein–B\'ezier basis functions are supported on single simplices, the invalid-IV updates decouple across simplices, while the causal-effect update can be assembled from simplexwise quantities, so the algorithm parallelizes across workers with communication limited to shared-interface coefficients. The triangulation-based representation applies naturally to both 2D and 3D imaging domains, and the domain decomposition allows tuning parameters to adapt across subdomains, accommodating spatial heterogeneity in the underlying IV signals.  Together, these features make Hi-FIVE suitable for statistically robust and computationally scalable analysis of high-resolution imaging data. 

We then develop statistical inference for the spatially varying causal effect. Unlike classical variable-selection settings \citep{Li:Wang:Wang:21}, which typically assume sparsity among a large number of candidate variables,  we allow invalid-IV effects to occupy nonvanishing portions of the spatial domain as the sample size increases. Furthermore, we accommodate spatially correlated observations through the error covariance structure and region-specific effective sample sizes. Building on the pointwise results, we construct simultaneous confidence regions (SCRs) by recalibrating the pointwise intervals with a wild bootstrap. These regions provide uncertainty quantification for the entire causal effect surface rather than isolated pointwise intervals, which is essential for interpretable inference in imaging studies.

Finally, we evaluate Hi-FIVE through simulation studies and an application to the Alzheimer's Disease Neuroimaging Initiative (ADNI) dataset. The simulations compare Hi-FIVE with pointwise robust IV methods, post-smoothed pointwise estimators, highlighting the role of spatial information in invalid-IV causal learning. In the ADNI analysis, we study the causal effect of phosphorylated tau (p-tau) on FDG-PET imaging responses, illustrating how Hi-FIVE can estimate spatially varying causal effects while identifying region-specific invalid instruments.

The rest of this article is organized as follows. Section \ref{section:method} introduces the image-on-scalar IV model with invalid instruments, presents the spatial identifiability results, and describes the Hi-FIVE estimation procedure. Section \ref{section:theory} establishes the large-sample properties of Hi-FIVE and develops a bootstrap procedure for SCRs. Section \ref{sec:simulation} reports simulation studies, and Section \ref{section:application} presents the ADNI application. Section \ref{sec:discussion} concludes with a discussion. Technical proofs, implementation details, and additional numerical results are provided in the Supplementary Material.

\section{Image-on-scalar Instrumental Variable Regression} 
\label{section:method}

This section develops an image-on-scalar instrumental variable regression framework for estimating spatially varying causal effects when candidate instruments may be invalid over parts of the image domain.

Let $\Omega$ denote the image domain, assumed without loss of generality to be a compact subset of $\mathbb{R}^\kappa$ with $\kappa \in \{2, 3\}$.  Let $\bm{s}\in \Omega$ denote a location point in the domain $\Omega$. Let $Y(\bm{s})$ be the imaging outcome at $\bm{s} \in \Omega$, $D$ be the exposure, and $\mathbf Z=(Z_{1},\ldots,Z_{p})^{\top}$ be a vector of candidate IVs.
Consider the following image-on-scalar IV regression model:
    \begin{equation}
    \label{Model:IV}
        Y(\bm{s}) = D\beta^\ast(\bm{s}) + \mathbf{Z}^{\top} \bm{\phi}^\ast(\bm{s}) +  U(\bm{s}) + \varepsilon(\bm{s}), 
        \qquad
        D = \mathbf{Z}^\top \bm{\gamma}^\ast + \xi, 
    \end{equation}
for $\bm{s}\in \Omega$, where $\beta^\ast(\cdot)$ is the spatially varying causal effect function of primary interest, $\bm\phi^\ast(\cdot) = (\phi_1^\ast(\cdot),\ldots,\phi_p^\ast(\cdot))^{\top}$ represents the direct effects of the candidate instruments on the imaging response, and $\bm\gamma^\ast = (\gamma_1^\ast,\ldots,\gamma_{p}^\ast)^{\top}$ gives the first-stage effects of the instruments on the exposure. 
The process $U(\bm{s})$ is a mean-zero stochastic process that captures subject-level image variation, $\varepsilon(\bm{s})$ is the mean-zero measurement error at location $\bm s$ with variance $\sigma_\varepsilon^2(\bm{s})$, and $\xi$ is the mean-zero error term for exposure $D$. We assume that $\varepsilon(\cdot)$ is uncorrelated with $(D,\mathbf Z)$ and that $\xi$ is uncorrelated with $\mathbf Z$; nonetheless, $U(\cdot)$ and $\xi$ can be correlated because of subject-level unmeasured confounding, and $U(\cdot)$ may also be correlated with $\mathbf Z$. 

Provided that $\Sigma_{\mathbf Z}:=\E(\mathbf Z\mathbf Z^\top)$ is positive definite, we may write
\begin{equation}
\label{eq:eta}
U(\bm s) = \mathbf Z^\top \bm \varphi^\ast (\bm s)  + \eta(\bm s),
\end{equation}
where $\eta(\bm s)$ is mean zero and uncorrelated with $\mathbf Z$, and $\bm\varphi^\ast(\cdot) = (\varphi_1^\ast(\cdot),\ldots,\varphi^\ast_p(\cdot))^{\top}$ is the coefficient vector of regressing $U(\cdot)$ on $\mathbf Z$.
It yields the observed-data representation of \eqref{Model:IV}:
\begin{align}
Y(\bm{s}) = D\beta^\ast(\bm{s}) + \mathbf{Z}^{\top} \bm{\alpha}^\ast(\bm{s})  +\epsilon(\bm{s}),\qquad D = \mathbf{Z}^\top \boldsymbol{\gamma}^\ast + \xi,\label{EQU:obs}
\end{align}
where $\bm\alpha^\ast(\bm s) \equiv \bm\phi^\ast(\bm s) + \bm\varphi^\ast(\bm s)$ and $\epsilon(\bm s) \equiv \eta(\bm s) + \varepsilon(\bm s)$. As discussed in Section \ref{sec:Identifiability}, a nonzero $\alpha_\ell^\ast(\bm s)$ indicates that $Z_\ell$ is invalid at location $\bm s$, either because it has a direct effect on the imaging response through $\phi_\ell^\ast(\bm s)$, or because it is associated with unmeasured confounding through $\varphi_\ell^\ast(\bm s)$, or both. These two sources of invalidity are not separately identifiable from observed data and are therefore modeled jointly through $\alpha_\ell^\ast(\bm s)$.

Let $G_\eta(\bm s,\bm s')$ be the covariance function of $\eta(\cdot)$. Under the covariance regularity in Assumption~\ref{A:covariance} of the Supplementary Material,
$
G_\eta(\bm s,\bm s')=\sum_{t\ge1}\rho_t\psi_t(\bm s)\psi_t(\bm s'),
$
where $\rho_1\ge\rho_2\ge\cdots\ge0$, $\sum_{t\ge1}\rho_t<\infty$, and $\{\psi_t\}$ is an orthonormal set of eigenfunctions. Equivalently, $\eta$ admits the Karhunen--Lo\`eve expansion $\eta(\bm s)=\sum_{t\ge1}\sqrt{\rho_t}\,\zeta_t\psi_t(\bm s)$, where $\{\zeta_t\}_{t \ge 1}$ are mean-zero, mutually uncorrelated scores with unit variance.

\subsection{Causal Identifiability from Spatial Information}
\label{sec:Identifiability}

Because of endogeneity, the causal-effect function $\beta^\ast(\cdot)$ in \eqref{Model:IV} or \eqref{EQU:obs} is generally non-identifiable from the observed data alone. 
Formally, non-identifiability implies that two different causal functions, $\beta^\ast(\cdot)$ and $\widetilde{\beta}(\cdot)$, can be observationally equivalent, in the sense that, along with their respective nuisance parameters, both can generate the exact same probability distribution for the observed variables $\{ Y(\bm s), D, \mathbf Z \}$. Consequently, the data provide no basis for distinguishing between these two different causal functions.

Following \cite{angrist1996identification}, however, if an instrument $Z_{\ell}$ is known to be a valid IV at location $\bm s$, then $\beta^\ast(\bm s)$ becomes identifiable. 
Specifically, a \textit{valid} IV ${Z}_{\ell}$ at location $\bm s$ must satisfy three IV assumptions: (V1) ${Z}_{\ell}$ is (conditionally) associated with ${D}$, i.e., $\gamma^\ast_{\ell}\neq 0$; (V2) ${Z}_{\ell}$ does not directly affect the imaging outcome $Y(\bm s)$, i.e., $\phi^\ast_{\ell}(\bm s)=0$; and (V3) ${Z}_{\ell}$ is uncorrelated with the unmeasured confounders, i.e., $\varphi^\ast_\ell(\bm s) = 0$.

In practice, the mechanisms underlying candidate instruments are rarely fully known, and it is often unclear which instruments are valid at which locations. Recent seminal work \citep[etc.]{kang2016instrumental,guo2018confidence,lin2024instrumental} has developed data-driven approaches to select valid IVs based on the conjecture that valid IVs are of plurality. To accommodate our setup, we formally introduce the following definition.

\begin{definition}
For a fixed location $\bm s\in \Omega$, let $\mathcal V(\bm s) \equiv \{ \ell : \alpha^\ast_{\ell}(\bm s) = 0, \ \gamma^\ast_\ell\neq 0 \}$ denote the set of valid IVs at $\bm s$. We say that the \textit{pointwise plurality condition} holds at $\bm s$, if 
\begin{align}\label{eq:pointwise-plurality}
    |\mathcal V(\bm s)| > \max_{c\neq 0} |\{ \ell : \alpha^\ast_\ell(\bm s) / \gamma^\ast_\ell = c, \ \gamma^\ast_\ell \neq 0 \}|.
\end{align}
Further, we say that the \textit{everywhere plurality condition} holds, if \eqref{eq:pointwise-plurality} holds for every $\bm s\in\Omega$.
\label{def:plurality}
\end{definition}

When scalar robust-IV methods are applied pixelwise to imaging MR data, they require the everywhere plurality condition. This assumption is often unrealistic for imaging outcomes, which aggregate signals from multiple pathways and are therefore especially prone to pleiotropy \citep{taschler2022causmrreview}. This motivates new identifiability conditions for IV regression with imaging or functional responses that explicitly exploit spatial structure rather than treating each location as an isolated scalar outcome.

Functional responses, including imaging phenotypes, often exhibit intrinsic spatial structure, such as smoothness or local coherence, that can be exploited to improve causal identifiability.
We next show how this structure can be used to identify the spatially varying causal effect $\beta^\ast(\cdot)$ by taking into account the spatial information in images while relaxing the everywhere plurality condition.
To proceed, we introduce the following stability condition.

\begin{definition}
    A candidate IV $Z_{\ell}$ is said to be \textit{stable} at location $\bm s\in\Omega$, if there exists an $\varepsilon$-neighborhood $\mathbb N_{\varepsilon}(\bm s)=\{ \bm s'\in\Omega : \|\bm s'-\bm s\| < \varepsilon \}$ such that either (a) $\ell\in\mathcal V(\bm s')$ for every $\bm s'\in\mathbb N_{\varepsilon}(\bm s)$, or (b) $\ell\notin\mathcal V(\bm s')$ for every $\bm s'\in\mathbb N_{\varepsilon}(\bm s)$. 
    Otherwise, $Z_{\ell}$ is said to be \textit{unstable}.
\end{definition}

Let $\Omega\subset\mathbb R^{\kappa}$ be compact and connected with nonempty interior and \(0<{\rm Vol}(\Omega)<\infty\). Suppose \(\Sigma_{\mathbf Z}:=\E(\mathbf Z\mathbf Z^\top)\) is positive definite. Define \(\bm\gamma^\ast=\Sigma_{\mathbf Z}^{-1}\E(\mathbf ZD)\) and \(\mathcal I=\{\ell:\gamma^\ast_\ell\neq0\}\). For $\bm s\in \Omega$, let \(\bm\varpi(\bm s)=\Sigma_{\mathbf Z}^{-1}\E\{\mathbf ZY(\bm s)\}\), and for \(\ell\in\mathcal I\), define ratio \(r_\ell(\bm s)=(\gamma^\ast_\ell)^{-1}{\varpi_\ell(\bm s)}\) . Consider parameter pairs \((\beta^{\ast},\bm \alpha^{\ast})\) satisfying \eqref{EQU:obs} with \(\E\{\mathbf Z\epsilon(\bm s)\}=0\) and \(\E(\mathbf Z\xi)=0\). 
Let $\Omega_0\equiv \Omega_0(\beta^\ast,\bm\alpha^\ast)\subseteq\Omega$ denote the region where the pointwise plurality condition holds. We have the following result. 

\begin{theorem}\label{theorem:identifiability}
Let $\beta^\ast(\cdot)\in C^r(\Omega)$, $r \geq 0$ and $\Sigma_{\mathbf Z}$ be positive definite. Suppose
    \begin{itemize}
        \item [(A1)] There exists a stable, valid IV at each $\bm s\in\Omega$.
        \item [(A2)] Either (a)
        there is a Lebesgue-measurable subset $\Omega_\oplus \subseteq \Omega_0$ such that $\operatorname{Vol}(\Omega_\oplus) >  \operatorname{Vol}(\Omega) / 2$, where $\operatorname{Vol}(\cdot)$ is the Lebesgue measure.\\
        Or (b) there exists a nonempty open set $\Omega_{\bigodot}\subseteq \Omega$ such that \(r_\ell(\bm s)=r_k(\bm s)\) for all \(\ell,k\in\mathcal I\) and all \(\bm s\in\Omega_{\bigodot}\). (Under (A1), (b) is equivalent to \(\alpha_\ell^\ast(\bm s)=0\) for all \(\ell\in\mathcal I\) and \(\bm s\in\Omega_{\bigodot}\).)
        \item [(A3)] For any $\ell=1,\ldots,p$ and every $\bm s\in {\rm cl}(\Omega_0^c)$, if $\alpha_{\ell}^\ast(\cdot)$ is $C^r$-continuous in the neighborhood of $\bm s$, then $Z_{\ell}$ is stable at $\bm s$.    
    \end{itemize}
    {Then $\beta^\ast(\cdot)$ is identifiable on $\Omega$ within the class of parameter pairs $(\beta,\bm\alpha)$ that satisfy \eqref{EQU:obs}, $\beta\in C^r(\Omega)$ and (A1)--(A3): any two such pairs that induce the same distribution of $\{Y(\cdot),D,\mathbf Z\}$ have the same causal effect function.}
    Moreover, if any of (A1)--(A3) fails, then $\beta^\ast(\cdot)$ is generally not identifiable.
\end{theorem}

Theorem \ref{theorem:identifiability} establishes a new path to causal identifiability by using the spatial information inherent to functional data.
As a result, it becomes possible to identify $\beta^\ast(\cdot)$ on $\Omega$ even if $\Omega_0$ is much smaller than $\Omega$. Theorem \ref{theorem:identifiability} can potentially be extended to a general domain and other functional-response settings.

\begin{figure}
    \centering
    \includegraphics[width=0.75\linewidth]{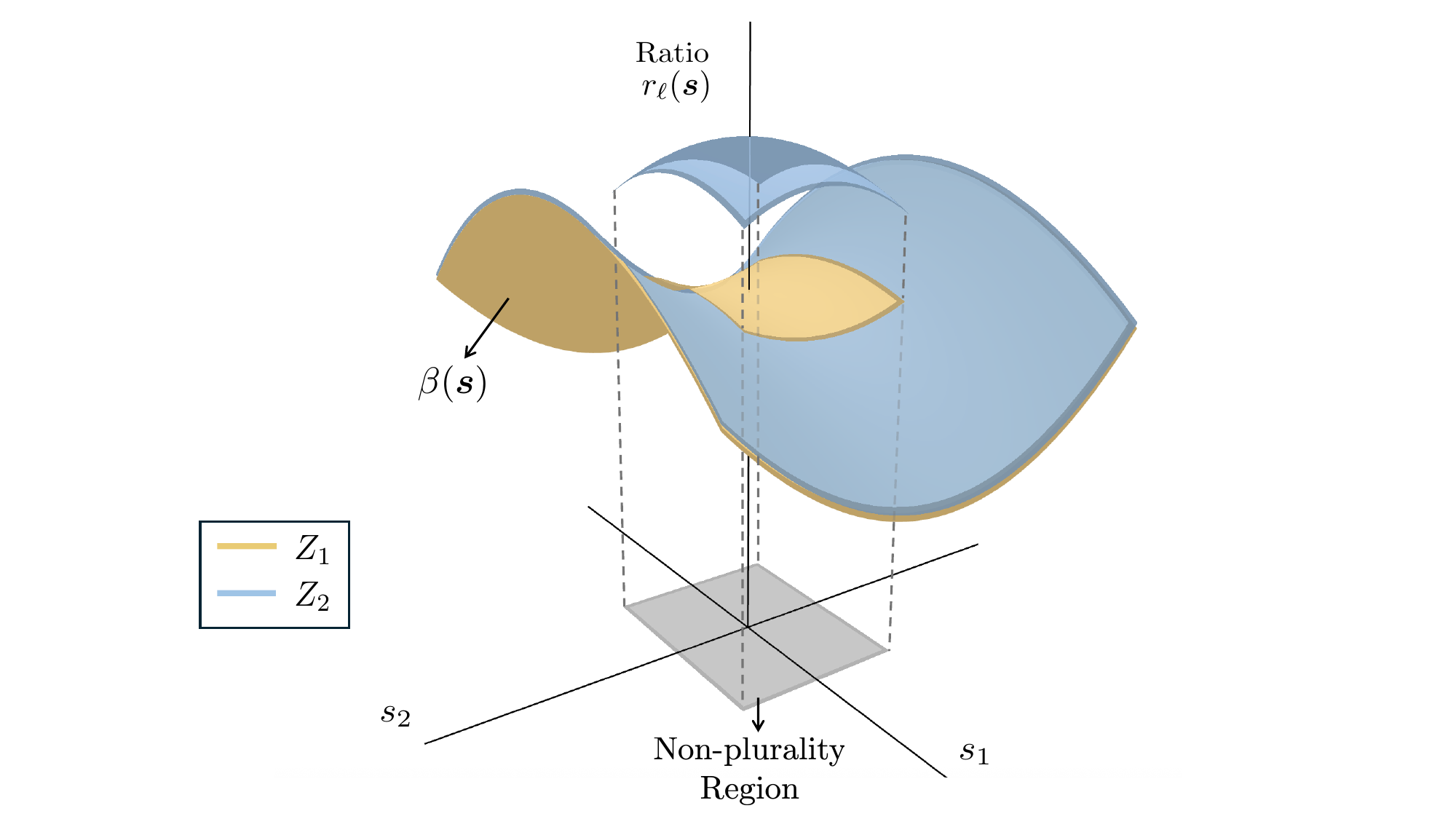}
    \caption{An illustration of the identifiability of $\beta(\bm{s})$: Suppose $p=2$, the yellow and blue surfaces represent the ratios $r_1(\bm s)$ and \(r_2(\bm s)\), corresponding to $Z_1$ and $Z_2$. The gray region is where the pointwise plurality condition does not hold.}
    \label{fig:identifiability}
\end{figure}

Figure~\ref{fig:identifiability} provides an intuitive illustration of Theorem~\ref{theorem:identifiability} with $p=2$.  Instrument $Z_1$ is valid and stable over the whole region, and \(r_1(\bm s)=\beta^\ast(\bm s)\) for all $\bm s\in \Omega$, satisfying $\textit{(A1)}$. 
The pointwise plurality condition fails in the gray non-plurality region, since with only two IVs there may be one valid IV at each location $\bm s$, while for the rest of the region, $\textit{(A2)}$ is satisfied, since in this nonempty and open region, \(r_\ell(\bm s)=r_k(\bm s)\) for all \(\ell,k\in\mathcal I\). 
At every boundary point of the non-plurality region, if $\alpha^\ast(\cdot)$ is $C^r$ locally, then the corresponding IV $Z_\ell$ is stable in a neighborhood of that point. Then $\textit{(A3)}$ is satisfied, and we can extend the identification to the non-plurality region. Therefore, the causal effect $\beta^\ast(\bm s)$ is identifiable.

\subsection{Finite-sample Estimation using Hi-FIVE}
\label{sec:estimation}

To estimate the spatially varying causal effect $\beta(\cdot)$, we construct a two-stage IV estimator tailored to imaging responses. Suppose we observe an independent sample:
$\{D_i,\mathbf Z_i,Y_i(\bm{s}_j): i=1,\ldots,n,\ j=1,\ldots,N\}$
from \eqref{Model:IV}, where $i=1,\ldots,n$ indexes subjects, $j=1,\ldots, N$ indexes spatial locations, $\bm{s}_j$ denotes the $j$th observed pixel or voxel location in $\Omega$.

In the first stage, we project the exposure onto the candidate instruments by fitting $D_i=\mathbf Z_i^\top\bm\gamma+\xi_i$, $i=1,\ldots,n$, yielding fitted exposure values $\widehat D_i=\mathbf Z_i^\top\widehat{\bm\gamma}$. In the second stage, the imaging response is modeled as a function of the fitted exposure and the candidate instruments, allowing the spatially varying causal effect $\beta(\bm{s})$ and the invalid-IV effect functions $\alpha_\ell(\bm{s})$ to be estimated jointly.

The second-stage estimator is constructed using multivariate splines over triangulations. We therefore first introduce the triangulation and spline notation needed to define the penalized image-on-scalar regression criterion.

Triangulations provide an effective way to represent data distributed over irregular domains with complex boundaries \citep{lai2013bpst}. This approach is well suited for image processing, including both two- and three-dimensional brain modeling. Let $\kappa\in\{2,3\}$ denote the spatial dimension. We denote by $T$ a $\kappa$-simplex, that is, the convex hull of $\kappa+1$ affinely independent points. In particular, $T$ is a triangle when $\kappa=2$ and a tetrahedron when $\kappa=3$. A collection $\triangle=(T_1,\ldots,T_K)$ of $K$ simplices is called a triangulation of $\Omega=\bigcup_{k=1}^K T_k$ if the intersection of any two simplices in $\triangle$ is either empty or a common lower-dimensional face, such as a vertex, an edge, or, in three dimensions, a triangular face.

For an integer $r\geq 0$, let $C^r(\Omega)$ be the space of all $r$th continuously differentiable functions over $\Omega$. Given a triangulation $\triangle$, define the spline space of degree $d$ and smoothness $r$ as
$
\mathcal{S}^r_d(\triangle) = 
\left\{
s\in C^r(\Omega): s|_{T}\in \mathcal{P}_d(T),\ T\in\triangle
\right\},
$
where $s|_{T}$ denotes the polynomial piece of $s$ restricted to simplex $T$, and $\mathcal{P}_d(T)$ is the space of all polynomials on $T$ with total degree no greater than $d$. We represent multivariate splines using Bernstein--Bézier basis polynomials on simplices. Let
$
\mathbf{B}(\bm{s})=\{B_m(\bm{s}), m\in\mathcal{M}\}^{\top}
$
denote the collection of Bernstein basis polynomials associated with $\mathcal{S}^r_d(\triangle)$. Then any function $g\in \mathcal{S}^r_d(\triangle)$ can be written as
$
g(\bm{s})=\mathbf{B}^{\top}(\bm{s})\bm{\theta},
~\text{subject to } \mathbf{\Phi}\bm{\theta}=\mathbf{0},
$
where $\bm{\theta}=(\theta_m,m\in\mathcal{M})^\top$ is the spline coefficient vector. The matrix $\mathbf{\Phi}$ collects the smoothness constraints across shared interfaces of adjacent simplices (shared edges in 2D and shared triangular faces in 3D). Its dimension is $J\times |\mathcal{M}|$, where $J$ depends on the triangulation structure and the smoothness order $r$. See \cite{Zhou:Pan:14} and \cite{li2024nonparametric} for examples of such constraint matrices.

To borrow information across neighboring pixels or voxels while accommodating spatially localized invalid-IV effects, we represent $\beta(\cdot)$ in $\mathcal{S}^{r_1}_{d_1}(\triangle_1)$ and each $\alpha_\ell(\cdot)$ in the broken polynomial space
$
\mathcal S^{-1}_{d_2}(\triangle_2)
=\{f:f|_{T_k}\in\mathcal P_{d_2}(T_k),\ k=1,\ldots,K_2\}.
$
The invalid-effect coefficients on different simplices are unconstrained, so jumps across their interfaces are allowed. Each observation on an interface is assigned to one adjacent simplex. The causal effect satisfies the continuity constraints defining $\mathcal S^{r_1}_{d_1}(\triangle_1)$. For the theory, $d_1=d_2=d$, $r_1=r\geq0$, and $\triangle_2$ is a fixed-level refinement of $\triangle_1$. 

Denote $\|\alpha_{\ell}\|_{T_k}  = \{\int_{T_k} \alpha_{\ell}^2(\bm{s}) d \bm{s}\}^{1/2}$ as the norm of function $\alpha_{\ell}$ on the simplex $T_k$,  $k=1,\ldots, K_2$.
We consider the following objective function with region-wise penalization:
\begin{align}
\notag
\argmin_{\beta \in \mathcal{S}^{r_1}_{d_1}(\triangle_1), \alpha_{\ell} \in \mathcal{S}^{-1}_{d_2}(\triangle_2)} &  \frac{1}{nN} \sum_{i=1}^n \sum_{j=1}^N \left\{Y_{i}(\bm{s}_j) - (\widehat{\bm{\gamma}}^{\top}\mathbf{Z}_{i})\beta(\bm s_j) - \sum_{\ell=1}^p Z_{i\ell} \alpha_{\ell}(\bm s_j) \right\}^2 \\
\label{EQU:objective-IV}
& + \frac{\lambda^{\rm R}}{nN}\mathcal{E}(\beta)  + \sum_{\ell = 1}^p  \sum_{k=1}^{K_2} p_{\lambda^{\rm S}}(\|\alpha_{\ell}\|_{T_k}),
\end{align}
where the roughness penalty is defined as $\mathcal{E}(\beta) = \int_{\Omega} \sum_{a=1}^{\kappa}\sum_{b=1}^{\kappa} \left( \nabla_{s_a}\nabla_{s_b}\beta \right)^2 d\bm{s}$. The parameter $\lambda^{\text{R}}$ controls the roughness and spatial complexity of the causal-effect function $\beta(\bm{s})$, while $p_{\lambda^{\textrm{S}}}(\cdot)$ denotes the SCAD penalty \citep{fan2001variable},  which performs region-wise selection of invalid-IV effects.

Let $\mathbf{B}_1(\bm s)$ and $\mathbf{B}_2(\bm s)$ denote the vectors of the Bernstein--B\'ezier basis polynomials for $\mathcal{S}^{r_1}_{d_1}(\triangle_1)$ and $\mathcal{S}^{-1}_{d_2}(\triangle_2)$, respectively. For a simplex $T_k$ of $\triangle_2$, let $\mathbf{B}_{2,k}(\bm s)$ collect the degree-$d_2$ Bernstein basis functions supported on $T_k$, and let $\bm{\theta}_{\alpha,\ell,k}$ be their coefficient vector, so that $\alpha_\ell(\bm s)=\mathbf{B}_{2,k}^{\top}(\bm s)\bm{\theta}_{\alpha,\ell,k}$ for $\bm s\in T_k$. 
Let $\mathbf{P}_1$ denote the roughness matrix satisfying $\bm{\theta}_\beta^{\top}\mathbf{P}_1\bm{\theta}_\beta=\mathcal{E}(\mathbf{B}_1^{\top}\bm{\theta}_\beta)$. 
By the stability property of the spline basis,
$
\|\alpha_{\ell}\|_{T_k}
\asymp
|\triangle_2|^{\kappa/2}
\|\bm{\theta}_{\alpha,\ell,k}\|_2.
$
This norm equivalence gives the scaling of the blocks, but does not make the two SCAD penalties identical after a scalar change of the tuning parameter. We define Hi-FIVE using the Euclidean coefficient-block penalty in the following criterion:
\begin{align}
\argmin_{\bm \theta_\beta, \bm \theta_{\alpha, \ell}}& \frac{1}{nN}\sum_{i=1}^n \sum_{j=1}^N \left\{Y_{i}(\bm{s}_j) - (\widehat{\bm{\gamma}}^{\top}\mathbf{Z}_{i})\mathbf{B}_1^{\top}(\bm{s}_j)\bm{\theta}_{\beta} - \sum_{\ell=1}^p Z_{i\ell} \mathbf{B}_2^{\top}(\bm{s}_j)\bm{\theta}_{\alpha,\ell} \right\}^2 \notag \\
& \quad + \frac{\lambda^{\rm R}}{nN}\bm{\theta}_{\beta}^{\top}\mathbf{P}_1\bm{\theta}_{\beta}  +\sum_{\ell = 1}^p  \sum_{k=1}^{K_2} p_{\lambda^{\rm S}}(\|\bm{\theta}_{\alpha,\ell,k}\|), \label{EQU:objective-IV-2}
\end{align}
subject to $\mathbf{\Phi}_1\bm{\theta}_{\beta} = \mathbf{0}$. We use the standard SCAD penalty with $p_\lambda(0)=0$, a fixed shape parameter $a>2$, and
$
p_\lambda'(t)=\lambda\mathbf1\{0<t\leq\lambda\}
+\frac{(a\lambda-t)_+}{a-1}\mathbf1\{t>\lambda\}.
$
In particular, $p_\lambda(t)=(a+1)\lambda^2/2$ for $t\geq a\lambda$.

The two penalty terms in \eqref{EQU:objective-IV-2} play distinct but complementary roles in causal identification. The roughness penalty on \(\beta(\mathbf s)\) exploits the spatial smoothness of the causal effect and favors a spatially regular causal-effect surface. In contrast, the region-wise sparsity penalty on \(\alpha_\ell(\mathbf s)\) exploits the local validity/plurality structure of the IVs by identifying regions with nonzero direct IV effects. Together, these penalties help distinguish the spatial causal effect \(\beta(\mathbf s)\) from spatially heterogeneous IV effects \(\alpha_\ell(\mathbf s)\).

\chg{Theorem~\ref{theorem:identifiability} requires a valid instrument at every location (A1). Thus, we minimize \eqref{EQU:objective-IV-2} subject to the \emph{anchor constraint}
\begin{equation}
\label{eq:anchor}
\text{for every }k\in\{1,\ldots,K_2\},\qquad \bm\theta_{\alpha,\ell,k}=\mathbf0\ \text{ for at least one }\ell\in\widehat{\mathcal I}_{\tau_\gamma} ,
\end{equation}
where $\widehat{\mathcal I}_{\tau_\gamma}=\{\ell:|\widehat\gamma_\ell|>\tau_\gamma\}$, for some $\tau_\gamma > 0$, collects the instruments with a nonnegligible effect. The constraint requires that on every simplex at least one relevant instrument is treated as valid. The anchor constraint ties the two penalties together: it prevents the sparsity penalty from trading a few flagged blocks for an arbitrary shift of $\beta$, because some instrument always remains unflagged and its ratio must be matched. Section~\ref{SEC:anchor-separation} of the Supplementary Material shows the requirements without a valid-IV plurality.
}
Let $\widehat{\bm{\theta}} = (\widehat{\bm{\theta}}_{\beta}^{\top}, \widehat{\bm{\theta}}_{\alpha,1}^{\top}, \ldots, \widehat{\bm{\theta}}_{\alpha,p}^{\top})^{\top}$ be a measurable choice of a global minimizer of \eqref{EQU:objective-IV-2} subject to $\mathbf\Phi_1\bm\theta_\beta=\mathbf0$ and \eqref{eq:anchor}. Proposition~\ref{S-Prop:Selection} in the Supplementary Material proves that there exists a unique minimizer with probability tending to 1. The Hi-FIVE estimators of $\beta$ and $\alpha_{\ell}$ are:
\begin{equation}
\widehat{\beta}(\bm{s}) = \mathbf{B}_1(\bm{s})^{\top}\widehat{\bm{\theta}}_{\beta},\qquad \widehat{\alpha}_{\ell}(\bm{s}) = \mathbf{B}_2(\bm{s})^{\top}\widehat{\bm{\theta}}_{\alpha,\ell}, \quad \ell = 1, \ldots, p.
\label{EQ:beta_l_hat}
\end{equation}

\subsection{Spatially Distributed Adaptive Learning} \label{sec:distributed-optimization}

\begin{figure}[H]
    \centering
    \includegraphics[width=0.855\linewidth]{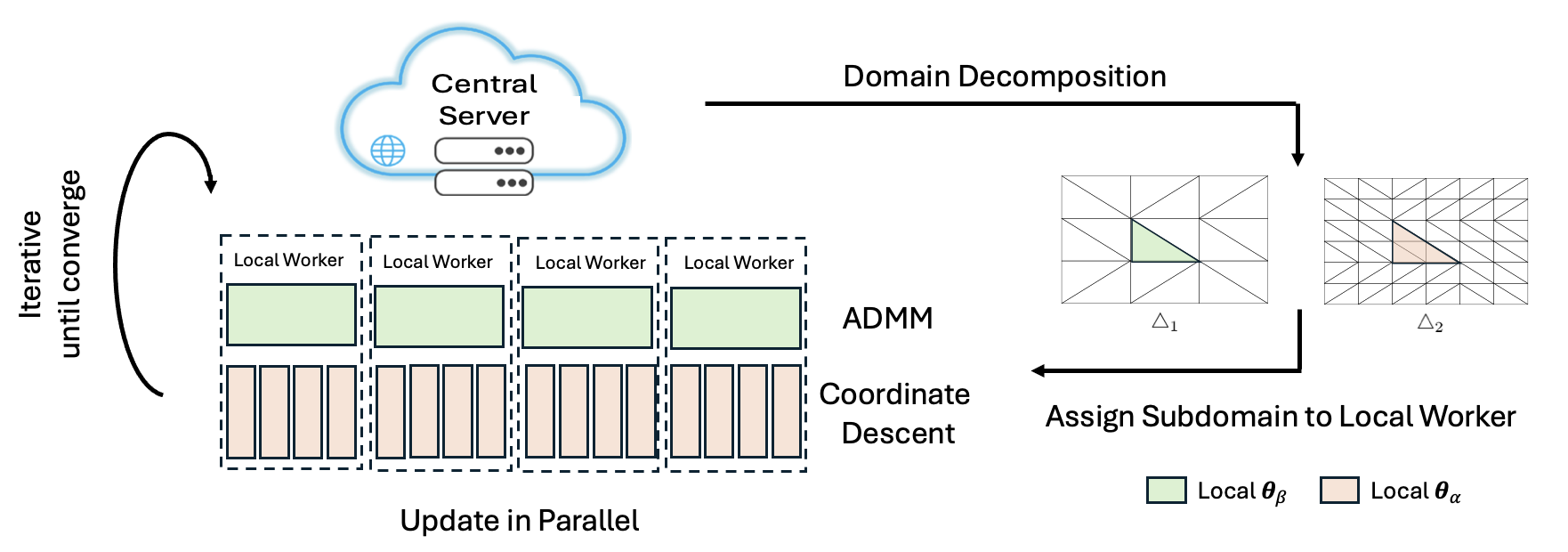}
    \caption{An illustration of the distributed optimization framework based on domain decomposition. The coarse triangulation $\triangle_1$ is used for the causal-effect surface, whereas the refined triangulation $\triangle_2$ supports more localized invalid-IV effect updates.} 
    \label{fig:distributed-overview}
\end{figure}

The optimization framework illustrated in Figure~\ref{fig:distributed-overview} provides an overview of the proposed spatially distributed adaptive learning procedure for Hi-FIVE. The local support of the spline basis enables parallel computation over spatial subdomains, substantially improving computational scalability for large-scale imaging data. 

In addition, domain decomposition provides a flexible framework for modeling heterogeneous spatial effects over large and complex domains. The main motivation is that spatial smoothness, signal strength, and patterns of local IV validity may vary across different subdomains. By partitioning the full domain into several smaller subdomains, the proposed method can adaptively select subdomain-specific hyperparameters within each subdomain, rather than imposing a single set of global tuning parameters over the entire domain. This allows the Hi-FIVE estimators to better accommodate spatial heterogeneity while also reducing the computational burden for large-scale imaging data.

\vspace{-0.5cm}
\subsubsection{Distributed algorithm}

We seek a minimizer of \eqref{EQU:objective-IV-2} by a distributed scheme that combines three ingredients: local linear approximation (LLA) of the nonconvex SCAD penalty, an alternating direction method of multipliers (ADMM) that enforces the smoothness constraint $\mathbf{\Phi}_1\bm{\theta}_\beta=\mathbf{0}$ across worker boundaries, and coordinate descent with group soft-thresholding for the region-wise invalid-IV blocks. Because the Bernstein--B\'{e}zier bases are locally supported, each update below is assembled from simplexwise contributions and parallelizes across workers.

For the current iterate $(\bm{\theta}_\beta,\{\bm{\theta}_{\alpha,\ell}\})$, write $\widehat{D}_i=\widehat{\bm{\gamma}}^{\top}\mathbf{Z}_i$ and define the partial residuals
\begin{equation*}\label{eq:partial-residuals}
	R_{ij}^{\beta}=Y_i(\bm{s}_j)-\sum_{\ell=1}^{p}Z_{i\ell}\mathbf{B}_2^{\top}(\bm{s}_j)\bm{\theta}_{\alpha,\ell},~
	R_{ij}^{\ell}=Y_i(\bm{s}_j)-\widehat{D}_i\mathbf{B}_1^{\top}(\bm{s}_j)\bm{\theta}_\beta
	-\sum_{\ell'\neq\ell}Z_{i\ell'}\mathbf{B}_2^{\top}(\bm{s}_j)\bm{\theta}_{\alpha,\ell'}.
\end{equation*}

\noindent\textbf{Causal-effect update.} Holding $\{\bm{\theta}_{\alpha,\ell}\}$ fixed, $\bm{\theta}_\beta$ solves the constrained penalized least-squares problem
\begin{equation}\label{obj:admm}
	\bm{\theta}_\beta^{+}=\argmin_{\bm{\theta}_\beta:\,\mathbf{\Phi}_1\bm{\theta}_\beta=\mathbf{0}}\
	\frac{1}{nN}\sum_{i=1}^{n}\sum_{j=1}^{N}\bigl\{R_{ij}^{\beta}-\widehat{D}_i\mathbf{B}_1^{\top}(\bm{s}_j)\bm{\theta}_\beta\bigr\}^{2}
	+\frac{\lambda^{\rm R}}{nN}\bm{\theta}_\beta^{\top}\mathbf{P}_1\bm{\theta}_\beta.
\end{equation}
Let $\mathbf{H}_\beta=\sum_{i,j}\widehat{D}_i^{2}\mathbf{B}_1(\bm{s}_j)\mathbf{B}_1^{\top}(\bm{s}_j)$ and $\mathbf{c}_\beta=\sum_{i,j}\widehat{D}_i R_{ij}^{\beta}\mathbf{B}_1(\bm{s}_j)$. Let $\mathbf{Q}_2$ be an orthonormal basis of the null space $\{\bm{\theta}:\mathbf{\Phi}_1\bm{\theta}=\mathbf{0}\}$, so that $\bm{\theta}_\beta=\mathbf{Q}_2\bm{\vartheta}$. Then \eqref{obj:admm} is a quadratic program with solution $\bm{\vartheta}^{+}=\{\mathbf{Q}_2^{\top}(\mathbf{H}_\beta+\lambda^{\mathrm{R}}\mathbf{P}_1)\mathbf{Q}_2\}^{-1}\mathbf{Q}_2^{\top}\mathbf{c}_\beta$. Since $\mathbf{H}_\beta,\mathbf{c}_\beta$ are sums of simplexwise terms, they can be assembled in parallel. The interface coupling induced by $\mathbf{\Phi}_1$ is handled by consensus ADMM with local copies $\bm{\theta}_\beta^{(w)}$, consensus variable $\bm{z}$, scaled dual $\bm{u}^{(w)}$, and penalty $\rho>0$:
\begin{align} \notag
	&\bm{\theta}_\beta^{(w)}\!\leftarrow\!\argmin_{\bm{\theta}}\Bigl\{f_w(\bm{\theta})+\tfrac{\rho}{2}\|\bm{\theta}-\bm{z}+\bm{u}^{(w)}\|^2\Bigr\},\quad
 \bm{z}\leftarrow\Pi_{\mathbf{\Phi}_1}\!\Bigl\{\tfrac1W\textstyle\sum_w(\bm{\theta}_\beta^{(w)}+\bm{u}^{(w)})\Bigr\},\\ 
	&\bm{u}^{(w)}\leftarrow\bm{u}^{(w)}+\bm{\theta}_\beta^{(w)}-\bm{z}, \label{obj:admm-updates}
\end{align}
where $f_w$ collects the least-squares and roughness contributions of the simplices assigned to worker $w$, so that the objective in \eqref{obj:admm} equals $\sum_{w=1}^{W}f_w$, and $\Pi_{\mathbf{\Phi}_1}=\mathbf{I}-\mathbf{\Phi}_1^{\top}(\mathbf{\Phi}_1\mathbf{\Phi}_1^{\top})^{-1}\mathbf{\Phi}_1$. The reparametrized closed form is used when a central solve is affordable; \eqref{obj:admm-updates} is its distributed counterpart and exchanges only the shared-interface coefficients between workers.

\noindent\textbf{Invalid-IV update.} The nonconvex SCAD penalty is handled by LLA: at the current iterate the weight of block $(\ell,k)$ is $w_{\ell,k}=p'_{\lambda^{\mathrm{S}}}(\|\bm{\theta}_{\alpha,\ell,k}\|)$, and each block then solves the weighted group-lasso subproblem
\begin{equation}\label{obj:coord}
	\bm{\theta}_{\alpha,\ell,k}^{+}=\argmin_{\bm{\theta}} \frac{1}{nN}
	\sum_{i=1}^{n}\sum_{\bm{s}_j\in T_k}\!\bigl\{R_{ij}^{\ell}-Z_{i\ell}\mathbf{B}_{2,k}^{\top}(\bm{s}_j)\bm{\theta}\bigr\}^{2}
	+w_{\ell,k}\,\|\bm{\theta}\|.
\end{equation}
Let $\mathbf{H}_{\ell,k}=\sum_{i=1}^{n}\sum_{\bm s_j\in T_k}Z_{i\ell}^{2}\mathbf{B}_{2,k}(\bm s_j)\mathbf{B}_{2,k}^{\top}(\bm s_j)$ and 
$
\bm{r}_{\ell,k}= \frac{2}{nN}\sum_{i=1}^{n}\sum_{\bm s_j\in T_k}Z_{i\ell}\bigl(Z_{i\ell}\mathbf{B}_{2,k}^{\top}(\bm{s}_j)\bm{\theta}_{\alpha,\ell,k}-R_{ij}^{\ell}\bigr)\mathbf{B}_{2,k}(\bm{s}_j)
$
as the gradient of the least-squares term of \eqref{obj:coord}. With $\tau_{\ell,k}\le nN\{2\lambda_{\max}(\mathbf{H}_{\ell,k})\}^{-1}$, the proximal-gradient step is
\begin{equation}\label{obj:coord-update}
	\bm{v}_{\ell,k}=\bm{\theta}_{\alpha,\ell,k}-\tau_{\ell,k}\bm{r}_{\ell,k},
	\qquad
	\bm{\theta}_{\alpha,\ell,k}^{+}=\Bigl(1-\frac{\tau_{\ell,k}\,w_{\ell,k}}{\|\bm{v}_{\ell,k}\|}\Bigr)_{\!+}\bm{v}_{\ell,k}.
\end{equation}
\noindent\textbf{Anchor step.} After the invalid-IV update, the anchor constraint \eqref{eq:anchor} is enforced simplex by simplex. If every block $\bm\theta_{\alpha,\ell,k}$ with $\ell\in\widehat{\mathcal I}_{\tau_\gamma}$ is nonzero on $T_k$, we set to zero the block whose removal increases the local objective least:
\begin{equation}\label{obj:anchor-step}
\ell_k^\dagger=\argmin_{\ell\in\widehat{\mathcal I}_{\tau_\gamma}}\ \min_{\bm\theta_{\alpha,\cdot,k}:\ \bm\theta_{\alpha,\ell,k}=\mathbf0}\ \frac{1}{nN}\sum_{i=1}^{n}\sum_{\bm s_j\in T_k}\Bigl\{\widetilde R_{ij}-\sum_{\ell'=1}^pZ_{i\ell'}\mathbf B_{2,k}^\top(\bm s_j)\bm\theta_{\alpha,\ell',k}\Bigr\}^2+\sum_{\ell'\neq\ell}w_{\ell',k}\|\bm\theta_{\alpha,\ell',k}\|,
\end{equation}
where $\widetilde R_{ij}=Y_i(\bm s_j)-\widehat D_i\mathbf B_1^\top(\bm s_j)\bm\theta_\beta$, and then refit the remaining blocks on $T_k$ by \eqref{obj:coord-update}. Each local problem involves only the $p$ blocks of one simplex, so the anchor step parallelizes across simplices and adds a cost of order $p$ local refits per simplex.

Algorithm~\ref{ALG:hifive} in the Supplementary Material alternates the updates \eqref{obj:admm-updates}, \eqref{obj:coord-update} and \eqref{obj:anchor-step} in an inner block-coordinate loop nested within the outer LLA loop, until convergence.

\subsubsection{Triangulation and tuning-parameter selection}

In practice, we recommend using a finer triangulation to estimate the instrumental effects $\alpha_\ell$ than for the causal effect $\beta$. A finer triangulation resolves the boundary of an invalid region more sharply, thereby offering greater flexibility in identifying spatially varying invalid instruments. We employ a hierarchical triangulation scheme to facilitate distributed computation. Starting from an initial triangulation $\triangle_1$, we generate a refined triangulation $\triangle_2$ through $L$ levels of uniform refinement, with each level subdividing every simplex into $2^{\kappa}$ smaller ones. 
Throughout we take $L$ fixed, so that $|\triangle_1|$ and $|\triangle_2|$ are of the same order and $K_2 = 2^{\kappa L}K_1$; unless stated otherwise we use $L=1$. 
Figure \ref{fig:distributed-overview} illustrates an example of $\triangle_1$ and $\triangle_2$, with the green region highlighting a simplex assigned to a single worker and the orange regions representing the refined simplices within each worker.

For tuning parameter selection, we use a nested procedure. First, we fix a candidate value of the roughness penalty parameter $\lambda^{\rm R}$. Conditional on this value, we select the $\lambda^{\rm S}$ by minimizing the modified BIC (mBIC). We then repeat this procedure over a grid of candidate values for $\lambda^{\rm R}$ and choose the $\lambda^{\rm R}$ using $K$-fold cross-validation with folds formed at the subject level, so that within-image dependence is preserved. 

We now introduce the mBIC criterion used to select $\lambda^{\rm S}$. Let
$\widehat{\mathcal A}_{\lambda^S}=\{(\ell,k):\|\widehat\alpha_\ell\|_{T_k}\neq0\}$
be the estimated active set of invalid-IV blocks, among the $pK_2$ candidate blocks. Write
$
\operatorname{pISE}_{\lambda^S}=\operatorname{ISE}_{\lambda^S}
+\frac{\lambda^R}{nN}\widehat{\bm\theta}_\beta^\top\mathbf P_1\widehat{\bm\theta}_\beta ,
$
where 
$
\text{ISE}_{\lambda^S}=\frac{1}{nN}\sum_{i=1}^n \sum_{j = 1}^N \Bigl\{Y_{i}(\bm{s}_j)-  (\widehat{\bm{\gamma}}^{\top}\mathbf{Z}_{i})\mathbf{B}_1^{\top}(\bm{s}_j)\bm{\widehat{\theta}}_{\beta} - \sum_{\ell=1}^p Z_{i\ell} \mathbf{B}_2^{\top}(\bm{s}_j)\bm{\widehat{\theta}}_{\alpha,\ell} \Bigr\}^2.
$ 
To account for spatial dependence within simplices, consider the total effective sample size
$
\widehat q_n=n\sum_{k=1}^{K_2}\widehat N_{\mathrm{eff},k},
$
where $\widehat N_{\mathrm{eff},k}$ is the effective number of independent spatial observations within $T_k$; see Section~\ref{SEC:eff-sample} for estimation details. When the pixels are nearly independent, $\widehat q_n$ is close to $nN$; under strong spatial correlation, it is of order $nK_2$.

The mBIC is defined as:
\[
\operatorname{mBIC}(\lambda^S)=\log\bigl(\operatorname{pISE}_{\lambda^S}\bigr)
+\frac{|\widehat{\mathcal A}_{\lambda^S}|}{\widehat q_n}
\bigl\{\log \widehat q_n+2a_0\log(pK_2)\bigr\},\qquad a_0\in(0,1),
\]
and $\widehat\lambda^{\rm S}$ minimizes this criterion over a finite
deterministic grid. The first term measures fit and roughness; the second
penalizes the number of active invalid-IV blocks. The additional term
$2a_0\log(pK_2)$ accounts for the number of candidate supports
\citep{chen2008extended}. Section~\ref{sec:mBIC} of the Supplementary Material investigates the theoretical properties of the proposed mBIC. Propositions~\ref{Prop:mBIC} and~\ref{Prop:mBIC-size} in the Supplementary Material show that the same oracle equality holds with probability tending to one when $\lambda^{\rm S}$ is selected by mBIC over a finite grid, each under its own additional conditions.

\section{Asymptotics and Simultaneous Inference}
\label{section:theory}

\subsection{Properties of the Hi-FIVE Estimator}


We next establish the asymptotic properties of Hi-FIVE. Our primary inferential target is the spatially varying causal effect $\beta^\ast(\cdot)$, while the regionwise invalid-IV effects $\alpha_\ell(\cdot)$ are treated as nuisance components that capture local violations of the instrumental variable assumptions. Unlike classical variable-selection settings \citep{Li:Wang:Wang:21}, which typically assume sparsity among a large number of candidate variables, our domain-selection framework does not impose a sparsity assumption on the spatial regions affected by invalid IVs. Consequently, the invalid-IV effects are allowed to occupy nonvanishing portions of the spatial domain as the sample size increases. Moreover, observations within each region may be spatially correlated. Our analysis accommodates this dependence through the error covariance structure and region-specific effective sample sizes. For the asymptotic results below, we use the shorthand $\triangle\equiv\triangle_1$. With fixed $L$, we have $|\triangle_2|\asymp|\triangle|$ and $K_2\asymp|\triangle|^{-\kappa}$.

{In the following, we partition the coefficient vector of \eqref{EQU:objective-IV-2} as $\bm\theta=(\bm\theta_1^{\top},\bm\theta_2^{\top})^{\top}$, where $\bm\theta_1=(\bm\theta_\beta^{\top},\{\bm\theta_{\alpha,\ell,k}^{\top}:(\ell,k)\in\mathcal A^\ast\})^{\top}$ collects the causal-effect coefficients together with the coefficients of the truly invalid IV--simplex blocks, $\mathcal A^\ast=\{(\ell,k):\|\alpha^\ast_\ell\|_{T_k}>0\}$, and $\bm\theta_2=\{\bm\theta_{\alpha,\ell,k}:(\ell,k)\notin\mathcal A^\ast\}$ collects the remaining blocks, so that the true value satisfies $\bm\theta^\ast_2=\mathbf 0$. We define the oracle estimator $\widehat{\bm\theta}^{\,o}_1$ as the minimizer of \eqref{EQU:objective-IV-2} over $\bm\theta_1$ subject to $\bm\theta_2 = \mathbf 0$ and omitting the SCAD penalty, with the corresponding oracle estimator given by $\widehat\beta^{\,o}(\cdot) = \mathbf B_1(\cdot)^{\top}\widehat{\bm\theta}^{\,o}_\beta$.}

Proposition~\ref{S-Prop:Selection} in the Supplementary Material shows that, with probability tending to one, the oracle estimator is the unique global minimizer of the nonconvex criterion \eqref{EQU:objective-IV-2} under the anchor constraint \eqref{eq:anchor}. Consequently, the convergence rate (Theorem~\ref{THM:beta-convergence}) and the pointwise asymptotic normality (Theorem~\ref{THM:beta-normality}) carry over from the oracle estimator to Hi-FIVE. 

\begin{theorem}
\label{THM:beta-convergence}
Under Assumptions~\ref{A:tri}--\ref{AssumpC:signal} in the Supplementary Material, with probability tending to one, criterion \eqref{EQU:objective-IV-2} under the anchor constraint \eqref{eq:anchor} has a unique global minimizer $\widehat{\bm\theta}=(\widehat{\bm\theta}_1^{\top},\widehat{\bm\theta}_2^{\top})^{\top}$. This minimizer has the following properties:
(a) \textup{(Recovery of zero blocks}) $\widehat{\bm\theta}_2 = \mathbf{0}$;
(b) \textup{(Oracle Equivalence)} $\widehat{\bm\theta}_1 = \widehat{\bm\theta}_1^{\,o}$, so that the resulting Hi-FIVE estimator satisfies $\widehat\beta(\cdot) = \widehat\beta^{\,o}(\cdot)$.
Furthermore, the estimator $\widehat{\beta}$ is $L^2$-consistent with convergence rate
		$$\Vert \widehat{\beta}- \beta^{\ast}\Vert_{L^2(\Omega)} = \OP\left\{\left(1 + \frac{\lambda^{\text{R}}}{nN|\triangle|^{4}}\right)|\triangle|^{d +1}\|\beta^\ast\|_{d+1,\infty} + \frac{1}{\sqrt{n}} +\frac{\lambda^{\text{R}}}{nN|\triangle|^2}\|\beta^\ast\|_{2,\infty} \right\}.$$
\end{theorem}

The convergence rate in Theorem~\ref{THM:beta-convergence} explicitly captures the trade-off between the triangulation resolution $|\triangle|$, sample size $n$, and spatial resolution $N$. A finer partition ($|\triangle| \to 0$) reduces spline approximation bias but increases the effective parameter dimension, requiring $n$ and $N$ to grow at compatible rates to maintain variance control.

\begin{theorem}
\label{THM:beta-normality}
Suppose Assumptions~\ref{A:tri}--\ref{AssumpC:signal} and~\ref{A:normality} in the Supplementary Material hold at a fixed location $\bm s\in\Omega$. For the Hi-FIVE estimator of Theorem~\ref{THM:beta-convergence},
\[
\sigma_{\beta}^{-1}(\bm{s}) \left\{\widehat{\beta}(\bm{s}) -\beta^\ast(\bm{s})\right\}\overset{\mathcal{L}}{\longrightarrow}N\left(0, 1\right),
\]
as $n\to\infty$, where $\sigma_\beta^2(\bm s)$ is the conditional oracle asymptotic variance defined in \eqref{eq:var_orac_est} of the Supplementary Material. Consequently, for $0<\tau<1$ and this fixed $\bm s$,
\[
\Pr\left\{\beta^*(\bm{s})\in\left(\widehat{\beta}(\bm{s})-z_{1-\tau / 2} \sigma_{\beta}(\bm{s}), \widehat{\beta}(\bm{s})+z_{1-\tau / 2}\sigma_{\beta}(\bm{s})\right)\right\} \rightarrow 1-\tau.
\]

\end{theorem}

The asymptotic variance $\sigma^2_{\beta}(\bm{s})$, given in \eqref{eq:var_orac_est} of the Supplementary Material, depends on three unknown quantities: the population first stage $\mathbf Z_i^\top\bm\gamma^\ast$, the true support $\mathcal A^\ast$ of the invalid-IV blocks, and the covariance of the within-subject error $\epsilon(\cdot)$. We replace them, respectively, by the fitted exposure $\widehat D_i$, the support $\widehat{\mathcal A}$ selected by Hi-FIVE, and the empirical covariance of the Hi-FIVE residuals $Y_i(\bm s_j)-D_i\widehat\beta(\bm s_j)-\mathbf Z_i^\top\widehat{\bm\alpha}(\bm s_j)$. This gives the plug-in covariance $\widehat{\bm{\Xi}}$ in \eqref{EQU:var-k-estimate} and hence $\widehat{\sigma}^2_{\beta}(\bm{s})$ in \eqref{eq:est_varbeta}; Supplementary Section~\ref{sec:inferences} gives the complete procedure, including the smoothed subject-level residual processes $\widehat\eta_i(\cdot)$ and the measurement-error residuals $\widehat\varepsilon_{ij}$ used in the bootstrap below. Corollary~\ref{COR:variance-consistency} in that section shows that, the plug-in estimator is ratio-consistent, $\widehat\sigma_\beta(\bm s)/\sigma_\beta(\bm s)\to1$ in probability.

\begin{remark}\label{rem:anchoring}
 Since the fitted exposure is a linear combination of the candidate instruments, $\widehat D_i=\widehat{\bm\gamma}^{\top}\mathbf Z_i =\sum_{\ell=1}^{p}\widehat\gamma_\ell Z_{i\ell}$, the least-squares term in \eqref{EQU:objective-IV} depends on $(\beta,\{\alpha_\ell\})$ only
through the $p$ combined effect functions
$\Gamma_\ell(\bm s)=\widehat\gamma_\ell\,\beta(\bm s)+\alpha_\ell(\bm s)$,
$\ell=1,\ldots,p$, because $\widehat D_i\,\beta(\bm s)+\sum_{\ell}Z_{i\ell}\alpha_\ell(\bm s) =\sum_{\ell}Z_{i\ell}\Gamma_\ell(\bm s)$. Hence, for any shift function
$c(\cdot)$, the reparametrization
\begin{equation}
\label{eq:invariance}
(\beta,\alpha_1,\ldots,\alpha_p)\longmapsto
\bigl(\beta-c,\ \alpha_1+\widehat\gamma_1 c,\ \ldots,\ \alpha_p+\widehat\gamma_p c\bigr)
\end{equation}
leaves the entire least-squares term unchanged.
In such case, the fit alone does not identify $\beta$:
\eqref{eq:invariance} is the finite-sample counterpart of the population
non-identifiability that the plurality condition resolves (Definition~\ref{def:plurality} and Theorem~\ref{theorem:identifiability}). What separates the true representation from a shifted one is therefore not the fit but the penalties and the anchor constraint: a shift $c\not\equiv0$ changes which invalid-IV blocks are nonzero, and hence the SCAD penalty, and the anchor constraint forbids any simplex on which every instrument is flagged. Theorems~\ref{THM:beta-convergence} and~\ref{THM:beta-normality} formalize this through a joint separation condition Assumption~\ref{AssumpC:signal}(i): 
 every incorrect anchored support with no more invalid-IV blocks than $\mathcal A^\ast$ must fit the reduced form worse, by a margin $\Delta_n^{\rm G}$ per block saved. When pointwise plurality holds everywhere, this condition is verified in Example~\ref{EX:global-compatible} of the Supplementary Material. Section~\ref{SEC:anchor-separation} of the Supplementary Material also shows that Hi-FIVE applies to regions where plurality fails: the anchor constraint \eqref{eq:anchor} makes the condition hold there as well.    
\end{remark}

\vspace{-0.5cm}
\subsection{Bootstrap-calibrated Simultaneous Confidence Regions}
Now, we use a wild bootstrap approach \citep{hall2013simple} to recalibrate pointwise confidence intervals (PCIs) and construct simultaneous confidence regions (SCRs). 
Given point estimates $\widehat{\beta}(\cdot)$ and $\widehat{\sigma}_{\beta}(\cdot)$, and nominal level $\tau_0$, the procedure is described as follows:
\begin{enumerate}

\item {\it Bootstrap Estimates.} 
For $b=1,\ldots,B$, generate bootstrap samples $\{ Y^{*(b)}_i(\bm s_j), D_i^{*(b)}, \mathbf Z_i \}$, $i=1,\ldots,n$, $j=1,\ldots,N$, using
$
Y_{i}^{\ast(b)}(\bm s_j)=D_i^{\ast(b)}\widehat{\beta}(\bm{s}_j)+ \mathbf Z_{i}^\top\widehat{\bm \alpha}(\bm{s}_j)+\nu_{i}^{(b)} \widehat{\eta}_i(\bm{s}_j)+\nu_{ij}^{(b)} \widehat{\varepsilon}_{ij}, 
D_i^{\ast(b)}=\widehat{\bm{\gamma}}^{\top} \mathbf{Z}_i +\nu_{i}^{(b)}\widehat{\xi}_i,
$
where $\nu_{i}^{(b)}$ and $\nu_{ij}^{(b)}$ are independent Rademacher variables (taking values $\pm 1$ with probability 0.5 each) and $\widehat{\xi}_i = D_i -  \widehat{\bm{\gamma}}^{\top} \mathbf{Z}_i$. 
Accordingly, compute $\widehat{\beta}^{*(b)}(\cdot)$ from \eqref{EQU:objective-IV-2} under \eqref{eq:anchor} and $\widehat{\sigma}_{\beta}^{\ast(b)}(\cdot)$ from \eqref{eq:est_varbeta}. 

\item {\it Recalibration.} 
Solve the equation $\widehat{\iota}(\tau) = 1 - \tau_0$ for $\tau$, and denote the root by $\widehat{\tau}:=\widehat{\iota}^{-1}(1-\tau_0)$, where 
$\widehat{\iota}(\tau) = B^{-1}\sum_{b=1}^{B} \mathbb I\big\{\widehat{\beta}(\bm{s}_j) \in \big[\widehat{\beta}^{*(b)}(\bm{s}_j) \pm \widehat{\sigma}_{\beta}^{*(b)}(\bm{s}_j) z_{1-\tau/2}\big], j = 1, \ldots, N \big\}$.
We construct $\widehat{\rm SCR}(\bm s)=\big[\widehat{\beta}(\bm{s}) \pm \widehat{\sigma}_{\beta}(\bm{s}) z_{1-\widehat\tau/2}\big]$.
\end{enumerate}
The same multiplier $\nu_i^{(b)}$ is applied to both $\widehat\eta_i$ and $\widehat\xi_i$ so that the bootstrap preserves the dependence between the subject-level image process and the exposure error. Theorem~\ref{THM:beta-normality} and Corollary~\ref{COR:variance-consistency} in the supplementary materials provide theoretical justification for the pointwise intervals. The recalibration step is a finite-sample adjustment whose simultaneous coverage is assessed through simulation in Section~\ref{sec:simulation}.

\section{Simulation Studies}
\label{sec:simulation}
We evaluate the estimation, 
simultaneous inference, robustness, and computational performance of Hi-FIVE through simulations. Because no existing functional-data method allows spatially varying invalid instruments, we compare Hi-FIVE with pointwise sisVIVE applied at each image location and with sisVIVE+ which smooths the pointwise sisVIVE estimates.

\vspace{-0.5cm}
\paragraph{Simulation Design}

For subject $i=1,\ldots,n$ and spatial location $\bm s_j\in\Omega$, we generate
\begin{align*}
Y_i(\bm s_j)
&=
D_i\beta^\ast(\bm s_j)
+
\mathbf Z_i^\top\bm\alpha^\ast(\bm s_j)
+
\eta_i(\bm s_j)
+
\varepsilon_i(\bm s_j), \quad 
D_i
=
\mathbf Z_i^\top\bm\gamma^\ast+\xi_i,
\label{eq:sim-dgm}
\end{align*}
where $\beta^\ast(\cdot)$ is the causal-effect function and
$\bm\alpha^\ast(\cdot)$ is the spatially varying invalid-IV effect. The spatial process $\eta_i(\cdot)$ is correlated with $\xi_i$ to induce
unmeasured confounding. We consider $n\in\{100,300\}$, $p\in\{3,10\}$, and low- and high-noise
settings. The two-dimensional (2D) images contain $95\times79$ pixels, and the
three-dimensional (3D) images contain $67^3$ voxels.

The invalid-IV configuration is constructed to violate pointwise plurality.
Let $q=\lfloor p/2\rfloor+1$. The first $q$ instruments share a common nonzero
invalid-IV effect over a localized region $\mathcal R_\alpha$, while the
remaining instruments are valid throughout the domain. Hence, the invalid instruments form the largest ratio group within $\mathcal R_\alpha$, although
the spatial identifiability conditions in Section~\ref{sec:Identifiability}
continue to hold.

We consider three scenarios: a smooth 2D causal effect, a 2D causal effect with a jump discontinuity, and a smooth 3D causal effect. The exact coefficient functions, invalid-IV regions $\mathcal R_\alpha$, covariance parameters, signal-to-noise ratios (SNR), triangulations, and tuning specifications are provided in Sections \ref{sec:supp-dgm}--\ref{sec:supp-snr} in the Supplementary Material. Specifically, the first-stage SNR, defined in Equation \eqref{EQU:snr-1} as a measure of the candidate IV strength, ranges from approximately 0.2 to 0.4. The second-stage SNR, which measures the strength of the imaging outcome signal, ranges from 1 to 2 under the high-noise setting and from 4 to 8 under the low-noise setting. For each configuration, we generate 200 Monte Carlo replications and report the mean integrated squared error 
${\rm MISE} = R^{-1} \sum^R_{r=1} \text{ISE}^{(r)}$ and $\text{SD}_{\text{ISE}}= \{(R-1)^{-1}\sum^R_{r=1} \{ \text{ISE}^{(r)}-\text{MISE}\}^2\}^{1/2}$, where \(\text{ISE}^{(r)}=N^{-1}\sum_{j=1}^N
\{\beta^\ast(\bm{s}_j)-\widehat{\beta}^{(r)}(\bm{s}_j)\}^2\).
 Simultaneous inference is assessed using the Empirical Coverage Rates \(\text{ECR} = R^{-1}\sum_{r=1}^R I\{\beta^\ast(\bm{s}_j)\in\widehat{\rm SCR}^{(r)}_{1-\tau}(\bm{s}_j) \text{ for all } j\}\) and average confidence region widths. 

\begin{figure}[H]
    \centering
    \includegraphics[width=0.85\linewidth]{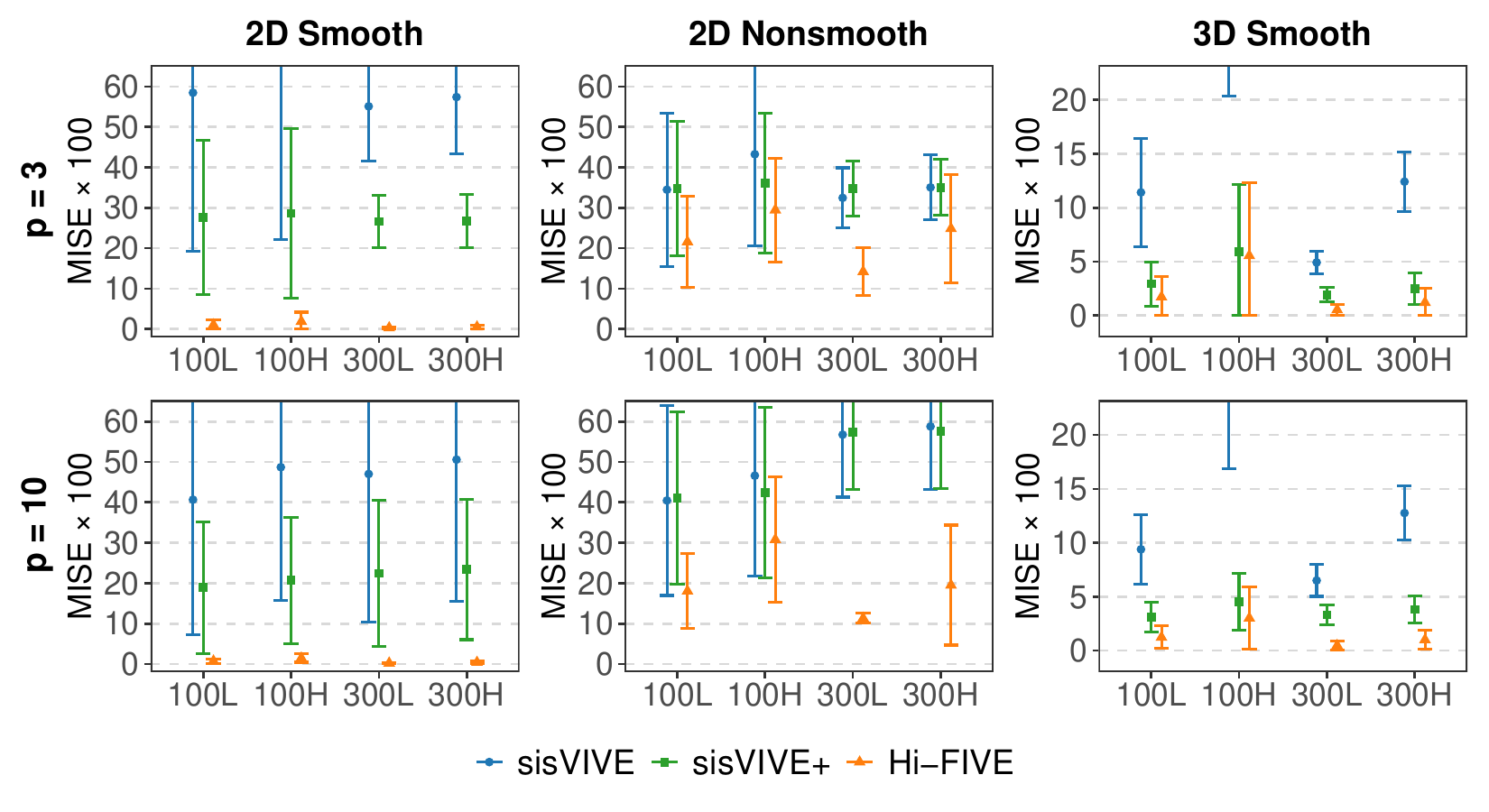}
    \vspace{-0.5cm}
    \caption{Estimation results: MISE and $\text{SD}_{\text{ISE}}$ (L: low noise; H: high noise).}
    \label{fig:mise}
\end{figure}

\vspace{-0.5cm}
\paragraph{Causal-Effect Estimation}
Figure~\ref{fig:mise} summarizes the estimation results across all settings. Tables~\ref{tab:2dcont-mise-ecr}--\ref{tb:mise-3d} in the Supplementary Material report the corresponding numerical values. Under the common-ratio invalid group used here, pointwise plurality fails on $\mathcal R_\alpha$. In the two-dimensional (2D) settings, Hi-FIVE attains the smallest MISE in every configuration, with particularly large improvements where the pointwise plurality
condition fails. Increasing the sample size reduces both the estimation error
and Monte Carlo variability of Hi-FIVE. 
In contrast, sisVIVE remains substantially biased because it performs
identification independently at each location. Although sisVIVE+ reduces local
variability through post-smoothing, it cannot remove the bias caused by failed pointwise identification.  

\begin{figure}[H]
    \centering
    \def\plotheight{2.8cm}

    \newcommand{\cropthenrotate}[2]{%
        \rotatebox[origin=c]{270}{%
            \includegraphics[
                height=\plotheight,
                trim={#1},
                clip
            ]{#2}%
        }%
    }

    \def\gapAB{1.1cm}
    \def\gapBC{1.1cm}
    \def\gapCD{0.96cm}

    \captionsetup[subfigure]{skip=0pt}

    \makebox[\textwidth][c]{%
        \begin{minipage}[t]{0.85\textwidth}
            \centering
            \makebox[0.18\linewidth][c]{True}%
            \hspace{\gapAB}%
            \makebox[0.18\linewidth][c]{Hi-FIVE}%
            \hspace{\gapBC}%
            \makebox[0.18\linewidth][c]{sisVIVE}%
            \hspace{\gapCD}%
            \makebox[0.18\linewidth][c]{sisVIVE+}%
        \end{minipage}%
        \rlap{\hspace{0mm}\begin{minipage}[t]{0.12\textwidth}\mbox{}\end{minipage}}%
    }

    \vspace{-1.1cm}
    \makebox[\textwidth][c]{%
        \llap{\raisebox{-2.6em}{\small (a)}\hspace{1mm}}%
        \begin{minipage}[t]{0.85\textwidth}
            \centering
            \begin{subfigure}[t]{0.18\linewidth}
                \vspace{0pt}
                \centering
                \makebox[\linewidth][c]{%
                    \cropthenrotate
                        {0cm 1cm 2.5cm 1cm}
                        {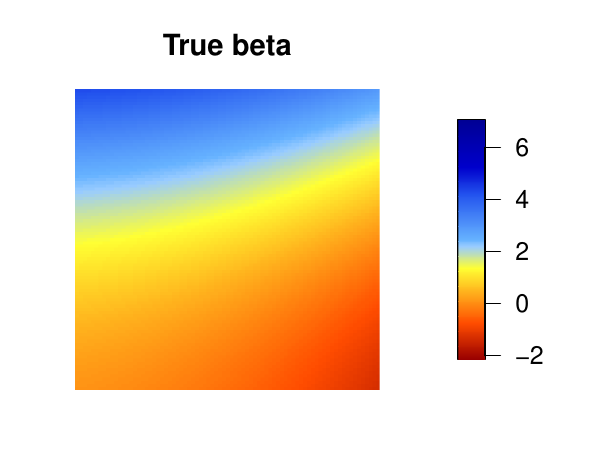}
                }
            \end{subfigure}%
            \hspace{\gapAB}%
            \begin{subfigure}[t]{0.18\linewidth}
                \vspace{0pt}
                \centering
                \makebox[\linewidth][c]{%
                    \cropthenrotate
                        {0cm 1cm 2.5cm 1cm}
                        {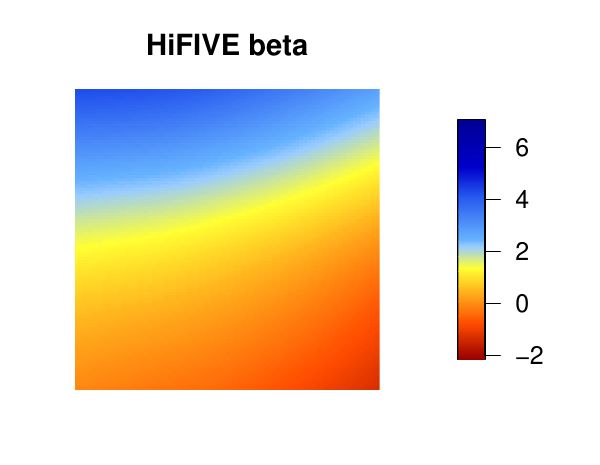}
                }
            \end{subfigure}%
            \hspace{\gapBC}%
            \begin{subfigure}[t]{0.18\linewidth}
                \vspace{0pt}
                \centering
                \makebox[\linewidth][c]{%
                    \cropthenrotate
                        {0cm 1cm 2.5cm 1cm}
                        {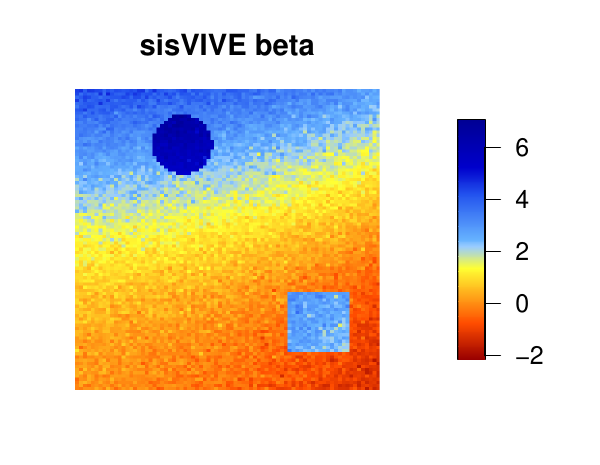}
                }
            \end{subfigure}%
            \hspace{\gapCD}%
            \begin{subfigure}[t]{0.18\linewidth}
                \vspace{0pt}
                \centering
                \makebox[\linewidth][c]{%
                    \cropthenrotate
                        {0cm 1cm 2.5cm 1cm}
                        {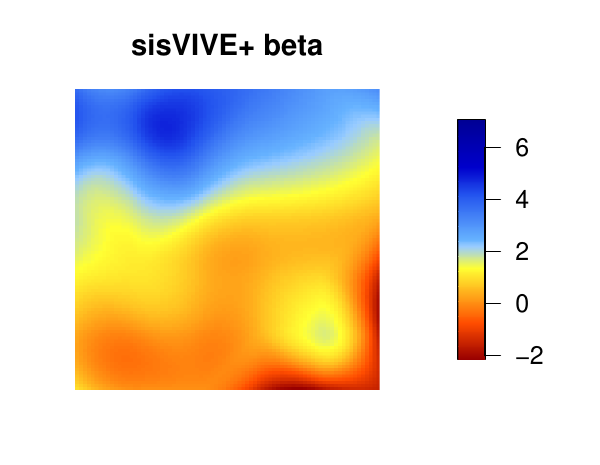}
                }
            \end{subfigure}%
        \end{minipage}%
        \rlap{%
            \hspace{0mm}%
            \begin{minipage}[t]{0.12\textwidth}
                \vspace{0pt}
                \centering
                \includegraphics[
                    height=\plotheight,
                    trim={7.5cm 1.3cm 0cm 0.1cm},
                    clip
                ]{figures/sim_plot/continuous/beta_comparison_cont_sisviveplus.pdf}
            \end{minipage}%
        }%
    }

    \vspace{-0.8cm}
    \makebox[\textwidth][c]{%
        \llap{\raisebox{-2.6em}{\small (b)}\hspace{1mm}}%
        \begin{minipage}[t]{0.85\textwidth}
            \centering
            \begin{subfigure}[t]{0.18\linewidth}
                \vspace{0pt}
                \centering
                \makebox[\linewidth][c]{%
                    \cropthenrotate
                        {0cm 1cm 2.5cm 1cm}
                        {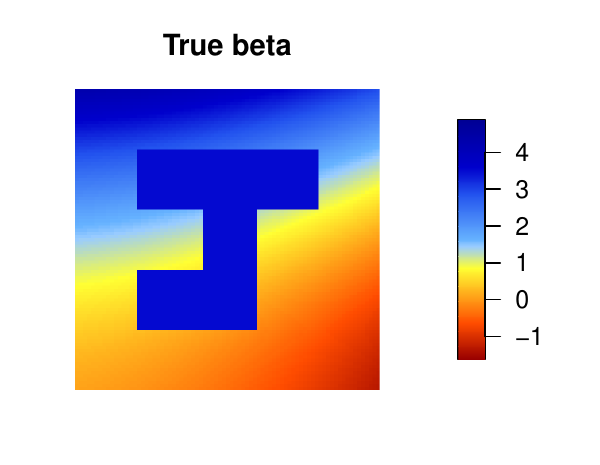}
                }
            \end{subfigure}%
            \hspace{\gapAB}%
            \begin{subfigure}[t]{0.18\linewidth}
                \vspace{0pt}
                \centering
                \makebox[\linewidth][c]{%
                    \cropthenrotate
                        {0cm 1cm 2.5cm 1cm}
                        {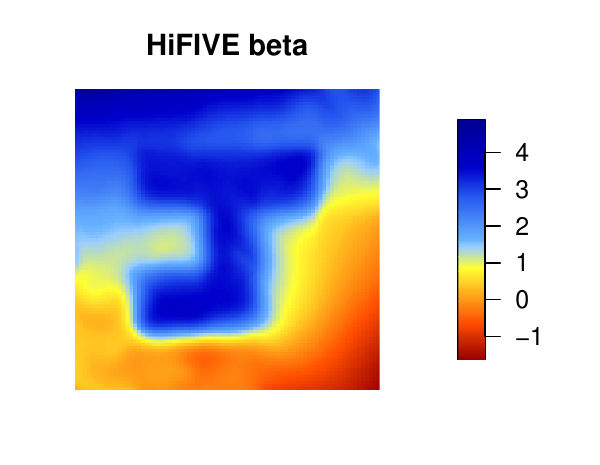}
                }
            \end{subfigure}%
            \hspace{\gapBC}%
            \begin{subfigure}[t]{0.18\linewidth}
                \vspace{0pt}
                \centering
                \makebox[\linewidth][c]{%
                    \cropthenrotate
                        {0cm 1cm 2.5cm 1cm}
                        {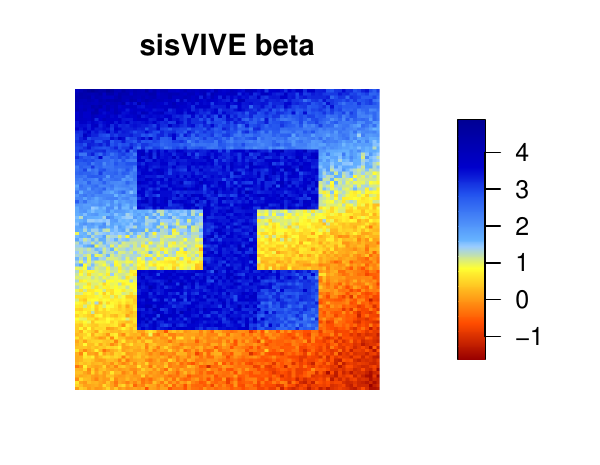}
                }
            \end{subfigure}%
            \hspace{\gapCD}%
            \begin{subfigure}[t]{0.18\linewidth}
                \vspace{0pt}
                \centering
                \makebox[\linewidth][c]{%
                    \cropthenrotate
                        {0cm 1cm 2.5cm 1cm}
                        {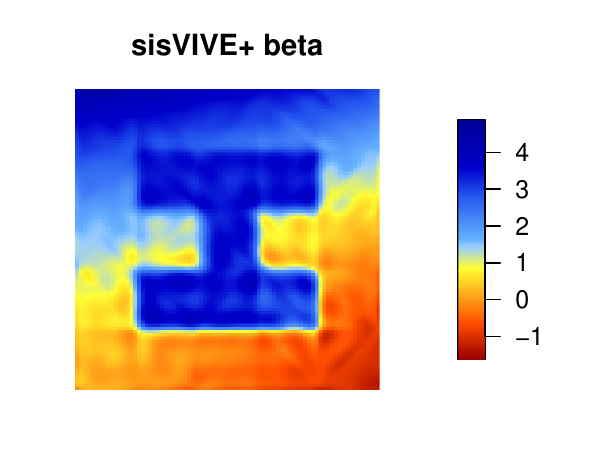}
                }
            \end{subfigure}%
        \end{minipage}%
        \rlap{%
            \hspace{0mm}%
            \begin{minipage}[t]{0.12\textwidth}
                \vspace{0pt}
                \centering
                \includegraphics[
                    height=\plotheight,
                    trim={7.5cm 1.3cm 0cm 0.1cm},
                    clip
                ]{figures/sim_plot/jump/beta_comparison_jump_sisviveplus.pdf}
            \end{minipage}%
        }%
    }

    \vspace{-0.6cm}
    \caption{Estimated $\beta^\ast(\cdot)$ in the 2D settings with $n=100$, $p=10$ candidate IVs, and high noise: (a) smooth causal effect; (b) causal effect with a jump discontinuity. Each row shares the color scale shown on its right.}
    \label{fig:sim-beta-est}
\end{figure}

Figure~\ref{fig:sim-beta-est}(a) shows representative estimated surfaces in the smooth 2D setting. Hi-FIVE closely recovers the true causal effect. In contrast, the sisVIVE estimate is noisy, and its error is dominated by bias in the invalid-IV regions. Smoothing in sisVIVE+ removes much of the noise, but this bias remains clearly visible.

\vspace{-0.5cm}
We next consider a causal-effect function with a jump discontinuity. {This design violates the continuity assumption of Theorem~\ref{theorem:identifiability} and the smoothness assumption of Theorems~\ref{THM:beta-convergence}--\ref{THM:beta-normality}, so it is a misspecification study rather than a setting covered by the theory; in particular, a jump of $\beta^\ast$ across a region boundary is not distinguishable from an invalid-IV effect by the identification argument alone.} 
The true signal forms a digital-font ``4,'' while the invalid-IV effect occupies the additional region required to form an ``H.'' Thus, a method that fails to separate the causal and invalid-IV effects tends to recover an H-shaped coefficient function. As shown in Figure~\ref{fig:mise}, under low noise, the Hi-FIVE estimator improves clearly as the sample size increases, with substantial decreases in both the MISE and its standard deviation. Under high noise, Hi-FIVE still achieves the lowest MISE across all nonsmooth settings. 
Figure~\ref{fig:sim-beta-est}(b) shows that the comparison methods tend to mix the invalid-IV signal with the true causal effect and estimate an H-shaped jump in the causal-effect function. In contrast, the Hi-FIVE estimator mostly preserves the ``4'' pattern, suggesting that it can distinguish the causal-effect structure from the invalid-IV pattern. Although Hi-FIVE is designed for a continuous causal-effect function and therefore does not perfectly recover the discontinuity, it still captures the main causal pattern and remains robust in nonsmooth settings.
Its MISE in this design is, however, an order of magnitude larger than in the smooth design (Tables~\ref{tab:2dcont-mise-ecr} and \ref{tab:jump-mise}), which quantifies the cost of the misspecification.

We further evaluate Hi-FIVE for three-dimensional (3D) volumetric outcomes. Figure \ref{S-fig:bias_allmethod_cont} displays estimates from Hi-FIVE and the comparison methods for selected 3D slices. This setting is more challenging because of the substantially larger number of spatial locations, the increased computational burden, and the more complex geometry of spatially localized invalid-IV effects. Nevertheless, Hi-FIVE consistently attains the smallest MISE across all configurations. Its advantage becomes more pronounced as the sample size increases, accompanied by a substantial reduction in the standard deviation of the MISE and a wider performance gap relative to sisVIVE+. These results show that the proposed spatial identification and estimation framework performs effectively in 3D settings.

\vspace{-0.5cm}
\paragraph{Inferences and Computational Cost} Table \ref{tab:cont-ecr-scr} evaluates the inference performance of Hi-FIVE in the 2D smooth case at significance level $\tau \in \{0.1, 0.05, 0.01\}$. Across most settings, empirical coverage rates are close to their nominal levels \(1-\tau\). As expected, the SCR widths increase when the nominal level becomes lower, from \(\tau=0.10\) to \(\tau=0.01\), and they are generally wider under high noise. Increasing the sample size from \(n=100\) to \(n=300\) reduces the SCR widths, reflecting improved estimation precision. In addition, the SCRs are also narrower for \(p=10\) than for \(p=3\), suggesting that the additional IV information improves the inference efficiency.

Table \ref{tab:computation-time} reports the average computational time for model estimation on a standard computing platform. Hi-FIVE is substantially faster than sisVIVE, particularly in the 3D setting, while its computational time increases moderately with sample size, number of instruments, and image dimension. Moreover, we assess triangulation sensitivity by considering coarse, default, and fine triangulations; see Section \ref{sec:supp-triangulation} in the Supplementary Material. The estimated causal effects and invalid-IV regions remain stable, while finer triangulations improve boundary resolution at additional computational cost.

\begin{table}[H]
\centering
\caption{Empirical simultaneous coverage rates and average widths of the SCRs based on 200 replications. Columns are indexed by the nominal significance level $\tau$.}
\label{tab:cont-ecr-scr}
\resizebox{\textwidth}{!}{
\begingroup
\setlength{\aboverulesep}{0.05ex}
\setlength{\belowrulesep}{0.05ex}

\begin{tabular}{ccccccccccccccc}
\toprule
\multirow{3}{*}{$\kappa$} 
& \multirow{3}{*}{$n$} 
& \multirow{3}{*}{Noise}
& \multicolumn{6}{c}{ECR} 
& \multicolumn{6}{c}{SCR Width}  \\
\cmidrule(lr){4-9} \cmidrule(lr){10-15}
&&& \multicolumn{3}{c}{$p=3$} 
& \multicolumn{3}{c}{$p=10$}
& \multicolumn{3}{c}{$p=3$} 
& \multicolumn{3}{c}{$p=10$}  \\
\cmidrule(lr){4-6} \cmidrule(lr){7-9} 
\cmidrule(lr){10-12} \cmidrule(lr){13-15}
&&& 0.10 & 0.05 & 0.01 
& 0.10 & 0.05 & 0.01 
& 0.10 & 0.05 & 0.01 
& 0.10 & 0.05 & 0.01 \\
\midrule
\multirow{4}{*}{2D}&
\multirow{2}{*}{100} & Low & 0.860 & 0.920 & 0.980 & 0.885 & 0.920 & 0.970 & 0.582 & 0.631 & 0.716 & 0.504 & 0.537 & 0.592 \\
& & High & 0.880 & 0.915 & 0.975 & 0.855 & 0.920 & 0.970 & 0.761 & 0.830 & 0.943 & 0.632 & 0.701 & 0.799 \\
\cmidrule(lr){2-15}
&\multirow{2}{*}{300} & Low & 0.915 & 0.960 & 0.985 & 0.925 & 0.965 & 0.985 & 0.465 & 0.483 & 0.518 & 0.420 & 0.440 & 0.472 \\
& & High & 0.910 & 0.965 & 0.985 & 0.930 & 0.970 & 0.975 & 0.599 & 0.621 & 0.673 & 0.516 & 0.546 & 0.603 \\
\midrule
\multirow{4}{*}{3D}
& \multirow{2}{*}{100} 
& Low  & 0.880 & 0.930 & 0.975 & 0.870 & 0.935 & 0.980 & 1.607 & 1.990 & 2.632 & 1.439 & 1.712 & 2.199   \\
&& High & 0.865 & 0.925 & 0.970 & 0.875 & 0.930 & 0.975 & 1.907 & 2.280 & 3.078 & 1.789 & 2.094 & 2.621 \\
\cmidrule(lr){2-15}
& \multirow{2}{*}{300} 
& Low  & 0.885 & 0.950 & 0.990 & 0.890 & 0.945 & 0.990 & 0.793 & 0.945 & 1.286 & 0.725 & 0.876 & 1.164 \\
&& High & 0.890 & 0.935 & 0.990 & 0.890 & 0.935 & 0.990 & 0.934 & 1.113 & 1.524 & 0.870 & 1.033 & 1.354 \\
\bottomrule
\end{tabular}
\endgroup
}
\end{table}

\begin{table}[H]
\centering
\caption{Average computation time (seconds) per replication.}
\label{tab:computation-time}

\renewcommand{\arraystretch}{0.7}
\setlength{\tabcolsep}{10pt}

\resizebox{0.9\textwidth}{!}{%
\begin{tabular}{c cccc cccc}
\toprule
& \multicolumn{4}{c}{2D ($N=7505$)}
& \multicolumn{4}{c}{3D ($N=300763$)} \\
\cmidrule(lr){2-5}
\cmidrule(lr){6-9}
& \multicolumn{2}{c}{$n=100$}
& \multicolumn{2}{c}{$n=300$}
& \multicolumn{2}{c}{$n=100$}
& \multicolumn{2}{c}{$n=300$} \\
\cmidrule(lr){2-3}
\cmidrule(lr){4-5}
\cmidrule(lr){6-7}
\cmidrule(lr){8-9}
$p$
& sisVIVE & Hi-FIVE
& sisVIVE & Hi-FIVE
& sisVIVE & Hi-FIVE
& sisVIVE & Hi-FIVE \\
\midrule
3  & 90  & 15 & 125 & 40 & 4420 & 1090 & 5110 & 1415 \\
10 & 145 & 25 & 210 & 60 & 6490 & 1145 & 7800 & 1940 \\
\bottomrule
\end{tabular}%
}

\vspace{2pt}
\begin{minipage}{0.9\textwidth}
\footnotesize
\textit{Note:} Computation times are rounded to the nearest 5 seconds.
\end{minipage}

\end{table}

\section{Application}
\label{section:application}
Hi-FIVE is applied to data from the Alzheimer’s Disease Neuroimaging Initiative (ADNI). 
The imaging responses are spatially normalized fluorodeoxyglucose positron emission tomography (FDG-PET) images of size 79$\times$95$\times$68, which measure regional glucose metabolism, a widely used proxy for neuronal activity. We consider phosphorylated tau (p-tau) as the exposure, and five SNPs (rs744373, rs3820757, rs11136000, rs6656401, rs3851179), chosen from prior genetic and biomarker studies of tau pathology and CSF AD biomarkers \citep{franzmeier2019bin1taupet, franzmeier2022rs744373, xu2020picalmbiomarkers}, as candidate instrumental variables. Age, sex, and diagnosis are partialled out of the imaging outcome, exposure, and instruments before the two-stage fit. 
The complete-case sample with baseline FDG-PET, p-tau, genotype, and covariates comprises 105 subjects: 29 cognitively normal, 39 early MCI, and 37 MCI participants; none had dementia, because the matched genotype–imaging subset was concentrated among cognitively normal and MCI participants.
We have no evidence that any of these SNPs violates the IV assumptions, but their pointwise validity cannot be guaranteed since their biological functions and potential pleiotropic pathways are not fully understood, a common issue in MR. Because some SNPs are involved in the broader pathway linking A$\beta$, p-tau, and cognition, they may affect FDG-PET through mechanisms other than p-tau, which motivates our analysis allowing for invalid IVs.
As a key biomarker of tau-related AD pathology closely associated with downstream neurodegeneration, p-tau lets us assess whether tau burden causally affects neurodegeneration, as reflected by FDG-PET across brain regions. To accommodate potential differences in spatial effects between hemispheres, we partition the image domain into left- and right-hemisphere subdomains.

\begin{center}
    \begin{figure}[!htbp]
        \centering
\includegraphics[width=0.9\linewidth]{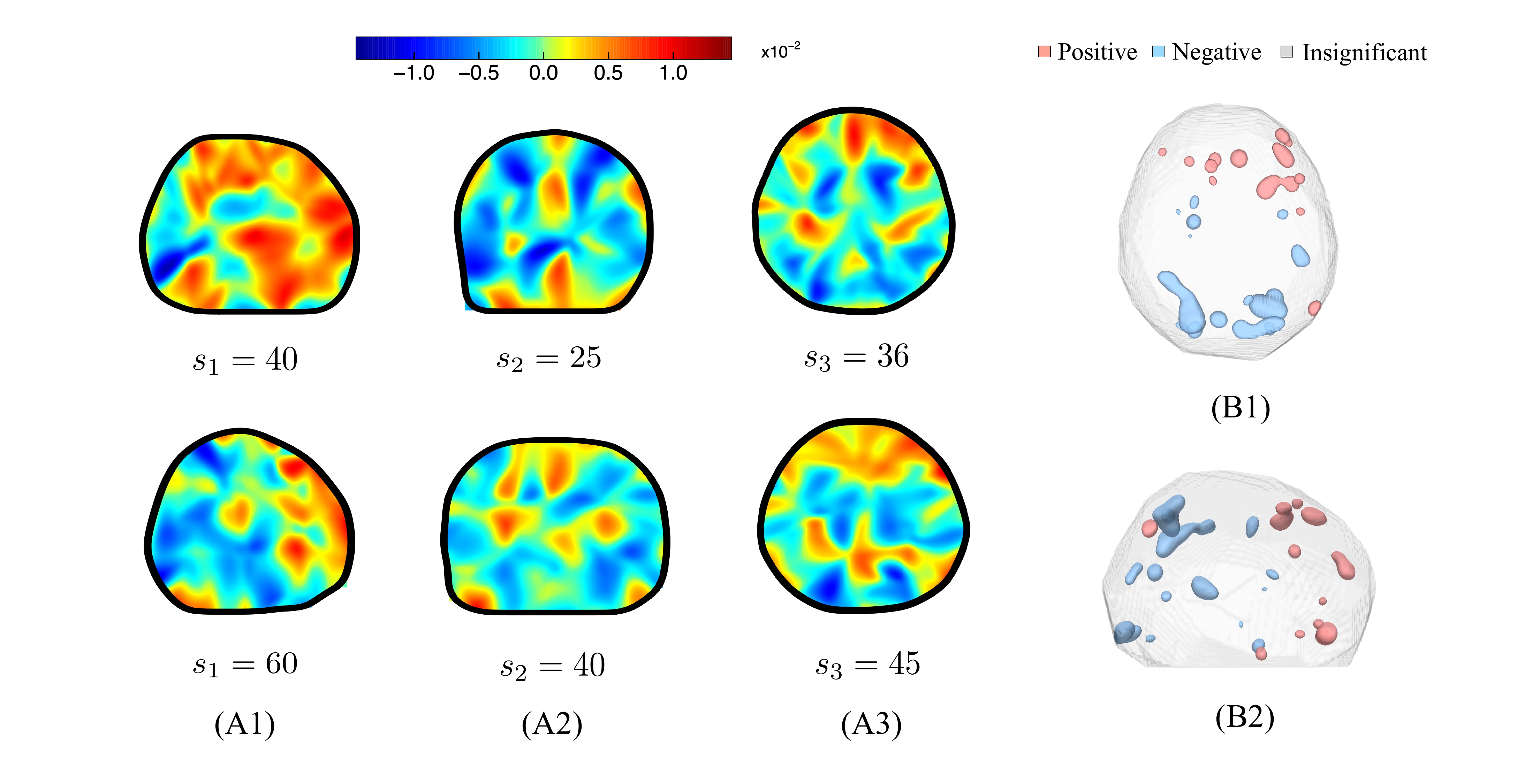}
        \vspace{-0.5cm}
        \caption{Causal effects of p-tau on FDG-PET imaging by Hi-FIVE: (A1) sagittal view; (A2) coronal view; (A3) axial view. 
        Significance map at a 5\% significance level of Hi-FIVE causal effect: (B1) Superior view (top = anterior); (B2) Right lateral view (right = anterior). }
        \label{fig:3drealdata-est-ref}
    \end{figure}
\end{center}

Figure \ref{fig:3drealdata-est-ref} (A1)-(A3) shows the estimated causal effects of p-tau on FDG-PET metabolism. Blue regions represent negative causal effects, suggesting that higher p-tau leads to reduced metabolic activity under the fitted IV model, whereas red regions indicate positive causal effects, and green regions correspond to estimates close to zero. In the 3D causal-effect maps, $s_1$, $s_2$, and $s_3$ denote sagittal, coronal, and axial slice indices, respectively. 

Across these views, the estimated causal effect of p-tau on FDG-PET is negative in localized regions, most prominently in the left-hemispheric sagittal slice ($s_1=60$), posterior coronal slices ($s_2=25$ and $s_2=40$), and mid-to-superior axial slices ($s_3=36$ and $s_3=45$). Under the fitted model, these negative patterns indicate that higher p-tau lowers FDG-PET uptake there. Their spatial distribution is consistent with previous ADNI studies linking elevated plasma or CSF pTau181 to lower FDG metabolism in cognitively impaired individuals, particularly in temporal, cingulate, precuneus, and orbitofrontal regions \citep{shen2021ptaupet}. The axial slices also cover posterior cingulate, precuneus, and parietotemporal regions, which are commonly hypometabolic on FDG-PET in MCI and AD \citep{bailly2015fdgpet}. Although these prior findings are associational, they support the biological plausibility of our causal estimates. Some positive estimates are observed, particularly in the near-midline sagittal slice ($s_1=40$). Because prior evidence most consistently links p-tau to reduced glucose metabolism, we interpret these positive regions as localized effects in this nondemented cohort rather than as a typical AD-related pattern.

\vspace{-0.5cm}

The 3D significance map in Figure \ref{fig:3drealdata-est-ref} (B1)-(B2) further indicates significant heterogeneous Hi-FIVE causal effects of p-tau on FDG-PET, with significant positive clusters primarily in anterior regions and significant negative clusters in posterior and inferior regions. The posterior negative pattern is consistent with AD-related FDG-PET hypometabolism in the posterior cingulate, precuneus, and temporoparietal cortex \citep{wilson2019petdementia,hanseeuw2017fdgtauamyloid}. The anterior positive clusters may reflect localized metabolic increases that differ from the typical posterior hypometabolic pattern, consistent with prior evidence linking p-tau181 to FDG-PET abnormalities and reports of tau-related FDG-PET hypermetabolism in the MCI group \citep{shen2021ptaupet,rubinski2020fdgpethypermetabolism}.

\section{Discussion}
\label{sec:discussion}

We propose Hi-FIVE, a functional data approach for causal estimation with imaging outcomes under unmeasured confounding. Hi-FIVE leverages spatial structure to relax the plurality requirement for valid IVs, enabling identification and inference in settings where scalar-outcome methods fail. It also uses domain decomposition and local basis representations to improve computational scalability while preserving localized spatial features. Simulations show that Hi-FIVE achieves lower estimation error and more accurate invalid-IV selection than pixelwise competitors and their smoothed variants.

The framework suggests several extensions. First, the current theory treats the number
of candidate IVs as fixed; accommodating a very large instrument set is an important open
problem. Second, weak instruments remain a serious obstacle in genetic applications, and how instrument strength interacts with the relaxed plurality condition is not yet understood \citep{lin2024instrumental,ye2024genius}. Third, we have assumed throughout that the causal effect function is continuous; adapting both the estimator and the theory to effects that jump across anatomical boundaries is an open problem. Fourth, many genome-wide studies release only summary statistics computed on separate exposure and outcome samples, and a two-sample version of Hi-FIVE would widen its applicability considerably \citep{ye2021debiased,dai2024inference,zilinskas2024inferring}. Finally, the identification argument is not specific to IVs and may prove useful for other functional causal estimands, particularly mediation analysis with imaging mediators.

\section*{Data Availability Statement}

The data that support the findings of this study are available from the Alzheimer's Disease Neuroimaging Initiative (ADNI). Restrictions apply to the availability of these data, which were used under the ADNI Data Use Agreement for this study. Data are available through  \url{https://adni.loni.usc.edu} with the permission of ADNI.

\if1\blind
\section*{Acknowledgment}
This research was partially supported by NSF grants DMS-2515788, 
DMS-2515789, and DMS-2426173, and by NIH grants 
1R01AG085616 and R01HL184065. Data used in the preparation of this paper were obtained from the ADNI database (adni.loni.usc.edu). As such, the investigators within the ADNI contributed to the design and implementation of ADNI and/or provided data but did not participate in the analysis or writing of this report. A complete listing of ADNI investigators can be found at \url{http://adni.loni.usc.edu/wp-content/uploads/how_to_apply/ADNI_Acknowledgement_List.pdf}. 
\fi

\section*{Declaration of Generative AI Use}
The authors used generative AI tools to polish the language and improve the readability of the text, and to assist with writing and debugging code for the simulations and figures. The authors conceived the research question, developed the methodology, formulated the identification conditions and theoretical results, designed the proof strategies, and verified all proofs, numerical results, and references. All AI-generated text and code were reviewed and edited by the authors, who take full responsibility for the content of the manuscript.

\spacingset{1}
\bibliographystyle{asa}
\bibliography{references}

\clearpage
\setcounter{page}{1}
\spacingset{1}
\begin{center}
{\Large\bf Supplementary Materials for ``Causal Learning with Invalid
Instruments for High-Dimensional Imaging Responses''}
\end{center}
	
	\spacingset{1.65} 

	\renewcommand{\suppanchor}{A}
	\section*{Section A. Additional Lemmas and Proof Details}
	\renewcommand{\thedefinition}{A.\arabic{definition}}
	\renewcommand{\theequation}{A.\arabic{equation}}
	\renewcommand{\thesubsection}{A.\arabic{subsection}}
	\renewcommand{\thetheorem}{{\sc A.\arabic{theorem}}}
	\renewcommand{\thelemma}{{\sc A.\arabic{lemma}}} %
	\renewcommand{\theproposition}{{\sc A.\arabic{proposition}}}
	\renewcommand{\thecorollary}{\sc A.\arabic{corollary}}
	\renewcommand{\thefigure}{A.\arabic{figure}}
	\renewcommand{\thetable}{A.\arabic{table}}
	\renewcommand{\theremark}{A.\arabic{remark}}
	\renewcommand{\theexample}{A.\arabic{example}}
	\setcounter{equation}{0}
	\setcounter{definition}{0}
	\setcounter{theorem}{0}
	\setcounter{lemma}{0}
	\setcounter{figure}{0}
	\setcounter{remark}{0}
	\setcounter{proposition}{0}
	\setcounter{corollary}{0}
	\setcounter{example}{0}
	\setcounter{subsection}{0}
	\setcounter{subsubsection}{0}

	\subsection{Identifiability: Proof of Theorem 1}

	\begin{proof}
		Let \(P\) denote the observed distribution of \((Y(\cdot),D,\mathbf Z)\). Since \(\Sigma_{\mathbf Z}\), \(\bm\gamma^\ast\), \(\bm\varpi(\cdot)\), \(\mathcal I\), and \(r_\ell(\cdot)\) have been defined from \(P\) before the theorem, we only record the implication needed below. For any parameter vector $\bm\alpha$, define $\mathcal V_{\bm\alpha}(\bm s):=\{\ell\in\mathcal I:\alpha_\ell(\bm s)=0\}$. For any admissible parameter pair \((\beta,\bm\alpha)\) satisfying the compact model and the orthogonality conditions, we have
		\begin{equation}\label{eq:ratio_ident}
		\begin{aligned}
			\bm\varpi(\bm s)&=\bm\gamma^\ast\beta(\bm s)+\bm\alpha(\bm s),\\
			r_\ell(\bm s)&=\beta(\bm s)+\frac{\alpha_\ell(\bm s)}{\gamma^\ast_\ell},
			\qquad \ell\in\mathcal I.
		\end{aligned}
		\end{equation}

		To prove identifiability, suppose that two parameter pairs \((\beta,\bm\alpha)\) and \((\widetilde\beta,\widetilde{\bm\alpha})\) both satisfy the compact model, the orthogonality conditions, and assumptions (A1)--(A3), and suppose that they induce the same observed distribution \(P\). Since \(\bm\gamma^\ast\) and \(\bm\varpi(\cdot)\) are determined by \(P\), both pairs satisfy, for every \(\ell\in\mathcal I\),
		\begin{equation}\label{eq:ratio_pair}
			r_\ell(\bm s)=\beta(\bm s)+\frac{\alpha_\ell(\bm s)}{\gamma^\ast_\ell}=\widetilde\beta(\bm s)+\frac{\widetilde\alpha_\ell(\bm s)}{\gamma^\ast_\ell},\qquad \bm s\in\Omega.
		\end{equation}
		We show that \(\beta(\bm s)=\widetilde\beta(\bm s)\) for all \(\bm s\in\Omega\).
		
		\smallskip
		\noindent\textbf{Step 1: Existence of a coincidence point.} 
		
		We first show that the two causal effect functions coincide at some point of $\Omega$.
		Since the ratio surfaces \(r_\ell(\cdot)\), \(\ell\in\mathcal I\), are determined by the observed distribution, condition (A2)(b) is not pair-specific: either it holds for both observationally equivalent parameter pairs, or it holds for neither. In the latter case, since each pair satisfies (A2), both pairs must satisfy (A2)(a).
		
		\textit{Suppose that (A2)(b) holds.} Then
		there exists a nonempty open set \(\Omega_{\bigodot}\subseteq\Omega\) such that \(r_\ell(\bm s)=r_k(\bm s)\) for all \(\ell,k\in\mathcal I\) and all \(\bm s\in\Omega_{\bigodot}\). Fix any \(\bm s\in\Omega_{\bigodot}\). By (A1), applied to \((\beta,\bm\alpha)\), there exists at least one relevant IV \(\ell\in\mathcal I\) such that \(\alpha_\ell(\bm s)=0\). Then \eqref{eq:ratio_ident} gives \(r_\ell(\bm s)=\beta(\bm s)\). Since all ratio values are equal on \(\Omega_{\bigodot}\), their common value equals \(\beta(\bm s)\). Similarly, by (A1), applied to \((\widetilde\beta,\widetilde{\bm\alpha})\), there exists \(k\in\mathcal I\) such that \(\widetilde\alpha_k(\bm s)=0\), and hence \(r_k(\bm s)=\widetilde\beta(\bm s)\). Therefore \(\beta(\bm s)=\widetilde\beta(\bm s)\), so \(U:=\Omega_{\bigodot}\) is a nonempty open set on which the two causal effect functions coincide.
		
		\textit{Suppose that (A2)(a) holds.}
		There exist Lebesgue-measurable sets $\Omega_\oplus\subseteq\Omega_0(\beta,\bm\alpha)$ and $\widetilde\Omega_\oplus\subseteq\Omega_0(\widetilde\beta,\widetilde{\bm\alpha})$ such that ${\rm Vol}(\Omega_\oplus)>{\rm Vol}(\Omega)/2$ and ${\rm Vol}(\widetilde\Omega_\oplus)>{\rm Vol}(\Omega)/2$. Therefore ${\rm Vol}(\Omega_\oplus\cap\widetilde\Omega_\oplus)>0$, so $U:=\Omega_\oplus\cap\widetilde\Omega_\oplus$ is nonempty. For any $\bm s\in\Omega_\oplus$, if $\ell\in\mathcal V_{\bm\alpha}(\bm s)$, then $\alpha_\ell(\bm s)=0$, and hence \eqref{eq:ratio_ident} gives $r_\ell(\bm s)=\beta(\bm s)$. By the pointwise plurality condition,
        \[
        \left|\{\ell\in\mathcal I:r_\ell(\bm s)=\beta(\bm s)\}\right|
        >\max_{v\neq\beta(\bm s)}
          \left|\{\ell\in\mathcal I:r_\ell(\bm s)=v\}\right|,
        \]
        so $\beta(\bm s)$ is the unique mode of the observable multiset $\{r_\ell(\bm s):\ell\in\mathcal I\}$. Similarly, for every $\bm s\in\widetilde\Omega_\oplus$, $\widetilde\beta(\bm s)$ is the unique mode of the same observable multiset. Hence, for every $\bm s\in U=\Omega_\oplus\cap\widetilde\Omega_\oplus$, both $\beta(\bm s)$ and $\widetilde\beta(\bm s)$ equal the same unique mode, and therefore $\beta(\bm s)=\widetilde\beta(\bm s)$.

		\smallskip
		\noindent\textbf{Step 2: Extension of the coincidence region.} 
		
		Define \(S:=\{\bm s\in\Omega:\beta(\bm s)=\widetilde\beta(\bm s)\}\). Since \(\beta,\widetilde\beta\in C^r(\Omega)\), both functions are continuous, and therefore \(S\) is closed in \(\Omega\). By Step 1, $S$ is nonempty. We show that \(S=\Omega\).
		
		Suppose, for a contradiction, that \(S\neq\Omega\). Because \(\Omega\) is connected and \(S\) is a nonempty proper closed subset, its boundary \(\partial_\Omega S\) is nonempty. Choose \(\bm t\in\partial_\Omega S\). Since \(S\) is closed in \(\Omega\), we have \(\bm t\in S\), and hence \(\beta(\bm t)=\widetilde\beta(\bm t)\). Moreover, \(\bm t\) is not an interior point of \(S\); equivalently, every neighborhood of \(\bm t\) in \(\Omega\) contains points outside \(S\).
		
		Let \(\Omega_0:=\Omega_0(\beta,\bm\alpha)\) and \(\widetilde\Omega_0:=\Omega_0(\widetilde\beta,\widetilde{\bm\alpha})\). We first claim that
		\begin{equation}\label{eq:t_ship}
			\bm t\in {\rm cl}_\Omega(\Omega_0^c)\cup {\rm cl}_\Omega(\widetilde\Omega_0^c).
		\end{equation}
		To prove the claim, suppose instead that \(\bm t\notin{\rm cl}_\Omega(\Omega_0^c)\) and \(\bm t\notin{\rm cl}_\Omega(\widetilde\Omega_0^c)\). Equivalently, \(\bm t\in{\rm int}_\Omega(\Omega_0)\cap{\rm int}_\Omega(\widetilde\Omega_0)\), so there exists an open neighborhood \(W\subseteq\Omega\) of \(\bm t\) such that \(W\subseteq\Omega_0\cap\widetilde\Omega_0\). For every \(\bm s\in W\), the pointwise plurality condition holds for both parameter pairs. Therefore, \(\beta(\bm s)\) and \(\widetilde\beta(\bm s)\) are both the unique mode of the same observable multiset \(\{r_\ell(\bm s):\ell\in\mathcal I\}\), and hence \(\beta(\bm s)=\widetilde\beta(\bm s)\) for all \(\bm s\in W\). Thus \(W\subseteq S\), which implies that \(\bm t\in{\rm int}_\Omega(S)\). This contradicts \(\bm t\in\partial_\Omega S\), since every neighborhood of a boundary point must contain points outside \(S\). Therefore, the claim \eqref{eq:t_ship} holds.
		
		By \eqref{eq:t_ship}, there are two cases. First suppose that \(\bm t\in{\rm cl}_\Omega(\widetilde\Omega_0^c)\). By (A1), applied to the first pair \((\beta,\bm\alpha)\) at \(\bm t\), there exist \(\ell\in\mathcal I\) and \(\varepsilon>0\) such that \(\alpha_\ell(\bm s)=0\) for all \(\bm s\in N:=\mathbb N_\varepsilon(\bm t)\). Hence \(r_\ell(\bm s)=\beta(\bm s)\) for all \(\bm s\in N\). Combining this with \eqref{eq:ratio_pair} for the second pair gives
		\begin{equation}\label{eq:alpha_form}
			\widetilde\alpha_\ell(\bm s)=\gamma^\ast_\ell\{\beta(\bm s)-\widetilde\beta(\bm s)\},\qquad \bm s\in N.
		\end{equation}
		Since \(\beta,\widetilde\beta\in C^r(\Omega)\), \eqref{eq:alpha_form} implies that \(\widetilde\alpha_\ell(\cdot)\) is \(C^r\) in a neighborhood of \(\bm t\). Also, because \(\bm t\in S\), \eqref{eq:alpha_form} gives \(\widetilde\alpha_\ell(\bm t)=0\), so \(Z_\ell\) is valid at \(\bm t\) under the second pair. However, every neighborhood of \(\bm t\) contains points outside \(S\). For any \(\bm s'\in N\setminus S\), \eqref{eq:alpha_form} gives \(\widetilde\alpha_\ell(\bm s')\neq0\). Thus every neighborhood of \(\bm t\) contains both a valid point and an invalid point for \(Z_\ell\) under the second pair, so \(Z_\ell\) is unstable at \(\bm t\). This contradicts (A3), applied to the second pair, because \(\bm t\in{\rm cl}_\Omega(\widetilde\Omega_0^c)\) and \(\widetilde\alpha_\ell(\cdot)\) is \(C^r\) in a neighborhood of \(\bm t\).
		
		If \(\bm t\notin{\rm cl}_\Omega(\widetilde\Omega_0^c)\), then by \eqref{eq:t_ship} we must have \(\bm t\in{\rm cl}_\Omega(\Omega_0^c)\). Similarly, by (A1), applied to the second pair \((\widetilde\beta,\widetilde{\bm\alpha})\), there exist \(k\in\mathcal I\) and \(\delta>0\) such that \(\widetilde\alpha_k(\bm s)=0\) for all \(\bm s\in N:=\mathbb N_\delta(\bm t)\). Hence \(r_k(\bm s)=\widetilde\beta(\bm s)\) for all \(\bm s\in N\), and combining this with \eqref{eq:ratio_pair} for the first pair gives
		\[\alpha_k(\bm s)=\gamma^\ast_k\{\widetilde\beta(\bm s)-\beta(\bm s)\},\qquad \bm s\in N.\]
		Thus \(\alpha_k(\cdot)\) is \(C^r\) in \(N\) and \(\alpha_k(\bm t)=0\), while every neighborhood of \(\bm t\) contains points \(\bm s\notin S\), for which \(\alpha_k(\bm s)\neq0\). Therefore \(Z_k\) is unstable at \(\bm t\) under the first pair. This contradicts (A3), because \(\bm t\in{\rm cl}_\Omega(\Omega_0^c)\) and \(\alpha_k(\cdot)\) is \(C^r\) in a neighborhood of \(\bm t\). 
		
		Both cases lead to contradictions. Therefore the assumption \(S\neq\Omega\) is false, and hence \(S=\Omega\). Consequently, \(\beta(\bm s)=\widetilde\beta(\bm s)\) for every \(\bm s\in\Omega\). This proves identifiability of \(\beta^\ast(\cdot)\) on \(\Omega\).

		\smallskip
		\noindent\textbf{Step 3: Counterexamples when one assumption fails.} 
		
		Take $\Omega=[0,1]^2$, a compact, connected domain with nonempty interior. The following two-dimensional examples show that failure of each condition can lead to non-identifiability. To construct the common observed distribution, take mean-zero $\xi$
        and a mean-zero process $u(\cdot)$ with $(\xi,u(\cdot))$ independent of $\mathbf Z$, and set
        $D=\mathbf Z^\top\bm\gamma+\xi$ and
        $Y(\bm s)=\mathbf Z^\top\bm\varpi(\bm s)+u(\bm s)$.
        For each pair satisfying $\bm\gamma\beta+\bm\alpha=\bm\varpi$,
        define $\epsilon(\bm s)=u(\bm s)-\xi\beta(\bm s)$.
        Then \eqref{EQU:obs} and both orthogonality conditions hold,
        while the observed distribution is the same for every such pair.
        Dependence between $\xi$ and $\epsilon$ is allowed in the
        identifiability theorem.
		
		\emph{Failure of (A1):} Let \(p=2\), \(\bm\gamma=(1,1)^\top\), and let \(r_1(\bm s)=r_2(\bm s)=0\) for all \(\bm s\in\Omega\). Consider Model A with \(\beta_A(\bm s)=0\) and \(\bm\alpha_A(\bm s)=(0,0)^\top\), and Model B with \(\beta_B(\bm s)=1\) and \(\bm\alpha_B(\bm s)=(-1,-1)^\top\). Then
		\[\bm\gamma\beta_A(\bm s)+\bm\alpha_A(\bm s)=(0,0)^\top=\bm\gamma\beta_B(\bm s)+\bm\alpha_B(\bm s),\]
		so the two models are observationally equivalent, but \(\beta_A(\bm s)\neq\beta_B(\bm s)\). In Model B, no relevant IV is valid anywhere, because both components of \(\bm\alpha_B\) are equal to \(-1\). Hence (A1) fails. However, (A2)(b) holds because all ratio surfaces are equal on \(\Omega\), and (A3) holds because the components of \(\bm\alpha_B\) are constant and both IVs are invalid everywhere, hence stable everywhere. Therefore, without (A1), \(\beta(\cdot)\) is generally not identifiable.
		
		\emph{Failure of (A2):} Let \(p=2\), \(\bm\gamma=(1,1)^\top\), and let \(r_1(\bm s)=0\) and \(r_2(\bm s)=1\) for all \(\bm s\in\Omega\). Consider Model A with \(\beta_A(\bm s)=0\) and \(\bm\alpha_A(\bm s)=(0,1)^\top\), and Model B with \(\beta_B(\bm s)=1\) and \(\bm\alpha_B(\bm s)=(-1,0)^\top\). Then
		\[\bm\gamma\beta_A(\bm s)+\bm\alpha_A(\bm s)=(0,1)^\top=\bm\gamma\beta_B(\bm s)+\bm\alpha_B(\bm s),\]
		so the two models are observationally equivalent, but \(\beta_A(\bm s)\equiv0\) and \(\beta_B(\bm s)\equiv1\). In Model A, \(Z_1\) is valid everywhere and \(Z_2\) is invalid everywhere, while in Model B, \(Z_2\) is valid everywhere and \(Z_1\) is invalid everywhere. Thus (A1) holds for both models. The components of \(\bm\alpha_A\) and \(\bm\alpha_B\) are constant, so every IV has a stable validity status, and (A3) holds. However, (A2) fails. At every \(\bm s\), the observable ratio multiset is \(\{0,1\}\), so there is no strict plurality value, and there is no nonempty open set on which all relevant ratio surfaces are equal. Therefore, without (A2), \(\beta(\cdot)\) is generally not identifiable.
		
		\emph{Failure of (A3):} Let \(p=3\), \(\bm\gamma=(1,1,1)^\top\), and define \(g(\bm s)=(s_1-2/3)_+^{r+1}\) for \(\bm s=(s_1,s_2)\in[0,1]^2\), where \((x)_+=\max(x,0)\). Then \(g\in C^r(\Omega)\), \(g(\bm s)=0\) for \(s_1\le2/3\), and \(g(\bm s)>0\) for \(s_1>2/3\). Let the observable ratio surfaces be \(r_1(\bm s)=0\), \(r_2(\bm s)=0\), and \(r_3(\bm s)=g(\bm s)\). Consider Model A with \(\beta_A(\bm s)=0\) and \(\bm\alpha_A(\bm s)=(0,0,g(\bm s))^\top\), and Model B with \(\beta_B(\bm s)=g(\bm s)\) and \(\bm\alpha_B(\bm s)=(-g(\bm s),-g(\bm s),0)^\top\). Then
		\[
		\bm\gamma\beta_A(\bm s)+\bm\alpha_A(\bm s)=(0,0,g(\bm s))^\top=\bm\gamma\beta_B(\bm s)+\bm\alpha_B(\bm s),
		\]
		so the two models are observationally equivalent, but \(\beta_A(\bm s)\equiv0\) whereas \(\beta_B(\bm s)=g(\bm s)\), and these differ on \(\{s_1>2/3\}\). Assumption (A1) holds: in Model A, \(Z_1\) and \(Z_2\) are valid everywhere, while in Model B, \(Z_3\) is valid everywhere. Assumption (A2) also holds, because on the nonempty open set \(\Omega_{\bigodot}:=\{\bm s\in[0,1]^2:s_1<2/3\}\), $r_1=r_2=r_3=0$. However, (A3) fails. For Model B, when \(s_1\le2/3\), all three IVs are valid; when \(s_1>2/3\), only \(Z_3\) is valid. Hence the pointwise plurality region for Model B is \(\Omega_0(\beta_B,\bm\alpha_B)=\{\bm s\in\Omega:s_1\le2/3\}\), so \({\rm cl}_\Omega\{\Omega_0(\beta_B,\bm\alpha_B)^c\}=\{\bm s\in\Omega:s_1\ge2/3\}\). Take any boundary point \(\bm t=(2/3,t_2)\). Then \(\bm t\in{\rm cl}_\Omega\{\Omega_0(\beta_B,\bm\alpha_B)^c\}\), and \(\alpha_{B,1}(\bm s)=\alpha_{B,2}(\bm s)=-g(\bm s)\) are \(C^r\) in a neighborhood of \(\bm t\). However, \(Z_1\) and \(Z_2\) are valid on the side \(s_1\le2/3\) and invalid on the side \(s_1>2/3\), so every neighborhood of \(\bm t\) contains both valid and invalid points for these IVs. Thus \(Z_1\) and \(Z_2\) are unstable at \(\bm t\), contradicting (A3). Therefore, if (A3) is dropped, two different causal effect functions yield the same observed distribution, and \(\beta(\cdot)\) is not identifiable.
		
		Combining Steps 1 and 2 proves identifiability under (A1)--(A3), and the three examples in Step 3 show that failure of any one of these assumptions can generally lead to non-identifiability.
	\end{proof}

	\subsection{Model Estimation}
	
	\subsubsection{Preliminaries}
	\label{SEC:Pre}
	
We use the main-paper notation for the triangulations, spline bases,
first-stage fit, active supports and oracle estimator. For the rate
statements, put $\triangle=\triangle_1$, write $|\triangle|$ for its
maximum simplex diameter, and rescale spatial coordinates
so that $\operatorname{Vol}(\Omega)=1$.

\paragraph{Supports and design matrices.}
For a support $\mathcal A\subseteq[p]\times[K_2]$, write
\[
 \mathcal A_\ell=\{k:(\ell,k)\in\mathcal A\},\qquad
 \mathcal A(k)=\{\ell:(\ell,k)\in\mathcal A\},\qquad
 s_n^\ast=|\mathcal A^\ast|.
\]
Let $\Theta_{\mathcal A}$ contain the pairs $(\beta,\bm\alpha)$ in
$\mathcal S_d^r(\triangle_1)\times\{\mathcal S_d^{-1}(\triangle_2)\}^p$
with $\alpha_\ell|_{T_k}=0$ whenever $(\ell,k)\notin\mathcal A$.
\chg{\emph{Anchored supports.} Let $\mathcal I=\{\ell:\gamma^\ast_\ell\neq0\}$ and, for the threshold $\tau_\gamma>0$ in the anchor constraint \eqref{eq:anchor} of the main paper, $\widehat{\mathcal I}_{\tau_\gamma}=\{\ell:|\widehat\gamma_\ell|>\tau_\gamma\}$. A support $\mathcal A$ is \emph{anchored} if $\mathcal I\setminus\mathcal A(k)\neq\emptyset$ for every $k\in[K_2]$, that is, if on every simplex of $\triangle_2$ at least one relevant instrument carries no invalid-IV block. On the event $\{\widehat{\mathcal I}_{\tau_\gamma}=\mathcal I\}$, whose probability tends to one under Assumption~\ref{AssumpC:signal}(iii) because $\widehat{\bm\gamma}-\bm\gamma^\ast=\OP(n^{-1/2})$, the anchor constraint restricts the Hi-FIVE criterion to coefficient vectors with anchored supports. Unless stated otherwise, the supports compared below are anchored.}
Throughout the proofs, we use reduced causal-effect coefficients:
\[
 \widetilde{\mathbf B}_1=\mathbf Q_2^\top\mathbf B_1,\qquad
 \beta=\widetilde{\mathbf B}_1^\top\bm\theta_\beta.
\]
Thus the constrained coefficient vector in the main-paper criterion is
$\mathbf Q_2\bm\theta_\beta$. On a restricted support, inactive
invalid-effect blocks are omitted from coefficient vectors and matrices;
their full-space versions fill those blocks with zeros.

Let $\mathbf B_{2,\mathcal A_\ell}$ concatenate the vectors
$\mathbf B_{2,k}$ for $k\in\mathcal A_\ell$, and set
\[
 \mathbf U_{ij,\mathcal A}
 =\bigl(\widehat D_i\widetilde{\mathbf B}_1^\top(\bm s_j),
 Z_{i1}\mathbf B_{2,\mathcal A_1}^\top(\bm s_j),\ldots,
 Z_{ip}\mathbf B_{2,\mathcal A_p}^\top(\bm s_j)\bigr)^\top.
\]
The matrix $\mathbf U_{i,\mathcal A}$ stacks
$\mathbf U_{ij,\mathcal A}^\top$ over locations, and $\mathbf U_{\mathcal A}$
stacks $\mathbf U_{i,\mathcal A}$ over subjects. We omit the
support subscript for $\mathcal A=\mathcal A^\ast$, and write
$\mathbf U^{\rm f}$ for the design containing all invalid-IV blocks.
The full sampled broken-spline matrix is
$\mathbb B_2=\{\mathbf B_2(\bm s_j)^\top\}_{j=1}^N$.
Put
\[
 \rho_n=\frac{\lambda^{\rm R}}{nN},\qquad
 \mathbf W_1=\mathbf Q_2^\top\mathbf P_1\mathbf Q_2,\qquad
 \mathbf W=\operatorname{diag}(\mathbf W_1,\mathbf0),
\]
where the zero block has the dimension of the retained invalid effects.
Define
\begin{equation}
\label{DEF:Gamma}
 \bm\Gamma_{\mathcal A,0}
 =\frac{\mathbf U_{\mathcal A}^\top\mathbf U_{\mathcal A}}{nN},\qquad
 \bm\Gamma_{\mathcal A,\lambda^{\rm R}}
 =\bm\Gamma_{\mathcal A,0}+\rho_n\mathbf W.
\end{equation}
We abbreviate $\bm\Gamma_{\mathcal A}=\bm\Gamma_{\mathcal A,\lambda^{\rm R}}$.
Let $D^0=\mathbf Z^\top\bm\gamma^\ast$ be the population linear
projection of $D$ onto the instruments, and write $D_i^0=\mathbf Z_i^\top\bm\gamma^\ast$.
A superscript $0$ on a design matrix replaces $\widehat D_i$ by $D_i^0$.
Define the corresponding penalized Gram matrix by
\[
 \bm\Gamma_{\mathcal A}^{0}
 :=\bm\Gamma_{\mathcal A,\lambda^{\rm R}}^{0}
 =\frac{(\mathbf U_{\mathcal A}^{0})^\top\mathbf U_{\mathcal A}^{0}}{nN}
  +\rho_n\mathbf W.
\]

\paragraph{Approximants, losses and rates.}
Let $\widetilde\beta=Q_{\triangle_1}\beta^\ast$, where
$Q_{\triangle_1}$ is the quasi-interpolant in
Assumption~\ref{A:true-coef}. On each active simplex,
$\widetilde\alpha_\ell$ is the $L^2$ projection of $\alpha_\ell^\ast$
onto $\mathcal P_d(T_k)$; it is zero on inactive simplices.
Their coefficients satisfy
$\widetilde\beta=\widetilde{\mathbf B}_1^\top\widetilde{\bm\theta}_\beta$
and $\widetilde\alpha_\ell=\mathbf B_2^\top\widetilde{\bm\theta}_{\alpha,\ell}$.
Write $\widetilde{\bm\theta}_\alpha$ for the concatenated invalid-effect
coefficients and $\widetilde{\bm\theta}$ for the entire coefficient vector.

Stack responses and errors in subject order:
\begin{equation}
\label{DEF:epsilon}
 \mathbf Y=(Y_{11},\ldots,Y_{1N},\ldots,Y_{nN})^\top,\qquad
 \bm\epsilon=(\epsilon_{11},\ldots,\epsilon_{1N},\ldots,\epsilon_{nN})^\top.
\end{equation}
For $b=(\ell,k)$, $\bm\theta_b$ denotes an invalid-IV coefficient block.
The smooth loss and its support-restricted minimum are
\begin{equation}
\label{eq:global-refit}
\begin{aligned}
 F(\bm\theta)&=\frac{\|\mathbf Y-\mathbf U^{\rm f}\bm\theta\|^2}{nN}
              +\rho_n\bm\theta_\beta^\top\mathbf W_1\bm\theta_\beta,\\
 F_{\mathcal A}&=\min_{\bm\theta:\,\bm\theta_b=\mathbf0\ (b\notin\mathcal A)}
                 F(\bm\theta).
\end{aligned}
\end{equation}
The minimum is attained for every support: appending roughness rows
gives a finite-dimensional least-squares problem, even when its
coefficient design is singular.
For the SCAD shape parameter $a$ from the main paper, put
$c_a=(a+1)/2$ and $k_a=(a-1)/(2a^2)$. The common rate abbreviations are
\begin{equation}
\label{eq:proof-rates}
\begin{aligned}
 r_{n,N}&=|\triangle|^{d+1}+n^{-1/2}
       +\rho_n|\triangle|^{-2}\{1+|\triangle|^{d-1}\},\\
 e_{n,N}&=n^{-1}+|\triangle|^{2d+2}+\rho_n.
\end{aligned}
\end{equation}

\paragraph{Covariance and effective sample size.}
Let $G_\epsilon(\bm s,\bm s')=\operatorname{Cov}\{\epsilon_i(\bm s),\epsilon_i(\bm s')\}$
and $\mathbf G_\epsilon=[G_\epsilon(\bm s_j,\bm s_{j'})]_{j,j'=1}^N$.
The error correlation is
\begin{equation}
\label{EQ:correlation}
 \rho_\epsilon(\bm s,\bm s')
 =\frac{G_\epsilon(\bm s,\bm s')}
        {\{G_\epsilon(\bm s,\bm s)G_\epsilon(\bm s',\bm s')\}^{1/2}}.
\end{equation}
With $m_k=\sum_{j=1}^N I(\bm s_j\in T_k)$, define
\[
 N_{\mathrm{eff},k}
 =\frac{m_k^2}{\sum_{\bm s_j,\bm s_{j'}\in T_k}
                       \rho_\epsilon(\bm s_j,\bm s_{j'})},\qquad
 \bar N_{\mathrm{eff}}=\frac1{K_2}\sum_{k=1}^{K_2}N_{\mathrm{eff},k},
\]
\[
 N^-_{\mathrm{eff}}=\min_kN_{\mathrm{eff},k},\qquad
 N^+_{\mathrm{eff}}=\max_kN_{\mathrm{eff},k},\qquad
 q_n=nK_2\bar N_{\mathrm{eff}}.
\]
For a fitted pair $(\widehat\beta,\widehat{\bm\alpha})$, write
\begin{equation}
\label{DEF:pilot}
 \varrho_n=\|\widehat\beta-\beta^\ast\|_\infty
       +\max_{1\le\ell\le p}\|\widehat\alpha_\ell-\alpha_\ell^\ast\|_\infty.
\end{equation}

	\paragraph{Assumptions.}

	\begin{assumpB}[Triangulations and observation locations]
		\label{A:tri}
		The simplicial partition $\triangle$ is $\pi$-quasi-uniform:
		$(\min_{T\in\triangle}\varrho_T)^{-1}|\triangle|\leq\pi$
		for a constant $\pi>0$, where $\varrho_T$ is the inradius of $T$.
		Moreover, $\triangle_2$ is obtained from $\triangle_1$ by $L$ levels of uniform refinement with $L$ fixed and independent of $n$ 
		and
		$|\triangle_1|/|\triangle_2| = 2^{L}$. Consequently,
		$|\triangle_2|\asymp|\triangle|$ and $K_2\asymp|\triangle|^{-\kappa}$.
		{The observation locations are quasi-uniform with respect to $\triangle_2$: there are constants $0<c_m\le C_m<\infty$ such that
		\[
		c_mN|\triangle_2|^{\kappa}\le m_k\le C_mN|\triangle_2|^{\kappa}
		\qquad\text{for all }k,
		\]
		and
		\[
		\frac1N\sum_{\bm s_j\in T_k}f(\bm s_j)
		=\int_{T_k}f(\bm s)\,d\bm s\,\{1+o(1)\}
		\]
		uniformly over $k$ and over polynomials $f\ge0$ of degree at most $2d$ on $T_k$.}
	\end{assumpB}

	\begin{assumpB}[Causal-effect smoothness]
		\label{A:true-coef}
		The true causal effect function
		$\beta^\ast(\cdot)\in\mathcal W^{d+1,\infty}(\Omega)$ for an integer $d\ge1$. {The degree $d$ and smoothness $r$ of $\mathcal S^r_d(\triangle_1)$ are such that the space has full approximation power \citep[Chapters 10 and 18]{lai2007spline}: there is a quasi-interpolation operator $Q_{\triangle_1}$ onto $\mathcal S^r_d(\triangle_1)$ with $\|D^{m}(f-Q_{\triangle_1}f)\|_{\infty}\le C|\triangle|^{d+1-m}\|f\|_{d+1,\infty}$ for $0\le m\le d$ and all $f\in\mathcal W^{d+1,\infty}(\Omega)$, with $C$ depending only on $d$, $\kappa$ and $\pi$ (for $\kappa=2$ this holds whenever $d\ge3r+2$).}
	\end{assumpB}

	\begin{assumpB}[Invalid-effect structure and signal size]
		\label{A:true-alpha}
		At every location $\bm s\in\Omega$, at least one relevant instrument is
		locally valid; that is,
		$
		\{\ell:\gamma_\ell^\ast\neq0,\ \alpha_\ell^\ast(\bm s)=0\}
		\neq\emptyset.
		$ In addition, we assume the simplexwise version of this condition: the true support $\mathcal A^\ast$ is anchored, so that for every $k$ some relevant instrument satisfies $\alpha^\ast_\ell\equiv0$ on $T_k$. This local relevance is uniform: for some constant $c_\gamma>0$,
		$
		\inf_{\bm s\in\Omega}
		\sum_{\ell:\alpha_\ell^\ast(\bm s)=0}(\gamma_\ell^\ast)^2
		\ge c_\gamma.
		$
		Moreover, the number of locally invalid instruments is uniformly bounded:
		for some fixed integer $s_0$,
		$
		\max_{1\le k\le K_2}|\mathcal A^\ast(k)|\le s_0.
		$
		Consequently,
		$
		|\mathcal A^\ast|\le s_0K_2=O(K_2)
		=O(|\triangle|^{-\kappa}).
		$
		
		Each invalid-IV coefficient is piecewise smooth on its active simplices,
		namely
		$
		\alpha_\ell^\ast|_{T_k}\in \mathcal W^{d+1,\infty}(T_k), (\ell,k)\in\mathcal A^\ast.
		$ Their elementwise smoothness norms are uniformly bounded: $\max_{(\ell,k)\in\mathcal A^\ast}\|\alpha_\ell^\ast\|_{\mathcal W^{d+1,\infty}(T_k)}\le C_\alpha<\infty$.
		Moreover, there exists a constant $c_\alpha>0$ such that
		\[
		\min_{(\ell,k)\in\mathcal A^\ast}
		\operatorname{Vol}(T_k)^{-1/2}
		\|\alpha_\ell^\ast\|_{T_k}
		\ge c_\alpha n^{-c},\qquad
		0<c<{\tfrac12-\frac{\kappa}{4(d+1)}}.
		\]
	\end{assumpB}

	\begin{assumpB}[Error distribution]
		\label{A:error-subgaussian}
		The variables $\zeta_{it}$ are independent, have zero mean and unit variance, and are sub-Gaussian with parameter $\sigma_\zeta^2$. The variables $e_{ij}$ are also independent, have zero mean and unit variance, and are sub-Gaussian with parameter $\sigma_e^2$. The measurement error is $\varepsilon_i(\bm s_j)=\sigma_\varepsilon(\bm s_j)e_{ij}$.
		That is, there exist finite constants $\sigma_\zeta,\sigma_e>0$ such
		that
		\[
		\mathbb E\{\exp(\lambda\zeta_{it})\}
		\le
		\exp\left(\frac{\lambda^2\sigma_\zeta^2}{2}\right),
		\qquad
		\mathbb E\{\exp(\lambda e_{ij})\}
		\le
		\exp\left(\frac{\lambda^2\sigma_e^2}{2}\right)
		\]
		for all $\lambda\in\mathbb R$.
		The subject-level arrays $\{(\mathbf Z_i,\xi_i,\{\zeta_{it}\}_{t\ge1},\{e_{ij}\}_{j\le N})\}_{i=1}^n$ are independent and identically distributed across $i${, and $\mathbf Z_i$, $\{\zeta_{it}\}_{t\ge1}$ and $\{e_{ij}\}_{j\le N}$ are mutually independent; in particular $\E\{\mathbf Z_i\epsilon_i(\bm s)\}=\mathbf 0$ for every $\bm s$, and the conditional covariance of $\epsilon_i(\cdot)$ given $\mathbf Z_i$ equals $G_\epsilon$}.
	\end{assumpB}

	\begin{assumpB}[Spatial covariance and effective sample size]
		\label{A:covariance}
		The function $\sigma_\varepsilon$ belongs to $\mathcal C^1(\Omega)$ and satisfies
		$0<c_\sigma\le\sigma_\varepsilon(\bm s)\le C_\sigma<\infty$ for all $\bm s\in\Omega$.
		For every $t$, $\psi_t\in\mathcal C^1(\Omega)$. The covariance kernel satisfies $G_\eta\in L^2(\Omega\times\Omega)$ and
		$0<c_G\le G_\eta(\bm s,\bm s)\le C_G<\infty$ for all $\bm s\in\Omega$.
		The Karhunen--Lo\`eve expansion is uniformly summable:
		$
		\left\|\sum_{t=1}^{\infty}\rho_t^{1/2}|\psi_t|\right\|_\infty<\infty.
		$
		Recall that $\epsilon (\bm s) = \eta(\bm s) + \varepsilon(\bm s)$. The within-subject error covariance has the smooth-plus-nugget form
		$
		G_\epsilon(\bm s,\bm s')
		=G_\eta(\bm s,\bm s')
		+\sigma_\varepsilon^2(\bm s)\mathbf 1\{\bm s=\bm s'\}.
		$
		The covariance kernel $G_\epsilon$ is symmetric and positive semidefinite,
		and its variance function satisfies $0<c_\epsilon
		\leq G_\epsilon(\bm s,\bm s)
		\leq C_\epsilon<\infty$ for all $\bm s\in\Omega$.
		
		Moreover, the local correlation satisfies
		\[
		\max_{1\le k\le K_2}
		\sum_{j,j'=1}^{N}
		I(\bm s_j\in T_k,\bm s_{j'}\in T_k)\,\rho_\epsilon(\bm s_j,\bm s_{j'})
		\;\le\;
		C\,\frac{N^{2}|\triangle|^{2\kappa}}{N^{-}_{\mathrm{eff}}} .
		\]
		The effective within-simplex sample sizes are comparable across simplices: there exists a constant $c_0\in(0,1]$, not depending on $n$ or $N$, such that $N^{-}_{\mathrm{eff}}\;\ge\;c_0\,\bar N_{\mathrm{eff}}$. For the common pilot in \eqref{DEF:pilot}, as $n,N\to\infty$,
		\[
		\log N=o(n),
		\qquad
		N^{+}_{\mathrm{eff}}\Bigl(\sqrt{\log N/n}+\varrho_n\Bigr)\to0 .
		\]
		Finally, $q_n\to\infty$ and $\log(pK_2)/q_n\to0$.
	\end{assumpB}

	\begin{assumpB}[Covariates]
		\label{A:covariates}
		Let $\mathbf X=(D,Z_1,\ldots,Z_p)^\top$. The eigenvalues of
		$\bm\Sigma_{\mathbf X}=\mathbb E(\mathbf X\mathbf X^\top)$ are bounded away
		from zero and infinity. The number of candidate instruments $p$ is fixed
		and $\mathbb E(\mathbf Z\xi)=\mathbf0$. For a constant $C_Z<\infty$,
		\[
		\max_{1\le i\le n}\max_{1\le\ell\le p}|Z_{i\ell}|\le C_Z
		\quad\text{almost surely}.
		\]
	\end{assumpB}

	\begin{assumpB}[Sample sizes and roughness tuning]
		\label{A:size}
		As $n,N\to\infty$, for some $\iota\in(0,1)$,
		\[
		n^{1/2}|\triangle|^{d+1}\to0,\qquad
		N^{1/\kappa}|\triangle|\to\infty,\qquad
		N^{-1}n^{(\kappa+1)/(2d+2)+\iota}\to0.
		\]
		The roughness penalty parameter $\lambda^{\rm R}$ satisfies
		\[
		(\lambda^{\rm R})^{-1/2}N^{-1}|\triangle|^{-(\kappa+1)}\to0,
		\qquad \lambda^{\rm R}(nN)^{-1}|\triangle|^{-4}\to0.
		\]
	\end{assumpB}


	\begin{assumpB}[Joint signal separation and selection tuning]
		\label{AssumpC:signal}
		\label{A:global}
		\emph{(i) Joint signal separation.} There exist deterministic sequences $\upsilon_n\to0$ and
		$\Delta_n^{\rm G}>0$, with $\sqrt n\upsilon_n\to\infty$, such that,
		for every anchored $\mathcal A\neq\mathcal A^\ast$ with $|\mathcal A|\le s_n^\ast$,
		\begin{align}
		\label{eq:global-separation}
		\inf_{\|\bm\gamma-\bm\gamma^\ast\|\le\upsilon_n}
		\inf_{(\beta,\bm\alpha)\in\Theta_{\mathcal A}}
		\bigg[&\int_\Omega\E\!\left[\Big\{
		\mathbf Z^\top\big[\bm\gamma\{\widetilde\beta(\bm s)-\beta(\bm s)\}
		+\widetilde{\bm\alpha}(\bm s)-\bm\alpha(\bm s)\big]\Big\}^2\right]\,d\bm s
		+\rho_n\mathcal E(\beta)\bigg]\notag\\
		&\hspace{15mm}\ge (1+s_n^\ast-|\mathcal A|)\Delta_n^{\rm G}.
		\end{align}
		\emph{(ii) {Selection tuning.}} The tuning parameters satisfy
		\begin{equation}
		\label{eq:global-tuning}
		e_{n,N}=o\{(\lambda^{\rm S})^2\},
		\qquad (\lambda^{\rm S})^2=o(\Delta_n^{\rm G}).
		\end{equation}
		If $s_n^\ast=0$, the family in \eqref{eq:global-separation} is empty;
		only the first condition in \eqref{eq:global-tuning} is required.
		\emph{(iii) Anchor threshold.} The threshold in \eqref{eq:anchor} satisfies $0<\tau_\gamma<\min_{\ell\in\mathcal I}|\gamma^\ast_\ell|$.
	\end{assumpB}

	\begin{assumpB}[Pointwise normality]
		\label{A:normality}
		{For the fixed location $\bm s\in\Omega$, as $n,N\to\infty$,
		\begin{itemize}
			\item[(U1)] $n^{1/2}|\triangle|^{d+1-\kappa/2}\to0$;
			\item[(U2)] $n^{1/2}\lambda^{\rm R}(nN)^{-1}|\triangle|^{-2-\kappa/2}\to0$;
			\item[(U3)] $n|\triangle|^{\kappa}\to\infty$;
			\item[(V)] the asymptotic variance $\sigma_\beta^2(\bm s)$, defined in \eqref{eq:var_orac_est} below, satisfies $\Pr\{n\sigma_\beta^2(\bm s)\ge c_V\}\to1$ for some constant $c_V>0$.
		\end{itemize}}
	\end{assumpB}

\paragraph{Remarks on the assumptions.}

\begin{remark}[Spline approximation and invalid instruments]
\label{REM:spline-assumptions}
Assumptions~\ref{A:tri}, \ref{A:true-coef} and \ref{A:size} are
standard in spline estimation over triangulations; see
\citet{lai2007spline} and \citet{Yu:etal:21}. Quasi-uniformity excludes
degenerate simplices, while smoothness and the rate conditions control
the approximation and roughness-penalty biases.

The fixed refinement level gives
$\mathcal S_d^r(\triangle_1)\subseteq\mathcal S_d^{-1}(\triangle_2)$
and comparable mesh sizes. The elementwise $L^2$ projection error of
each $\alpha_\ell^\ast$ is orthogonal to the causal spline space in the
population inner product. Lemma~\ref{LEM:U-mu-diff} bounds its empirical
contribution; population orthogonality does not make that contribution
zero. Nesting also places all fitted spatial vectors in a common broken
spline space, as used in Lemma~\ref{LEM:global-support}.

Assumption~\ref{A:true-alpha} allows validity to vary spatially but
requires a relevant valid instrument at each location. Its lower bound
on nonzero effects is a signal-size condition for support recovery
\citep{fan2001variable,wainwright2009sharp}.
\end{remark}

\begin{remark}[Error covariance and effective sample size]
\label{REM:covariance-assumptions}
Assumptions~\ref{A:error-subgaussian} and \ref{A:covariates} provide
independent subjects, exponential concentration and a well-conditioned
population design \citep{vershynin2018high,wainwright2009sharp}.
Within-subject spatial dependence is described by
Assumption~\ref{A:covariance}. Since $\rho_\epsilon\le1$,
$N_{\mathrm{eff},k}\ge1$; if correlations are nonnegative, then also
$N_{\mathrm{eff},k}\le m_k$. Under nonnegative correlations, the effective sample size therefore ranges
from one observation under perfect dependence to $m_k$ under independence.
Quasi-uniformity gives comparable pixel counts across simplices, while
$N^-_{\mathrm{eff}}\ge c_0\bar N_{\mathrm{eff}}$ additionally requires
comparable effective sample sizes. This condition permits the two
effective sample sizes to be interchanged up to a constant.

The pilot condition
$N^+_{\mathrm{eff}}\{\sqrt{\log N/n}+\varrho_n\}\to0$
is used in Lemma~\ref{LEM:Neff-consistency} to estimate $q_n$.
It is not used by Theorems~\ref{THM:beta-convergence}
and~\ref{THM:beta-normality} or Proposition~\ref{S-Prop:Selection}.
\end{remark}

	\begin{remark}[Global signal separation and selection]
		\label{REM:global-needed}

        Assumption~\ref{AssumpC:signal} compares the joint causal and
        invalid-effect fit with every incorrect anchored support containing no more
        blocks than the true support. Both $\beta$ and the retained
        $\alpha$ blocks may change. The neighborhood of $\bm\gamma^\ast$
        accommodates the estimated first stage. The first bound in
        \eqref{eq:global-tuning} makes the cost of an additional block
        exceed the total stochastic gain in fit; the second prevents
        the savings from deleting blocks from exceeding their loss of fit.
        The first bound is sufficient and is not asserted to be sharp.

        According to the Hi-FIVE criterion, the comparison family contains only anchored supports. This matters when pointwise plurality fails: the invariance \eqref{eq:invariance} produces representations that reproduce the reduced form exactly and, when valid instruments are a minority on a region of fixed size, use fewer blocks than $\mathcal A^\ast$ once the mesh is fine; such representations must flag every instrument on a transition band and are therefore not anchored. Anchored supports keep an unflagged instrument on every simplex, so a transition band incurs a loss of fit. Section~\ref{SEC:anchor-separation} verifies part~(i) in this regime with $\Delta_n^{\rm G}\asymp b_n^2|\triangle|^{\kappa+1}$, where $b_n$ is the size of the invalid-IV effect.


        Lemma~\ref{LEM:mbic-refit-ordering} uses this joint gap to compare
        incorrect supports of no larger size. Proposition~\ref{Prop:mBIC}
        excludes larger supports from the global SCAD path through
        \eqref{eq:mbic-grid-lower}. Proposition~\ref{Prop:mBIC-size}
        instead compares all anchored supports under its additional conditions.
	\end{remark}

\begin{remark}[Conditions for pointwise normality]
\label{REM:normality-assumptions}
Assumption~\ref{A:normality} is used by the oracle and Hi-FIVE
pointwise limit laws and by Corollary~\ref{COR:variance-consistency}.
It is not needed for the convergence rates or support recovery.
Conditions (U1)--(U2) make the spline approximation and roughness biases
$o(n^{-1/2})$ at the location under study. The available pointwise
bounds lose a factor $|\triangle|^{-\kappa/2}$ relative to the $L^2$
bounds, so Assumption~\ref{A:size} alone does not imply (U1)--(U2).
Condition (U3) makes the number of basis functions smaller in order than
the number of subjects and gives the required Lindeberg bound.

Condition (V) separately requires a nondegenerate pointwise variance.
The heuristic $n\sigma_\beta^2(\bm s)\approx G_\eta(\bm s,\bm s)$
comes from spline reproduction of smooth functions, but is not proved
here. In particular, the rate examples below do not verify (V).
Sharper bounds on the inverse Gram matrix in the locally supported basis
could reduce the undersmoothing requirements: the intended relaxations
replace (U1) by the approximation rate in Assumption~\ref{A:size},
replace (U2) by $n^{1/2}\lambda^{\rm R}(nN)^{-1}\to0$, and remove
(U3). Such equivalent-kernel bounds are not established here and are
not used in the stated results.
\end{remark}

\begin{remark}[Compatibility of the rate conditions]
\label{REM:rate-compatibility}
For $|\triangle|\asymp n^{-a}$, the rate restrictions in
Assumptions~\ref{A:true-alpha} and \ref{A:size}, combined with
Assumption~\ref{AssumpC:signal} when
$\Delta_n^{\rm G}\asymp|\triangle|^\kappa n^{-2c}$ as in
Example~\ref{EX:global-compatible} (so that $n^{-1}$ and
$|\triangle|^{2d+2}$, both bounded by $e_{n,N}$, are
$o(\Delta_n^{\rm G})$), require
\[
 \frac{1}{2(d+1)}<a<\frac{1-2c}{\kappa},
 \qquad a(d+1-\kappa/2)>c.
\]
For suitable $N$ and $\lambda^{\rm R}$, these inequalities can hold
if and only if $\kappa<2(d+1)$ and
$c<1/2-\kappa/\{4(d+1)\}$, explaining the upper bound in
Assumption~\ref{A:true-alpha}.
Condition (U1) of Assumption~\ref{A:normality} further requires
\[
 \frac{1}{2(d+1)-\kappa}<a<\frac{1-2c}{\kappa}.
\]
This interval is nonempty if and only if
$c<1/2-\kappa/\{4(d+1)-2\kappa\}$.
These comparisons concern the rate conditions. Joint signal separation
and the variance lower bound remain separate requirements.
\end{remark}

\paragraph{Compatibility examples.}

	\begin{example}[Compatibility of the global conditions]
		\label{EX:global-compatible}
		Take $p=3$, $\bm\gamma^\ast=(1,1,1)^\top$,
		$\bm\Sigma_{\mathbf Z}=\mathbf I_3$,
		$\alpha_1^\ast=\alpha_2^\ast=0$ and $\alpha_3^\ast=b_n=n^{-c}$
		on all of $\Omega$, with a fixed smooth $\beta^\ast$.
		Then $s_n^\ast=K_2$. For an incorrect support $\mathcal A$ with $|\mathcal A|\le K_2$, call a
		simplex deficient if its retained instrument set is
		$\emptyset$, $\{1\}$ or $\{2\}$. Even if $\beta$ is allowed to vary
		freely on each simplex, the squared reduced-form discrepancy on each
		deficient simplex is at least $C |\triangle|^\kappa b_n^2$, uniformly for
		$\bm\gamma$ near $(1,1,1)^\top$. For example, with retained set $\{1\}$,
		the two unadjusted components have discrepancies $\gamma_2 u$ and
		$b_n+\gamma_3u$; minimizing their squared sum over $u\in\mathbb R$
		still leaves $b_n^2\gamma_2^2/(\gamma_2^2+\gamma_3^2)$.
		If $B$ is the number of deficient simplices, counting retained blocks
		gives $B\ge K_2-|\mathcal A|$. Also $B\ge1$, since a nondeficient
		model with at most $K_2$ blocks must retain exactly $\{3\}$ everywhere.
		Thus $B\ge(1+K_2-|\mathcal A|)/2$, which proves
		\eqref{eq:global-separation} with
		$\Delta_n^{\rm G}\asymp |\triangle|^\kappa n^{-2c}$.
		For instance, $\kappa=3$, $d=5$, $|\triangle|\asymp n^{-1/7}$, $N\asymp n$,
		$\lambda^{\rm R}=1$, $c=1/20$ and $\lambda^{\rm S}=n^{-1/3}$ give
		$e_{n,N}\asymp n^{-1}=o(n^{-2/3})$ and
		$n^{-2/3}=o(\Delta_n^{\rm G})$, where $\Delta_n^{\rm G}\asymp n^{-37/70}$.
		These rates also satisfy Assumption~\ref{A:size}
		and conditions (U1)--(U3) of Assumption~\ref{A:normality}.
		\chg{The argument bounds the loss of fit of every competing support and therefore applies to the family of anchored supports. In this example plurality holds everywhere; Proposition~\ref{S-Prop:anchor-separation} treats the case in which it fails.}
	\end{example}

\begin{example}[Rate conditions for pointwise normality]
\label{EX:normality-rates}
Take $\kappa=3$, $d=5$, $|\triangle|\asymp n^{-1/7}$,
$N\asymp n$, $\lambda^{\rm R}\asymp1$,
$\lambda^{\rm S}\asymp n^{-1/10}$ and $0<c<1/10$.
These choices satisfy Assumption~\ref{A:size}
and the rate restriction in Assumption~\ref{A:true-alpha}, with any
fixed $0<\iota<2/3$ in Assumption~\ref{A:size}.
The quantities in (U1)--(U3) of Assumption~\ref{A:normality} have orders
\[
 n^{1/2}|\triangle|^{d+1-\kappa/2}\asymp n^{-1/7},\qquad
 n^{1/2}\lambda^{\rm R}(nN)^{-1}|\triangle|^{-2-\kappa/2}
 \asymp n^{-1},\qquad
n|\triangle|^\kappa\asymp n^{4/7}\to\infty.
\]
Also $n|\triangle|^{2\kappa}\asymp n^{1/7}\to\infty$, verifying
(C1) of Corollary~\ref{COR:variance-consistency}.
This example verifies rate compatibility. The structural conditions in
Assumption~\ref{A:true-alpha}, the joint separation and tuning in
Assumption~\ref{AssumpC:signal}, the variance condition (V), and the
covariance-estimation condition (C2) require separate verification.
\end{example}

\subsubsection{Joint Separation under Anchoring when Plurality Fails}
\label{SEC:anchor-separation}

\begin{figure}[htbp]
	\centering
	\includegraphics[width=0.8\linewidth]{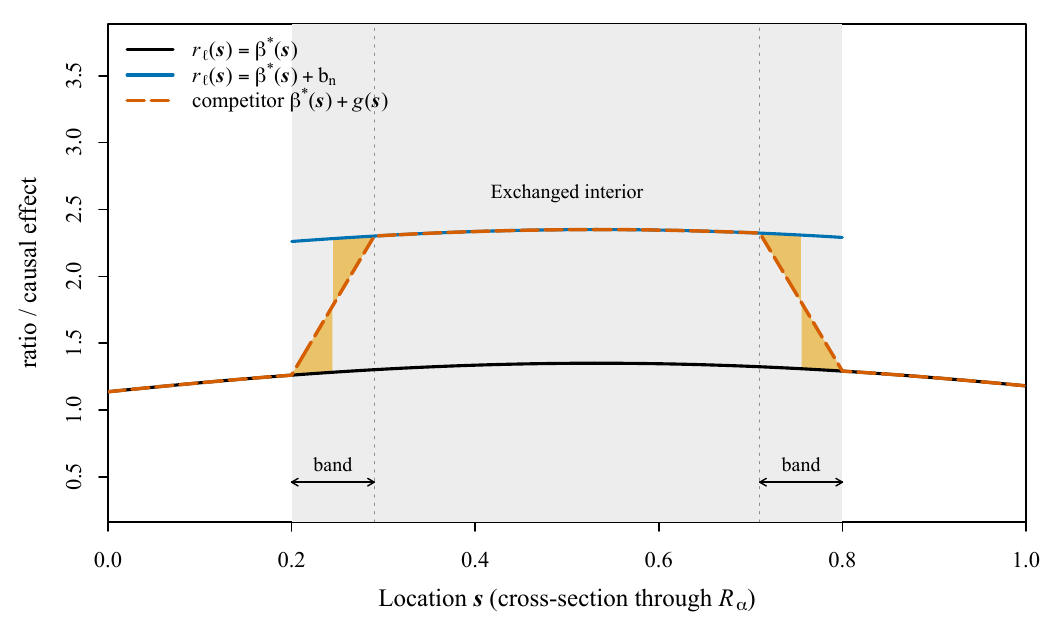}
	\caption{\chg{The anchor constraint rules out the observationally equivalent competitor: a one-dimensional cross-section through $\mathcal R_\alpha$ in the setting of \eqref{eq:anchor-setting} ($p=3$, $q=2$, constant shift $b_n$ on a fixed region). The black curve is the common ratio $r_\ell(\bm s)=\beta^\ast(\bm s)$ of the valid instruments, the blue curve is the ratio $r_\ell(\bm s)=\beta^\ast(\bm s)+b_n$ shared by the invalid group $G$ on $\mathcal R_\alpha$, and the dashed orange curve is the competitor $\beta^\ast(\bm s)+g(\bm s)$. The gray area is the exchanged interior, where the competitor saves $2q-p$ blocks per simplex. The anchor constraint keeps at least one instrument unflagged on every simplex, so the fitted causal effect must track that instrument's ratio: $\beta^\ast$ on standard simplices and $\beta^\ast+b_n$ on exchanged ones. Being a continuous spline, it must move between the two ratios across a transition band of width $\asymp|\triangle|$ (arrows and dotted lines), and the shaded gap on the band is a loss of fit of order $b_n^2$ times the volume of the band (Lemma~\ref{LEM:no-jump}). Under the tuning window $(\lambda^{\rm S})^2=o(b_n^2|\triangle|^{\kappa+1})$, this loss exceeds the penalty saved (Proposition~\ref{S-Prop:anchor-separation}).}}
	\label{fig:anchor-mechanism}
\end{figure}

\begin{chgblock}
This section supports Remark~\ref{rem:anchoring} of the main paper. 
It verifies Assumption~\ref{AssumpC:signal}(i) for the Hi-FIVE criterion under the anchor constraint \eqref{eq:anchor} in a mesh-aligned idealization of the simulation designs, with an invalid region of fixed size, which is the regime in which valid instruments may be a minority on a region of positive volume. The critical competitors are generated by the invariance \eqref{eq:invariance}: for $g\in\mathcal S^r_d(\triangle_1)$, the representation $(\beta^\ast+g,\ \alpha^\ast_1-\gamma^\ast_1g,\ldots,\alpha^\ast_p-\gamma^\ast_pg)$ has the same combined effects $\gamma^\ast_\ell\beta+\alpha_\ell$ as $(\beta^\ast,\bm\alpha^\ast)$ and reproduces the population reduced form exactly, so it can be ruled out only through the block count and the anchor constraint. Throughout, blocks are counted at the population first stage $\bm\gamma^\ast$. Figure~\ref{fig:anchor-mechanism} illustrates the mechanism in a one-dimensional cross-section. On every simplex, anchoring leaves some instrument unflagged and thereby ties the fitted causal effect to that instrument's ratio $r_\ell=\beta^\ast+\alpha^\ast_\ell/\gamma^\ast_\ell$; the two lemmas below quantify the loss of fit this forces on a competitor that exchanges the valid and invalid groups on part of $\mathcal R_\alpha$, and Proposition~\ref{S-Prop:anchor-separation} compares that loss with the number of blocks saved.

\paragraph{Setting.}
Let $\Omega$ be a bounded polyhedral Lipschitz domain with connected interior and $\operatorname{Vol}(\Omega)=1$. Let $\triangle^{(0)}$ be a fixed triangulation of $\Omega$, and let each $\triangle_1$ be obtained from $\triangle^{(0)}$ by uniform refinement (for $\kappa=3$, by a stable red-refinement rule such as that of \citet{bey1995tetrahedral}, which produces finitely many similarity classes and children of equal volume), so that $\triangle_1$ is $\pi$-quasi-uniform for a fixed $\pi$, and $\triangle_2$ is obtained from $\triangle_1$ as in Assumption~\ref{A:tri}. Let $\mathcal R_\alpha$ be a nonempty union of simplices of $\triangle^{(0)}$ with $\operatorname{Vol}(\Omega\setminus\mathcal R_\alpha)>0$; it is then a union of simplices of $\triangle_1$ and of $\triangle_2$ at every level, and it does not shrink as the mesh is refined. Let $G\subseteq[p]$ with $|G|=q$ and $1\le q\le p-1$, let $\gamma^\ast_\ell\neq0$ for every $\ell$, so that $\mathcal I=[p]$, and let $b_n>0$. Suppose that
\begin{equation}
\label{eq:anchor-setting}
\alpha^\ast_\ell=\gamma^\ast_\ell b_n\mathbf 1_{\mathcal R_\alpha}\quad(\ell\in G),
\qquad
\alpha^\ast_\ell\equiv0\quad(\ell\notin G),
\end{equation}
with $\beta^\ast$ as in Assumption~\ref{A:true-coef}. The configuration covers plurality ($2q<p$), ties ($2q=p$) and strict majorities of invalid instruments ($2q>p$) on $\mathcal R_\alpha$. Because $b_n>0$, the invalid-IV effects jump by $\gamma^\ast_\ell b_n$ at the boundary of $\mathcal R_\alpha$, a quantitative form of (A3). Write $\gamma_{\min}=\min_\ell|\gamma^\ast_\ell|$, $\lambda_{\mathbf Z}=\lambda_{\min}(\bm\Sigma_{\mathbf Z})>0$, and $M=2^{\kappa L}$ for the number of children of a simplex of $\triangle_1$ in $\triangle_2$.
\end{chgblock}

\begin{lemma}[Continuous splines cannot switch between ratios]
\label{LEM:no-jump}
\begin{chgblock}
Under the quasi-uniformity and refinement in Assumption~\ref{A:tri}, there is a constant $c_J>0$, depending only on $\kappa$, $d$, $L$ and $\pi$, such that the following hold for every $b>0$ and every $T\in\triangle_1$.
\begin{enumerate}
\item[(a)] If $T'$ and $T''$ are distinct children of $T$, then
$\int_{T'}(P-\vartheta b)^2+\int_{T''}P^2\ge c_Jb^2\operatorname{Vol}(T)$
for every $P\in\mathcal P_d(T)$ and every $\vartheta\in[1/3,3]$.
\item[(b)] If $\widetilde T\in\triangle_1$ shares a facet $F$ with $T$, then
$\sum_{T'}\int_{T'}(P-\vartheta_{T'}b)^2+\int_{\widetilde T}\widetilde P^2\ge c_Jb^2\operatorname{Vol}(T)$
for all $P\in\mathcal P_d(T)$ and $\widetilde P\in\mathcal P_d(\widetilde T)$ with $P=\widetilde P$ on $F$ and all $\vartheta_{T'}\in[1/3,3]$, where the sum is over the children $T'$ of $T$.
\end{enumerate}
\end{chgblock}
\end{lemma}

\begin{proof}
\begin{chgblock}
Let $\widehat T$ be a reference simplex and $A_T:\widehat T\to T$ an affine bijection. Uniform refinement is defined through edge midpoints and, for $\kappa=3$, a choice among finitely many interior diagonals, so $A_T^{-1}$ maps the children of $T$ onto one of finitely many fixed refinement patterns of $\widehat T$. By quasi-uniformity, $c_\pi|\triangle|^\kappa\le\operatorname{Vol}(T)\le|\triangle|^\kappa$, and every facet $F$ of $T$ satisfies $c_\pi|\triangle|^{\kappa-1}\le|F|\le|\triangle|^{\kappa-1}$, with $c_\pi>0$ depending only on $\kappa$ and $\pi$. Integrals transform as $\int_Tf=\{\operatorname{Vol}(T)/\operatorname{Vol}(\widehat T)\}\int_{\widehat T}f\circ A_T$, and similarly on facets with the ratio of facet areas.

(a) Fix a refinement pattern and two distinct children $\widehat T',\widehat T''$ of $\widehat T$, and put $\mu_a=\inf\{\int_{\widehat T'}(P-\vartheta)^2+\int_{\widehat T''}P^2:P\in\mathcal P_d,\ \vartheta\in[1/3,3]\}$. Along a minimizing sequence, $\int_{\widehat T''}P^2$ stays bounded; since $\widehat T''$ has nonempty interior and $\mathcal P_d$ is finite-dimensional, $P$ stays in a compact set. The infimum is therefore attained. If it were zero, $P$ would equal the nonzero constant $\vartheta$ on $\widehat T'$ and vanish on $\widehat T''$, which is impossible for a polynomial. Hence $\mu_a>0$. Applying this bound to $P\circ A_T/b$ and changing variables gives (a) with constant $\mu_a/\operatorname{Vol}(\widehat T)$. There are finitely many patterns and pairs of children, so the smallest constant is positive.

(b) Fix a pattern and a facet $\widehat F$ of $\widehat T$, and put $\mu_b=\inf\{\sum_{T'}\int_{\widehat T'}(P-\vartheta_{T'})^2+\int_{\widehat F}P^2:P\in\mathcal P_d,\ \vartheta_{T'}\in[1/3,3]\}$. The same compactness argument shows that the infimum is attained. If it were zero, $P$ would be constant on each child and hence a constant $\vartheta\ge1/3$ on $\widehat T$, while vanishing on $\widehat F$; this is impossible, so $\mu_b>0$. Changing variables, and using $\operatorname{Vol}(T)/|F|\le|\triangle|/c_\pi$, gives
\[
\sum_{T'}\int_{T'}(P-\vartheta_{T'}b)^2+C_1|\triangle|\int_FP^2\ge c_2b^2\operatorname{Vol}(T),
\]
with $C_1,c_2>0$ depending only on $\kappa$, $d$, $L$ and $\pi$. Next, norm equivalence on $\mathcal P_d(\widehat T)$ gives $\int_{\widehat F_i}\widehat P^2\le C_d\int_{\widehat T}\widehat P^2$ for each facet $\widehat F_i$ of $\widehat T$. Mapping $\widetilde T$ onto $\widehat T$ and using quasi-uniformity yields the discrete trace inequality $\int_F\widetilde P^2\le C_3|\triangle|^{-1}\int_{\widetilde T}\widetilde P^2$. Since $P=\widetilde P$ on $F$,
\[
\sum_{T'}\int_{T'}(P-\vartheta_{T'}b)^2+C_1C_3\int_{\widetilde T}\widetilde P^2\ge c_2b^2\operatorname{Vol}(T),
\]
which gives (b) with $c_J=c_2/\max\{1,C_1C_3\}$ after taking the minimum with the constant in (a).
\end{chgblock}
\end{proof}

\begin{lemma}[Counting interfaces]
\label{LEM:interface-count}
\begin{chgblock}
In the setting of \eqref{eq:anchor-setting}, there is a constant $c_I>0$, independent of the refinement level, such that every nonempty set $\mathfrak E$ of simplices of $\triangle_1$ contained in $\mathcal R_\alpha$ satisfies
\[
|\partial\mathfrak E|\ \ge\ \max\{1,\ c_I|\triangle|\,|\mathfrak E|\},
\]
where $\partial\mathfrak E$ is the set of facets shared by a member of $\mathfrak E$ and a simplex of $\triangle_1$ not in $\mathfrak E$.
\end{chgblock}
\end{lemma}

\begin{proof}
\begin{chgblock}
Let $U=\bigcup_{T\in\mathfrak E}T$. Since $U$ is polyhedral, its perimeter relative to $\Omega$ is the total area of the facets in $\partial\mathfrak E$; facets on $\partial\Omega$ do not contribute. Each facet has area at most $|\triangle|^{\kappa-1}$, so $\operatorname{Per}(U;\Omega)\le|\triangle|^{\kappa-1}|\partial\mathfrak E|$. Quasi-uniformity gives $\operatorname{Vol}(U)\ge c_\pi|\triangle|^\kappa|\mathfrak E|$. Because $\Omega$ is a bounded Lipschitz domain with connected interior, the relative isoperimetric inequality \citep[Chapter~3]{ambrosio2000functions} gives a constant $C_\Omega$ with
\[
\min\{\operatorname{Vol}(U),\operatorname{Vol}(\Omega\setminus U)\}^{(\kappa-1)/\kappa}\le C_\Omega\operatorname{Per}(U;\Omega).
\]
Put $w_0=\operatorname{Vol}(\Omega\setminus\mathcal R_\alpha)\in(0,1]$. Since $U\subseteq\mathcal R_\alpha$, $\operatorname{Vol}(\Omega\setminus U)\ge w_0$, and $\operatorname{Vol}(U)\le1$. If $\operatorname{Vol}(U)\le\operatorname{Vol}(\Omega\setminus U)$, the left-hand side is at least $\operatorname{Vol}(U)\ge w_0\operatorname{Vol}(U)$; otherwise it is at least $\operatorname{Vol}(\Omega\setminus U)\ge w_0\ge w_0\operatorname{Vol}(U)$. Hence
\[
|\partial\mathfrak E|\ge\frac{w_0\operatorname{Vol}(U)}{C_\Omega|\triangle|^{\kappa-1}}\ge\frac{w_0c_\pi}{C_\Omega}|\triangle|\,|\mathfrak E|.
\]
Finally, $\operatorname{Vol}(U)>0$ and $\operatorname{Vol}(\Omega\setminus U)>0$, so $\operatorname{Per}(U;\Omega)>0$ and $|\partial\mathfrak E|\ge1$.
\end{chgblock}
\end{proof}

\begin{proposition}[Joint separation under anchoring]
\label{S-Prop:anchor-separation}
\begin{chgblock}
In the setting of \eqref{eq:anchor-setting}, let $r\ge0$ and $\upsilon_n\le\gamma_{\min}/2$. There is a constant $\kappa_0>0$, depending only on $p$, $\lambda_{\mathbf Z}$, $\gamma_{\min}$, $\kappa$, $d$, $L$, $\pi$, $\Omega$ and $\mathcal R_\alpha$, such that, whenever $|\triangle|\le1$, inequality \eqref{eq:global-separation} holds for every anchored support $\mathcal A\neq\mathcal A^\ast$ with $|\mathcal A|\le s_n^\ast$, with
\[
\Delta_n^{\rm G}=\kappa_0\,b_n^2\,|\triangle|^{\kappa+1}.
\]
In particular, Assumption~\ref{AssumpC:signal}(i) holds when valid instruments are a strict minority ($2q>p$) on a region of fixed positive volume.
\end{chgblock}
\end{proposition}

\begin{proof}
\begin{chgblock}
Fix $\bm\gamma$ with $\|\bm\gamma-\bm\gamma^\ast\|\le\upsilon_n$, an anchored support $\mathcal A\neq\mathcal A^\ast$ with $|\mathcal A|\le s_n^\ast$, and $(\beta,\bm\alpha)\in\Theta_{\mathcal A}$. Then $|\gamma_\ell|\ge\gamma_{\min}/2$ and $\vartheta_\ell=\gamma^\ast_\ell/\gamma_\ell\in[2/3,2]$ for every $\ell$. On each active simplex, $\alpha^\ast_\ell$ is constant, so $\widetilde{\bm\alpha}=\bm\alpha^\ast$. Put $g=\beta-\widetilde\beta\in\mathcal S^r_d(\triangle_1)$, $u_\ell=\gamma_\ell(\widetilde\beta-\beta)+\widetilde\alpha_\ell-\alpha_\ell$, and let $\mathfrak L$ denote the bracket in \eqref{eq:global-separation}.

\emph{Step 1: anchor lower bound.} Since $\rho_n\mathcal E(\beta)\ge0$ and $\bm\Sigma_{\mathbf Z}\succeq\lambda_{\mathbf Z}\mathbf I$,
\[
\mathfrak L\ge\lambda_{\mathbf Z}\sum_{\ell=1}^p\int_\Omega u_\ell^2
\ge\lambda_{\mathbf Z}\sum_{T\in\triangle_2}\ \sum_{\ell\notin\mathcal A(T)}\int_Tu_\ell^2 .
\]
If $\ell\notin\mathcal A(T)$, then $\alpha_\ell=0$ on $T$, and
$u_\ell=-\gamma_\ell\{g-\vartheta_\ell b_n\mathbf 1(\ell\in G,\,T\subseteq\mathcal R_\alpha)\}$ on $T$. Put $c_u=\lambda_{\mathbf Z}\gamma_{\min}^2/4$. Each unflagged instrument therefore contributes at least $c_u\int_T(g-t)^2$, where $t=\vartheta_\ell b_n$ if $\ell\in G$ and $T\subseteq\mathcal R_\alpha$, and $t=0$ otherwise. Anchoring enters here and in Step~2 through $\mathcal U(T)\neq\emptyset$ below: it guarantees an unflagged instrument, and hence such a term, on every simplex.

\emph{Step 2: classification and block count.} Every simplex of $\triangle_2$ lies in $\mathcal R_\alpha$ or in the closure of its complement. For $T\in\triangle_2$, let $\mathcal U(T)=[p]\setminus\mathcal A(T)$, which is nonempty because $\mathcal A$ is anchored. Call $T$ \emph{mixed} if $T\subseteq\mathcal R_\alpha$ and $\mathcal U(T)$ meets both $G$ and $G^c$; \emph{exchanged} if $T\subseteq\mathcal R_\alpha$ and $\mathcal U(T)\subseteq G$; and \emph{standard} otherwise. Let $X$ and $E$ be the sets of mixed and exchanged simplices. Choose one instrument in $\mathcal U(T)$ for each $T$, from $G$ if $T\in E$. Step~1 then gives
\[
\mathfrak L\ge c_u\sum_{T\in E}\int_T(g-\vartheta_Tb_n)^2+c_u\sum_{T\notin E\cup X}\int_Tg^2+c_u\sum_{T\in X}\int_T\{(g-\vartheta_Tb_n)^2+g^2\},
\]
with $\vartheta_T\in[2/3,2]$. Since $(x-a)^2+x^2\ge a^2/2$, each mixed simplex contributes at least $(2c_u/9)b_n^2\operatorname{Vol}(T)$.

For the block count, $|\mathcal A^\ast(T)|=q$ if $T\subseteq\mathcal R_\alpha$ and $0$ otherwise. If $T$ is standard and $T\subseteq\mathcal R_\alpha$, then $\mathcal U(T)\subseteq G^c$, so $|\mathcal A(T)|\ge q$; if $T$ is exchanged, then $G^c\subseteq\mathcal A(T)$, so $|\mathcal A(T)|\ge p-q$. Hence
\[
d_{\mathcal A}=s_n^\ast-|\mathcal A|=\sum_T\{|\mathcal A^\ast(T)|-|\mathcal A(T)|\}\le(2q-p)_+|E|+q|X|.
\]
Moreover, $E\cup X\neq\emptyset$. Otherwise, every term in this sum is nonpositive while the sum is nonnegative, so $|\mathcal A(T)|=|\mathcal A^\ast(T)|$ for every $T$. On $\mathcal R_\alpha$ this means $|\mathcal U(T)|=p-q$ with $\mathcal U(T)\subseteq G^c$, so $\mathcal A(T)=G$; outside $\mathcal R_\alpha$ it means $\mathcal A(T)=\emptyset$. Thus $\mathcal A=\mathcal A^\ast$, a contradiction.

\emph{Step 3: cost of the interfaces.} Let $\mathfrak E$ be the set of simplices of $\triangle_1$ with at least one exchanged child; they lie in $\mathcal R_\alpha$, and $|E|\le M|\mathfrak E|$. On each $T\in\triangle_1$, $g$ is a polynomial $P_T\in\mathcal P_d(T)$, and since $r\ge0$, $P_T=P_{\widetilde T}$ on a facet shared by $T$ and $\widetilde T$. Consider a facet in $\partial\mathfrak E$ shared by $T\in\mathfrak E$ and $\widetilde T\notin\mathfrak E$. Exactly one of the following holds.
\begin{enumerate}
\item[(i)] $T$ has a standard child. Then $T$ has an exchanged and a standard child, and Lemma~\ref{LEM:no-jump}(a) bounds the contribution of $T$ below by $c_uc_Jb_n^2\operatorname{Vol}(T)$.
\item[(ii)] $T$ has no standard child but a mixed child, or $T$ has only exchanged children and $\widetilde T$ has a mixed child. A mixed child has volume at least $c_\pi|\triangle|^\kappa/M$ and therefore contributes at least $(2c_u/9)c_\pi b_n^2|\triangle|^\kappa/M$.
\item[(iii)] $T$ has only exchanged children and $\widetilde T$ has only standard children. Then Lemma~\ref{LEM:no-jump}(b) bounds the contribution of $T\cup\widetilde T$ below by $c_uc_Jb_n^2\operatorname{Vol}(T)$.
\end{enumerate}
In every case the contribution of $T\cup\widetilde T$ is at least $c_\ast b_n^2|\triangle|^\kappa$, with $c_\ast=c_uc_\pi\min\{c_J,2/(9M)\}$. A simplex has $\kappa+1$ facets, so summing over $\partial\mathfrak E$ counts each simplex at most $2(\kappa+1)$ times. Each mixed simplex also contributes at least $c_\ast b_n^2|\triangle|^\kappa$ on its own. Averaging the two resulting bounds gives
\[
\mathfrak L\ \ge\ \frac{c_\ast}{4(\kappa+1)}\,b_n^2|\triangle|^\kappa\bigl(|\partial\mathfrak E|+|X|\bigr),
\]
where $|\partial\mathfrak E|=0$ if $\mathfrak E=\emptyset$.

\emph{Step 4: comparison with the saving.} If $\mathfrak E\neq\emptyset$, Lemma~\ref{LEM:interface-count} gives $|\partial\mathfrak E|\ge\max\{1,c_I|\triangle||\mathfrak E|\}$; if $\mathfrak E=\emptyset$, then $X\neq\emptyset$ by Step~2. In both cases $|\partial\mathfrak E|+|X|\ge\frac13(1+c_I|\triangle||\mathfrak E|+|X|)$. On the other hand, with $|\triangle|\le1$,
\[
(1+d_{\mathcal A})\Delta_n^{\rm G}\le\kappa_0b_n^2|\triangle|^{\kappa}\bigl(|\triangle|+pM|\triangle||\mathfrak E|+p|\triangle||X|\bigr)
\le\kappa_0C_4\,b_n^2|\triangle|^\kappa\bigl(1+c_I|\triangle||\mathfrak E|+|X|\bigr),
\]
with $C_4=\max\{1,pM/c_I,p\}$. Choosing $\kappa_0=c_\ast/\{12(\kappa+1)C_4\}$ gives $\mathfrak L\ge(1+d_{\mathcal A})\Delta_n^{\rm G}$. The constants do not depend on $\bm\gamma$, $\mathcal A$, $(\beta,\bm\alpha)$ or the refinement level, which proves the claim.
\end{chgblock}
\end{proof}

\begin{remark}[Interpretation of the gap]
\label{REM:anchor-gap}
\begin{chgblock}
Proposition~\ref{S-Prop:anchor-separation} covers plurality, ties and strict majorities of invalid instruments on $\mathcal R_\alpha$ in a single statement. Its gap is smaller than the gap $|\triangle|^\kappa n^{-2c}$ of Example~\ref{EX:global-compatible} by a factor $|\triangle|$. This factor is the price of comparing a perimeter with a volume. A competitor that exchanges the groups on a region of fixed size saves of order $|\triangle|^{-\kappa}$ blocks but pays a loss of fit only on a transition band of order $|\triangle|^{-(\kappa-1)}$ simplices. When $2q>p$, this order is attained. Take $g=Q_{\triangle_1}\varphi$ with $\varphi=b_n$ on a fixed ball $B\subset\mathcal R_\alpha$ and $\varphi=0$ outside its $C|\triangle|$-neighborhood, and leave one valid instrument unflagged on each band simplex. The band contains of order $|\triangle|^{-(\kappa-1)}$ simplices, so its loss of fit is of order $b_n^2|\triangle|$. The roughness term satisfies $\rho_n\mathcal E(\widetilde\beta+g)=O(\rho_n+\rho_nb_n^2|\triangle|^{-3})=o(b_n^2|\triangle|)$, because $\rho_n|\triangle|^{-4}\to0$ by Assumption~\ref{A:size} and $\rho_n\le e_{n,N}=o(b_n^2|\triangle|^{\kappa+1})$ by the tuning window. Each band simplex costs at most $p$ additional blocks, so the support still saves of order $|\triangle|^{-\kappa}$ blocks. Hence Assumption~\ref{AssumpC:signal}(i) can hold only with $\Delta_n^{\rm G}=O(b_n^2|\triangle|^{\kappa+1})$, and the exponent $\kappa+1$ cannot be improved in this configuration. Two features of the setting are essential. Anchoring is used in Steps~1 and~2: it guarantees an unflagged instrument on every band simplex, and this is what converts the transition band into a loss of fit rather than a change in the block count alone. The jump of size $b_n$ at $\partial\mathcal R_\alpha$ is used in Lemma~\ref{LEM:no-jump}. If the invalid-IV effects switched on smoothly, a smooth $\beta$ could follow the invalid ratio without loss, which is the non-identifiability described by Theorem~\ref{theorem:identifiability} when (A3) fails. For general invalid-IV configurations, Assumption~\ref{AssumpC:signal}(i) remains an assumption. The proposition verifies it in a mesh-aligned idealization of the configuration used by the simulation designs, without restricting the size of $\mathcal R_\alpha$ relative to the mesh.
\end{chgblock}
\end{remark}

\begin{example}[Rates for the anchored criterion]
\label{EX:anchor-rates}
\begin{chgblock}
With $\Delta_n^{\rm G}=\kappa_0b_n^2|\triangle|^{\kappa+1}$, the tuning condition \eqref{eq:global-tuning} becomes
\[
e_{n,N}=o\{(\lambda^{\rm S})^2\},\qquad(\lambda^{\rm S})^2=o\bigl(b_n^2|\triangle|^{\kappa+1}\bigr).
\]
Take $b_n=n^{-c}$ and $|\triangle|\asymp n^{-a}$. Since $e_{n,N}=n^{-1}+|\triangle|^{2d+2}+\rho_n$, this window is nonempty when $2c+a(\kappa+1)<1$, $a(2d+1-\kappa)>2c$, and $\rho_n=o(n^{-2c}|\triangle|^{\kappa+1})$. Combined with Remark~\ref{REM:rate-compatibility}, the admissible range of $a$ becomes $\max[1/\{2(d+1)\},\,2c/(2d+1-\kappa)]<a<(1-2c)/(\kappa+1)$, and condition (U1) of Assumption~\ref{A:normality} further requires $a>1/\{2(d+1)-\kappa\}$. These ranges are nonempty if and only if $c<(2d+1-\kappa)/\{4(d+1)\}$ and, with (U1), $c<(2d+1-2\kappa)/\{2(2d+2-\kappa)\}$; both bounds are stricter than the one in Assumption~\ref{A:true-alpha}, and for $\kappa=3$, $d=5$ they equal $1/3$ and $5/18$. For instance, with the choices of Example~\ref{EX:global-compatible}, namely $\kappa=3$, $d=5$, $|\triangle|\asymp n^{-1/7}$, $N\asymp n$, $\lambda^{\rm R}=1$ and $c=1/20$, the gap is $\Delta_n^{\rm G}\asymp n^{-47/70}$. The value $\lambda^{\rm S}=n^{-1/3}$ used there gives $(\lambda^{\rm S})^2=n^{-2/3}$, which is not $o(n^{-47/70})$. The choice $\lambda^{\rm S}=n^{-2/5}$ satisfies both conditions, since $e_{n,N}\asymp n^{-1}=o(n^{-4/5})$ and $n^{-4/5}=o(n^{-47/70})$. The remaining conditions of Assumptions~\ref{A:size} and~\ref{A:normality} hold as in Example~\ref{EX:global-compatible}.
\end{chgblock}
\end{example}

\paragraph{Auxiliary lemmas.}
For the next two lemmas, define
\[
 \mathbb G_{\mathcal A}
 =\left\{g(\mathbf Z,t,\bm s)=t\beta(\bm s)
       +\sum_{\ell=1}^pZ_\ell\alpha_\ell(\bm s):
       (\beta,\bm\alpha)\in\Theta_{\mathcal A}\right\}.
\]
For $g^{(1)},g^{(2)}\in\mathbb G_{\mathcal A}$, the empirical inner product is
\begin{equation}
\label{DEF:DDC-empirical_product}
 \langle g^{(1)},g^{(2)}\rangle_{n,N}
 =\frac1{nN}\sum_{i=1}^n\sum_{j=1}^N
 g^{(1)}(\mathbf Z_i,\widehat D_i,\bm s_j)
 g^{(2)}(\mathbf Z_i,\widehat D_i,\bm s_j).
\end{equation}
	The population inner product is
	\begin{equation}
		\langle g^{(1)}, g^{(2)}\rangle = \E \int g^{(1)}(\mathbf{Z}, D^0, \bm{s})g^{(2)}(\mathbf{Z}, D^0, \bm{s}) d \bm{s}.
		\label{DEF:DDC-theoretical_product}
	\end{equation}
	The associated norms are $\|g\|_{n,N}^2=\langle g,g\rangle_{n,N}$ and $\|g\|^2=\langle g,g\rangle$.
	
	\begin{lemma}
		\label{LEM:norm}
		Let $\mathcal A\subseteq[p]\times[K_2]$ be a block support {satisfying the following condition for a constant $c^{\prime}_\gamma>0$:}
		\[
		{\text{(S)}\qquad \min_{1\le k\le K_2}\ \sum_{\ell\notin\mathcal A(k)}(\gamma^\ast_\ell)^2\ \ge\ c^{\prime}_\gamma .}
		\]
		{Under Assumption~\ref{A:covariates}, for} $g(\bm s) \in \mathbb{G}_{\mathcal{A}}$, $\|g\|^2 \asymp \|\beta\|_{L^2(\Omega)}^2 + \sum_{\ell = 1}^p \|\alpha_{\ell}\|_{L^2(\Omega)}^2${, with constants depending only on $c^{\prime}_\gamma$, $\bm\gamma^\ast$ and $\bm\Sigma_{\mathbf X}$. Under Assumptions~\ref{A:true-alpha}, \ref{A:covariates} and \ref{A:size}, the true support $\mathcal A=\mathcal A^\ast$ satisfies (S) for all sufficiently large $n$, with $c^{\prime}_\gamma=\min_{\ell:\gamma^\ast_\ell\neq0}(\gamma^\ast_\ell)^2$}.
	\end{lemma}
	
	\begin{proof}
		Write $\mathcal R(\mathcal A_\ell)=\bigcup_{k\in\mathcal A_\ell}T_k$, $\mathbf X^0=(D^0,Z_1,\ldots,Z_p)^\top$, $\bm\Sigma_{\mathbf X^0}=\E(\mathbf X^0\mathbf X^{0\top})$, and $\bm g(\bm s)=(\beta(\bm s),\alpha_1(\bm s),\ldots,\alpha_p(\bm s))^\top$. By \eqref{DEF:DDC-theoretical_product},
		\begin{align*}
			\|g\|^2 
			&= \int_{\Omega} \bm{g}^{\top}(\bm{s}) \, \bm\Sigma_{\mathbf X^0} \, \bm{g}(\bm{s}) \, d\bm{s} = \int_{\Omega} \widetilde{\bm{g}}^{\top}(\bm{s}) \, 
			\widetilde{\bs{\Sigma}}_{X}(\bm{s}) \, 
			\widetilde{\bm{g}}(\bm{s}) \, d\bm{s}.
		\end{align*}
		Here, the local covariance structure is given by
		\[
		\widetilde{\bs{\Sigma}}_{X}(\bm{s}) 
		= 
		\begin{pmatrix}
			{\E\{(D^0)^2\}} & {\E(D^0\mathbf{Z}_{\bm s}^{\top})} \\
			{\E(\mathbf{Z}_{\bm s}D^0)} & {\E(\mathbf{Z}_{\bm s}\mathbf{Z}_{\bm s}^{\top})}
		\end{pmatrix},
		\]
		where $\mathbf{Z}_{\bm s} = (Z_\ell : \bm{s} \in \mathcal R(\mathcal{A}_{\ell}))^{\top}$ 
		contains the instruments whose active regions include $\bm s$. 
		The corresponding subvector of $\bm{g}(\bm{s})$ is
		$\widetilde{\bm{g}}(\bm{s}) 
		= \big(\beta(\bm{s}), \, \alpha_{\ell}(\bm{s}) : \bm{s} \in \mathcal R(\mathcal{A}_{\ell}) \big)^{\top}$. Write $S(\bm s)=\{\ell:\bm s\in\mathcal R(\mathcal A_\ell)\}$, so that $\mathbf Z_{\bm s}=\mathbf Z_{S(\bm s)}$ and $S(\bm s)=\mathcal A(k)$ for $\bm s$ in the interior of $T_k$. Since $\bm\Sigma_{\mathbf Z}$ is a principal submatrix of $\bm\Sigma_{\mathbf X}$, it is positive definite by Assumption~\ref{A:covariates}, so the components of $\mathbf Z$ are linearly independent in $L^2$, and $\widetilde{\bs{\Sigma}}_{X}(\bm{s})=\E\{(D^0,\mathbf Z_{S(\bm s)}^{\top})^{\top}(D^0,\mathbf Z_{S(\bm s)}^{\top})\}$ is singular if and only if $D^0=\mathbf Z^{\top}\bm\gamma^\ast$ lies in the span of $\{Z_\ell:\ell\in S(\bm s)\}$, that is, if and only if $\gamma^\ast_\ell=0$ for every $\ell\notin S(\bm s)$. Condition (S) excludes this for every $k$, so $\widetilde{\bs{\Sigma}}_{X}(\bm{s})$ is positive definite for every $\bm s$ in the interior of a simplex of $\triangle_2$, hence for almost every $\bm s\in\Omega$, which is all the integral requires. Moreover, $S(\bm s)$ ranges over at most $2^p$ subsets of $\{1,\ldots,p\}$ with $p$ fixed, so $\operatorname{\,inf}_{\bm s}\lambda_{\min}\{\widetilde{\bs{\Sigma}}_{X}(\bm{s})\}>0$ and $\sup_{\bm s}\lambda_{\max}\{\widetilde{\bs{\Sigma}}_{X}(\bm{s})\}\le\lambda_{\max}(\bm\Sigma_{\mathbf X^0})<\infty$, with bounds depending only on $c^{\prime}_\gamma$, $\bm\gamma^\ast$ and $\bm\Sigma_{\mathbf X}$.
		
		For any given location $\bm s$, we have
		\[
		\widetilde{\bm{g}}^{\top}(\bm{s}) \, \widetilde{\bs{\Sigma}}_{X}(\bm{s}) \, \widetilde{\bm{g}}(\bm{s}) 
		\asymp \beta^2(\bm{s}) + \sum_{\ell : \bm{s} \in \mathcal R(\mathcal{A}_{\ell})} \alpha_{\ell}^2(\bm{s}).
		\]
		Therefore,
\[
\begin{aligned}
 \int_\Omega\widetilde{\bm g}^\top(\bm s)
 \widetilde{\bs\Sigma}_X(\bm s)\widetilde{\bm g}(\bm s)\,d\bm s
 &\asymp\int_\Omega\left\{\beta^2(\bm s)
       +\sum_{\ell:\bm s\in\mathcal R(\mathcal A_\ell)}
             \alpha_\ell^2(\bm s)\right\}\,d\bm s=\|\beta\|_{L^2(\Omega)}^2
       +\sum_{\ell=1}^p\|\alpha_\ell\|_{L^2(\Omega)}^2.
\end{aligned}
\]

		{It remains to verify that $\mathcal A^\ast$ satisfies (S) for all large $n$ under Assumptions~\ref{A:true-alpha}, \ref{A:covariates} and \ref{A:size}. Fix $k$ and suppose that every $\ell$ with $\gamma^\ast_\ell\neq0$ belongs to $\mathcal A^\ast(k)$. Since $\mathcal W^{d+1,\infty}(T_k)\subset C^{d}(T_k)$ on the convex set $T_k$, the sets $\mathcal N_\ell=\{\bm s\in T_k:\alpha^\ast_\ell(\bm s)=0\}$, over $\ell$ with $\gamma^\ast_\ell\neq0$, are closed and, by the pointwise validity clause of Assumption~\ref{A:true-alpha}, cover $T_k$. Since these are finitely many closed sets, one of them contains a nonempty open subset $U$ of $T_k$. For that $\ell$ we have $(\ell,k)\in\mathcal A^\ast$, so $\alpha^\ast_\ell|_{T_k}\in\mathcal W^{d+1,\infty}(T_k)$ with norm at most $C_\alpha$; it vanishes on $U$ together with all its derivatives of order at most $d$, and Taylor's theorem along segments of the convex set $T_k$ starting from a point in the interior of $U$ gives $|\alpha^\ast_\ell(\bm s)|\le C_{d,\kappa}C_\alpha|\triangle_2|^{d+1}$ for all $\bm s\in T_k$, with $C_{d,\kappa}$ depending only on $d$ and $\kappa$. Hence $\operatorname{Vol}(T_k)^{-1/2}\|\alpha^\ast_\ell\|_{T_k}\le C_{d,\kappa}C_\alpha|\triangle_2|^{d+1}=o(n^{-1/2})$ by Assumption~\ref{A:size}, which contradicts the lower bound $c_\alpha n^{-c}$ with $c<1/2$ in Assumption~\ref{A:true-alpha} once $n$ is large. Therefore, for all large $n$ and every $k$, some $\ell\notin\mathcal A^\ast(k)$ has $\gamma^\ast_\ell\neq0$, and (S) holds with $c^{\prime}_\gamma=\min_{\ell:\gamma^\ast_\ell\neq0}(\gamma^\ast_\ell)^2>0$.}
		This proves the lemma.
	\end{proof}
	
	\begin{lemma}
        Suppose that Assumptions~\ref{A:tri}, \ref{A:error-subgaussian}
        and \ref{A:covariates} hold, and let $\mathcal A$ satisfy
        condition (S) of Lemma~\ref{LEM:norm}. In particular,
        $\mathcal A=\mathcal A^\ast$ satisfies (S) under
        Assumptions~\ref{A:true-alpha} and \ref{A:size}.
        If $N^{1/2}|\triangle|^{\kappa/2}\to\infty$ as $N\to\infty$,
        there exist constants $0<c_\Gamma<C_\Gamma<\infty$ such that
        the following bounds hold with probability approaching one as
        $n\to\infty$:
        \begin{itemize}
        \item[(a)]
        \[
        c_\Gamma|\triangle|^\kappa
        \le\lambda_{\min}(\bm\Gamma_{\mathcal A,0})
        \le\lambda_{\max}(\bm\Gamma_{\mathcal A,0})
        \le C_\Gamma|\triangle|^\kappa.
        \]
        \item[(b)]
        \[
        \begin{aligned}
        c_\Gamma|\triangle|^\kappa
        &\le\lambda_{\min}(\bm\Gamma_{\mathcal A,\lambda^{\rm R}})
        \le\lambda_{\max}(\bm\Gamma_{\mathcal A,\lambda^{\rm R}})\le C_\Gamma\left(|\triangle|^\kappa
             +\frac{\lambda^{\rm R}}{nN|\triangle|^{4-\kappa}}\right).
        \end{aligned}
        \]
        \end{itemize}
        The probability is taken over $(\mathbf Z_i,\xi_i)_{i\le n}$,
        with the observation locations fixed. The Gram matrices are
        defined in \eqref{DEF:Gamma}.
		\label{LEM:Gamma_0}
	\end{lemma}
	
	\begin{proof}
		For any vector $\bm \theta 
		= \big(\bm \theta_\beta^\top, \, \bm \theta_{\alpha,1}^\top, \, \ldots, \, \bm \theta_{\alpha,p}^\top \big)^\top$, we have
		\[
		(nN)^{-1}\bm \theta^{\top}\mathbf{U}_{\mathcal{A}}^\top\mathbf{U}_{\mathcal{A}} \bm \theta 
		= \|g_{\bm \theta}\|_{n,N}^2,
		\]
        where $
        g_{\bm\theta}(\bm s)
        =\widehat D\beta(\bm s)
+\sum_{\ell=1}^pZ_\ell\alpha_\ell(\bm s),
        \beta(\bm s)=\widetilde{\mathbf B}_1^\top(\bm s)\bm\theta_\beta,
        \alpha_\ell(\bm s)
        =\mathbf B_{2,\mathcal A_\ell}^\top(\bm s)\bm\theta_{\alpha,\ell}.$
        
		Let $\delta_N$ be the uniform relative quadrature error in
		Assumption~\ref{A:tri}, so $\delta_N=o(1)$. Since $p$ is fixed,
		independent subjects and bounded instruments give
		$\|n^{-1}\mathbf Z^\top\mathbf Z-\bm\Sigma_{\mathbf Z}\|_{\rm op}
		=\OP(n^{-1/2})$ under Assumptions~\ref{A:error-subgaussian}
		and~\ref{A:covariates}. Applied to the nonnegative elementwise
		polynomial $\E\{g(\mathbf Z,D^0,\bm s)^2\}$, the quadrature
		bound therefore gives
		\[
		 \big|\|g\|_{n,N,0}^2-\|g\|^2\big|
		 \le\{\OP(n^{-1/2})+O(\delta_N)\}\|g\|^2
		\]
		uniformly over $\mathbb G_{\mathcal A}$, where the subscript $0$
		uses $D_i^0$ in place of $\widehat D_i$.
		Also $\widehat{\bm\gamma}-\bm\gamma^\ast=\OP(n^{-1/2})$ by
		$\E(\mathbf Z\xi)=\mathbf0$, bounded instruments and
		$\E\xi^2<\infty$. Thus
		$\max_i|\widehat D_i-D_i^0|=\OP(n^{-1/2})$.
		Lemma~\ref{LEM:norm} and the same quadrature bound imply that
		replacing $D_i^0$ by $\widehat D_i$ changes the squared norm by
		at most $\OP(n^{-1/2})\|g\|^2$, uniformly over this space.
		Polarization now gives
		\begin{equation}
		\label{EQU:Rn}
        \begin{aligned}
        R_{n,N}
        &:=\sup_{\substack{g^{(1)},g^{(2)}\in\mathbb G_{\mathcal A}\\
                          \|g^{(1)}\|\,\|g^{(2)}\|>0}}
        \frac{|\langle g^{(1)},g^{(2)}\rangle_{n,N}
                 -\langle g^{(1)},g^{(2)}\rangle|}
             {\|g^{(1)}\|\,\|g^{(2)}\|}\\
        &=\OP(n^{-1/2})+O(\delta_N)=o_{\rm P}(1).
        \end{aligned}
		\end{equation}

		Moreover, by the stability of spline functions,
		for $\beta=\widetilde{\mathbf B}_1^{\top}\bm\theta_\beta\in\mathcal S^r_d(\triangle_1)$ and for $\alpha_\ell$ with $\alpha_\ell|_{T_k}=\mathbf B_{2,k}^{\top}\bm\theta_{\alpha,\ell,k}\in\mathcal P_d(T_k)$, we have
		\begin{equation}\label{eq:spline-stability}
		\begin{aligned}
		c_B|\triangle|^{\kappa}\|\bm\theta_\beta\|^2
		&\le\|\beta\|^2_{L^2(\Omega)}
		\le C_B|\triangle|^{\kappa}\|\bm\theta_\beta\|^2,\\
		c_B|\triangle|^{\kappa}\|\bm\theta_{\alpha,\ell,k}\|^2
		&\le\|\alpha_\ell\|^2_{T_k}
		\le C_B|\triangle|^{\kappa}\|\bm\theta_{\alpha,\ell,k}\|^2,
		\end{aligned}
		\end{equation}
		for constants $0<c_B\le C_B<\infty$ depending only on $d$, $\kappa$ and $\pi$ \citep{lai2007spline}. The second pair of inequalities follows from the stability of the Bernstein--B\'ezier basis on a single simplex of a $\pi$-quasi-uniform partition, and the first pair follows from the same result on $\triangle_1$ because $\mathbf Q_2$ has orthonormal columns, so that the Gram matrix $\mathbf Q_2^{\top}\{\int_\Omega\mathbf B_1\mathbf B_1^{\top}\}\mathbf Q_2$ has its spectrum inside that of $\int_\Omega\mathbf B_1\mathbf B_1^{\top}$. The bound $\|g_{\bm\theta}\|^2\asymp\|\beta\|^2_{L^2(\Omega)}+\sum_\ell\|\alpha_\ell\|^2_{L^2(\Omega)}$ is Lemma~\ref{LEM:norm}, and $\sum_\ell\|\alpha_\ell\|^2_{L^2(\Omega)}=\sum_{(\ell,k)\in\mathcal A}\|\alpha_\ell\|^2_{T_k}$. Therefore, for some constants $c,C>0$,
		\[
		c \,(1-R_{n,N})\,|\triangle|^{\kappa}\,\|\bm \theta\|^{2}   
		\;\leq\; (1-R_{n,N}) \,\|g_{\bm \theta}\|^{2}
		\;\leq\; \|g_{\bm \theta}\|_{n,N}^{2},
		\]
		and
		\[
		\|g_{\bm \theta}\|_{n,N}^{2} 
		\;\leq\; (1+R_{n,N})\,\|g_{\bm \theta}\|^{2}
		\;\leq\; C \,(1+R_{n,N})\,|\triangle|^{\kappa}\,\|\bm \theta\|^{2}.
		\]
		By \eqref{EQU:Rn}, $R_{n,N}=o_{\rm P}(1)$, and part (a) follows. Since $\mathbf W\succeq\mathbf 0$, $\lambda_{\min}(\bm{\Gamma}_{\mathcal{A}, \lambda^{\rm R}})\ge\lambda_{\min}(\bm{\Gamma}_{\mathcal{A}, 0})\ge c_\Gamma|\triangle|^{\kappa}$, which is the lower bound in part (b). It remains to bound $\lambda_{\max}(\mathbf W_1)$.
		
		Using Markov's inequality for piecewise polynomials on a $\pi$-quasi-uniform partition and \eqref{eq:spline-stability}, we have
		\[
		\bm \theta_\beta^{\top} \mathbf{W}_1 \bm \theta_\beta=\mathcal E(\widetilde{\mathbf B}_1^{\top}\bm\theta_\beta) \leq \frac{C}{|\triangle|^{4}}\bigl\Vert \widetilde{\mathbf B}_1^{\top}\bm\theta_\beta\bigr\Vert_{L^2}^{2}\leq \frac{C}{|\triangle|^{4-\kappa}}\Vert\bm{\theta}_\beta\Vert^{2}.
		\]
        Hence
        \[
        \lambda_{\max}(\bm\Gamma_{\mathcal A,\lambda^{\rm R}})
        \le C\left\{(1+R_{n,N})|\triangle|^\kappa
                +\frac{\lambda^{\rm R}}{nN|\triangle|^{4-\kappa}}\right\}.
        \]
        Since $R_{n,N}=o_{\rm P}(1)$, this gives the upper bound in
        part (b) with probability approaching one.
	\end{proof}

\subsubsection{Theoretical Properties of the Oracle Estimator}

The oracle estimator from the main paper minimizes the smooth loss on
$\mathcal A^\ast$. In the reduced coordinates of Section~\ref{SEC:Pre},
\begin{equation}\label{EQU:objective-oracle}
\begin{gathered}
 \widehat{\bm\theta}^{o}\in\operatorname*{arg\,min}_{\bm\theta}
 \left\{\frac{\|\mathbf Y-\mathbf U\bm\theta\|^2}{nN}
       +\rho_n\bm\theta_\beta^\top\mathbf W_1\bm\theta_\beta\right\},\\
 \widehat\beta^o=\widetilde{\mathbf B}_1^\top\widehat{\bm\theta}_\beta^o.
\end{gathered}
\end{equation}
On the event of Lemma~\ref{LEM:Gamma_0}, its normal equations give
\begin{equation}
\label{equ:norm-equation}
 \widehat{\bm\theta}^{o}
 =\bm\Gamma_{\mathcal A^\ast}^{-1}\frac{\mathbf U^\top\mathbf Y}{nN}.
\end{equation}

Partition $\mathbf U_{ij}=(\mathbf U_{ij}^{(1)\top},
\mathbf U_{ij}^{(2)\top})^\top$ into its causal-effect and invalid-effect
blocks, and put
\[
\begin{gathered}
 \mathbf U_{ab}=\sum_{i=1}^n\sum_{j=1}^N
                    \mathbf U_{ij}^{(a)}\mathbf U_{ij}^{(b)\top},
 \quad a,b\in\{1,2\},\\
 \mathbf E=\frac1{nN}\sum_{i,j}\mathbf U_{ij}^{(1)}Y_{ij},\qquad
 \mathbf F=\frac1{nN}\sum_{i,j}\mathbf U_{ij}^{(2)}Y_{ij}.
\end{gathered}
\]
For the approximation terms, let
\[
 \mu^\ast_{ij}=D_i\beta^\ast(\bm s_j)
                    +\mathbf Z_i^\top\bm\alpha^\ast(\bm s_j),\qquad
 \widetilde\mu_{ij}=D_i\widetilde\beta(\bm s_j)
                    +\mathbf Z_i^\top\widetilde{\bm\alpha}(\bm s_j),
\]
and stack these quantities in subject order:
\begin{align}
\label{DEF:mu}
 \bm\mu^\ast&=(\mu^\ast_{11},\ldots,\mu^\ast_{1N},\ldots,\mu^\ast_{nN})^\top,\\
\label{DEF:mu-star}
 \widetilde{\bm\mu}&=(\widetilde\mu_{11},\ldots,\widetilde\mu_{1N},\ldots,\widetilde\mu_{nN})^\top.
\end{align}
Conditionally on $\mathbf Z_1,\ldots,\mathbf Z_n$, the covariance of the
leading oracle term computed with the population first stage is
	\begin{equation}
		\label{EQU:var-theory-estimate}
		\mathbf{\Xi}(\mathcal{A}^\ast) = \sum_{i=1}^n {\boldsymbol{\Gamma}_{\mathcal{A}^\ast}^{0\,-1}} \!\left\{\frac{1}{n^2N^2}\sum_{j,j' =1}^N {\mathbf{U}^{0}_{ij,\mathcal{A}^\ast} \mathbf{U}^{0\top}_{ij',\mathcal{A}^\ast}} G_{\epsilon}(\bm{s}_j, \bm{s}_{j^{\prime}})\right\}{\boldsymbol{\Gamma}_{\mathcal{A}^\ast}^{0\,-1}}. 
	\end{equation}
	{The asymptotic variance of $\widehat\beta^{o}(\bm s)$ is}
	\begin{equation}
		\sigma_{\beta}^2(\bm s) = \widetilde{\mathbf{B}}_1(\bm s)^{\top}\mathbf{\Xi}_\beta(\mathcal{A}^\ast)\widetilde{\mathbf{B}}_1(\bm s),
		\label{eq:var_orac_est}
	\end{equation}
	where $\mathbf{\Xi}_\beta(\mathcal{A}^\ast)$ is {the $\bm\theta_\beta$-block of $\mathbf{\Xi}(\mathcal{A}^\ast)$. Note that $\sigma_\beta^2(\bm s)$ is a function of $\mathbf Z_1,\ldots,\mathbf Z_n$, that is, a random normaliser; the proof establishes the limit law conditionally on $\mathbf Z_1,\ldots,\mathbf Z_n$, and the unconditional statement follows}.

	\begin{theorem}
		\label{THM:beta-convergence-oracle}
		Under Assumptions~\ref{A:tri}--\ref{A:size}, the oracle estimator
        $\widehat{\bm\theta}^{o}$ in \eqref{equ:norm-equation} exists and is
        unique with probability tending to one. With $r_{n,N}$ as defined
        in \eqref{eq:proof-rates}, it satisfies
        \[
        \begin{aligned}
        \|\widehat{\bm\theta}^{o}-\widetilde{\bm\theta}\|
            &=\OP(|\triangle|^{-\kappa/2}r_{n,N}),\\
        \sum_{\ell=1}^p\|\widehat\alpha_\ell^{o}-\alpha_\ell^\ast\|_{L^2(\Omega)}
            &=\OP(r_{n,N}).
        \end{aligned}
        \]
        Moreover, $\widehat\beta^o$ is consistent and satisfies
        \[
        \begin{aligned}
        \Vert\widehat\beta^o-\beta^\ast\Vert_{L^2(\Omega)}
        =\OP\bigg\{&\frac{\lambda^{\rm R}}{nN|\triangle|^2}
                    \|\beta^\ast\|_{2,\infty}+\frac1{\sqrt n}+\left(1+\frac{\lambda^{\rm R}}{nN|\triangle|^4}\right)
          |\triangle|^{d+1}\|\beta^\ast\|_{d+1,\infty}\bigg\}.
        \end{aligned}
        \]
	\end{theorem}
	
	Theorem~\ref{THM:beta-convergence-oracle} gives the $L^2$ convergence rate of the oracle estimator $\widehat{\beta}^{o}$. The roughness penalty bias is of order $\OP\{\lambda^{\text{R}}({nN|\triangle|^2})^{-1}\|\beta^\ast\|_{2,\infty}\}$. The spline approximation bias of $\widehat{\beta}^{o}$ is of order $\OP\{\left(1 + \lambda^{\text{R}}/{nN|\triangle|^4}\right)|\triangle|^{d +1}\|\beta^\ast\|_{d+1,\infty} \}$. The random error is of order $\OP(n^{-1/2})$.

	\begin{theorem}
		\label{LEM:beta-normality-oracle}
		Suppose Assumptions~\ref{A:tri}--\ref{A:size} and \ref{A:normality} hold for a fixed $\bm s\in\Omega$. Then, as $n \rightarrow \infty$,
		\begin{equation*}
			\sigma_{\beta}^{-1}(\bm{s}) \left\{\widehat{\beta}^{o}(\bm{s}) -\beta^\ast(\bm{s})\right\}\overset{\mathcal{L}}{\longrightarrow}N\left(0, 1\right).
		\end{equation*}
	\end{theorem}

The approximants in Section~\ref{SEC:Pre} satisfy
\[
 \|D^m(\beta^\ast-\widetilde\beta)\|_\infty
 \le C|\triangle|^{d+1-m}\|\beta^\ast\|_{d+1,\infty},\qquad 0\le m\le d,
\]
and, on each active simplex,
\[
 \|\alpha^\ast_\ell-\widetilde\alpha_\ell\|_{L^\infty(T_k)}
 \le C|\triangle|^{d+1}
             \|\alpha^\ast_\ell\|_{\mathcal W^{d+1,\infty}(T_k)}
 \le CC_\alpha|\triangle|^{d+1}.
\]
The first bound follows from Assumption~\ref{A:true-coef}. The second
uses the $L^\infty$ stability of the $L^2$ projection onto polynomials
of fixed degree on a simplex, with a constant depending only on
$d$, $\kappa$ and $\pi$. The error is zero on inactive simplices.
Consequently, the response has the decomposition
\[
 Y_{ij}=\mathbf U_{ij}^\top\widetilde{\bm\theta}
       +(D_i-\widehat D_i)\widetilde\beta(\bm s_j)
       +(\mu^\ast_{ij}-\widetilde\mu_{ij})+\epsilon_{ij}.
\]

	\begin{lemma}
		\label{LEM:U-mu-diff}
		Under Assumptions~\ref{A:tri}{--\ref{A:error-subgaussian} and \ref{A:covariates}}, we have $\|(nN)^{-1}(\mathbf{U}^\top \bm{\mu}^\ast - \mathbf{U}^\top \widetilde{\bm{\mu}})\| = \OP(|\triangle|^{d+1+\kappa/2})$.
	\end{lemma}
	
	\begin{proof}
		By Lemma S1.2 in \cite{Yu:etal:21},
		\begin{align*}
			&\bigl\|(nN)^{-1}\mathbf U^\top(\bm\mu^\ast-\widetilde{\bm\mu})\bigr\|^2\\
            &\quad=\left\|\frac1{nN}\sum_{i=1}^n\sum_{j=1}^N
            \widehat D_i\widetilde{\mathbf B}_1(\bm s_j)
                 (\mu_{ij}^\ast-\widetilde\mu_{ij})\right\|^2+\sum_{\ell=1}^p\left\|\frac1{nN}
               \sum_{i=1}^n\sum_{j=1}^N
               Z_{i\ell}\mathbf B_{2,\mathcal A_\ell^\ast}^{\top}(\bm s_j)
                 (\mu_{ij}^\ast-\widetilde\mu_{ij})\right\|^2\\
            &\quad=\OP\left\{|\triangle|^{-\kappa}
                (|\triangle|^\kappa\cdot|\triangle|^{d+1})^2
                +C_2\sum_{\ell=1}^p|\mathcal A_\ell^\ast|
                (|\triangle|^\kappa\cdot|\triangle|^{d+1})^2\right\}\\
            &\quad=\OP(|\triangle|^{2d+2+\kappa}).
        \end{align*}
		{Here, for the first term, write $\widetilde{\mathbf B}_1=\mathbf Q_2^{\top}\mathbf B_1$ and use $\|\mathbf Q_2^{\top}\bm v\|\le\|\bm v\|$, where}
		\[
		{\bm v=(nN)^{-1}\sum_{i=1}^n\sum_{j=1}^N\widehat D_i\,\mathbf B_1(\bm s_j)\,(\mu^\ast_{ij}-\widetilde\mu_{ij}) .}
		\]
		{Each of the $O(|\triangle|^{-\kappa})$ coordinates of $\bm v$ involves a basis function bounded by one and supported on $O(1)$ simplices, hence on $O(N|\triangle|^{\kappa})$ locations by Assumption~\ref{A:tri}, and}
		\[
		{|\mu^\ast_{ij}-\widetilde\mu_{ij}|\ \le\ C|\triangle|^{d+1}\bigl(\|\beta^\ast\|_{d+1,\infty}+C_\alpha\bigr)\bigl(|D_i|+\|\mathbf Z_i\|_1\bigr)}
		\]
		{by the preceding sup-norm approximation bounds (Assumptions~\ref{A:true-coef} and \ref{A:true-alpha}), while $n^{-1}\sum_i|\widehat D_i|(|D_i|+\|\mathbf Z_i\|_1)=\OP(1)$ by Assumption~\ref{A:covariates}. The second term is bounded in the same way, each $\mathbf B_{2,k}$ being supported on a single simplex, with $\sum_\ell|\mathcal A^\ast_\ell|=|\mathcal A^\ast|\le s_0K_2=O(|\triangle|^{-\kappa})$ coordinates (Assumption~\ref{A:true-alpha}). The constant in the $\OP$ bound is proportional to $\|\beta^\ast\|_{d+1,\infty}+C_\alpha$.}
		Taking square roots yields the desired result.
	\end{proof}
	
	\begin{lemma}
		\label{LEM:D-cancel}
		
		For the first-stage OLS fit,
		\[
		\left\|\frac1{nN}\sum_{i=1}^n\sum_{j=1}^N
		\mathbf U_{ij}(D_i-\widehat D_i)
		\widetilde{\mathbf B}_1(\bm s_j)^\top
		\widetilde{\bm\theta}_\beta\right\|=0.
		\]
	\end{lemma}
	
	\begin{proof}
		Let $u_i := D_i - \widehat{D}_i$ be the first-stage OLS residuals from regressing $D$ on the instrument matrix (whose columns include $Z_\ell$). 
		By OLS orthogonality, $ \sum_{i=1}^n Z_{i\ell} u_i = 0$ for each $\ell$, and since $\widehat D \in \mathrm{span}\{Z\}$ we also have $\sum_{i=1}^n \widehat D_i\, u_i = 0$.
		
		Note that the sums over $j$ do not depend on $i$, so we may factor them out:
		\[
		\begin{aligned}
        &\frac1{nN}\sum_{i=1}^n Z_{i\ell}u_i\sum_{j=1}^N
          \mathbf B_{2,\mathcal A_\ell^\ast}(\bm s_j)
          \widetilde{\mathbf B}_1(\bm s_j)^\top\widetilde{\bm\theta}_\beta\\
        &\quad=\left\{\frac1n\sum_{i=1}^n Z_{i\ell}u_i\right\}
          \left\{\frac1N\sum_{j=1}^N
          \mathbf B_{2,\mathcal A_\ell^\ast}(\bm s_j)
          \widetilde{\mathbf B}_1(\bm s_j)^\top\widetilde{\bm\theta}_\beta\right\}
          =\mathbf0,
        \end{aligned}
        \]
        and similarly
        \[
        \begin{aligned}
        &\frac1{nN}\sum_{i=1}^n\widehat D_i u_i\sum_{j=1}^N
          \widetilde{\mathbf B}_1(\bm s_j)
          \widetilde{\mathbf B}_1(\bm s_j)^\top\widetilde{\bm\theta}_\beta\\
        &\quad=\left\{\frac1n\sum_{i=1}^n\widehat D_i u_i\right\}
          \left\{\frac1N\sum_{j=1}^N
          \widetilde{\mathbf B}_1(\bm s_j)
          \widetilde{\mathbf B}_1(\bm s_j)^\top\widetilde{\bm\theta}_\beta\right\}
          =\mathbf0.
        \end{aligned}
        \]
        Both identities hold because the first factor in each product is zero by the OLS normal equations.
	\end{proof}

	\begin{lemma} 
		\label{lem:Ueps-order}
		Under Assumptions~\ref{A:tri}{, \ref{A:error-subgaussian}, \ref{A:covariance} and \ref{A:covariates}}, we have
		\begin{align}
        \label{lem:Ueps-order-1}
        \|(nN)^{-1}\mathbf U^\top\bm\epsilon\|
        &=\OP\{n^{-1/2}|\triangle|^{\kappa/2}+(nN)^{-1/2}\},\\
        \label{lem:Ueps-order-2}
        \|(nN)^{-1}\mathbf U^\top\bm\epsilon\|_\infty
        &=\OP\Big[\{n^{-1/2}|\triangle|^{\kappa/2}+(nN)^{-1/2}\}
\times\{\log n+\log(pK_2)\}^{1/2}\Big].
        \end{align}
		
	\end{lemma}
	
	\begin{proof}
        Assumption~\ref{A:covariance} gives
        \[
        \E\{\epsilon_i(\bm s)\epsilon_i(\bm s')\}
        =G_\eta(\bm s,\bm s')+
          \sigma_\varepsilon^2(\bm s)\mathbf1\{\bm s=\bm s'\},
        \]
		with $G_\eta$ square-integrable on $\Omega\times\Omega$ and $\sigma_\varepsilon^2$ bounded. Let $\bm{X}_i=({D^0_i}, Z_{i1},\cdots, Z_{ip})^\top$, $\mathbf{b}_1(\bm s_j) = \widetilde{\mathbf B}_1(\bm s_j)$ and $\mathbf{b}_{\ell+1}(\bm s_j) = \mathbf B_{2,\mathcal A_\ell^\ast}(\bm s_j)$ for $\ell = 1, \ldots, p$.
		{We first replace $\widehat D_i$ by $D^0_i$. Writing $\widehat D_i=D^0_i+\mathbf Z_i^{\top}\bm\delta$ with $\bm\delta=\widehat{\bm\gamma}-\bm\gamma^\ast=\OP(n^{-1/2})$ (Assumption~\ref{A:covariates}), the first block of $(nN)^{-1}\mathbf U^{\top}\bm\epsilon$ equals $(nN)^{-1}\sum_{i,j}D^0_i\mathbf b_1(\bm s_j)\epsilon_{ij}+\sum_{\ell=1}^p\delta_\ell\,(nN)^{-1}\sum_{i,j}Z_{i\ell}\mathbf b_1(\bm s_j)\epsilon_{ij}$. Each of these $p+1$ sums is of the form treated below, with an $i$-factor built from $(D^0_i,\mathbf Z_i)$, which is independent of $\bm\epsilon_i$ and independent across $i$ by Assumption~\ref{A:error-subgaussian}; hence the second term is $\OP(n^{-1/2})$ times the bound obtained for the first, and it suffices to prove the lemma with $D^0_i$ in place of $\widehat D_i$. Below, $\mathbf U$ is understood accordingly. (The factorisation of expectations used next fails for $\widehat D_i$, which depends on the whole sample and is correlated with $\epsilon_{ij}$ through $\xi_i$; this is why the split is needed.)}
		The pairs $(\bm X_i,\bm\epsilon_i)$ are independent across $i$,
        and $\bm X_i$ is independent of $\bm\epsilon_i$ by
        Assumption~\ref{A:error-subgaussian}. The basis vectors are
        deterministic, and
        $\mathbf U_{ij}^\top\mathbf U_{ij'}
        =\sum_{\ell=1}^{p+1}X_{i\ell}^2
          \mathbf b_\ell(\bm s_j)^\top\mathbf b_\ell(\bm s_{j'})$.
        Therefore,
        \begin{align*}
        &\E\bigl\|(nN)^{-1}\mathbf U^\top\bm\epsilon\bigr\|^2\\
        &\quad=(nN)^{-2}\sum_{i=1}^n\sum_{\ell=1}^{p+1}\E(X_{i\ell}^2)
          \sum_{j,j'=1}^N\mathbf b_\ell(\bm s_j)^\top
             \mathbf b_\ell(\bm s_{j'})G_\epsilon(\bm s_j,\bm s_{j'})\\
        &\quad=(nN)^{-2}\sum_{i=1}^n\sum_{\ell=1}^{p+1}\E(X_{i\ell}^2)
          \Bigg\{\sum_{j,j'}\mathbf b_\ell(\bm s_j)^\top
             \mathbf b_\ell(\bm s_{j'})G_\eta(\bm s_j,\bm s_{j'})+\sum_j\|\mathbf b_\ell(\bm s_j)\|^2
                \sigma_\varepsilon^2(\bm s_j)\Bigg\}.
        \end{align*}
		{We bound the two sums for each $\ell$:}
		\[
		\begin{aligned}
        \sum_{j,j'}\mathbf b_\ell(\bm s_j)^\top\mathbf b_\ell(\bm s_{j'})
           G_\eta(\bm s_j,\bm s_{j'})&\lesssim N^2|\triangle|^\kappa,\\
        \sum_j\|\mathbf b_\ell(\bm s_j)\|^2\sigma_\varepsilon^2(\bm s_j)
           &\lesssim N.
        \end{aligned}
		\]
		{These bounds hold uniformly over $\ell$. For $\ell\ge2$, $|G_\eta(\bm s,\bm s')|\le\|\sum_t\rho_t^{1/2}|\psi_t|\|_\infty^2$ and $\sigma_\varepsilon^2\le C_\sigma^2$ (Assumption~\ref{A:covariance}), each $B_{2,m}$ is bounded by one and supported on a single simplex containing $O(N|\triangle|^{\kappa})$ locations (Assumption~\ref{A:tri}), and there are $O(|\triangle|^{-\kappa})$ coordinates, so that}
		\[
		\begin{aligned}
        &\sum_{j,j'}|\mathbf b_\ell(\bm s_j)^\top\mathbf b_\ell(\bm s_{j'})|
                |G_\eta(\bm s_j,\bm s_{j'})|\lesssim\sum_m\Bigl\{\sum_j|B_{2,m}(\bm s_j)|\Bigr\}^2
          \lesssim|\triangle|^{-\kappa}(N|\triangle|^\kappa)^2.
        \end{aligned}
		\]
		{For $\ell=1$, $\mathbf b_1=\mathbf Q_2^{\top}\mathbf B_1$ is not locally supported, but $\sum_{j,j'}\mathbf b_1(\bm s_j)^{\top}\mathbf b_1(\bm s_{j'})G_\eta(\bm s_j,\bm s_{j'})=\operatorname{tr}(\mathbf Q_2^{\top}\mathbf A\mathbf Q_2)\le\operatorname{tr}(\mathbf A)$ with $\mathbf A=\sum_{j,j'}\mathbf B_1(\bm s_j)\mathbf B_1(\bm s_{j'})^{\top}G_\eta(\bm s_j,\bm s_{j'})\succeq\mathbf 0$ (as $G_\eta$ is a positive semidefinite kernel), and $\sum_j\|\mathbf Q_2^{\top}\mathbf B_1(\bm s_j)\|^2\le\sum_j\|\mathbf B_1(\bm s_j)\|^2$, so the same bounds hold. The moments $\E(X_{i\ell}^2)$ are bounded by Assumption~\ref{A:covariates}.}
		Therefore,
		\[
		\E\Big\|(nN)^{-1}\mathbf{U}^\top \bm{\epsilon}\Big\|^2
		\;\lesssim\; (nN)^{-2}\, n \big(N^2|\triangle|^{\kappa} + N\big)
		\;=\; n^{-1}|\triangle|^{\kappa} \;+\; (nN)^{-1},
		\]
		and
		\[
		\Big\|(nN)^{-1}\mathbf{U}^\top \bm{\epsilon}\Big\|
		= \OP\!\left( \sqrt{n^{-1}|\triangle|^{\kappa} + (nN)^{-1}} \right)
		= \OP\!\left(n^{-1/2}|\triangle|^{\kappa/2} \;+\; (nN)^{-1/2}\right),
		\]
		which proves the claim in (\ref{lem:Ueps-order-1}). {For \eqref{lem:Ueps-order-2}, condition on $\mathbf Z_1,\ldots,\mathbf Z_n$, so that $\bm X_i$ is fixed. Each coordinate of $(nN)^{-1}\mathbf U^{\top}\bm\epsilon$ is then a sum over $i$ of independent mean-zero variables $(nN)^{-1}X_{i\ell}\sum_j b_{\ell m}(\bm s_j)\epsilon_{ij}$, each sub-Gaussian with parameter proportional to its variance (Assumption~\ref{A:error-subgaussian}: $\eta_i(\bm s)=\sum_t\rho_t^{1/2}\zeta_{it}\psi_t(\bm s)$ and $\varepsilon_i(\bm s_j)=\sigma_\varepsilon(\bm s_j)e_{ij}$ are linear in independent sub-Gaussian variables), with $|X_{i\ell}|\le C_Z\max(1,\|\bm\gamma^\ast\|_1)$; the sum of these variances is bounded by the diagonal term computed above, of order at most $n^{-1}|\triangle|^{\kappa}+(nN)^{-1}$. A sub-Gaussian tail bound and a union bound over the $O(K_1+|\mathcal A^\ast|)=O(pK_2)$ coordinates give \eqref{lem:Ueps-order-2}, the factor $\{\log n+\log(pK_2)\}^{1/2}$ coming from the union bound; note that $\log(pK_2)$ is not $O(\log n)$ under Assumption~\ref{A:size} alone, so it cannot be dropped.}
	\end{proof}
	
	\begin{lemma}
		\label{LEM:roughness}
		Under Assumptions~\ref{A:tri} and \ref{A:true-coef},
        \[
        \|\lambda^{\rm R}\mathbf W\widetilde{\bm\theta}\|
        \le C\lambda^{\rm R}|\triangle|^{\kappa/2-2}
        \bigl\{\|\beta^\ast\|_{2,\infty}
              +|\triangle|^{d-1}\|\beta^\ast\|_{d+1,\infty}\bigr\}.
        \]
    \end{lemma}
	\begin{proof}
	The block structure of $\mathbf W$ gives
	$\|\mathbf W\widetilde{\bm\theta}\|=\|\mathbf W_1\widetilde{\bm\theta}_{\beta}\|$. Since $\mathbf W_1$ is symmetric and positive semidefinite,
	\begin{align}
		\|\mathbf W_1\widetilde{\bm\theta}_{\beta}\|^2
		&=
		\widetilde{\bm\theta}_{\beta}^{\top}
		\mathbf W_1^2
		\widetilde{\bm\theta}_{\beta} \leq
		\lambda_{\max}(\mathbf W_1)
		\widetilde{\bm\theta}_{\beta}^{\top}
		\mathbf W_1
		\widetilde{\bm\theta}_{\beta}.
		\label{eq:Ptheta-bound}
	\end{align}
	
	We first bound $\lambda_{\max}(\mathbf W_1)$. For an arbitrary
	coefficient vector $\bm v$, let
	$
	g_{\bm v}(\bm s)
	=
	\widetilde{\mathbf B}_1^{\top}(\bm s)\bm v.
	$
	By the definition of the roughness penalty,
	$
	\bm v^{\top}\mathbf W_1\bm v
	=
	\|g_{\bm v}\|_{\mathcal{E}}^2.
	$
	The spline inverse inequality on a quasi-uniform simplicial partition gives
	$\|g_{\bm v}\|_{\mathcal{E}}
	\leq
	C|\triangle|^{-2}
	\|g_{\bm v}\|_{L^2(\Omega)}$.
	
	Moreover, by the $L^2$ stability of the spline basis,
	$\|g_{\bm v}\|_{L^2(\Omega)}^2
	\leq C|\triangle|^{\kappa}\|\bm v\|^2$. 
	Therefore,
	\[
	\bm v^{\top}\mathbf W_1\bm v
	\leq
	C|\triangle|^{-4}
	\|g_{\bm v}\|_{L^2(\Omega)}^2
	\leq
	C|\triangle|^{\kappa-4}\|\bm v\|^2.
	\]
	Taking the supremum over $\|\bm v\|=1$ yields
	$\lambda_{\max}(\mathbf W_1)
	\leq C|\triangle|^{\kappa-4}$.

	Next,
    $\widetilde{\bm\theta}_\beta^\top\mathbf W_1\widetilde{\bm\theta}_\beta
    =\|\widetilde\beta\|_{\mathcal E}^2$.
    For $d\ge2$, the derivative bound in Assumption~\ref{A:true-coef}
    with $m=2$ gives
    $\|\widetilde\beta-\beta^\ast\|_{\mathcal E}
      \le C|\triangle|^{d-1}\|\beta^\ast\|_{d+1,\infty}$.
    Also,
    $\|\beta^\ast\|_{\mathcal E}
      \le\kappa\operatorname{Vol}(\Omega)^{1/2}\|\beta^\ast\|_{2,\infty}$.
    The triangle inequality therefore gives
    \begin{align*}
		\|\widetilde\beta\|_{\mathcal{E}}
		&\leq
		\|\beta^\ast\|_{\mathcal{E}}
		+
		\|\widetilde\beta-\beta^\ast\|_{\mathcal{E}}\leq
		C\left\{
		\|\beta^\ast\|_{2,\infty}
		+
		|\triangle|^{d-1}
		\|\beta^\ast\|_{d+1,\infty}
		\right\}.
	\end{align*}
	For $d=1$, $\widetilde\beta$ is piecewise linear and
    $\mathcal E(\widetilde\beta)=0$, so the same bound holds directly.
    Consequently,
    \begin{equation}
		\left(
		\widetilde{\bm\theta}_{\beta}^{\top}
		\mathbf W_1
		\widetilde{\bm\theta}_{\beta}
		\right)^{1/2}
		\leq
		C\left\{
		\|\beta^\ast\|_{2,\infty}
		+
		|\triangle|^{d-1}
		\|\beta^\ast\|_{d+1,\infty}
		\right\}.
		\label{eq:P-quadratic-bound}
	\end{equation}
	
	Combining \eqref{eq:Ptheta-bound}--\eqref{eq:P-quadratic-bound},
	we obtain
	\begin{align*}
        \|\mathbf W\widetilde{\bm\theta}\|
        &=\|\mathbf W_1\widetilde{\bm\theta}_\beta\|\le\lambda_{\max}^{1/2}(\mathbf W_1)
           \bigl(\widetilde{\bm\theta}_\beta^\top\mathbf W_1
                    \widetilde{\bm\theta}_\beta\bigr)^{1/2}\\
        &\le C|\triangle|^{\kappa/2-2}
           \bigl\{\|\beta^\ast\|_{2,\infty}
                 +|\triangle|^{d-1}\|\beta^\ast\|_{d+1,\infty}\bigr\}.
        \end{align*}
        Multiplying by $\lambda^{\rm R}$ proves the lemma.
	\end{proof}
	
	\begin{proof}[Proof of Theorem \ref{THM:beta-convergence-oracle}]
		Throughout this proof, $\mathcal A=\mathcal A^\ast$.
        Lemma~\ref{LEM:D-cancel} removes the first-stage residual term
        from the normal equations. Hence,
        \[
        \begin{aligned}
        \widehat{\bm\theta}^o-\widetilde{\bm\theta}
        =\bm\Gamma_{\mathcal A,\lambda^{\rm R}}^{-1}
          \bigg\{&\frac{\mathbf U^\top(\bm\mu^\ast-\widetilde{\bm\mu})}{nN}
                  +\frac{\mathbf U^\top\bm\epsilon}{nN}-\frac{\lambda^{\rm R}}{nN}
                    \mathbf W\widetilde{\bm\theta}\bigg\}.
        \end{aligned}
        \]
        By {Lemma~\ref{LEM:Gamma_0}(b), applied with $\mathcal A=\mathcal A^\ast$ (which satisfies condition (S) of Lemma~\ref{LEM:norm} for all large $n$), there is an event $\mathcal E_n$ with $\Pr(\mathcal E_n)\to1$ on which $\bm\Gamma_{\mathcal A^\ast,\lambda^{\rm R}}$ is invertible, so that $\widehat{\bm\theta}^{o}$ exists and is unique, and} we have
		$\lambda_{\max}((\bm{\Gamma}_{\mathcal{A},\lambda^{\rm R}})^{-1}) \;{\le\; c_\Gamma^{-1}}|\triangle|^{-\kappa}$.
		Lemmas~\ref{LEM:U-mu-diff}, \ref{lem:Ueps-order} and
        \ref{LEM:roughness} bound the approximation, noise and roughness
        terms, respectively. Combining these bounds gives
        \begin{align*}
        \|\widehat{\bm\theta}^o-\widetilde{\bm\theta}\|
        =\OP\bigg\{&|\triangle|^{d+1-\kappa/2}
          +n^{-1/2}|\triangle|^{-\kappa/2}
          +(nN)^{-1/2}|\triangle|^{-\kappa}\\
          &+\frac{\lambda^{\rm R}}{nN}|\triangle|^{-2-\kappa/2}
          \big(\|\beta^\ast\|_{2,\infty}
             +|\triangle|^{d-1}\|\beta^\ast\|_{d+1,\infty}\big)\bigg\}.
        \end{align*}

		By \eqref{eq:spline-stability} and the approximation bounds preceding
        Lemma~\ref{LEM:U-mu-diff}, there are constants $C_1,C_2>0$ such that
        \[
		\begin{aligned}
        &\|\widehat\beta^o-\beta^\ast\|
          +\sum_{\ell=1}^p\|\widehat\alpha_\ell^o-\alpha_\ell^\ast\|\\
        &\quad\le C_1|\triangle|^{\kappa/2}
          \bigg\{\|\widehat{\bm\theta}_\beta^o-\widetilde{\bm\theta}_\beta\|
            +\sum_{\ell=1}^p
             \|\widehat{\bm\theta}_{\alpha,\ell}^o
                         -\widetilde{\bm\theta}_{\alpha,\ell}\|\bigg\}
          +C_2|\triangle|^{d+1}.
        \end{aligned}
		\]
        Here, $C_2|\triangle|^{d+1}$ bounds
        $\|\widetilde\beta-\beta^\ast\|
          +\sum_\ell\|\widetilde\alpha_\ell-\alpha_\ell^\ast\|$,
        since $\sum_{k\in\mathcal A_\ell^\ast}\operatorname{Vol}(T_k)
              \le\operatorname{Vol}(\Omega)$.
        Substituting the coefficient bound gives
        \begin{align*}
        &\|\widehat\beta^o-\beta^\ast\|
+\sum_{\ell=1}^p\|\widehat\alpha_\ell^o-\alpha_\ell^\ast\|\\
        &\quad=\OP\bigg\{|\triangle|^{d+1}+n^{-1/2}
        +(nN)^{-1/2}|\triangle|^{-\kappa/2}+\frac{\lambda^{\rm R}}{nN}|\triangle|^{-2}
        \big(\|\beta^\ast\|_{2,\infty}
             +|\triangle|^{d-1}\|\beta^\ast\|_{d+1,\infty}\big)\bigg\}.
        \end{align*}
		{The term $(nN)^{-1/2}|\triangle|^{-\kappa/2}=n^{-1/2}(N|\triangle|^{\kappa})^{-1/2}=o(n^{-1/2})$ because $N^{1/\kappa}|\triangle|\to\infty$ by Assumption~\ref{A:size}. Thus, the displayed bound gives the claimed rate. The same bound holds for $\sum_{\ell}\|\widehat\alpha^{o}_\ell-\alpha^\ast_\ell\|$, and by the same absorption the coefficient bound displayed above is $\OP(|\triangle|^{-\kappa/2}r_{n,N})$ with $r_{n,N}$ as in \eqref{eq:proof-rates}, which proves the remaining assertions.}
	\end{proof}
	
	We next derive the pointwise asymptotic distribution of
    $\widehat\beta^o$. Applying the Schur complement to the block normal equations
    \[
	\begin{pmatrix}
		\mathbf U_{11} + \lambda^{\rm R}\mathbf W_1 & \mathbf U_{12}\\
		\mathbf U_{21} & \mathbf U_{22}
	\end{pmatrix}
	\begin{pmatrix}\widehat{\bm\theta}_\beta^{o}\\ \widehat{\bm\theta}_\alpha^{o}\end{pmatrix}
	=
	{nN}\begin{pmatrix}\mathbf E\\ \mathbf F\end{pmatrix},
	\]
	we obtain
	\begin{equation}\label{eq:theta-beta-Schur}
\begin{aligned}
 \widehat{\bm\theta}_\beta^o
 &=\bigl(\mathbf U_{11}+\lambda^{\rm R}\mathbf W_1
       -\mathbf U_{12}\mathbf U_{22}^{-1}\mathbf U_{21}\bigr)^{-1}\times nN\bigl(\mathbf E-\mathbf U_{12}\mathbf U_{22}^{-1}\mathbf F\bigr).
\end{aligned}
\end{equation}
	
	We now analyze the right-hand side. First, plug the component
	\(\mathbf{U}_{ij}^{(2)\top}\widetilde{\bm{\theta}}_\alpha\) into
	\(\mathbf{E}-\mathbf{U}_{12}\mathbf{U}_{22}^{-1}\mathbf{F}\):
	\begin{align*}
 &\sum_{i=1}^n\sum_{j=1}^N
       \mathbf U_{ij}^{(1)}\mathbf U_{ij}^{(2)\top}\widetilde{\bm\theta}_\alpha-\mathbf U_{12}\mathbf U_{22}^{-1}
       \sum_{i=1}^n\sum_{j=1}^N
       \mathbf U_{ij}^{(2)}\mathbf U_{ij}^{(2)\top}\widetilde{\bm\theta}_\alpha
       =\mathbf0.
 \end{align*}
    The first-stage residual term also vanishes by Lemma~\ref{LEM:D-cancel}.
    The remaining components of $Y_{ij}$ in the decomposition preceding
    Lemma~\ref{LEM:U-mu-diff} are
    $\mathbf U_{ij}^{(1)\top}\widetilde{\bm\theta}_\beta$,
    the approximation error $\mu_{ij}^\ast-\widetilde\mu_{ij}$, and
    the noise $\epsilon_{ij}$. On the event $\mathcal E_n$ of
    Lemma~\ref{LEM:Gamma_0}(b), both $\mathbf U_{22}$ and
    $\bm\Gamma_{\mathcal A^\ast,\lambda^{\rm R}}$ are invertible, and
    \begin{equation}\label{eq:EF-exact}
\begin{aligned}
 nN\bigl(\mathbf E-\mathbf U_{12}\mathbf U_{22}^{-1}\mathbf F\bigr)
 &=\bigl(\mathbf U_{11}-\mathbf U_{12}\mathbf U_{22}^{-1}\mathbf U_{21}\bigr)
        \widetilde{\bm\theta}_\beta+\bm\delta_n^u+\bm\varsigma_n^u,
\end{aligned}
\end{equation}
	{where}
	\begin{align*}
 \bm\delta_n^u
 &=\sum_{i=1}^n\sum_{j=1}^N
   \bigl(\mathbf U_{ij}^{(1)}
       -\mathbf U_{12}\mathbf U_{22}^{-1}\mathbf U_{ij}^{(2)}\bigr)
          (\mu_{ij}^\ast-\widetilde\mu_{ij}),\\
 \bm\varsigma_n^u
 &=\sum_{i=1}^n\sum_{j=1}^N
   \bigl(\mathbf U_{ij}^{(1)}
       -\mathbf U_{12}\mathbf U_{22}^{-1}\mathbf U_{ij}^{(2)}\bigr)
          \epsilon_{ij}.
 \end{align*}
	{Write $\mathbf S=(nN)^{-1}(\mathbf U_{11}-\mathbf U_{12}\mathbf U_{22}^{-1}\mathbf U_{21})$ for the normalised Schur complement and}
	\begin{equation}\label{eq:Lambda-beta}
		{\bm\Lambda_\beta=\Big(\mathbf S+\frac{\lambda^{\rm R}}{nN}\mathbf W_1\Big)^{-1}\begin{pmatrix}\mathbf I & -\mathbf U_{12}\mathbf U_{22}^{-1}\end{pmatrix},}
	\end{equation}
	{so that $\bm\Lambda_\beta=(\mathbf I\ \ \mathbf 0)\,\bm\Gamma_{\mathcal A^\ast,\lambda^{\rm R}}^{-1}$ is the $\bm\theta_\beta$-block of rows of $\bm\Gamma_{\mathcal A^\ast,\lambda^{\rm R}}^{-1}$ (block inversion), and in particular $\|\bm\Lambda_\beta\|\le\|\bm\Gamma_{\mathcal A^\ast,\lambda^{\rm R}}^{-1}\|\le c_\Gamma^{-1}|\triangle|^{-\kappa}$ on $\mathcal E_n$. Substituting \eqref{eq:EF-exact} into \eqref{eq:theta-beta-Schur} gives the exact decomposition}
	\begin{equation}\label{eq:theta-beta-exact}
\begin{aligned}
 \widehat{\bm\theta}_\beta^o-\widetilde{\bm\theta}_\beta
 &=-\left(\mathbf S+\frac{\lambda^{\rm R}}{nN}\mathbf W_1\right)^{-1}
       \frac{\lambda^{\rm R}}{nN}\mathbf W_1\widetilde{\bm\theta}_\beta+\bm\Lambda_\beta\frac{\mathbf U^\top(\bm\mu^\ast-\widetilde{\bm\mu})}{nN}
       +\bm\Lambda_\beta\frac{\mathbf U^\top\bm\epsilon}{nN},
\end{aligned}
\end{equation}
	{where $(\mathbf I,-\mathbf U_{12}\mathbf U_{22}^{-1})$ acting on the two blocks of $\mathbf U^{\top}\bm v$ reproduces $\bm\delta^{u}_n$ and $\bm\varsigma^{u}_n$. The covariance objects are, with the same normalisation,}
	\begin{equation}\label{eq:Psi-def}
\begin{aligned}
 \bm\Psi_\eta&=\frac1{n^2N^2}\sum_{i=1}^n\sum_{j,j'=1}^N
      \mathbf U_{ij}\mathbf U_{ij'}^\top G_\eta(\bm s_j,\bm s_{j'}),\\
 \bm\Psi_\varepsilon&=\frac1{n^2N^2}\sum_{i=1}^n\sum_{j=1}^N
      \mathbf U_{ij}\mathbf U_{ij}^\top\sigma_\varepsilon^2(\bm s_j).
\end{aligned}
\end{equation}
	{Set $\bm\Psi_\epsilon=\bm\Psi_\eta+\bm\Psi_\varepsilon$.}
	{The quantities $\bm\Lambda^{0}_\beta$, $\bm\Psi^{0}_\epsilon$, $\bm\Gamma^{0}_{\mathcal A^\ast,\lambda^{\rm R}}$ denote the corresponding objects computed with the population first stage; these are functions of $\mathbf Z_1,\ldots,\mathbf Z_n$ only. By block inversion, $\bm\Lambda^{0}_\beta\bm\Psi^{0}_\epsilon\bm\Lambda^{0\top}_\beta$ is the $\bm\theta_\beta$-block of $\bm\Xi(\mathcal A^\ast)$ in \eqref{EQU:var-theory-estimate}, so that}
	\begin{equation}\label{eq:sigma-quadratic}
\begin{gathered}
 \sigma_\beta^2(\bm s)=\bm a_0^\top\bm\Psi_\epsilon^0\bm a_0, \qquad
 \bm a_0:=\bm\Lambda_\beta^{0\top}\widetilde{\mathbf B}_1(\bm s),
 \qquad\bm a:=\bm\Lambda_\beta^\top\widetilde{\mathbf B}_1(\bm s).
\end{gathered}
\end{equation}
	{Lemma~\ref{LEM:Gamma_0} applies to $\bm\Gamma^{0}_{\mathcal A^\ast,\lambda^{\rm R}}$ as well (its proof only simplifies, the reconciliation of $\widehat D_i$ with $D^0_i$ being unnecessary), so $\|\bm\Lambda^{0}_\beta\|\le c_\Gamma^{-1}|\triangle|^{-\kappa}$ on the event $\mathcal E^{0}_n=\{\lambda_{\min}(\bm\Gamma^{0}_{\mathcal A^\ast,\lambda^{\rm R}})\ge c_\Gamma|\triangle|^{\kappa}\}\cap\{n\sigma^2_\beta(\bm s)\ge c_V\}$, which is a function of $\mathbf Z_1,\ldots,\mathbf Z_n$ only and has probability tending to one by Lemma~\ref{LEM:Gamma_0} and Assumption~\ref{A:normality}(V). From now on $\mathcal E_n$ denotes the intersection of $\mathcal E^{0}_n$ with the event of Lemma~\ref{LEM:Gamma_0}(b) for $\bm\Gamma_{\mathcal A^\ast,\lambda^{\rm R}}$; it also has probability tending to one, but it depends on $\widehat{\bm\gamma}$ and is therefore not $\mathbf Z$-measurable. Finally, $\|\widetilde{\mathbf B}_1(\bm s)\|\le\|\mathbf B_1(\bm s)\|\le1$ for every $\bm s$, because $\mathbf Q_2$ has orthonormal columns and the Bernstein--B\'ezier basis functions are nonnegative and sum to one on each simplex; hence $\|\bm a_0\|\le c_\Gamma^{-1}|\triangle|^{-\kappa}$ on $\mathcal E^{0}_n$ and $\|\bm a\|\le c_\Gamma^{-1}|\triangle|^{-\kappa}$ on $\mathcal E_n$.}

	\begin{proof}[Proof of Theorem \ref{LEM:beta-normality-oracle}]
		{Throughout, $\mathcal A=\mathcal A^\ast$; the bounds of Steps~1 and~2 are unconditional $\OP$ statements established on $\mathcal E_n$, while Step~3 conditions on $\mathbf Z_1,\ldots,\mathbf Z_n$ and uses only the $\mathbf Z$-measurable event $\mathcal E^{0}_n$. Multiplying \eqref{eq:theta-beta-exact} by $\widetilde{\mathbf B}_1(\bm s)^{\top}$ and adding $\widetilde\beta(\bm s)-\beta^\ast(\bm s)$,}
		\begin{equation}\label{eq:beta-point-decomp}
			{\widehat\beta^{o}(\bm s)-\beta^\ast(\bm s)=b_{\rm a}(\bm s)+b_{\rm p}(\bm s)+b_{\rm s}(\bm s)+T_n(\bm s),}
		\end{equation}
		The four terms are the spline approximation bias,
        the roughness penalty bias, the projected approximation bias,
        and the noise term, respectively:
        \[
        \begin{aligned}
        b_{\rm a}(\bm s)&=\widetilde\beta(\bm s)-\beta^\ast(\bm s),\\
        b_{\rm p}(\bm s)&=-\widetilde{\mathbf B}_1(\bm s)^\top
          \left(\mathbf S+\frac{\lambda^{\rm R}}{nN}\mathbf W_1\right)^{-1}
          \frac{\lambda^{\rm R}}{nN}\mathbf W_1\widetilde{\bm\theta}_\beta,\\
        b_{\rm s}(\bm s)&=\bm a^\top
          \frac{\mathbf U^\top(\bm\mu^\ast-\widetilde{\bm\mu})}{nN},\\
        T_n(\bm s)&=\bm a^\top\frac{\mathbf U^\top\bm\epsilon}{nN}.
        \end{aligned}
        \]

        \medskip\noindent{\emph{Step 1: the bias terms are $o_{\rm P}(n^{-1/2})$.} By the quasi-interpolant bound in Assumption~\ref{A:true-coef}, $|b_{\rm a}(\bm s)|\le C|\triangle|^{d+1}\|\beta^\ast\|_{d+1,\infty}=o(n^{-1/2})$ by Assumption~\ref{A:size}. Since $(\mathbf S+\lambda^{\rm R}(nN)^{-1}\mathbf W_1)^{-1}\bm v$ is the $\bm\theta_\beta$-block of $\bm\Gamma^{-1}_{\mathcal A^\ast,\lambda^{\rm R}}(\bm v^{\top},\mathbf 0^{\top})^{\top}$, Lemma~\ref{LEM:roughness} gives}
		\begin{align*}
		{|b_{\rm p}(\bm s)|}&{\le\|\widetilde{\mathbf B}_1(\bm s)\|\,\|\bm\Gamma^{-1}_{\mathcal A^\ast,\lambda^{\rm R}}\|\,\Big\|\frac{\lambda^{\rm R}}{nN}\mathbf W\widetilde{\bm\theta}\Big\|}\\
		&{\le C\,\frac{\lambda^{\rm R}}{nN}|\triangle|^{-2-\kappa/2}\big\{\|\beta^\ast\|_{2,\infty}+|\triangle|^{d-1}\|\beta^\ast\|_{d+1,\infty}\big\}=o(n^{-1/2})}
		\end{align*}
		{by Assumption~\ref{A:normality}(U2), and Lemma~\ref{LEM:U-mu-diff} gives}
		\begin{align*}
		|b_{\rm s}(\bm s)|
        &\le\|\bm a\|\,\left\|\frac{\mathbf U^\top(\bm\mu^\ast-\widetilde{\bm\mu})}{nN}\right\|=\OP(|\triangle|^{-\kappa}\cdot|\triangle|^{d+1+\kappa/2})
          =\OP(|\triangle|^{d+1-\kappa/2})
          =o_{\rm P}(n^{-1/2})
        \end{align*}
		{by Assumption~\ref{A:normality}(U1). Together with (V), $b_{\rm a}(\bm s)+b_{\rm p}(\bm s)+b_{\rm s}(\bm s)=o_{\rm P}\{\sigma_\beta(\bm s)\}$.}

		\medskip\noindent{\emph{Step 2: replacing the estimated first stage in the noise term.} The weights $\bm a$ and the design $\mathbf U$ in $T_n(\bm s)$ depend on $\widehat{\bm\gamma}$, hence on $\xi_1,\ldots,\xi_n$, which are correlated with $\eta_1,\ldots,\eta_n$; conditioning on $\widehat D_1,\ldots,\widehat D_n$ is therefore not admissible, and we pass to the population first stage instead. Write $\widehat D_i=D^0_i+\mathbf Z_i^{\top}\bm\delta$ with $\bm\delta=\widehat{\bm\gamma}-\bm\gamma^\ast=\OP(n^{-1/2})$ (Assumption~\ref{A:covariates} and $\E(\mathbf Z\xi)=\mathbf 0$), so that $\max_i|\widehat D_i-D^0_i|\le C_Z\|\bm\delta\|_1=\OP(n^{-1/2})$. Let $\mathbf U^{0}$ be the design with $D^0_i$ in place of $\widehat D_i$ and $T^{0}_n(\bm s)=\bm a_0^{\top}(nN)^{-1}\mathbf U^{0\top}\bm\epsilon$. Then}
		\[
		\begin{aligned}
        T_n(\bm s)-T_n^0(\bm s)
        &=(\bm a-\bm a_0)^\top\frac{\mathbf U^{0\top}\bm\epsilon}{nN}+\bm a^\top\frac{(\mathbf U-\mathbf U^0)^\top\bm\epsilon}{nN}
          =:R_{n,1}+R_{n,2}.
        \end{aligned}
		\]
		{For $R_{n,2}$, the only nonzero block of $\mathbf U_{ij}-\mathbf U^{0}_{ij}$ is $(\mathbf Z_i^{\top}\bm\delta)\widetilde{\mathbf B}_1(\bm s_j)$, so}
		\[
		{R_{n,2}=\sum_{\ell=1}^p\delta_\ell\,\bm a_\beta^{\top}\,\frac{1}{nN}\sum_{i=1}^n\sum_{j=1}^N Z_{i\ell}\widetilde{\mathbf B}_1(\bm s_j)\epsilon_{ij},}
		\]
		{and the argument of Lemma~\ref{lem:Ueps-order} applied with the $\mathbf Z$-measurable weights $Z_{i\ell}$ gives, using $N|\triangle|^{\kappa}\to\infty$,}
		\begin{align*}
		{\Big\|\frac{1}{nN}\sum_{i=1}^n\sum_{j=1}^N Z_{i\ell}\widetilde{\mathbf B}_1(\bm s_j)\epsilon_{ij}\Big\|}&{=\OP\big\{n^{-1/2}|\triangle|^{\kappa/2}+(nN)^{-1/2}\big\},}\\
		{|R_{n,2}|}&{=\OP\big\{n^{-1/2}\cdot|\triangle|^{-\kappa}\cdot\big(n^{-1/2}|\triangle|^{\kappa/2}+(nN)^{-1/2}\big)\big\}}\\
		&{=\OP\big\{n^{-1}|\triangle|^{-\kappa/2}\big\}.}
		\end{align*}
		{For $R_{n,1}$, $\bm a-\bm a_0=(\bm\Gamma^{-1}-\bm\Gamma^{0-1})(\widetilde{\mathbf B}_1(\bm s)^{\top},\mathbf 0^{\top})^{\top}$ and $\bm\Gamma^{-1}-\bm\Gamma^{0-1}=\bm\Gamma^{-1}(\bm\Gamma^{0}-\bm\Gamma)\bm\Gamma^{0-1}$, where $\bm\Gamma-\bm\Gamma^{0}=(nN)^{-1}\sum_{i,j}(\mathbf U_{ij}\mathbf U_{ij}^{\top}-\mathbf U^{0}_{ij}\mathbf U^{0\top}_{ij})$ has only its $\bm\theta_\beta$-row and $\bm\theta_\beta$-column blocks nonzero, each of the form $n^{-1}\sum_i c_i\{N^{-1}\sum_j\widetilde{\mathbf B}_1(\bm s_j)\mathbf b_\ell(\bm s_j)^{\top}\}$, where $\mathbf b_\ell$ is the basis vector of the $\ell$th block of $\mathbf U^{0}_{ij}$ ($\widetilde{\mathbf B}_1$ or $\mathbf B_{2,\mathcal A^\ast_\ell}$) and $c_i$ is a product of $\mathbf Z_i^{\top}\bm\delta$ with $Z_{i\ell}$ or with $\widehat D_i+D^0_i$, so that $\max_i|c_i|=\OP(n^{-1/2})$ by Assumption~\ref{A:covariates}. For the Gram factor, $\|N^{-1}\sum_j\bm u_j\bm v_j^{\top}\|\le\|N^{-1}\sum_j\bm u_j\bm u_j^{\top}\|^{1/2}\|N^{-1}\sum_j\bm v_j\bm v_j^{\top}\|^{1/2}$, and each factor is $O(|\triangle|^{\kappa})$ by the Riemann-sum clause of Assumption~\ref{A:tri} and \eqref{eq:spline-stability} (the bound for $\mathbf B_1$ transfers to $\widetilde{\mathbf B}_1=\mathbf Q_2^{\top}\mathbf B_1$ because $\mathbf Q_2$ has orthonormal columns); hence $\|N^{-1}\sum_j\widetilde{\mathbf B}_1(\bm s_j)\mathbf b_\ell(\bm s_j)^{\top}\|\le C|\triangle|^{\kappa}$. Therefore, using Lemma~\ref{lem:Ueps-order} for $\|(nN)^{-1}\mathbf U^{0\top}\bm\epsilon\|$ (its proof is written for the $\mathbf Z$-measurable design $\mathbf U^{0}$),}
		\begin{align*}
		\|\bm\Gamma-\bm\Gamma^0\|
        &=\OP(n^{-1/2}|\triangle|^\kappa),\\
        \|\bm a-\bm a_0\|
        &=\OP(|\triangle|^{-2\kappa}\cdot n^{-1/2}|\triangle|^\kappa)
          =\OP(n^{-1/2}|\triangle|^{-\kappa}),\\
        |R_{n,1}|
        &=\OP\bigl\{n^{-1/2}|\triangle|^{-\kappa}
                   (n^{-1/2}|\triangle|^{\kappa/2}+(nN)^{-1/2})\bigr\}\\
        &=\OP\{n^{-1}|\triangle|^{-\kappa/2}\}.
        \end{align*}
		{Therefore}
		\[
		{\frac{T_n(\bm s)-T^{0}_n(\bm s)}{\sigma_\beta(\bm s)}=\OP\Big\{\frac{n^{-1}|\triangle|^{-\kappa/2}}{n^{-1/2}}\Big\}=\OP\{(n|\triangle|^{\kappa})^{-1/2}\}=o_{\rm P}(1)}
		\]
		{by Assumption~\ref{A:normality}(U3) and (V).}

		\medskip\noindent{\emph{Step 3: conditional central limit theorem for $T^{0}_n(\bm s)$.} Condition on $\mathbf Z_1,\ldots,\mathbf Z_n$ and work on $\mathcal E^{0}_n$; since $\mathcal E^{0}_n$ is $\mathbf Z$-measurable and $\bm\epsilon$ is independent of $(\mathbf Z_1,\ldots,\mathbf Z_n)$ by Assumption~\ref{A:error-subgaussian}, the conditional law of $\bm\epsilon$ is unaffected by this restriction. The weights $w_{ij}=(nN)^{-1}\bm a_0^{\top}\mathbf U^{0}_{ij}$ are then fixed, and $T^{0}_n(\bm s)=\sum_{i=1}^n\chi_{n,i}$ with $\chi_{n,i}=\sum_{j=1}^Nw_{ij}\epsilon_{ij}$. By Assumption~\ref{A:error-subgaussian} (mutual independence of $\mathbf Z_i$, $\{\zeta_{it}\}$ and $\{e_{ij}\}$, and independence across $i$), the $\chi_{n,i}$ are independent with mean zero and variances}
		\begin{align*}
		v_i&=\sum_{j,j'=1}^N w_{ij}w_{ij'}G_\epsilon(\bm s_j,\bm s_{j'})
             =(a_i^\eta)^2+(a_i^\varepsilon)^2,\\
        (a_i^\eta)^2&=\sum_{j,j'=1}^Nw_{ij}w_{ij'}G_\eta(\bm s_j,\bm s_{j'}),\\
        (a_i^\varepsilon)^2&=\sum_{j=1}^Nw_{ij}^2\sigma_\varepsilon^2(\bm s_j),
        \end{align*}
		{and $\sum_{i=1}^nv_i=\bm a_0^{\top}\bm\Psi^{0}_\epsilon\bm a_0=\sigma_\beta^2(\bm s)$ by \eqref{eq:sigma-quadratic}. Moreover,}
		\[
		{\chi_{n,i}=\sum_{t=1}^{\infty}\Big\{\rho_t^{1/2}\sum_{j=1}^Nw_{ij}\psi_t(\bm s_j)\Big\}\zeta_{it}+\sum_{j=1}^Nw_{ij}\sigma_\varepsilon(\bm s_j)e_{ij}}
		\]
		{is a linear combination of the independent unit-variance sub-Gaussian variables $\zeta_{it}$ and $e_{ij}$, whose sub-Gaussian parameters are $\sigma_\zeta^2$ and $\sigma_e^2$; hence $\chi_{n,i}$ is sub-Gaussian with parameter at most $\max(\sigma_\zeta^2,\sigma_e^2)v_i$, and in particular $\E(|\chi_{n,i}|^3\mid\mathbf Z)\le C_3v_i^{3/2}$ for a constant $C_3$ depending only on $\sigma_\zeta$ and $\sigma_e$. By the Berry--Esseen inequality for independent, non-identically distributed summands,}
		\[
		\begin{aligned}
        &\sup_{x\in\mathbb R}\left|
          \Pr\left\{\frac{T_n^0(\bm s)}{\sigma_\beta(\bm s)}\le x
                    \mathrel{\Big|}\mathbf Z\right\}-\Phi(x)\right|\le C\frac{\sum_{i=1}^n\E(|\chi_{n,i}|^3\mid\mathbf Z)}
                           {\sigma_\beta^3(\bm s)}
          \le CC_3\left\{\frac{\max_{1\le i\le n}v_i}
                              {\sigma_\beta^2(\bm s)}\right\}^{1/2}.
        \end{aligned}
		\]
		{To bound the numerator, write $\bm a_0^{\top}\mathbf U^{0}_{ij}=\sum_{\ell=1}^{p+1}X^{0}_{i\ell}g_\ell(\bm s_j)$ with $\bm X^{0}_i=(D^0_i,Z_{i1},\ldots,Z_{ip})^{\top}$, $g_1=\bm a_{0,\beta}^{\top}\widetilde{\mathbf B}_1$ and $g_{\ell+1}=\bm a_{0,\alpha,\ell}^{\top}\mathbf B_{2,\mathcal A^\ast_\ell}$, so that $\max_i|X^{0}_{i\ell}|\le C_Z\max(1,\|\bm\gamma^\ast\|_1)$ (Assumption~\ref{A:covariates}). Then, by $\|G_\eta\|_\infty\le\|\sum_t\rho_t^{1/2}|\psi_t|\|_\infty^2$ and $\sigma_\varepsilon^2\le C_\sigma^2$ (Assumption~\ref{A:covariance}), the design regularity in Assumption~\ref{A:tri}, and \eqref{eq:spline-stability},}
		\begin{align*}
			{\max_i(a^{\eta}_i)^2} &{\le\frac{C}{n^2}\sum_{\ell,\ell'}\Big\{\frac1N\sum_j|g_\ell(\bm s_j)|\Big\}\Big\{\frac1N\sum_{j'}|g_{\ell'}(\bm s_{j'})|\Big\}}\\
			&{\le\frac{C}{n^2}\Big\{\sum_\ell\|g_\ell\|_{L^2(\Omega)}\Big\}^2\le\frac{C}{n^2}|\triangle|^{\kappa}\|\bm a_0\|^2\le Cn^{-2}|\triangle|^{-\kappa},}\\
			{\max_i(a^{\varepsilon}_i)^2} &{\le\frac{C}{n^2N}\sum_\ell\frac1N\sum_jg_\ell^2(\bm s_j)\le\frac{C}{n^2N}|\triangle|^{\kappa}\|\bm a_0\|^2\le Cn^{-2}N^{-1}|\triangle|^{-\kappa},}
		\end{align*}
		{where the middle inequalities use $N^{-1}\sum_j|g_\ell(\bm s_j)|\le\{N^{-1}\sum_jg_\ell^2(\bm s_j)\}^{1/2}\le C\|g_\ell\|_{L^2(\Omega)}$ ($g_\ell^2$ is a nonnegative piecewise polynomial, so the Riemann-sum clause of Assumption~\ref{A:tri} applies) and $\|g_\ell\|^2_{L^2(\Omega)}\le C_B|\triangle|^{\kappa}\|\bm a_{0,\ell}\|^2$. Hence, on $\mathcal E^{0}_n$,}
		\[
		{\frac{\max_iv_i}{\sigma_\beta^2(\bm s)}\le\frac{Cn^{-2}|\triangle|^{-\kappa}(1+N^{-1})}{c_V n^{-1}}=O\big\{(n|\triangle|^{\kappa})^{-1}\big\}\to0}
		\]
		{by Assumption~\ref{A:normality}(U3). The right-hand side is a deterministic sequence, so, writing $\Delta_n(\mathbf Z)=\sup_x|\Pr\{\sigma_\beta^{-1}(\bm s)T^{0}_n(\bm s)\le x\mid\mathbf Z\}-\Phi(x)|$, we have $\Delta_n(\mathbf Z)\mathbf 1_{\mathcal E^{0}_n}\to0$ uniformly: the conditional limit law holds on $\mathcal E^{0}_n$, an event of probability tending to one. Taking expectations, for every $x$,}
		\[
		\begin{aligned}
        &\big|\Pr\{\sigma_\beta^{-1}(\bm s)T_n^0(\bm s)\le x\}-\Phi(x)\big|\le\E\{\Delta_n(\mathbf Z)\mathbf1_{\mathcal E_n^0}\}
             +\Pr\{(\mathcal E_n^0)^c\}\longrightarrow0,
        \end{aligned}
		\]
		{so $\sigma_\beta^{-1}(\bm s)T^{0}_n(\bm s)\to N(0,1)$ unconditionally as well.}

		\medskip\noindent{\emph{Step 4: conclusion.} By \eqref{eq:beta-point-decomp}, Steps 1 and 2,}
		\[
		{\sigma_\beta^{-1}(\bm s)\{\widehat\beta^{o}(\bm s)-\beta^\ast(\bm s)\}=\sigma_\beta^{-1}(\bm s)T^{0}_n(\bm s)+o_{\rm P}(1).}
		\]
		{Slutsky's lemma and Step 3 complete the proof, both conditionally on $\mathbf Z_1,\ldots,\mathbf Z_n$ (in probability) and unconditionally.}
	\end{proof}
	
	\subsubsection{Theoretical Properties of the Hi-FIVE Estimator}

	Using the coefficient partition from the main paper, extend the
	oracle estimator in \eqref{equ:norm-equation} by zero on inactive blocks:
	\begin{equation}
	\label{EQU:scad-est}
	 \widehat{\bm\theta}^{\,\mathrm{orc}}
	 =(\widehat{\bm\theta}_1^{\,o\top},\mathbf0^\top)^\top.
	\end{equation}
	Proposition~\ref{S-Prop:Selection} shows that this candidate is the
	unique global minimizer of \eqref{EQU:objective-IV-2} \chg{under the anchor constraint \eqref{eq:anchor}} with probability
	tending to one. Throughout this subsubsection, write
	$\lambda=\lambda^{\rm S}$ and set
	\[
	 Q_\lambda(\bm\theta)
	 =F(\bm\theta)+\sum_b p_\lambda(\|\bm\theta_b\|),
	 \qquad
	 C_{\mathcal A}=F_{\mathcal A}+c_a\lambda^2|\mathcal A|,
	\]
	where the sum ranges over all invalid-IV blocks.

	\begin{lemma}[Reduction to a block-count penalty]
	\label{LEM:scad-global-reduction}
	Let $\mathbf U_b$ be the submatrix consisting of the columns of $\mathbf U^{\rm f}$
	corresponding to invalid-IV block $b$. If
	\[
	 \max_b\lambda_{\max}\{(nN)^{-1}\mathbf U_b^\top\mathbf U_b\}<k_a,
	\]
	then $Q_\lambda$ has a global minimizer \chg{over coefficient vectors with anchored support} and
	$\min_{\bm\theta}Q_\lambda(\bm\theta)=\min_{\mathcal A}C_{\mathcal A}$\chg{, where both minima are over anchored supports}.
	Every invalid-IV block of a global minimizer has norm either zero or
	at least $a\lambda$. If $C_{\mathcal A}$ has the unique minimizing
	\chg{anchored} support $\mathcal A^\ast$ and $F$ has a unique minimizer on that
	support, this minimizer is the unique global minimizer of $Q_\lambda$ \chg{under the anchor constraint}.
	\end{lemma}

	\begin{proof}
	The SCAD formula is
	\[
	p_\lambda(t)=\begin{cases}
	\lambda t,&0\le t\le\lambda,\\
	\{-t^2+2a\lambda t-\lambda^2\}/\{2(a-1)\},&\lambda<t<a\lambda,\\
	c_a\lambda^2,&t\ge a\lambda.
	\end{cases}
	\]
	It gives the following bound on the whole interval $[0,a\lambda]$:
	\begin{equation}\label{eq:scad-chord}
	 p_\lambda(t)-\frac{c_a\lambda t}{a}\ge k_a t(a\lambda-t).
	\end{equation}
	For $0<t\le\lambda$, the ratio of the left side to
	$t(a\lambda-t)$ is $(a-1)\lambda/\{2a(a\lambda-t)\}\ge k_a$.
	For $\lambda\le t<a\lambda$, this ratio is
	$(a-\lambda/t)/\{2a(a-1)\}\ge1/(2a)>k_a$.

	Fix all coefficients except a block $\bm\theta_b=t\bm u$ with
	$\|\bm u\|=1$ and $0<t<a\lambda$. As a device for comparing two
	deterministic choices, replace this block by zero with probability
	$1-t/(a\lambda)$ and by $a\lambda\bm u$ with probability
	$t/(a\lambda)$. Write $\bm\theta'$ for the result and
	$H=\bm u^\top\mathbf U_b^\top\mathbf U_b\bm u/(nN)$.
	The replacement preserves the mean coefficient vector, so
	\[
	 \E\{Q_\lambda(\bm\theta')\}-Q_\lambda(\bm\theta)
	 =Ht(a\lambda-t)+\frac{c_a\lambda t}{a}-p_\lambda(t)
	 \le(H-k_a)t(a\lambda-t)<0.
	\]
	The expectation here is only over the two replacements, with the data
	fixed. Thus, at least one replacement strictly decreases the objective.
	Apply it once to each block with norm in $(0,a\lambda)$. Other block
	norms do not change. \chg{Each replacement either sets a nonzero block to zero or keeps it nonzero, and zero blocks are untouched, so anchoring is preserved.} After finitely many steps the resulting vector
	$\overline{\bm\theta}$ has only zero or SCAD-flat blocks, and, writing
	$\mathcal A$ for its actual support,
	\[
	 Q_\lambda(\bm\theta)\ge Q_\lambda(\overline{\bm\theta})
	 =F(\overline{\bm\theta})+c_a\lambda^2|\mathcal A|
	 \ge C_{\mathcal A}\ge\min_{\mathcal B}C_{\mathcal B}.
	\]
	The first inequality is strict if any block was replaced.
	Conversely, take a minimizing support $\mathcal A_0$ in the finite
	collection of \chg{anchored} supports, and a quadratic refit $\bm\theta^0$ on it\chg{, which satisfies the anchor constraint}.
	Since $0\le p_\lambda(t)\le c_a\lambda^2$,
	$Q_\lambda(\bm\theta^0)\le C_{\mathcal A_0}$.
	This proves equality of the two minimum values and attainment.
	Every global minimizer is already zero or flat on each block and
	therefore attains $C_{\mathcal A}$ on its actual support. Unique
	minimization over supports, followed by uniqueness of the refit on
	that support, proves the last assertion.
	\end{proof}

	\begin{lemma}[Uniform comparison of supports]
	\label{LEM:global-support}
	Under Assumptions~\ref{A:tri}--\ref{AssumpC:signal},
	with probability tending to one,
	\[
	 C_{\mathcal A}>C_{\mathcal A^\ast}\qquad
	 \text{for every \chg{anchored} }\mathcal A\neq\mathcal A^\ast.
	\]
	\end{lemma}

	\begin{proof}
	\chg{We work on the event $\{\widehat{\mathcal I}_{\tau_\gamma}=\mathcal I\}$, whose probability tends to one; on it, the supports admitted by \eqref{eq:anchor} are exactly the anchored supports, and $\mathcal A^\ast$ is one of them by Assumption~\ref{A:true-alpha}.}

	\noindent\emph{Step 1: a common bound for the projected error.}
	Let $\mathcal T$ be the column space of
	$\mathbf Z\otimes\mathbb B_2$, with subjects stacked first.
	Nesting of the spline spaces implies that every fitted vector
	$\mathbf U^{\rm f}\bm\theta$ lies in $\mathcal T$.
	Let $\Pi_{\mathcal T}$ be its orthogonal projector and put
	\[
	 m_{ij}=\widehat D_i\widetilde\beta(\bm s_j)
	       +\mathbf Z_i^\top\widetilde{\bm\alpha}(\bm s_j),
	 \quad \mathbf w=\Pi_{\mathcal T}(\mathbf Y-\mathbf m),
	 \quad Z_n=\frac{\|\mathbf w\|^2}{nN}.
	\]
	Here $\mathbf m\in\mathcal T$. The model gives
	\[
	 \begin{aligned}
    Y_{ij}-m_{ij}
      &=(D_i-\widehat D_i)\widetilde\beta(\bm s_j)+b_{ij}+\epsilon_{ij},\\
    b_{ij}
      &=D_i(\beta^\ast-\widetilde\beta)(\bm s_j)
       +\mathbf Z_i^\top(\bm\alpha^\ast-\widetilde{\bm\alpha})(\bm s_j).
    \end{aligned}
	\]
	The first term is orthogonal to $\mathcal T$, since
	$\mathbf Z^\top(\mathbf D-\widehat{\mathbf D})=0$.
	The approximation bounds and the finite second moments of the
	covariates give $\|\mathbf b\|^2/(nN)=\OP(|\triangle|^{2d+2})$.
	For the remaining noise, write
	$\Pi_{\mathcal T}=\Pi_{\mathbf Z}\otimes\Pi_{\mathbb B_2}$.
	Conditional on $\mathbf Z$, the noise has mean zero and covariance
	$\mathbf I_n\otimes\mathbf G_\epsilon$, by
	Assumptions~\ref{A:error-subgaussian}--\ref{A:covariance}. Hence
	\begin{align*}
	 \E\left\{\frac{\|\Pi_{\mathcal T}\bm\epsilon\|^2}{nN}
	       \mathrel{\Big|}\mathbf Z\right\}
	 &=\frac{\tr(\Pi_{\mathbf Z})\tr(\Pi_{\mathbb B_2}\mathbf G_\epsilon)}{nN}\le\frac{p\,\tr(\mathbf G_\epsilon)}{nN}\le\frac{C}{n}.
	\end{align*}
	Markov's inequality and the bound for $\mathbf b$ show that
	\begin{equation}\label{eq:global-noise}
	 Z_n=\OP(n^{-1}+|\triangle|^{2d+2}).
	\end{equation}
	This argument conditions only on $\mathbf Z$, not on the estimated
	first stage, and controls all supports through the same projector.

	\medskip\noindent\emph{Step 2: compare the noiseless losses.}
	For every support define
	\[
	 D_{\mathcal A}=\inf_{(\beta,\bm\alpha)\in\Theta_{\mathcal A}}
	 \left[\frac1{nN}\sum_{i,j}
	 \{m_{ij}-\widehat D_i\beta(\bm s_j)-\mathbf Z_i^\top\bm\alpha(\bm s_j)\}^2
	 +\rho_n\mathcal E(\beta)\right].
	\]
	The first-stage estimator satisfies
	$\widehat{\bm\gamma}-\bm\gamma^\ast=\OP(n^{-1/2})$.
	Also, $n^{-1}\mathbf Z^\top\mathbf Z$ converges to the positive
	definite $\bm\Sigma_{\mathbf Z}$. On an event whose probability tends
	to one, $\|\widehat{\bm\gamma}-\bm\gamma^\ast\|\le\upsilon_n$ and
	$n^{-1}\mathbf Z^\top\mathbf Z\succeq\bm\Sigma_{\mathbf Z}/2$.
	For any competing coefficients, the corresponding reduced-form
	difference is polynomial of degree at most $d$ on each simplex.
	Apply the uniform quadrature condition in Assumption~\ref{A:tri} to
	its squared $\bm\Sigma_{\mathbf Z}$ norm. For all large $n$, the
	empirical squared norm is at least one fourth of its population
	integral, simultaneously over every support and coefficient vector.
	The roughness term is nonnegative and unchanged. Thus
	Assumption~\ref{AssumpC:signal} yields, on this same event,
	\begin{equation}\label{eq:global-gap-empirical}
	 D_{\mathcal A}\ge\tfrac14(1+s_n^\ast-|\mathcal A|)\Delta_n^{\rm G},
	 \qquad \mathcal A\neq\mathcal A^\ast\ \text{\chg{anchored}},\quad |\mathcal A|\le s_n^\ast.
	\end{equation}

	\medskip\noindent\emph{Step 3: compare all model sizes.}
	Let $B_n=\|(\mathbf I-\Pi_{\mathcal T})\mathbf Y\|^2/(nN)$.
	For every feasible coefficient vector, orthogonality gives
	\[
	 F(\bm\theta)=B_n+
	 \frac{\|\mathbf m-\mathbf U^{\rm f}\bm\theta+\mathbf w\|^2}{nN}
	 +\rho_n\mathcal E(\beta).
	\]
	Using $\|x+w\|^2\ge\tfrac12\|x\|^2-\|w\|^2$ and
	$\mathcal E(\beta)\ge0$, and evaluating the oracle-support loss at
	$(\widetilde\beta,\widetilde{\bm\alpha})$, we obtain
	\begin{equation}\label{eq:global-refit-bounds}
\begin{aligned}
 F_{\mathcal A}&\ge B_n+\tfrac12D_{\mathcal A}-Z_n,\\
 F_{\mathcal A}&\ge B_n,\\
 F_{\mathcal A^\ast}&\le B_n+Z_n+\rho_n\mathcal E(\widetilde\beta).
\end{aligned}
\end{equation}
	Lemma~\ref{LEM:roughness} gives $\mathcal E(\widetilde\beta)=O(1)$.
	Set $R_n=2Z_n+\rho_n\mathcal E(\widetilde\beta)=\OP(e_{n,N})$.
	By \eqref{eq:global-tuning}, $R_n=o_{\rm P}(\lambda^2)$.
	If $|\mathcal A|>s_n^\ast$, then
	\[
	 C_{\mathcal A}-C_{\mathcal A^\ast}
	 \ge c_a\lambda^2(|\mathcal A|-s_n^\ast)-R_n>0
	\]
	simultaneously for all such supports, with probability tending to one.
	If $\mathcal A\neq\mathcal A^\ast$ \chg{is anchored} and $|\mathcal A|\le s_n^\ast$,
	put $d_{\mathcal A}=s_n^\ast-|\mathcal A|\ge0$. Equations
	\eqref{eq:global-gap-empirical}--\eqref{eq:global-refit-bounds} give
	\[
	 C_{\mathcal A}-C_{\mathcal A^\ast}
	 \ge\tfrac18(1+d_{\mathcal A})\Delta_n^{\rm G}
	       -c_a\lambda^2d_{\mathcal A}-R_n>0
	\]
	simultaneously, since $\lambda^2=o(\Delta_n^{\rm G})$.
	Both bounds use the same random $R_n$; no union bound over models is
	needed. If $s_n^\ast=0$, only the first comparison is needed.
	\end{proof}

	\begin{proposition}
	\label{S-Prop:Selection}
	Suppose Assumptions~\ref{A:tri}--\ref{AssumpC:signal}
	hold. Then the oracle candidate in \eqref{EQU:scad-est} satisfies
	\[
	 \Pr\left\{\widehat{\bm\theta}^{\,\mathrm{orc}}
	 \text{ is the unique global minimizer of }
	 \eqref{EQU:objective-IV-2}\ \text{\chg{under \eqref{eq:anchor}}}\right\}\longrightarrow1.
	\]
	Consequently, every global Hi-FIVE estimator satisfies
	$\widehat{\mathcal A}=\mathcal A^\ast$,
	$\widehat{\bm\theta}_1=\widehat{\bm\theta}^{\,o}_1$ and
	$\widehat{\bm\theta}_2=\mathbf0$ on this event. In particular,
	$\widehat\beta(\bm s)=\widehat\beta^o(\bm s)$ for all $\bm s\in\Omega$.
	\end{proposition}

	\begin{proof}[Proof of Proposition~\ref{S-Prop:Selection}]
	For $b=(\ell,k)$ and a unit vector $\bm u$ in that block,
	\[
	 \frac{\|\mathbf U_b\bm u\|^2}{nN}
	 =\left(\frac1n\sum_i Z_{i\ell}^2\right)
	  \left(\frac1N\sum_{\bm s_j\in T_k}
	       \{\mathbf B_{2,k}(\bm s_j)^\top\bm u\}^2\right)
	 \le C |\triangle|^\kappa.
	\]
	Here, the constant is uniform over $b$ by bounded covariates,
	Assumption~\ref{A:tri}, and spline stability. There is no roughness
	penalty on an invalid-IV block. Since $|\triangle|\to0$, the curvature condition
	of Lemma~\ref{LEM:scad-global-reduction} holds for all large $n$.
	Lemma~\ref{LEM:global-support} makes $\mathcal A^\ast$ the unique
	minimizing \chg{anchored} support, with probability tending to one. \chg{By Assumption~\ref{A:true-alpha}, $\mathcal A^\ast$ is anchored, so the oracle candidate satisfies \eqref{eq:anchor} on the event $\{\widehat{\mathcal I}_{\tau_\gamma}=\mathcal I\}$.} On that event
	intersected with the event of Lemma~\ref{LEM:Gamma_0}(b), the refit on
	$\mathcal A^\ast$ is unique and equals the oracle estimator. Applying
	Lemma~\ref{LEM:scad-global-reduction} proves global uniqueness.
	Since $\mathcal A^\ast$ is the unique minimizing support, the global
	minimizer has actual support $\mathcal A^\ast$, so every active block of the
	oracle refit is nonzero and, by Lemma~\ref{LEM:scad-global-reduction}, lies on
	the flat part of SCAD; no separate beta-min or tuning condition is needed.
	The bijection through $\mathbf Q_2$ transfers uniqueness back
	to the original constrained coefficients.
	\end{proof}

	\begin{proof}[Proof of Theorems~\ref{THM:beta-convergence} and~\ref{THM:beta-normality}]
	By Proposition~\ref{S-Prop:Selection}, the global estimator equals the
	oracle with probability tending to one. Write
	$E_n=\{\widehat{\bm\theta}=\widehat{\bm\theta}^{\,\mathrm{orc}}\}$.
	For any $t>0$,
	\[
	 \Pr\{\|\widehat\beta-\beta^\ast\|_{L^2}>t\}
	 \le\Pr\{\|\widehat\beta^o-\beta^\ast\|_{L^2}>t\}+\Pr(E_n^c).
	\]
	The rate in Theorem~\ref{THM:beta-convergence-oracle} therefore
	transfers to $\widehat\beta$. At the fixed location of
	Assumption~\ref{A:normality}, the distribution functions of
	$\sigma_\beta^{-1}(\bm s)(\widehat\beta-\beta^\ast)(\bm s)$ and
	$\sigma_\beta^{-1}(\bm s)(\widehat\beta^o-\beta^\ast)(\bm s)$ differ
	by at most $\Pr(E_n^c)$, using the same oracle asymptotic variance.
	The limit law in Theorem~\ref{LEM:beta-normality-oracle} thus transfers
	as well.
	\end{proof}

	\subsubsection{Choosing the parameter for region-wise IV identification by mBIC}
\label{sec:mBIC}

Throughout this subsubsection, $\lambda^{\rm R}>0$ is a deterministic
sequence, held fixed when comparing {values of the parameter for region-wise IV identification}. \chg{Every fit is computed under the anchor constraint \eqref{eq:anchor}, so every fitted support and every support compared with $\mathcal A^\ast$ is anchored; auxiliary supersets such as $\mathcal A\cup\mathcal A^\ast$, which enter only through refit losses, need not be. We work on the event $\{\widehat{\mathcal I}_{\tau_\gamma}=\mathcal I\}$.}
We use the loss $F$ and its support-restricted minimum $F_{\mathcal A}$
from \eqref{eq:global-refit}. For the mBIC penalty with $a_0\in(0,1)$, write
\[
\begin{aligned}
 P_0&=pK_2,&
 c_n&=\log q_n+2a_0\log P_0,\\
 v_n^0&=\frac{c_n}{q_n},&
 v_n&=\frac{\log\widehat q_n+2a_0\log P_0}{\widehat q_n}.
\end{aligned}
\]
We use the same pilot estimate $\widehat q_n$ at every grid point. The
criterion at a quadratic refit is
\[
 \operatorname{pISE}_{\mathcal A}=F_{\mathcal A},\qquad
 \operatorname{mBIC}_{\mathcal A}=\log(F_{\mathcal A})+v_n|\mathcal A|.
\]
The results below also hold if $q_n$ replaces $\widehat q_n$ throughout.
For a fitted coefficient vector $\widehat{\bm\theta}_{\lambda^{\rm S}}$,
let $\mathcal A_{\lambda^{\rm S}}$ be its actual nonzero block support,
and define
\[
 \operatorname{pISE}_{\lambda^{\rm S}}
 =F(\widehat{\bm\theta}_{\lambda^{\rm S}}),\qquad
 \operatorname{mBIC}_{\lambda^{\rm S}}
 =\log(\operatorname{pISE}_{\lambda^{\rm S}})
   +v_n|\mathcal A_{\lambda^{\rm S}}|.
\]
These definitions distinguish the criterion at a quadratic refit from
the criterion at the fit returned at a grid point.

For the model-size comparisons below, write
$\mathcal G_n=\{\bm g:\|\bm g-\bm\gamma^\ast\|\le\upsilon_n\}$,
where $\upsilon_n\to0$ and $\sqrt n\upsilon_n\to\infty$.
When Assumption~\ref{AssumpC:signal} is imposed, use its sequence $\upsilon_n$.
For a support $\mathcal A$, let $\mathbf U_{\mathcal A}(\bm g)$ be its
design in reduced coordinates with fitted exposure $\mathbf Z\bm g$. Let
$\mathbf R_{\mathcal A}=(\mathbf W_1^{1/2},\mathbf0)$, using the same
number of roughness rows for every support, and set
\[
 \mathbf X_{\mathcal A}(\bm g)=
 \begin{pmatrix}\mathbf U_{\mathcal A}(\bm g)\\
 \sqrt{\lambda^{\rm R}}\mathbf R_{\mathcal A}\end{pmatrix},
 \quad \mathbf J\mathbf v=(\mathbf v^\top,\mathbf0^\top)^\top.
\]
Let $\mathbf P_{\mathcal A}(\bm g)$ be the orthogonal projector onto
$\operatorname{col}\{\mathbf X_{\mathcal A}(\bm g)\}$ and put
$\mathbf S_{\mathcal A}(\bm g)=\mathbf J^\top\mathbf P_{\mathcal A}(\bm g)\mathbf J$.
These projectors exist even when the coefficient design is singular.
Write $\bm\Sigma_\epsilon=\mathbf I_n\otimes\mathbf G_\epsilon$.

	\paragraph{Pilot residuals and effective sample size.}
	\label{SEC:eff-sample}
	Let
	$(\widehat\beta,
	\widehat{\bm\alpha})$
	denote a common pilot fit, held fixed over the selection grid.
	The residuals use the observed exposure $D_i$:

	\[
	\widehat R_i(\bm s_j)=Y_i(\bm s_j)-D_i\widehat\beta(\bm s_j)
	-\mathbf Z_i^\top\widehat{\bm\alpha}(\bm s_j),
	\qquad i=1,\ldots,n,\ j=1,\ldots,N.
	\]
	We estimate the covariance and correlation by
	\begin{equation}
		\label{EQU:G-epsilon}
        \begin{aligned}
        \widehat G_\epsilon(\bm s_j,\bm s_{j'})
        &=\frac1n\sum_{i=1}^n
          \widehat R_i(\bm s_j)\widehat R_i(\bm s_{j'}),\\
        \widehat\rho(\bm s_j,\bm s_{j'})
        &=\frac{\widehat G_\epsilon(\bm s_j,\bm s_{j'})}
          {\{\widehat G_\epsilon(\bm s_j,\bm s_j)
             \widehat G_\epsilon(\bm s_{j'},\bm s_{j'})\}^{1/2}}.
        \end{aligned}
	\end{equation}
	The estimates $\widehat N_{\mathrm{eff},k}$,
	$\widehat{\bar N}_{\mathrm{eff}}$, and $\widehat q_n$ are obtained from
	$N_{\mathrm{eff},k}$, $\bar N_{\mathrm{eff}}$, and $q_n$ by replacing $\rho_\epsilon$ with
	$\widehat\rho$. The pilot error $\varrho_n$ is given in
	\eqref{DEF:pilot}.
	
	\begin{corollary}
		\label{COR:pilot-sup}
		Under Assumptions~\ref{A:tri}--\ref{A:true-alpha}, suppose the pilot has $\widehat\beta\in\mathcal S_d^r(\triangle_1)$ and $\widehat\alpha_\ell\in\mathcal S_d^{-1}(\triangle_2)$ and satisfies
		$\|\widehat\beta-\beta^\ast\|_{L^2(\Omega)}
		+\max_{1\le\ell\le p}\|\widehat\alpha_\ell-\alpha^\ast_\ell\|_{L^2(\Omega)}
		=O_{\mathrm P}(a_n)$. Then
		$\varrho_n=O_{\mathrm P}\bigl\{|\triangle|^{-\kappa/2}\bigl(a_n+|\triangle|^{d+1}\bigr)\bigr\}$.
	\end{corollary}
	\begin{proof}
	Use the approximants $\widetilde\beta$ and $\widetilde\alpha_\ell$
	from Section~\ref{SEC:Pre}. Assumptions~\ref{A:true-coef}
	and~\ref{A:true-alpha} give sup-norm and $L^2$ approximation errors
	of order $|\triangle|^{d+1}$. Apply
	$\|g\|_\infty\le C|\triangle|^{-\kappa/2}\|g\|_{L^2(\Omega)}$
	to the spline differences $\widehat\beta-\widetilde\beta$ and
	$\widehat\alpha_\ell-\widetilde\alpha_\ell$, using quasi-uniformity
	and the fixed refinement level in Assumption~\ref{A:tri}.
	The triangle inequality gives the result.
	\end{proof}

	\begin{lemma}
		\label{LEM:Neff-consistency}
		Suppose Assumptions~\ref{A:tri}, \ref{A:error-subgaussian}--\ref{A:covariates} hold for the common pilot. Then
		\[
		\Delta_n:=\max_{1\le j,j'\le N}\bigl|\widehat\rho(\bm s_j,\bm s_{j'})
		-\rho_\epsilon(\bm s_j,\bm s_{j'})\bigr|
		=O_{\mathrm P}\Bigl(\sqrt{\log N/n}+\varrho_n\Bigr),
		\]
		and consequently
		$\widehat{\bar N}_{\mathrm{eff}}/\bar N_{\mathrm{eff}}\xrightarrow{\mathrm{P}}1$ and
		$\widehat q_n/q_n\xrightarrow{\mathrm{P}}1$.
	\end{lemma}
	
	\begin{proof}
		Write $\widehat R_i(\bm s_j)=\epsilon_i(\bm s_j)+b_i(\bm s_j)$ with
		\[
		b_i(\bm s_j)=D_i\{\beta^\ast(\bm s_j)-\widehat\beta(\bm s_j)\}
		+\mathbf Z_i^\top\{\bm\alpha^\ast(\bm s_j)-\widehat{\bm\alpha}(\bm s_j)\},
		\]
		and let $\bar G(\bm s_j,\bm s_{j'})=n^{-1}\sum_i\epsilon_i(\bm s_j)\epsilon_i(\bm s_{j'})$.
		Then
		\begin{equation}
			\label{EQU:Ghat-decomp}
            \begin{aligned}
            \widehat G_\epsilon-G_\epsilon
            ={}&(\bar G-G_\epsilon)+\frac1n\sum_{i=1}^n\bigl\{
              \epsilon_i(\bm s_j)b_i(\bm s_{j'})
              +b_i(\bm s_j)\epsilon_i(\bm s_{j'})
              +b_i(\bm s_j)b_i(\bm s_{j'})\bigr\}.
            \end{aligned}
		\end{equation}
		
		\emph{Step 1 (stochastic term).}
By Assumptions~\ref{A:error-subgaussian}--\ref{A:covariance}, write
$\epsilon_{ij}=\sum_{t\ge1}\rho_t^{1/2}\psi_t(\bm s_j)\zeta_{it}
+\sigma_\varepsilon(\bm s_j)e_{ij}$. Independence of the latent
variables gives, for every $u\in\mathbb R$,
\begin{equation}
\label{eq:mbic-subgaussian}
 \E\{\exp(u\epsilon_{ij})\}
 \le \exp\!\left[\frac{u^2}{2}
   \{\sigma_\zeta^2G_\eta(\bm s_j,\bm s_j)
       +\sigma_e^2\sigma_\varepsilon^2(\bm s_j)\}\right]
 \le \exp(u^2K_\epsilon^2/2),
\end{equation}
where $K_\epsilon^2=\sigma_\zeta^2C_G+\sigma_e^2C_\sigma^2$ is
independent of $i,j,n,N$. The first inequality follows for finite
Karhunen--Lo\`eve sums by multiplying their moment-generating bounds;
$L^2$ convergence and Fatou's lemma along an almost surely convergent
subsequence extend it to the full sum.
Consequently,
\[
 \Pr\{|\epsilon_{ij}|>t\}\le2\exp\{-t^2/(2K_\epsilon^2)\},
 \qquad \sup_{i,j}\E|\epsilon_{ij}|^4\le16K_\epsilon^4.
\]
Spatial independence within a subject is not required.

        Products $\epsilon_{ij}\epsilon_{ij'}$ are uniformly
        sub-exponential. Independence across subjects, Bernstein's
        inequality and a union bound over the $N^2$ pairs give
        \begin{equation}
        \label{eq:mbic-covariance-tail}
        \Pr\!\left\{\max_{j,j'}|\bar G(\bm s_j,\bm s_{j'})
                       -G_\epsilon(\bm s_j,\bm s_{j'})|>t\right\}
        \le2N^2\exp\{-c_1n\min(t^2,t)\},\quad t>0.
        \end{equation}
        Since $\log N=o(n)$ by Assumption~\ref{A:covariance},
        $\|\bar G-G_\epsilon\|_\infty=\OP(\sqrt{\log N/n})$.

		\emph{Step 2 (plug-in term).}
		By Assumption~\ref{A:covariates} and $|Z_{i\ell}|\le C_\ell$,
		$n^{-1}\sum_i D_i^2=O_{\mathrm P}(1)$ and
		$n^{-1}\sum_i\|\mathbf Z_i\|_1^2=O_{\mathrm P}(1)$, so
		\[
        \begin{aligned}
        \max_{1\le j\le N}\frac1n\sum_{i=1}^n b_i(\bm s_j)^2
        \le{}&2\Bigl(\frac1n\sum_i D_i^2\Bigr)
              \|\widehat\beta-\beta^\ast\|_\infty^2+2\Bigl(\frac1n\sum_i\|\mathbf Z_i\|_1^2\Bigr)
              \max_\ell\|\widehat\alpha_\ell-\alpha^\ast_\ell\|_\infty^2\\
        ={}&O_{\mathrm P}(\varrho_n^2).
        \end{aligned}
		\]
		Step 1 applied with $j=j'$ and Assumption~\ref{A:covariance} give
		$\max_j n^{-1}\sum_i\epsilon_i(\bm s_j)^2=O_{\mathrm P}(1)$. Cauchy--Schwarz then bounds
		each cross term in \eqref{EQU:Ghat-decomp} by $O_{\mathrm P}(\varrho_n)$ and the
		quadratic term by $O_{\mathrm P}(\varrho_n^2)$, uniformly over $(j,j')$. Hence
		$\|\widehat G_\epsilon-G_\epsilon\|_\infty=O_{\mathrm P}(\sqrt{\log N/n}+\varrho_n)$.
		
		\emph{Step 3 (covariance to correlation).}
		By Assumption~\ref{A:covariance}, $c_\epsilon\le G_\epsilon(\bm s,\bm s)\le C_\epsilon$, so on the
		event $\|\widehat G_\epsilon-G_\epsilon\|_\infty\le c_\epsilon/2$, which has probability
		tending to one, all diagonal entries of $\widehat G_\epsilon$ lie in
		$[c_\epsilon/2,\,2C_\epsilon]$. The map
		$(a,b,c)\mapsto a/\sqrt{bc}$ is Lipschitz on
		$[-2C_\epsilon,2C_\epsilon]\times[c_\epsilon/2,2C_\epsilon]^2$, whence
		\[
		\Delta_n\le L\|\widehat G_\epsilon-G_\epsilon\|_\infty
		=O_{\mathrm P}\Bigl(\sqrt{\log N/n}+\varrho_n\Bigr).
		\]
		
		\emph{Step 4 (transfer to $N_{\mathrm{eff}}$).}
		Let $S_k=\sum_{j,j'}I(\bm s_j,\bm s_{j'}\in T_k)\rho_\epsilon(\bm s_j,\bm s_{j'})$ and
		$\widehat S_k$ its estimate, so $N_{\mathrm{eff},k}=m_k^2/S_k$ and
		$\widehat N_{\mathrm{eff},k}=m_k^2/\widehat S_k$. Since $S_k$ has $m_k^2$ summands,
		\[
		|\widehat S_k-S_k|\le m_k^2\Delta_n,
		\qquad\text{hence}\qquad
		\Bigl|\frac{\widehat S_k}{S_k}-1\Bigr|\le\frac{m_k^2}{S_k}\Delta_n
		=N_{\mathrm{eff},k}\Delta_n\le N^{+}_{\mathrm{eff}}\Delta_n=:\delta_n .
		\]
		By Assumption~\ref{A:covariance}, $\delta_n\xrightarrow{\mathrm{P}}0$. On $\{\delta_n\le1/2\}$,
		\[
		\max_{1\le k\le K_2}
		\Bigl|\frac{\widehat N_{\mathrm{eff},k}}{N_{\mathrm{eff},k}}-1\Bigr|
		=\max_k\Bigl|\frac{S_k}{\widehat S_k}-1\Bigr|
		\le\frac{\delta_n}{1-\delta_n}\xrightarrow{\mathrm{P}}0 .
		\]
		Since $\bar N_{\mathrm{eff}}$ and $\widehat{\bar N}_{\mathrm{eff}}$ are averages of
		$N_{\mathrm{eff},k}$ and $\widehat N_{\mathrm{eff},k}$ over $k$, with all terms positive,
		\[
		\Bigl|\frac{\widehat{\bar N}_{\mathrm{eff}}}{\bar N_{\mathrm{eff}}}-1\Bigr|
		\le\max_k\Bigl|\frac{\widehat N_{\mathrm{eff},k}}{N_{\mathrm{eff},k}}-1\Bigr|
		\xrightarrow{\mathrm{P}}0,
		\]
		and $\widehat q_n/q_n=\widehat{\bar N}_{\mathrm{eff}}/\bar N_{\mathrm{eff}}\xrightarrow{\mathrm{P}}1$.
	\end{proof}

\paragraph{Loss bounds and comparison of supports.}
\begin{lemma}[Bounds for the refitted losses and the mBIC penalty]
\label{LEM:mbic-loss-bounds}
Under Assumptions~\ref{A:tri}--\ref{A:size}, there is a constant $b_0>0$
such that
\[
 \Pr\{\inf_{\bm\theta}F(\bm\theta)\ge b_0\}\longrightarrow1,
 \qquad
 \sup_{\mathcal A}F_{\mathcal A}\le F(\mathbf0)=\OP(1),
 \qquad
 v_n=o_{\rm P}(n^{-1}).
\]
In particular, all refitted and fitted mBIC criteria are well defined
with probability tending to one.
\end{lemma}

\begin{proof}
Let $H_N$ be the
orthogonal projector onto the complement of the span of the columns of $\mathbb B_2$
and the $p+1$ sampled true coefficient functions
$\beta^*,\alpha_1^*,\ldots,\alpha_p^*$. These are deterministic spatial
vectors. Since $d$ and $p$ are fixed,
\[
 \operatorname{rank}(H_N)\ge N-CK_2-(p+1),\qquad
 \frac{\operatorname{rank}(H_N)}{N}\longrightarrow1;
\]
the last statement follows from $N |\triangle|^\kappa\to\infty$ in
Assumption~\ref{A:size} and $K_2\asymp |\triangle|^{-\kappa}$. Nesting places
every fitted spatial vector in $\operatorname{col}(\mathbb B_2)$.
The true mean for subject $i$ is
$D_i\beta^*+\sum_\ell Z_{i\ell}\alpha_\ell^*$, so $H_N$ annihilates
it as well. Consequently, for every coefficient vector,
\[
 F(\bm\theta)\ge\frac1{nN}\sum_{i=1}^n
                   \|H_N\bm\epsilon_i\|^2.
\]
The independent measurement errors in
Assumptions~\ref{A:error-subgaussian}--\ref{A:covariance} give
$\mathbf G_\epsilon\succeq c_\sigma^2\mathbf I_N$. Thus
\[
 \E\left\{\frac{\|H_N\bm\epsilon_i\|^2}{N}\right\}
 =\frac{\operatorname{tr}(H_N\mathbf G_\epsilon)}{N}
 \ge c_\sigma^2\frac{\operatorname{rank}(H_N)}{N}.
\]
Equation~\eqref{eq:mbic-subgaussian} gives
$\sup_{i,j}\E|\epsilon_{ij}|^4\le16K_\epsilon^4$.
Since $H_N$ is a contraction,
\[
 \E\left\{\frac{\|H_N\bm\epsilon_i\|^4}{N^2}\right\}
 \le\frac1{N^2}\sum_{j,j'}
       \E\{\epsilon_i^2(\bm s_j)\epsilon_i^2(\bm s_{j'})\}\le C.
\]
Independence across subjects and Chebyshev's inequality now prove the
first assertion, for example with $b_0=c_\sigma^2/2$.
This argument does not condition on the estimated first stage.
The zero coefficient vector is feasible on every support, and the
bounded true coefficient functions and second moments of the
covariates and errors give $\E\{F(\mathbf0)\}\le C$. This proves the
second assertion.

For the last assertion, the covariance matrix is
positive semidefinite and the independent nugget makes each
within-simplex correlation sum positive. Since each correlation is
at most one, $N_{\mathrm{eff},k}\ge1$, and hence $q_n\ge nK_2$.
For all sufficiently large $n$, $x\mapsto(\log x)/x$ is decreasing
on $[nK_2,\infty)$, so
\[
 n v_n^0\le
 \frac{\log(nK_2)+2a_0\log(pK_2)}{K_2}\longrightarrow0.
\]
Indeed, $n^{1/2}|\triangle|^{d+1}\to0$ implies
$K_2/n^{\kappa/(2d+2)}\to\infty$, while
$\log K_2/K_2\to0$.
Lemma~\ref{LEM:Neff-consistency} gives
$\widehat q_n/q_n\to1$ in probability under the pilot condition in
Assumption~\ref{A:covariance}.
It follows that $v_n/v_n^0\to1$ in probability, proving the result.
\end{proof}

\begin{lemma}[Comparison with supports of no larger size]
\label{LEM:mbic-refit-ordering}
Suppose Assumptions~\ref{A:tri}--\ref{AssumpC:signal} hold at a
deterministic value of $\lambda^{\rm S}$. Then, with probability
tending to one,
\begin{equation}
\label{eq:mbic-refit-ordering}
 \operatorname{mBIC}_{\mathcal A^\ast}
 <\operatorname{mBIC}_{\mathcal A}
 \quad\text{for every \chg{anchored} }\mathcal A\ne\mathcal A^\ast
       \text{ with }|\mathcal A|\le s_n^\ast.
\end{equation}
\end{lemma}

\begin{proof}
If $s_n^\ast=0$, the comparison family is empty. Otherwise, let
$d_{\mathcal A}=s_n^\ast-|\mathcal A|\ge0$ and use the same random
$R_n=\OP(e_{n,N})$ as in Lemma~\ref{LEM:global-support}.
Equations~\eqref{eq:global-gap-empirical}--\eqref{eq:global-refit-bounds}
give, simultaneously for all supports in~\eqref{eq:mbic-refit-ordering},
\[
 F_{\mathcal A}-F_{\mathcal A^\ast}
 \ge\tfrac18(1+d_{\mathcal A})\Delta_n^{\rm G}-R_n.
\]
Assumption~\ref{AssumpC:signal} implies
$e_{n,N}=o(\Delta_n^{\rm G})$, so the right-hand side is at least
$(1+d_{\mathcal A})\Delta_n^{\rm G}/16$ on a single event whose
probability tends to one. By Lemma~\ref{LEM:mbic-loss-bounds}, on
another such event all these losses are positive. Write
$M_n=F(\mathbf0)=\OP(1)$. Since $F_{\mathcal A}\le M_n$, integrating
$1/x$ between the two losses gives
\begin{align*}
 \operatorname{mBIC}_{\mathcal A}
 -\operatorname{mBIC}_{\mathcal A^\ast}
 &\ge\frac{F_{\mathcal A}-F_{\mathcal A^\ast}}{M_n}
          -d_{\mathcal A}v_n\ge\frac{(1+d_{\mathcal A})\Delta_n^{\rm G}}{16M_n}
          -d_{\mathcal A}v_n>0.
\end{align*}
The last inequality holds uniformly with probability tending to one,
because $v_n=o_{\rm P}(n^{-1})$ and
$n^{-1}\le e_{n,N}=o(\Delta_n^{\rm G})$ imply
$M_nv_n=o_{\rm P}(\Delta_n^{\rm G})$.
There is no union bound over supports.
\end{proof}

\begin{proposition}[Consistency of mBIC on a global SCAD path]
\label{Prop:mBIC}
Fix a deterministic roughness-parameter sequence $\lambda^{\rm R}>0$ and let
$\Lambda_n\subset(0,\infty)$ be a finite deterministic grid. The grid
may depend on $n,N$. Suppose it contains a deterministic point
$\widetilde\lambda^{\rm S}$ at which
Assumptions~\ref{A:tri}--\ref{AssumpC:signal} hold, and that
\begin{equation}
\label{eq:mbic-grid-lower}
 e_{n,N}=o\{(\lambda_{\min}^{\rm S})^2\},
 \qquad \lambda_{\min}^{\rm S}=\min\Lambda_n.
\end{equation}
At every $\lambda^{\rm S}\in\Lambda_n$, let
$\widehat{\bm\theta}_{\lambda^{\rm S}}$ be a global minimizer of
\eqref{EQU:objective-IV-2} \chg{under \eqref{eq:anchor}}, with the same $\lambda^{\rm R}$ and pilot
estimate $\widehat q_n$ throughout. If
$\widehat\lambda^{\rm S}$ minimizes
$\operatorname{mBIC}_{\lambda^{\rm S}}$ over $\Lambda_n$, then
\[
 \Pr\left\{
   \mathcal A_{\widehat\lambda^{\rm S}}=\mathcal A^\ast,\quad
   \widehat{\bm\theta}_{\widehat\lambda^{\rm S}}
          =\widehat{\bm\theta}^{\,\mathrm{orc}}
       \right\}\longrightarrow1.
\]
There is no upper bound on the grid values and no restriction on the
rate at which the finite grid's cardinality grows.
\end{proposition}

\begin{proof}
The block-curvature bound in the proof of
Proposition~\ref{S-Prop:Selection} is $C|\triangle|^\kappa<k_a$ for all
sufficiently large $n$ and does not depend on $\lambda^{\rm S}$.
Lemma~\ref{LEM:scad-global-reduction} therefore applies simultaneously
to every positive grid value. Write $\lambda=\lambda^{\rm S}$ and
$C_{\mathcal A}(\lambda)=F_{\mathcal A}+c_a\lambda^2|\mathcal A|$.
Every global fit has only zero or SCAD-flat blocks. If $\mathcal A$
is its actual support, that lemma gives
\[
 \begin{aligned}
 \min_{\mathcal B}C_{\mathcal B}(\lambda)
 &=Q_\lambda(\widehat{\bm\theta}_{\lambda})
  =F(\widehat{\bm\theta}_{\lambda})+c_a\lambda^2|\mathcal A|\ge C_{\mathcal A}(\lambda)
  \ge\min_{\mathcal B}C_{\mathcal B}(\lambda).
 \end{aligned}
\]
All inequalities are equalities. In particular, $\mathcal A$ minimizes
$C_{\mathcal B}(\lambda)$ and
$F(\widehat{\bm\theta}_{\lambda})=F_{\mathcal A}$.

Next, the common refit bounds~\eqref{eq:global-refit-bounds} imply,
for every $|\mathcal A|>s_n^\ast$ and every grid value,
\[
 C_{\mathcal A}(\lambda)-C_{\mathcal A^\ast}(\lambda)
 \ge c_a(\lambda_{\min}^{\rm S})^2
           (|\mathcal A|-s_n^\ast)-R_n>0
\]
with probability tending to one, where the same
$R_n=\OP(e_{n,N})$ works for all supports and all grid values.
Condition~\eqref{eq:mbic-grid-lower} makes the final inequality strict
on a single event. Hence every grid fit has at most $s_n^\ast$ active
blocks on that event.

Proposition~\ref{S-Prop:Selection} gives the oracle fit at
$\widetilde\lambda^{\rm S}$ with probability tending to one.
Intersect this event with the preceding events and the event in
Lemma~\ref{LEM:mbic-refit-ordering}. The equality between each global
fit's loss and its refitted loss gives
\[
 \operatorname{mBIC}_{\mathcal A_{\widehat\lambda^{\rm S}}}
 =\operatorname{mBIC}_{\widehat\lambda^{\rm S}}
 \le\operatorname{mBIC}_{\widetilde\lambda^{\rm S}}
 =\operatorname{mBIC}_{\mathcal A^\ast}.
\]
Since $|\mathcal A_{\widehat\lambda^{\rm S}}|\le s_n^\ast$, the
strict ordering in~\eqref{eq:mbic-refit-ordering} forces
$\mathcal A_{\widehat\lambda^{\rm S}}=\mathcal A^\ast$.
Finally, the selected fit minimizes $F$ on this support. The oracle
quadratic refit is unique on the event of
Lemma~\ref{LEM:Gamma_0}(b), so the selected coefficient vector equals
$\widehat{\bm\theta}^{\,\mathrm{orc}}$. The intersection of these
finitely many common events has probability tending to one, regardless
of the number of grid points.
\end{proof}

The grid restriction is compatible with the existing global conditions.
For the rates in Example~\ref{EX:global-compatible},
$e_{n,N}\asymp n^{-1}$ and
$\widetilde\lambda^{\rm S}=n^{-1/3}$.
Any finite deterministic grid containing $n^{-1/3}$ and contained in
$[n^{-2/5},\infty)$ satisfies~\eqref{eq:mbic-grid-lower}, since
$n^{-1}=o(n^{-4/5})$. Grid points above
$\widetilde\lambda^{\rm S}$ need not satisfy the upper tuning bound in
Assumption~\ref{AssumpC:signal}; their fits may omit true blocks.
The comparison in Lemma~\ref{LEM:mbic-refit-ordering} excludes these
fits from the selected solution.

\paragraph{Model-size control.}
\label{sec:mbic-model-size}
We first bound the projection of the error onto one simplex. We then
add invalid-IV blocks one at a time and sum their residual projections.
Finally, we compare each support with its union with the true support.
All norms below are Euclidean norms. The normalization by $nN$ is
included when comparing a squared projection with the loss $F$.

Let $d_0=\binom{d+\kappa}{\kappa}$, and let $\bm\Pi_k$ project onto
the sampled degree-$d$ polynomials supported on $T_k$. Assign observations
on shared faces to one simplex, as for the broken basis. Put
\[
 \mathbf Q_k=\mathbf P_Z\otimes\bm\Pi_k,\qquad
 M_n=\max_{k\le K_2}\|\mathbf Q_k\bm\epsilon\|^2,
\]
where $\mathbf P_Z$ projects onto $\operatorname{col}(\mathbf Z)$.
The spaces $\operatorname{col}(\mathbf Q_k)$ are mutually orthogonal
and contain all instrument-polynomial columns on their simplices.

For any support $\mathcal B$, block $b\notin\mathcal B$ and
$\bm g\in\mathcal G_n$, define the augmented residual and its projector by
\[
 \mathbf R_{b\mid\mathcal B}(\bm g)
  =\{\mathbf I-\mathbf P_{\mathcal B}(\bm g)\}
          \mathbf J\mathbf U_{\alpha,b},\qquad
 \bm\Pi_{b\mid\mathcal B}(\bm g)
  =\operatorname{proj}\{\operatorname{col}(\mathbf R_{b\mid\mathcal B}(\bm g))\}.
\]
For an ordering $E=\{b_1,\ldots,b_s\}\subseteq\mathcal A^c$, set
$\mathcal A_j=\mathcal A\cup\{b_1,\ldots,b_j\}$ and
$\mathcal A_0=\mathcal A$. Then
\begin{equation}
\label{eq:mbic-nested-identity}
 \mathbf P_{\mathcal A\cup E}(\bm g)-\mathbf P_{\mathcal A}(\bm g)
 =\sum_{j=1}^s\bm\Pi_{b_j\mid\mathcal A_{j-1}}(\bm g),
\end{equation}
and the summands are mutually orthogonal projectors.

\begin{lemma}[Local noise and nested projections]
\label{LEM:mbic-simplex-projection}
\label{LEM:mbic-one-block-residual}
Under Assumptions~\ref{A:tri} and \ref{A:error-subgaussian}--\ref{A:covariates},
$M_n=\OP\{(N/K_2)\log(eP_0)\}$.
If also $|\triangle|\to0$, then there is a fixed constant $C_M$
such that, with probability tending to one,
\begin{equation}
\label{eq:mbic-local-noise}
 \frac{\|\mathbf Q_k\bm\epsilon\|^2}{nN}
 \le\frac{C_M\log(eP_0)}{nK_2N_{\mathrm{eff},k}}
 \quad(k\le K_2),\qquad
 \frac{M_n}{nN}\le\frac{C_M\log(eP_0)}{q_n},
\end{equation}
where $C_M$ may be enlarged for the second inequality.

Suppose the following locality bound holds on one event of probability
tending to one, uniformly over the comparisons under consideration
and $\bm g\in\mathcal G_n$. For each $b=(\ell,k)$, there are nested
unions $\omega_t(T_k)$ of at most $C_\omega(t+1)^\kappa$ fine simplices,
containing $T_k$, such that
\begin{equation}
\label{eq:mbic-nested-locality}
 \|\mathbf D_{\omega_t^c}\mathbf J^\top
        \bm\Pi_{b\mid\mathcal B}(\bm g)\|_{\rm op}
 \le C_0\vartheta^{t+1},\qquad t\ge0,
\end{equation}
with fixed $C_0,C_\omega<\infty$ and $0<\vartheta<1$.
Here $\mathbf D_{\omega_t^c}$ retains all subjects' data outside
$\omega_t(T_k)$. There is a fixed $C_L<\infty$, depending only on the locality
constants and $\kappa$, such that, on this event,
\begin{equation}
\label{eq:mbic-nested-pathwise}
 \|\{\mathbf P_{\mathcal A\cup E}(\bm g)
            -\mathbf P_{\mathcal A}(\bm g)\}\mathbf J\bm\epsilon\|^2
 \le C_L^2|E|M_n.
\end{equation}
Consequently, the normalized gain is at most
$C_1|E|\log(eP_0)/q_n$, simultaneously over those comparisons,
with probability tending to one and fixed $C_1$.
The same statements hold for ordinary projections by omitting the
roughness rows and taking $\mathbf J=\mathbf I$.
For ordinary projections with the unconstrained causal Bernstein
basis, no additional locality condition is needed.
\end{lemma}

\begin{proof}
\emph{Step 1: noise on a single simplex.}
Conditional on $\mathbf Z$, independence of the KL scores and nugget
variables gives, for every fixed $\mathbf v$ and $u\in\mathbb R$,
\begin{equation}
\label{eq:mbic-vector-subgaussian}
 \E\{e^{u\mathbf v^\top\bm\epsilon}\mid\mathbf Z\}
 \le\exp\{K_{\rm sg}^2u^2\mathbf v^\top
                  \bm\Sigma_\epsilon\mathbf v/2\},\qquad
 K_{\rm sg}^2=\max(\sigma_\zeta^2,\sigma_e^2).
\end{equation}
The infinite KL sum follows from its finite sums by $L^2$ convergence
and Fatou's lemma along an almost surely convergent subsequence.
On one simplex the covariance has trace at most $C_\epsilon m_k$.
Thus every unit vector in $\operatorname{col}(\mathbf Q_k)$ has
variance at most $C_\epsilon m_k$. That space has dimension at most
$pd_0$, a fixed number. A $1/2$-net of its unit sphere has at most
$5^{pd_0}$ points, and its largest absolute noise score is at least
$\|\mathbf Q_k\bm\epsilon\|/2$. The sub-Gaussian bound and a union
bound over the nets and simplices prove the first assertion.

To retain $N_{\mathrm{eff},k}$, use the fixed covariance kernel on
compact $\Omega$. Put
$G_M(\bm s,\bm s')=\sum_{t\le M}\rho_t\psi_t(\bm s)\psi_t(\bm s')$.
The increasing open sets
$\{\bm s:G_M(\bm s,\bm s)>c_G/2\}$ cover $\Omega$.
A finite subcover and continuity give fixed $M$ and $\delta>0$ such
that $G_M(\bm s,\bm s')\ge c_G/4$ whenever
$\|\bm s-\bm s'\|\le\delta$.
For sufficiently small simplices, positive semidefiniteness of the
remaining KL covariance and the nugget covariance yields
\[
 \begin{aligned}
 &S_k:=\sum_{j,j'\in T_k}\rho_\epsilon(\bm s_j,\bm s_{j'})
 \ge\sum_{j,j'\in T_k}
 \frac{G_M(\bm s_j,\bm s_{j'})}
      {\{G_\epsilon(\bm s_j,\bm s_j)G_\epsilon(\bm s_{j'},\bm s_{j'})\}^{1/2}}
 \ge\frac{c_Gm_k^2}{4C_\epsilon},\\
 &\sum_{j,j'\in T_k}|G_\epsilon(\bm s_j,\bm s_{j'})|
 \le C_\epsilon m_k^2\le\frac{4C_\epsilon^2}{c_G}S_k.
 \end{aligned}
\]
This argument allows signed correlations. It also shows
$1\le N_{\mathrm{eff},k}\le4C_\epsilon/c_G$ under the fixed-kernel
assumptions; the effective sizes are retained in the bounds below.

For a deterministic, uniformly bounded nonnegative polynomial
$f\in\mathcal P_d(T_k)$ with $\sum_{j\in T_k}f(\bm s_j)^2\ge c_fm_k$
for a fixed $c_f>0$, and a nonzero
$\mathbf Z$-measurable subject vector $\mathbf a$, set
$\mathbf u=\mathbf a\otimes(f(\bm s_j)I\{\bm s_j\in T_k\})_j$.
Then
\[
 \begin{aligned}
 \E\{(\mathbf u^\top\bm\epsilon)^2\mid\mathbf Z\}
 &=\|\mathbf a\|^2\sum_{j,j'\in T_k}
       f(\bm s_j)f(\bm s_{j'})G_\epsilon(\bm s_j,\bm s_{j'})
 \le C\|\mathbf a\|^2\frac{m_k^2}{N_{\mathrm{eff},k}},\\
 \E\left\{\frac{\|\mathbf P_u\bm\epsilon\|^2}{nN}
          \mathrel{\Big|}\mathbf Z\right\}
 &=\frac{\E\{(\mathbf u^\top\bm\epsilon)^2\mid\mathbf Z\}}
         {nN\|\mathbf u\|^2}
 \le\frac{C}{nK_2N_{\mathrm{eff},k}}.
 \end{aligned}
\]
The first equality concerns a conditional expectation, not a realized
squared score. The factor $\|\mathbf a\|^2$ is of order $n$ for a
nondegenerate instrument column and cancels from the projection ratio.

The bound extends to signed polynomial contrasts. By Assumption~\ref{A:tri} and polynomial
norm equivalence, $\max_{j\in T_k}|h(\bm s_j)|^2
\le C m_k^{-1}\sum_{j\in T_k}h(\bm s_j)^2$ for $h\in\mathcal P_d(T_k)$.
Applying this to each subject's polynomial shows, for every unit
$\mathbf v\in\operatorname{col}(\mathbf Q_k)$,
$\mathbf v^\top\bm\Sigma_\epsilon\mathbf v\le C m_k/N_{\mathrm{eff},k}$.
The same net argument now gives, for $x>0$,
\[
 \Pr\left\{\max_k\frac{N_{\mathrm{eff},k}}{m_k}
             \|\mathbf Q_k\bm\epsilon\|^2>C\{x+\log(eP_0)\}
           \mathrel{\Big|}\mathbf Z\right\}\le e^{-x}.
\]
Take $x=2\log(eP_0)$, use $m_k\asymp N/K_2$ and
$N^-_{\mathrm{eff}}\ge c_0\bar N_{\mathrm{eff}}$, and divide by $nN$.
This proves~\eqref{eq:mbic-local-noise} with a fixed constant.

\emph{Step 2: add blocks in a nested order.}
The enlarged design space is the orthogonal sum of the current
space and the span of the added residual. Hence
$\mathbf P_{\mathcal B\cup\{b\}}-\mathbf P_{\mathcal B}
=\bm\Pi_{b\mid\mathcal B}$, also for singular designs.
Successive increments have mutually orthogonal ranges. This proves
\eqref{eq:mbic-nested-identity} and
\[
 \|\{\mathbf P_{\mathcal A\cup E}-\mathbf P_{\mathcal A}\}
           \mathbf J\bm\epsilon\|^2
 =\sum_{j=1}^{|E|}\|\bm\Pi_{b_j\mid\mathcal A_{j-1}}
                         \mathbf J\bm\epsilon\|^2.
\]
The sign is enlarged projector minus smaller projector.

\emph{Step 3: apply the local bound to residuals.}
Let $\mathbf v$ be any unit vector in a one-block residual space.
Its data part $\mathbf v_0=\mathbf J^\top\mathbf v$ lies in
$\bigoplus_k\operatorname{col}(\mathbf Q_k)$ and has norm at most one.
On shell $\omega_t\setminus\omega_{t-1}$, its norm is at most
$\max(1,C_0)\vartheta^t$, with $\omega_{-1}=\emptyset$.
Cauchy--Schwarz on each shell gives
\[
 \sum_k\|\mathbf Q_k\mathbf v_0\|
 \le\max(1,C_0)\sqrt{C_\omega}
        \sum_{t\ge0}(t+1)^{\kappa/2}\vartheta^t=:C_L<\infty.
\]
The tail condition leaves no component outside the union of these shells.
Consequently, $|\mathbf v^\top\mathbf J\bm\epsilon|
\le C_L\sqrt{M_n}$. Taking the supremum over unit residual vectors,
and then summing the nested increments, proves
\eqref{eq:mbic-nested-pathwise}. The same event works for every
support, order and $\bm g$; no further union bound is needed.
In the ordinary unconstrained causal space, a residual is supported
on its coarse parent simplex, which contains only a fixed number of
fine simplices. The preceding bound then follows directly.
All stochastic conditioning was on $\mathbf Z$. In particular, the
result applies to $\widehat{\bm\gamma}$ when it belongs to $\mathcal G_n$,
even though it can depend on $\bm\epsilon$.
\end{proof}

\paragraph{Locality of the residuals.}
Positivity of an original spline column does not imply positivity
of its residual: projecting $\mathbf e_1$ off the constant vector
leaves $\mathbf e_1-\mathbf1/M$, with negative entries outside its
original support. The preceding argument needs only a bound on the
sizes of residual tails. The following lemma gives such a bound.

\begin{lemma}[Decay outside a simplex neighborhood]
\label{LEM:mbic-residual-locality}
For a one-block addition $b=(\ell,k)$, project out the current
invalid-IV columns first. Their projector is
\[
 \mathbf L_{\mathcal B}
 =\sum_j\mathbf P_{Z_{\mathcal B(j)}}\otimes\bm\Pi_j.
\]
For a sampled causal basis $\mathbf C=(c_1,\ldots,c_M)$, put
\[
 \mathbf H_0=(\mathbf I-\mathbf L_{\mathcal B})(\mathbf Z\bm g\otimes\mathbf C),
 \quad\mathbf V_b=(\mathbf I-\mathbf L_{\mathcal B})\mathbf U_{\alpha,b},
 \quad\mathbf H=\begin{pmatrix}\mathbf H_0\\\sqrt{\lambda^{\rm R}}\mathbf R_C\end{pmatrix},
\]
where $\mathbf R_C$ is the roughness factor in the original augmented
rows after the basis change, so $\mathbf R_C^\top\mathbf R_C=\mathbf W_C$.
Then
$\mathbf R_{b\mid\mathcal B}=(\mathbf I-\mathbf P_H)\mathbf J\mathbf V_b$.
Join basis indices whose supports meet a common fine simplex, and
let $\omega_t(T_k)$ consist of the fine simplices meeting supports
at graph distance at most $t$ from the supports meeting $T_k$.
For $\mathbf G=\mathbf H^\top\mathbf H\ne0$, let
$0<a_G\le b_G$ bound its positive spectrum and put
$q=(b_G-a_G)/(b_G+a_G)<1$. Then
\begin{equation}
\label{eq:mbic-residual-decay}
 \|\mathbf D_{\omega_t^c}\mathbf J^\top
                 \mathbf R_{b\mid\mathcal B}\bm c\|
 \le q^{t+1}\|\mathbf V_b\bm c\|.
\end{equation}
If, additionally,
\begin{equation}
\label{eq:mbic-residual-angle}
 \|\mathbf R_{b\mid\mathcal B}\bm c\|
 \ge\eta\|\mathbf V_b\bm c\|\quad\text{for every }\bm c,
 \qquad\eta>0,
\end{equation}
then the left side of~\eqref{eq:mbic-nested-locality} is at most
$\eta^{-1}q^{t+1}$. When $\mathbf G=0$, take $q=0$; the residual
has support on $T_k$. 
\end{lemma}

\begin{proof}
Let $\tau=2/(a_G+b_G)$ and
$p_t(\mathbf G)=\tau\sum_{j=0}^t(\mathbf I-\tau\mathbf G)^j$.
Singular-value decomposition gives
\[
 \|\mathbf P_H-\mathbf H p_t(\mathbf G)\mathbf H^\top\|_{\rm op}
 \le q^{t+1}.
\]
Indeed, the difference has eigenvalue $(1-\tau\lambda)^{t+1}$ on
an eigenvector with positive Gram eigenvalue $\lambda$, and zero
on the orthogonal complement of $\operatorname{col}(\mathbf H)$.
The data columns of $\mathbf H$ retain the support of their causal
basis functions. Both $\mathbf H_0^\top\mathbf H_0$ and $\mathbf W_C$
have zero entries between nonadjacent indices; for the latter this
follows from the integrals defining the roughness penalty.
Thus $p_t(\mathbf G)$ spreads $\mathbf H_0^\top\mathbf V_b$ by at
most $t$ graph steps. The data part of
$\mathbf H p_t(\mathbf G)\mathbf H^\top\mathbf J\mathbf V_b$
vanishes outside $\omega_t(T_k)$, as does $\mathbf V_b$ itself.
The operator bound proves~\eqref{eq:mbic-residual-decay}.
Dividing by the residual norm using~\eqref{eq:mbic-residual-angle}
proves the normalized bound. No local square root of $\mathbf W_C$
is required. The zero Gram case is immediate.
\end{proof}

For completeness, uniform constants follow on any family of enlarged
supports satisfying (S) in Lemma~\ref{LEM:norm} with a common margin,
provided the causal space has a basis of support diameter
$O(|\triangle|)$ satisfying
\begin{equation}
\label{eq:mbic-local-basis}
 cN|\triangle|^\kappa\|\bm v\|^2
 \le\|\mathbf C\bm v\|^2\le CN|\triangle|^\kappa\|\bm v\|^2.
\end{equation}
To see this, on the event where $\mathbf Z^\top\mathbf Z/n$ is
uniformly positive definite, write
$\mathbf w_j=(\mathbf I-\mathbf P_{Z_{\mathcal B(j)}})\mathbf Z\bm g$.
Condition (S) gives $cn\le\|\mathbf w_j\|^2\le Cn$ uniformly over
$\bm g\in\mathcal G_n$. Since
$\mathbf H_0^\top\mathbf H_0=\sum_j\|\mathbf w_j\|^2\mathbf C_j^\top\mathbf C_j$
and $\|\mathbf W_C\|_{\rm op}\le C|\triangle|^{\kappa-4}$ by the
polynomial inverse inequality, Assumption~\ref{A:size} gives
\[
 cnN|\triangle|^\kappa\mathbf I\preceq\mathbf G
 \preceq CnN|\triangle|^\kappa\mathbf I.
\]
Thus $q$ is bounded away from one. The empirical norm comparison
under (S) for the enlarged support gives
\[
 \|\mathbf R_{b\mid\mathcal B}\bm c\|^2
 \ge\inf_{\bm v}\|\mathbf U_{\alpha,b}\bm c
                  -\mathbf U_{\mathcal B}(\bm g)\bm v\|^2
 \ge cnN|\triangle|^\kappa\|\bm c\|^2,
 \qquad\|\mathbf V_b\bm c\|^2\le CnN|\triangle|^\kappa\|\bm c\|^2.
\]
This supplies a uniform $\eta>0$. Quasi-uniformity bounds the number
of simplices in $\omega_t$ by $C(t+1)^\kappa$.
These are sufficient conditions for~\eqref{eq:mbic-nested-locality}.
They are not automatic for every candidate: enlarging the true support
can violate (S), and approximation power in Assumption~\ref{A:true-coef} alone does not supply
\eqref{eq:mbic-local-basis}. For unrestricted support comparisons,
the required uniform locality or covariance bound must be stated.

\begin{lemma}[Gain from adding invalid-IV blocks]
\label{LEM:mbic-size-noise}
Suppose Assumptions~\ref{A:tri}--\ref{A:size} hold. If~\eqref{eq:mbic-nested-locality} holds uniformly
for every chain from $\mathcal A^\ast$ to $\mathcal A^\ast\cup E$,
$E\subseteq(\mathcal A^\ast)^c$, then, with probability tending to one,
\begin{equation}
\label{eq:mbic-size-loss-gain}
 0\le F_{\mathcal A^\ast}-F_{\mathcal A^\ast\cup E}
 \le C_g|E|\frac{\log(eP_0)}{q_n}+R_n^{\rm b}
 \quad\text{for every }E\subseteq(\mathcal A^\ast)^c,
\end{equation}
where $C_g$ is fixed and the same nonnegative remainder satisfies
$R_n^{\rm b}=\OP(|\triangle|^{2d+2}+\rho_n)$.

Alternatively, the same conclusion holds under the following
covariance bound, without a locality condition. Write
$\mathbf D_E(\bm g)=\mathbf P_{\mathcal A^\ast\cup E}(\bm g)
-\mathbf P_{\mathcal A^\ast}(\bm g)$.
On a $\mathbf Z$-measurable event of probability tending to one,
suppose there are $\mathbf Z$-measurable matrices $\mathbf H_b\succeq0$
such that, simultaneously for all $E$ and $\bm g\in\mathcal G_n$,
\begin{equation}
\label{eq:mbic-size-envelope}
 \mathbf J^\top\mathbf D_E(\bm g)\mathbf J
 \preceq\sum_{b\in E}\mathbf H_b,\qquad
 \frac{\operatorname{tr}(\bm\Sigma_\epsilon\mathbf H_b)}{nN}
 \le\frac{C_H}{q_n}.
\end{equation}
A sharper alternative replaces these matrices by a common
$\mathbf Z$-measurable envelope $\mathbf H_E\succeq0$ for each $E$:
\begin{equation}
\label{eq:mbic-size-envelope-sharp}
 \begin{gathered}
 \mathbf J^\top\mathbf D_E(\bm g)\mathbf J\preceq\mathbf H_E
 \quad(\bm g\in\mathcal G_n),\qquad
 \mathbf V_E=\frac{\bm\Sigma_\epsilon^{1/2}\mathbf H_E
                       \bm\Sigma_\epsilon^{1/2}}{nN},\\
 \operatorname{tr}(\mathbf V_E)\le\frac{C_H|E|}{q_n},\qquad
 \|\mathbf V_E\|_{\rm op}\le\frac{C_H}{q_n}.
 \end{gathered}
\end{equation}
Then $\log(eP_0)$ in~\eqref{eq:mbic-size-loss-gain} may be replaced
by $\log(eP_0/|E|)$ for $E\ne\emptyset$.
For example, this sharper condition holds if a common
$\mathbf Z$-measurable augmented projector $\mathbf Q_E$ contains
the ranges of $\mathbf D_E(\bm g)$, has rank at most $C|E|$, and
$\|\mathbf Q_E\mathbf J\bm\Sigma_\epsilon\mathbf J^\top\mathbf Q_E\|_{\rm op}
\le CN/(K_2\bar N_{\mathrm{eff}})$.
\end{lemma}

\begin{proof}
Under locality, the preceding nested bound supplies the common noise
event. For either covariance alternative we use the following
consequence of~\eqref{eq:mbic-vector-subgaussian}. For a
$\mathbf Z$-measurable $\mathbf H\succeq0$, let
$\mathbf V=\bm\Sigma_\epsilon^{1/2}\mathbf H\bm\Sigma_\epsilon^{1/2}/(nN)$.
Gaussian integration gives, for sufficiently small $t>0$,
\[
 \E\{\exp(t\bm\epsilon^\top\mathbf H\bm\epsilon/(nN))\mid\mathbf Z\}
 \le\det(\mathbf I-2K_{\rm sg}^2t\mathbf V)^{-1/2}.
\]
Expanding $-\log(1-x)$ and applying Chernoff's inequality gives
\[
 \Pr\left\{\frac{\bm\epsilon^\top\mathbf H\bm\epsilon}{nN}>
 K_{\rm sg}^2[\operatorname{tr}(\mathbf V)
 +2\sqrt{\operatorname{tr}(\mathbf V^2)x}+2\|\mathbf V\|_{\rm op}x]
 \mathrel{\Big|}\mathbf Z\right\}\le e^{-x}.
\]
Under~\eqref{eq:mbic-size-envelope}, take $x=3\log(eP_0)$, use
$\|\mathbf V_b\|_{\rm op}\le\operatorname{tr}(\mathbf V_b)$, and
sum the resulting bounds over $b\in E$.
For~\eqref{eq:mbic-size-envelope-sharp}, take
$x=3s\log(eP_0/s)$ for $s=|E|$ and use at most $(eP_0/s)^s$ sets
of that size. The total failure probability is at most
$\sum_{s=1}^{P_0}e^{-2s\log(eP_0/s)}=O(P_0^{-1})$.
The stated sufficient condition follows by taking
$\mathbf H_E=\mathbf J^\top\mathbf Q_E\mathbf J$ and using its
covariance and rank bounds. In every case, the event is uniform in
$\bm g$ and hence applies at $\widehat{\bm\gamma}$.

At the estimated first stage, $\mathbf D_E$ is an orthogonal projector
and the augmented least-squares identity gives
\[
 F_{\mathcal A^\ast}-F_{\mathcal A^\ast\cup E}
 =\frac{\|\mathbf D_E\mathbf J\mathbf Y\|^2}{nN}.
\]
Subtract the augmented approximating vector
$\mathbf X_{\mathcal A^\ast}\widetilde{\bm\theta}$, which
$\mathbf D_E$ annihilates. The first-stage residual with entries
$(D_i-\widehat D_i)\widetilde\beta(\bm s_j)$ and zero roughness
rows is also annihilated, since $\mathbf Z^\top(\mathbf D-\widehat{\mathbf D})=0$.
The remaining terms are $\mathbf J\bm\epsilon$ and the vector of
approximation errors and negative approximating roughness rows.
Thus the common noise bound and the contraction property give
\eqref{eq:mbic-size-loss-gain} with
\[
 R_n^{\rm b}=\frac{2\|\mathbf b\|^2}{nN}
               +2\rho_n\mathcal E(\widetilde\beta)
 =\OP(|\triangle|^{2d+2}+\rho_n),
\]
where $\mathbf b$ is as in Lemma~\ref{LEM:global-support}.
\end{proof}

\begin{proposition}[Consistency of mBIC with control by model size]
\label{Prop:mBIC-size}
Suppose Assumptions~\ref{A:tri}--\ref{AssumpC:signal} hold at a deterministic selection parameter
$\widetilde\lambda^{\rm S}$ and a deterministic roughness sequence
$\lambda^{\rm R}>0$. In addition, suppose the following conditions hold.

\smallskip\noindent
\emph{(M1) Residual noise.} The locality condition
\eqref{eq:mbic-nested-locality} holds uniformly over $\mathcal G_n$
and all chains from $\mathcal A^\ast$ to its supersets. Either
\eqref{eq:mbic-size-envelope} or~\eqref{eq:mbic-size-envelope-sharp}
may replace this condition. Let $C_g$ be the fixed constant in
\eqref{eq:mbic-size-loss-gain} under the chosen condition.

\smallskip\noindent
\emph{(M2) Approximation and roughness.}
$|\triangle|^{2d+2}+\rho_n=o(v_n^0)$.

\smallskip\noindent
\emph{(M3) Competing signals.} For a fixed $L_0\ge1$, whenever
\chg{$\mathcal A$ is anchored,} $|\mathcal A|>s_n^\ast$ and
$|\mathcal A^\ast\setminus\mathcal A|>L_0(|\mathcal A|-s_n^\ast)$,
require
\begin{equation}
\label{eq:mbic-size-separation}
 \begin{split}
 \inf_{\bm g\in\mathcal G_n}\inf_{(\beta,\bm\alpha)\in\Theta_{\mathcal A}}
 \bigg[\int_\Omega\E\left[\Big\{\mathbf Z^\top
 [\bm g(\widetilde\beta-\beta)+\widetilde{\bm\alpha}-\bm\alpha]
                (\bm s)\Big\}^2\right]\,d\bm s
       +\rho_n\mathcal E(\beta)\bigg]
 \ge\Delta_n^{\rm G}.
 \end{split}
\end{equation}
The sequences are those in Assumption~\ref{AssumpC:signal}. This condition is vacuous when
$s_n^\ast=0$; Assumption~\ref{AssumpC:signal} already treats supports of no larger size.

\smallskip\noindent
\emph{(M4) Penalty margin.} With $b_0$ from
Lemma~\ref{LEM:mbic-loss-bounds},
\[
 \liminf_{n,N\to\infty}\frac{c_n}{\log(eP_0)}
 >\frac{2C_g(L_0+1)}{b_0}.
\]
Then, with probability tending to one,
\begin{equation}
\label{eq:mbic-all-ordering}
 \operatorname{mBIC}_{\mathcal A}>\operatorname{mBIC}_{\mathcal A^\ast}
 \quad\text{for every \chg{anchored} }\mathcal A\ne\mathcal A^\ast.
\end{equation}
For all larger supports the gap is at least
$(|\mathcal A|-s_n^\ast)v_n^0/8$ on the same event.

Let $\Lambda_n\subset(0,\infty)$ be any finite deterministic grid
containing $\widetilde\lambda^{\rm S}$. At that point use a global
SCAD fit, and at the other points use any feasible fitted coefficients,
with the same $\lambda^{\rm R}$ and common pilot estimate of $q_n$.
The fitted-mBIC minimizer over this grid satisfies
\[
 \Pr\{\mathcal A_{\widehat\lambda^{\rm S}}=\mathcal A^\ast,
 \ \widehat{\bm\theta}_{\widehat\lambda^{\rm S}}
       =\widehat{\bm\theta}^{\,\mathrm{orc}}\}\longrightarrow1.
\]
There is no lower restriction on the other grid values or restriction
on the growth of the finite grid's cardinality.
\end{proposition}

\begin{proof}
For any $\mathcal A$, write
\[
 \mathcal U=\mathcal A\cup\mathcal A^\ast,\quad
 m=|\mathcal A^\ast\setminus\mathcal A|,\quad
 s=|\mathcal A\setminus\mathcal A^\ast|,\quad
 j=|\mathcal A|-s_n^\ast=s-m.
\]
The comparison through the union is
\begin{equation}
\label{eq:mbic-union-decomposition}
 \begin{aligned}
 \operatorname{mBIC}_{\mathcal A}-\operatorname{mBIC}_{\mathcal A^\ast}
 &=\{\operatorname{mBIC}_{\mathcal A}-\operatorname{mBIC}_{\mathcal U}\}
   +\{\operatorname{mBIC}_{\mathcal U}-\operatorname{mBIC}_{\mathcal A^\ast}\}\\
 &=\left\{\log\frac{F_{\mathcal A}}{F_{\mathcal U}}-mv_n\right\}
   +\left\{\log\frac{F_{\mathcal U}}{F_{\mathcal A^\ast}}+sv_n\right\}.
 \end{aligned}
\end{equation}
By (M1)--(M2) and Lemma~\ref{LEM:mbic-size-noise}, on a common event,
\[
 0\le H_{\mathcal A}:=F_{\mathcal A^\ast}-F_{\mathcal U}
 \le C_gs\frac{\log(eP_0)}{q_n}+R_n^{\rm b},\qquad
 R_n^{\rm b}=o_{\rm P}(v_n^0).
\]
This is the noise part of the comparison. Monotonicity also gives
$F_{\mathcal A}-F_{\mathcal U}\ge0$.

If $j\le0$ and $\mathcal A\ne\mathcal A^\ast$,
Lemma~\ref{LEM:mbic-refit-ordering} already proves strict ordering
using the signal gap of Assumption~\ref{AssumpC:signal}. Suppose henceforth that $j>0$.
If $m\le L_0j$, then $s\le(L_0+1)j$. All refitted losses are at
least $b_0$ with probability tending to one. The first logarithm
in~\eqref{eq:mbic-union-decomposition} is nonnegative, and integration
of $1/x$ bounds the second below by $-H_{\mathcal A}/b_0$.
Consequently,
\[
 \operatorname{mBIC}_{\mathcal A}-\operatorname{mBIC}_{\mathcal A^\ast}
 \ge jv_n-\frac{C_g(L_0+1)j\log(eP_0)}{b_0q_n}
            -\frac{R_n^{\rm b}}{b_0}.
\]
By (M4), the middle term is less than $jv_n^0/2$ eventually.
Intersect with $v_n\ge3v_n^0/4$ and
$R_n^{\rm b}/b_0\le v_n^0/8$, whose probabilities tend to one.
The resulting gap is at least $jv_n^0/8$ for all these supports.

If $m>L_0j$, apply the empirical covariance and quadrature comparison
in Lemma~\ref{LEM:global-support} to (M3). It gives
$D_{\mathcal A}\ge\Delta_n^{\rm G}/4$ uniformly. Therefore,
\eqref{eq:global-refit-bounds} gives
\[
 F_{\mathcal A}-F_{\mathcal A^\ast}
 \ge T_n:=\Delta_n^{\rm G}/8-R_n>0,\qquad
 R_n=\OP(e_{n,N})=o_{\rm P}(\Delta_n^{\rm G}),
\]
on a common event, by Assumption~\ref{AssumpC:signal}. Thus the signal part satisfies
\[
 \log\frac{F_{\mathcal A}}{F_{\mathcal U}}
 \ge\log\frac{F_{\mathcal A^\ast}}{F_{\mathcal U}}
       +\log\left(1+\frac{T_n}{F_{\mathcal A^\ast}}\right).
\]
The union loss cancels in~\eqref{eq:mbic-union-decomposition}.
The mBIC gap is consequently at least
$\log(1+T_n/F_{\mathcal A^\ast})+jv_n\ge3jv_n^0/4$.
The two union-model differences need not each be positive; retaining
their cancellation avoids any extra signal condition proportional
to the number of omitted blocks. This proves~\eqref{eq:mbic-all-ordering}.

Finally, Proposition~\ref{S-Prop:Selection} gives the unique oracle
fit at $\widetilde\lambda^{\rm S}$. A feasible fit on actual support
$\mathcal A$ has loss at least $F_{\mathcal A}$. Hence, on the
intersection of the preceding events,
\[
 \operatorname{mBIC}_{\mathcal A_{\widehat\lambda^{\rm S}}}
 \le\operatorname{mBIC}_{\widehat\lambda^{\rm S}}
 \le\operatorname{mBIC}_{\widetilde\lambda^{\rm S}}
 =\operatorname{mBIC}_{\mathcal A^\ast}.
\]
Strict ordering forces the selected support to be $\mathcal A^\ast$
and its loss to equal $F_{\mathcal A^\ast}$. Uniqueness of the
oracle quadratic refit then forces equality of the coefficient
vectors. Since all comparisons hold simultaneously over supports,
no union bound over the tuning grid is needed.
\end{proof}

The additional conditions have separate roles. Condition (M1) controls
residual noise; local support of the original basis alone does not
establish it. Condition (M2) excludes the mechanism by which a representation with many invalid-IV blocks can reproduce the fitted values of a smooth $\beta$ while saving the entire roughness penalty, at a complexity cost of only $O(K_2v_n)=o_{\rm P}(\rho_n)$. Condition (M3) supplies the
signal comparison for larger models outside the comparison family of Assumption~\ref{AssumpC:signal},
and (M4) makes the penalty dominate the noise gain per net added block.
The result uses a fixed deterministic roughness sequence; it does not
cover its later selection by cross-validation.

	\section*{Section B. Implementation Details}
	\renewcommand{\suppanchor}{B}
	
	\renewcommand{\thedefinition}{B.\arabic{definition}}
	\renewcommand{\theequation}{B.\arabic{equation}}
	\renewcommand{\thesubsection}{B.\arabic{subsection}}
	\renewcommand{\thetheorem}{{\sc B.\arabic{theorem}}}
	\renewcommand{\thelemma}{{\sc B.\arabic{lemma}}} %
	\renewcommand{\theproposition}{{\sc B.\arabic{proposition}}}
	\renewcommand{\thecorollary}{\sc B.\arabic{corollary}}
	\renewcommand{\thefigure}{B.\arabic{figure}}
	\renewcommand{\thetable}{B.\arabic{table}}
	\renewcommand{\theremark}{B.\arabic{remark}}
	\renewcommand{\theexample}{B.\arabic{example}}
	\setcounter{equation}{0}
	\setcounter{definition}{0}
	\setcounter{theorem}{0}
	\setcounter{lemma}{0}
	\setcounter{figure}{0}
	\setcounter{remark}{0}
	\setcounter{proposition}{0}
	\setcounter{corollary}{0}
	\setcounter{example}{0}
	\setcounter{subsection}{0}
	\setcounter{subsubsection}{0}

	\subsection{Hi-FIVE Optimization Algorithm}
	
	\begin{algorithm}[H]
		\caption{Hi-FIVE optimization algorithm}
		\label{ALG:hifive}
		\KwIn{
			$\{Y_i(\bm{s}_j),\mathbf Z_i,\widehat D_i\}$; bases $\mathbf B_1,\mathbf B_2$; matrices $\mathbf\Phi_1,\mathbf P_1,\mathbf Q_2,\{\mathbf M_{2,k}\}$; tuning parameters $\lambda^{\rm R},\lambda^{\rm S}$; ADMM parameter $\rho$; tolerances $\varepsilon_{\rm out},\varepsilon_{\rm in}$.
		}
		\KwOut{
			Estimated spline coefficients $\widehat{\bm\theta}_\beta$ and $\{\widehat{\bm\theta}_{\alpha,\ell}\}_{\ell=1}^p$.
		}
		
		Initialize $(\bm\theta_\beta,\{\bm\theta_{\alpha,\ell}\}_{\ell=1}^p)$ by the convex group-lasso fit with $w_{\ell,k}\equiv\lambda^{\rm S}$\;
		
		\While{outer LLA change $>\varepsilon_{\rm out}$}{
			\textit{LLA update:}
			$w_{\ell,k}\leftarrow p'_{\lambda^{\rm S}}\left(\|\bm\theta_{\alpha,\ell,k}\|\right)$,
			$\ell=1,\ldots,p$, $k=1,\ldots,K_2$\;
			
			\While{inner block-coordinate change $>\varepsilon_{\rm in}$}{
				\textit{Causal-effect update:}
				update $\bm\theta_\beta$ by consensus ADMM enforcing
				$\mathbf\Phi_1\bm\theta_\beta=\mathbf0$\;
				
				\ForEach{worker $w$ in parallel}{
					compute the local primal update in \eqref{obj:admm-updates}\;
				}
				The server performs the consensus projection and dual update in
				\eqref{obj:admm-updates}; set $\bm\theta_\beta\leftarrow\bm z$\;
				
				\textit{Invalid-IV update:}
				\ForEach{simplex $T_k$ of $\triangle_2$ and instrument $\ell$ in parallel}{
					update $\bm\theta_{\alpha,\ell,k}$ by the group soft-thresholding rule in
					\eqref{obj:coord-update}\;
				}
				\chg{\textit{Anchor step:}}
				\ForEach{\chg{simplex $T_k$ of $\triangle_2$ in parallel}}{
					\chg{if $\bm\theta_{\alpha,\ell,k}\neq\mathbf0$ for every $\ell\in\widehat{\mathcal I}_{\tau_\gamma}$, apply the anchor update \eqref{obj:anchor-step}}\;
				}
			}
		}
		
		\Return{$\widehat{\bm\theta}_\beta$ and $\{\widehat{\bm\theta}_{\alpha,\ell}\}_{\ell=1}^p$}\;
	\end{algorithm}

	\subsection{Inference}
	\paragraph{Theoretical Expression for $\sigma_{\beta}^2(\cdot)$}
	Let $\mathbf U^{0}_{ij,\mathcal A^\ast}$ and $\bm\Gamma^{0}_{\mathcal A^\ast}$ be as in Section~A (the oracle design and Gram matrix with the population fitted exposure $D^0_i=\mathbf Z_i^{\top}\bm\gamma^\ast$ in place of $\widehat D_i$), and let
	\begin{equation*}
		\label{eq:var-estimate}
		\mathbf{\Xi} = \sum_{i=1}^n \boldsymbol{\Gamma}_{\mathcal{A}^\ast}^{0\,-1} \!\left\{\sum_{j,j' =1}^N \frac{1}{n^2N^2}\mathbf{U}^{0}_{ij,\mathcal{A}^\ast} \mathbf{U}^{0\top}_{ij',\mathcal{A}^\ast} \mathbf{G}_{\epsilon}(\bm{s}_j, \bm{s}_{j^{\prime}})\right\}\boldsymbol{\Gamma}_{\mathcal{A}^\ast}^{0\,-1}, 
	\end{equation*}
	where $\boldsymbol{\Gamma}^{0}_{\mathcal{A}^\ast}=\frac{1}{nN}\mathbf{U}^{0\top}_{\mathcal{A}^\ast}\mathbf{U}^{0}_{\mathcal{A}^\ast}+\text{diag}\{\frac{\lambda^R}{nN}\mathbf{Q}_2^\top\mathbf{P}_1\mathbf{Q}_2,\mathbf{0},\cdots,\mathbf{0}\}$ and $\mathbf{U}^{0}_{\mathcal{A}^\ast} = (\mathbf{U}^{0}_{ij,\mathcal{A}^\ast})_{1\leq i \leq n, 1\leq j \leq N}$. By Proposition~\ref{S-Prop:Selection}, $\widehat\beta=\widehat\beta^{o}$ with probability tending to one, so the asymptotic variance of $\widehat\beta(\bm s)$ is that of $\widehat\beta^{o}(\bm s)$, given by \eqref{eq:var_orac_est} and repeated here for convenience,
	\begin{equation}\label{eq:est_varbeta_th}
		\sigma_{\beta}^2(\bm{s})
		=\widetilde{\mathbf{B}}_1(\bm{s})^\top  \mathbf{\Xi}_\beta\widetilde{\mathbf{B}}_1(\bm{s}),
	\end{equation}
	where $\mathbf{\Xi}_\beta$ is the $\bm\theta_\beta$-block of $\mathbf{\Xi}$; this is \eqref{eq:var_orac_est}, the conditional asymptotic variance given $\mathbf Z_1,\ldots,\mathbf Z_n$ of the oracle estimator computed with the population first stage. It depends on $\bm\gamma^\ast$ through $D^0_i$ and on $G_\epsilon$, both unknown; the estimator below replaces $D^0_i$ by $\widehat D_i$, $\mathcal A^\ast$ by the selected set $\widehat{\mathcal A}$, and $G_\epsilon$ by the residual-based estimator $\widehat G_\epsilon$. 
	
	\paragraph{Estimation of $\sigma_{\beta}^2(\cdot)$}
	\label{sec:inferences}
	We begin by estimating the variance of $\widehat{\beta}(\bm s)$. Define the estimated residual as 
	$\widehat {R}_{i}(\bm s_j) = Y_{i}(\bm s_j) -D_i\widehat{\beta}(\bm{s}_j)-  \mathbf Z_{i}^\top\widehat{\bm \alpha}(\bm{s}_j)$, $i=1,\ldots,n$, $j = 1,\ldots,N$ and the estimated active set \(\widehat{\mathcal A} = \{\,(\ell,k): \|\widehat\alpha_\ell\|_{T_k}\neq 0\}\).
	Then we estimate the $\mathbf{Z}$-orthogonalized subject-$i$ process $\eta_i(\cdot)$ in \eqref{eq:eta},  and the measurement error by
	\begin{equation}
		\widehat{\eta}_i(\bm{s}_j)=\argmin_{g_i\in \mathcal{S}^r_d(\triangle_\eta)}\sum_{j=1}^N\left\{\widehat{R}_{i}(\bm s_j)-g_i(\bm{s}_j)\right\}^2\;\text{ and }\; \widehat{\varepsilon}_{ij}=\widehat{R}_{i}(\bm s_j)-\widehat{\eta}_i(\bm{s}_j),
	\end{equation}
	where $\triangle_\eta$ is the triangulation for smoothing $\widehat\eta_i$.
	
	Denote $\widehat{\mathcal{A}}_{\ell}
	=
	\left\{
	k : (\ell,k)\in\widehat{\mathcal{A}}
	\right\}$. The covariance estimator $\widehat{\mathbf G}_{\epsilon}(\bm s_j,\bm s_{j'})$ and the effective-sample-size estimate $\widehat q_n$ are computed as in Section~\ref{SEC:eff-sample}, namely by \eqref{EQU:G-epsilon} and Lemma~\ref{LEM:Neff-consistency}, with the Hi-FIVE fit $(\widehat\beta,\widehat{\bm\alpha})$ used as the common pilot, so that $\widehat R_i(\bm s_j)$ above are the pilot residuals and $\varrho_n$ in \eqref{DEF:pilot} is the sup-norm error of this fit. Define
	$$
	\mathbf{U}_{ij,\widehat{\mathcal A}}
	=
	\left\{
	\widehat{D}_i\widetilde{\mathbf{B}}_1^{\top}(\bm{s}_j),
	Z_{i1}\mathbf{B}_{2,\widehat{\mathcal A}_1}^{\top}(\bm{s}_j),
	\ldots,
	Z_{ip}\mathbf{B}_{2,\widehat{\mathcal A}_p}^{\top}(\bm{s}_j)
	\right\}^{\top},
	$$
	where $\mathbf{B}_{2,\widehat{\mathcal A}_\ell}(\bm{s}_j)$ denotes the subvector of 
	$\mathbf{B}_2(\bm{s}_j)$ corresponding to the estimated active subregions for the 
	$\ell$-th candidate IV. Then, the estimator of the variance-covariance matrix of the spline coefficients $\widehat{\bm \theta}$ is 
	\begin{equation}
		\label{EQU:var-k-estimate}
		\widehat{\mathbf{\Xi}} = \sum_{i=1}^n \boldsymbol{\Gamma}_{\widehat{\mathcal{A}}}^{-1} \!\left\{\sum_{j,j' =1}^N \frac{1}{n^2N^2}\mathbf{U}_{ij,\widehat{\mathcal{A}}} \mathbf{U}_{ij',\widehat{\mathcal{A}}}^{\top} \widehat{\mathbf{G}}_{\epsilon}(\bm{s}_j, \bm{s}_{j^{\prime}})\right\}\boldsymbol{\Gamma}_{\widehat{\mathcal{A}}}^{-1}, 
	\end{equation}
	where $\boldsymbol{\Gamma}_{\widehat{\mathcal{A}}}=\frac{1}{nN}\mathbf{U}^\top_{\widehat{\mathcal{A}}}\mathbf{U}_{\widehat{\mathcal{A}}}+\text{diag}\{\frac{\lambda^R}{nN}\mathbf{Q}_2^\top\mathbf{P}_1\mathbf{Q}_2,\mathbf{0},\cdots,\mathbf{0}\}$ and $\mathbf{U}_{\widehat{\mathcal{A}}} = (\mathbf{U}_{ij,\widehat{\mathcal{A}}})_{1\leq i \leq n, 1\leq j \leq N}$. The estimator for $\sigma^2_{\beta}(\cdot)$ is given by 
	\begin{equation}\label{eq:est_varbeta}
		\widehat{\sigma}_{\beta}^2(\bm{s})
		=\widehat{\text{Var}}(\widehat{\beta}(\bm{s}))
		=\widetilde{\mathbf{B}}_1(\bm{s})^\top  \widehat{\mathbf{\Xi}}_\beta\widetilde{\mathbf{B}}_1(\bm{s}),
	\end{equation}
	where $\widehat{\mathbf{\Xi}}_\beta$ is a submatrix of $\widehat{\mathbf{\Xi}}$ corresponding to the estimated covariance matrix of spline coefficients $\bm{\theta}_{\beta}$. As a result, the $(1-\tau_0)$-pointwise confidence interval at $\bm s$ is 
	$\widehat{\rm PCI}(\bm s) = \big[\widehat{\beta}(\bm{s}) \pm  \widehat{\sigma}_{\beta}(\bm{s})z_{1-\tau_0 / 2}\big]$,
	where $z_{1-\tau_0/2}$ is the $(1-\tau_0/2)$-quantile of $\mathcal N(0,1)$. 
	
	\begin{corollary}[Consistency of the variance estimator]
		\label{COR:variance-consistency}
		Suppose Assumptions~\ref{A:tri}--\ref{A:normality} hold, let $\widehat{\bm\theta}$ be the global solution of Proposition~\ref{S-Prop:Selection} (so that $\Pr(\widehat{\mathcal A}=\mathcal A^\ast)\to1$), and suppose in addition that
		\begin{itemize}
			\item[(C1)] $n|\triangle|^{2\kappa}\to\infty$;
			\item[(C2)] $\max_{1\le j,j'\le N}|\widehat G_\epsilon(\bm s_j,\bm s_{j'})-G_\epsilon(\bm s_j,\bm s_{j'})|=o_{\rm P}(|\triangle|^{\kappa})$.
		\end{itemize}
		Then, at the location $\bm s$ of Assumption~\ref{A:normality},
		\[
		\frac{\widehat{\sigma}_{\beta}^2(\bm s)}
		{\sigma_{\beta}^2(\bm s)}
		\xrightarrow{\mathrm P}1,
		\qquad
		\frac{\widehat{\sigma}_{\beta}(\bm s)}
		{\sigma_{\beta}(\bm s)}
		\xrightarrow{\mathrm P}1.
		\]
		Consequently,
		\[
		\frac{\widehat{\beta}(\bm s)-\beta^\ast(\bm s)}
		{\widehat{\sigma}_{\beta}(\bm s)}
		\xrightarrow{d}\mathcal N(0,1),
		\]
		and the pointwise confidence interval
		$\widehat{\mathrm{PCI}}(\bm s)$ has asymptotic coverage
		$1-\tau_0$.
	\end{corollary}

	\begin{proof}
		On the event $\{\widehat{\mathcal A}=\mathcal A^\ast\}\cap\mathcal E_n$, whose probability tends to one, $\mathbf U_{ij,\widehat{\mathcal A}}=\mathbf U_{ij,\mathcal A^\ast}$ and $\bm\Gamma_{\widehat{\mathcal A}}=\bm\Gamma_{\mathcal A^\ast}$, so $\widehat\sigma^2_\beta(\bm s)=\bm a^{\top}\widehat{\bm\Psi}_\epsilon\bm a$ with $\bm a=\bm\Lambda_\beta^{\top}\widetilde{\mathbf B}_1(\bm s)$ as in \eqref{eq:sigma-quadratic} (estimated first stage) and $\widehat{\bm\Psi}_\epsilon$ defined as $\bm\Psi_\epsilon$ in \eqref{eq:Psi-def} with $\widehat G_\epsilon$ in place of $G_\epsilon$, whereas $\sigma^2_\beta(\bm s)=\bm a_0^{\top}\bm\Psi^{0}_\epsilon\bm a_0$ (population first stage). Write $\widehat\sigma^2_\beta(\bm s)-\sigma^2_\beta(\bm s)=\{\bm a^{\top}\bm\Psi_\epsilon\bm a-\bm a_0^{\top}\bm\Psi^{0}_\epsilon\bm a_0\}+\bm a^{\top}(\widehat{\bm\Psi}_\epsilon-\bm\Psi_\epsilon)\bm a=:\Delta_1+\Delta_2$.

		For $\Delta_1$, Step~2 of the proof of Theorem~\ref{LEM:beta-normality-oracle} gives $\|\bm a-\bm a_0\|=\OP(n^{-1/2}|\triangle|^{-\kappa})$ and $\|\bm a\|,\|\bm a_0\|\le c_\Gamma^{-1}|\triangle|^{-\kappa}$. Next, $\bm\Psi_\epsilon=(nN)^{-2}\sum_i\mathbf U_i^{\top}\mathbf G_\epsilon\mathbf U_i$ with $\mathbf U_i=(\mathbf U_{i1},\ldots,\mathbf U_{iN})^{\top}$ is positive semidefinite because $\mathbf G_\epsilon$ is (Assumption~\ref{A:covariance}), so $\|\bm\Psi_\epsilon\|\le\operatorname{tr}(\bm\Psi_\epsilon)=\OP\{n^{-1}(|\triangle|^{\kappa}+N^{-1})\}=\OP(n^{-1}|\triangle|^{\kappa})$ by the trace computation in the proof of Lemma~\ref{lem:Ueps-order} (with $\max_i|\widehat D_i|=\OP(1)$) and $N|\triangle|^{\kappa}\to\infty$. Finally, $\bm\Psi_\epsilon-\bm\Psi^{0}_\epsilon$ has nonzero entries only in the $\bm\theta_\beta$-row and $\bm\theta_\beta$-column blocks, each a sum over $i$ of $n^{-2}c_i$ times a matrix of the form $N^{-2}\sum_{j,j'}\widetilde{\mathbf B}_1(\bm s_j)\mathbf b_\ell(\bm s_{j'})^{\top}G_\epsilon(\bm s_j,\bm s_{j'})=(N^{-1}\mathbf G_\epsilon^{1/2}\mathbb B_1)^{\top}(N^{-1}\mathbf G_\epsilon^{1/2}\mathbb B_\ell)$, where $\mathbb B_1$ and $\mathbb B_\ell$ are the $N$-row matrices with rows $\widetilde{\mathbf B}_1(\bm s_j)^{\top}$ and $\mathbf b_\ell(\bm s_j)^{\top}$; its norm is at most the product of the two factor norms, each $O(|\triangle|^{\kappa/2})$ because $\|N^{-1}\mathbf G_\epsilon^{1/2}\mathbb B\|^2=\|N^{-2}\mathbb B^{\top}\mathbf G_\epsilon\mathbb B\|\le\operatorname{tr}(N^{-2}\mathbb B^{\top}\mathbf G_\epsilon\mathbb B)=O(|\triangle|^{\kappa})$ by the same trace computation; with $\max_i|c_i|=\OP(n^{-1/2})$ as in Step~2, $\|\bm\Psi_\epsilon-\bm\Psi^{0}_\epsilon\|=\OP(n^{-3/2}|\triangle|^{\kappa})$. Hence
		\[
		|\Delta_1|\le\|\bm a-\bm a_0\|\big(\|\bm a\|+\|\bm a_0\|\big)\|\bm\Psi_\epsilon\|+\|\bm a_0\|^2\|\bm\Psi_\epsilon-\bm\Psi^{0}_\epsilon\|=\OP\big(n^{-3/2}|\triangle|^{-\kappa}\big),
		\]
		and $|\Delta_1|/\sigma^2_\beta(\bm s)=\OP\{(n|\triangle|^{2\kappa})^{-1/2}\}=o_{\rm P}(1)$ by (C1) and Assumption~\ref{A:normality}(V). For $\Delta_2$, with $w_{ij}=(nN)^{-1}\bm a^{\top}\mathbf U_{ij}$,
		\begin{align*}
		|\Delta_2|&\le\sum_{i=1}^n\Big\{\sum_{j=1}^N|w_{ij}|\Big\}^2\max_{j,j'}|\widehat G_\epsilon(\bm s_j,\bm s_{j'})-G_\epsilon(\bm s_j,\bm s_{j'})|\\
		&\le Cn^{-1}|\triangle|^{-\kappa}\max_{j,j'}|\widehat G_\epsilon(\bm s_j,\bm s_{j'})-G_\epsilon(\bm s_j,\bm s_{j'})|,
		\end{align*}
		because $\sum_j|w_{ij}|\le Cn^{-1}|\triangle|^{\kappa/2}\|\bm a\|\le Cn^{-1}|\triangle|^{-\kappa/2}$ as in Step~3 of the proof of Theorem~\ref{LEM:beta-normality-oracle}; so $|\Delta_2|/\sigma^2_\beta(\bm s)=o_{\rm P}(1)$ by (C2) and (V). This proves $\widehat\sigma^2_\beta(\bm s)/\sigma^2_\beta(\bm s)\to_{\rm P}1$; the ratio of standard deviations follows by continuity, the limit law by Theorem~\ref{THM:beta-normality} and Slutsky's lemma, and the coverage statement by continuity of the normal distribution function.
	\end{proof}

	\begin{remark}
		Conditions (C1)--(C2) are the prices of estimating the first stage and the error covariance in the variance, with the bounds available here. Step~2 of the proof of Lemma~\ref{LEM:Neff-consistency} gives $\max_{j,j'}|\widehat G_\epsilon-G_\epsilon|=\OP(\sqrt{\log N/n}+\varrho_n)$, where $\varrho_n$ in \eqref{DEF:pilot} is the sup-norm error of $(\widehat\beta,\widehat{\bm\alpha})$; on the event $\{\widehat{\mathcal A}=\mathcal A^\ast\}$ these are the oracle estimators, so Theorem~\ref{THM:beta-convergence-oracle} and Corollary~\ref{COR:pilot-sup} with $a_n=r_{n,N}$ give $\varrho_n=\OP(|\triangle|^{-\kappa/2}r_{n,N})$. A sufficient condition for (C2) is therefore $\sqrt{\log N/n}+|\triangle|^{-\kappa/2}r_{n,N}=o(|\triangle|^{\kappa})$, which for $r_{n,N}\asymp n^{-1/2}$ reads $n|\triangle|^{3\kappa}\to\infty$. This is stronger than (C1), and for $\kappa=3$ it is compatible with (U1) only when $d\ge6$; for example, $\kappa=3$, $d=6$, $|\triangle|\asymp n^{-1/10}$, $N\asymp n$, $\lambda^{\rm R}\asymp1$, $\lambda^{\rm S}\asymp n^{-1/10}$ and $c<1/10$ satisfy Assumptions~\ref{A:true-alpha}, \ref{A:size}, \ref{A:normality}, (C1) and this sufficient condition for (C2), whereas in the regime given after Assumption~\ref{A:normality} ($d=5$, $|\triangle|\asymp n^{-1/7}$) condition (C2) does not follow from the bounds available and remains an assumption. Both (C1) and the factor $|\triangle|^{\kappa}$ in (C2) can be removed under a bound on the entries of $\bm\Gamma^{-1}_{\mathcal A^\ast,\lambda^{\rm R}}$ in the locally supported basis; see the remark after Assumption~\ref{A:normality}.
	\end{remark}
	
	\section*{Section C. Additional Simulation Results}
	\renewcommand{\suppanchor}{C}
	
	\renewcommand{\thedefinition}{C.\arabic{definition}}
	\renewcommand{\theequation}{C.\arabic{equation}}
	\renewcommand{\thesubsection}{C.\arabic{subsection}}
	\renewcommand{\thetheorem}{{\sc C.\arabic{theorem}}}
	\renewcommand{\thelemma}{{\sc C.\arabic{lemma}}} %
	\renewcommand{\theproposition}{{\sc C.\arabic{proposition}}}
	\renewcommand{\thecorollary}{\sc C.\arabic{corollary}}
	\renewcommand{\thefigure}{C.\arabic{figure}}
	\renewcommand{\thetable}{C.\arabic{table}}
	\renewcommand{\theremark}{C.\arabic{remark}}
	\renewcommand{\theexample}{C.\arabic{example}}
	\setcounter{equation}{0}
	\setcounter{definition}{0}
	\setcounter{theorem}{0}
	\setcounter{lemma}{0}
	\setcounter{figure}{0}
	\setcounter{remark}{0}
	\setcounter{proposition}{0}
	\setcounter{corollary}{0}
	\setcounter{example}{0}
	\setcounter{subsection}{0}
	\setcounter{subsubsection}{0}
	
	\label{sec:supp-simulation}
	
	This section provides the complete data-generating mechanisms, implementation details,
	and additional numerical summaries for the simulation studies in the main paper.
	
	\subsection{Data-generating mechanisms}
	\label{sec:supp-dgm}
	
		\subsubsection{Common model}
	
	For subject $i=1,\ldots,n$ and spatial location $\bm{s}_j\in\Omega$, the data are generated from
	\begin{align*}
		Y_i(\bm{s}_j)=D_i\beta^\ast(\bm{s}_j)
		+\bZ_i^\top\bm \alpha^\ast(\bm{s}_j)
		+\eta_i(\bm{s}_j)+\varepsilon_i(\bm{s}_j), \quad
		D_i=\bZ_i^\top\bm \gamma^\ast+\xi_i.
	\end{align*}
	We consider $n\in\{100,300\}$ and $p\in\{3,10\}$. The instruments are generated as
	\[
	\bZ_i\sim N_p(\bm 0,\sigma_Z^2\mathbf I_p),
	\qquad
	\bm\gamma^\ast=
	\begin{cases}
		\mathbf 1_p, & p=3,\\
		0.6\,\mathbf 1_p, & p=10.
	\end{cases}
	\]
	The smaller coefficient magnitude in the $p=10$ setting offsets the increase
	in the number of instruments, which controls the first-stage signal strength ($\mathrm{SNR}_D$) broadly comparable across $p=3$ and $p=10$. $\sigma_Z^2=0.075$ in the 2D design and  $\sigma_Z^2=0.108$ in the 3D design.
	
	Let $q=\lfloor p/2\rfloor+1$. 
	The first $q$ instruments are locally invalid and share the same nonzero effect over a
	specified region $\mathcal R_\alpha\subset\Omega$; the remaining instruments are valid over
	the full domain. Thus, $q=2$ for $p=3$ and $q=6$ for $p=10$. Because the locally invalid
	instruments form the largest ratio group on $\mathcal R_\alpha$, pointwise plurality fails
	on that region.
	
	The subject-level spatial error is generated from a finite Karhunen--Lo\`eve expansion,
	\[
	\eta_i(\bm{s})=\sum_{t=1}^{T}\rho_t^{1/2}\zeta_{it}\psi_t(\bm{s}),
	\]
	and the measurement errors are independent with
	$\varepsilon_i(\bm{s}_j)\overset{\mathrm{iid}}{\sim}N(0,\sigma^2)$.
	
	The variables $\{\varepsilon_i(\bm{s}_j)\}$ are independent of $(\bZ_i,\xi_i,\zeta_{i1},\ldots,
	\zeta_{iT})$. Dependence between $\eta_i(\cdot)$ and $\xi_i$ induces unmeasured confounding.
	
	\subsubsection{Smooth two-dimensional design}
	\label{sec:supp-dgm-2d-smooth}
	
\begin{figure}[H]
		\centering
		\rotatebox[origin=c]{270}{%
			\includegraphics[
			height=0.25\textwidth,
			trim={1cm 1cm 3.5cm 1.2cm},
			clip
			]{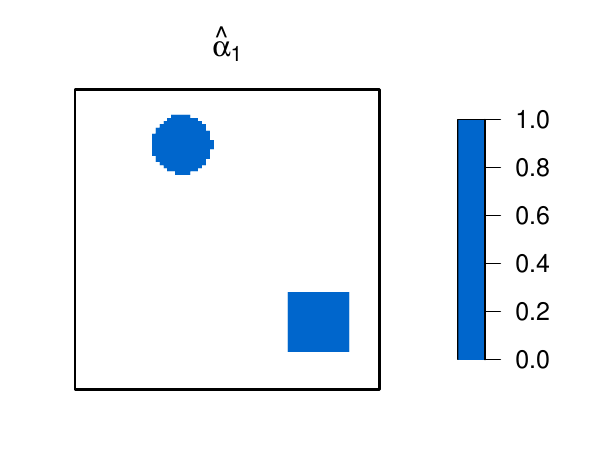}%
		}
		\caption{Design of the true IV effect $\alpha_\ell^\ast$ for
			$\ell\in\{1,\ldots,\lfloor p/2+1\rfloor\}$ in the 2D smooth simulations; the patterns are shown as invalid regions in the IV effects.}
		\label{fig:patt_2dcont}
	\end{figure}

	Let $\Omega=[0,1]^2$, with $N=95\times79=7{,}505$ equally spaced pixels, and write
	$\bm{s}=(s_1,s_2)^\top$. The causal-effect function is
	\[
	\beta^\ast(\bm{s})=3(s_1+0.2)^2-(s_2+0.2)^2.
	\]
	The spatial process contains two components,
	\[
	\eta_i(\bm{s})=\rho_1^{1/2}\zeta_{i1}\psi_1(\bm{s})
	+\rho_2^{1/2}\zeta_{i2}\psi_2(\bm{s}),
	\]
	where $\psi_1(\bm{s})=1.42\sin(2\pi s_1)$ and $\psi_2(\bm{s})=1.40\cos(2\pi s_2)$.
	We generate
	\[
	(\zeta_{i1},\zeta_{i2},\xi_i)^\top
	\sim N\!\left(
	\bm 0,
	\begin{pmatrix}
		1&0&0.3\\
		0&1&0.3\\
		0.3&0.3&1
	\end{pmatrix}
	\right).
	\]
	For $\ell=1,\ldots,q$,
	\[
	\alpha_\ell^\ast(\bm{s})=4I(\bm{s}\in\mathcal R_{\mathrm{sm}}),
	\qquad
	\mathcal R_{\mathrm{sm}}=\mathcal R_{\mathrm{sq}}\cup\mathcal R_{\mathrm{cir}},
	\]
	where
	$\mathcal R_{\mathrm{sq}}=[0.10,0.30]\times [0.68, 0.88]$ and $
	\mathcal R_{\mathrm{cir}}=\left\{\bm{s}\in\Omega:(s_1-0.82)^2+(s_2-0.35)^2\leq 0.10^2\right\}$.
	That is, $\mathcal R_{\mathrm{sq}}$ (shown in Figure~\ref{fig:patt_2dcont}) is a square centered at $(0.22,0.20)$ with side length $0.20$, whereas $\mathcal R_{\mathrm{cir}}$ is a disk centered at $(0.65,0.82)$ with radius $0.10$. For $\ell=q+1,\ldots,p$,
	$\alpha_\ell^\ast(\bm{s})=0$ on $\Omega$.
	
	The low- and high-noise settings are
	\[
	(\sigma,\rho_1,\rho_2)=
	\begin{cases}
		(0.8,0.1,0.1),&\text{low noise},\\
		(1.5,0.2,0.2),&\text{high noise}.
	\end{cases}
	\]
	We set the triangulation scheme as $K_1=50$, $K_2=200$. The degree of polynomials $d=2$, the smoothness parameter $r=1$.

	\subsubsection{Nonsmooth two-dimensional design}
	\label{sec:supp-dgm-2d-jump}
	
	The nonsmooth design uses the same instrument, first-stage, and error distributions as
	Section~\ref{sec:supp-dgm-2d-smooth}. Let $\mathcal Q\subset\Omega$ denote the pixel mask
	forming a digital-font ``4.'' The causal-effect function is
	\[
	\beta^\ast(\bm{s})
	=3.5\,I(\bm{s}\in\mathcal Q)
	+\{3(s_1+0.2)^2-(s_2+0.2)^2\}I(\bs s\notin\mathcal Q).
	\]
	where $
	\mathcal Q =
	\bigl([0.2,0.4]\times [0.2,0.6]\bigr)
	\cup\bigl([0.4,0.6]\times[0.42,0.6]\bigr)
	\cup\bigl([0.6,0.8]\times[0.2,0.8]\bigr)$.
	
\begin{figure}[H]
		\centering
		\includegraphics[width=0.3\textwidth,
		trim={0cm 1cm 2.5cm 1.2cm},
		clip]{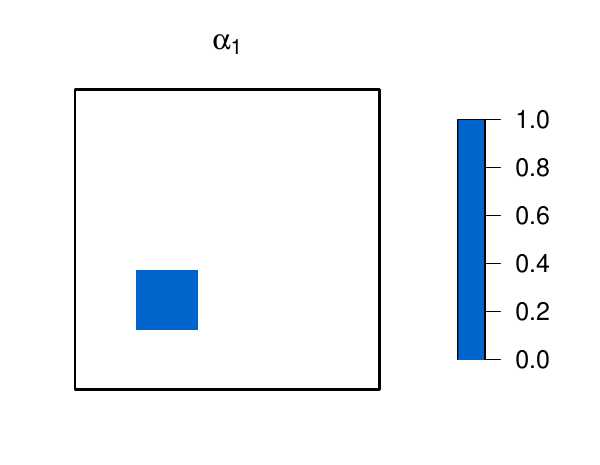}
		\caption{Design of the true IV effect $\alpha_\ell^\ast$ for $\ell\in \{1,\ldots,\lfloor p/2+1 \rfloor\}$ in the 2D nonsmooth simulations}
		\label{fig:patt_2djump}
	\end{figure}

	For $\ell=1,\ldots,q$, the invalid-IV effect is
	\[
	\alpha_\ell^\ast(\bm{s})
	=
	4I(\bm{s}\in\mathcal R_{\mathrm{jump}}),
	\qquad
	\mathcal R_{\mathrm{jump}}
	=
	[0.2,0.4]\times[0.6,0.8].
	\]
	For $\ell=q+1,\ldots,p$, $\alpha_\ell^\ast(\bm{s})=0$. The region $\mathcal R_{\mathrm{jump}}$ (shown in Figure~\ref{fig:patt_2djump}) continues the short left stem of the ``4,'' so that $\mathcal Q\cup\mathcal R_{\mathrm{jump}}$ forms a digital-font ``H.'' 
	
	For the spatially adaptive fit, the square domain is partitioned into the following four
	rectangular subdomains:
	\begin{align*}
		\Omega_1&=[0,0.6]\times[0,0.6],&
		\Omega_2&=[0.6,1.0]\times[0,0.6],\\
		\Omega_3&=[0,0.6]\times[0.6,1],&
		\Omega_4&=[0.6,1]\times[0.6,1.0].
	\end{align*}
	
	The triangulation representation is the same as the one in 2D smooth case, with $K_1=50$, $K_2=200$. The spline degree and smoothness pairs are $(d_1,r_1)=(5,1)$ for $\beta$ and $(d_2,r_2)=(5,0)$ for each $\alpha_\ell$.

    \subsubsection{Smooth three-dimensional design}
	\label{sec:supp-dgm-3d}
	
	Let $\Omega=[0,1]^3$, with $N=67^3$ equally spaced voxels, and write
	$\bm s=(s_1,s_2,s_3)^\top$. The causal-effect function is
	\[
	\beta^\ast(\bm{s})
	=3.75+0.75\exp\left[-12.5\left\{(s_1-0.5)^2+(s_2-0.5)^2+(s_3-0.5)^2\right\}\right].
	\]
	The spatial process is
	\[
	\eta_i(\bm{s})=\sum_{t=1}^3\rho_t^{1/2}\zeta_{it}\psi_t(\bm{s}),
	\]
	with
	$\psi_1(\bm{s})=\sin(\pi s_1)+0.576$, $\psi_2(\bm{s})=2.398\cos(\pi s_2)$,
	and $\psi_3(\bm{s})=6.391(s_3-1/2)$.
	
	Moreover, the score vector and first-stage error are jointly Gaussian with unit marginal variances, mutually uncorrelated scores, and \(\operatorname{Cov}(\zeta_{it},\xi_i)=0.02, t=1,2,3.\)
	
\begin{figure}[H]
		\centering
		\includegraphics[width=0.55\linewidth]{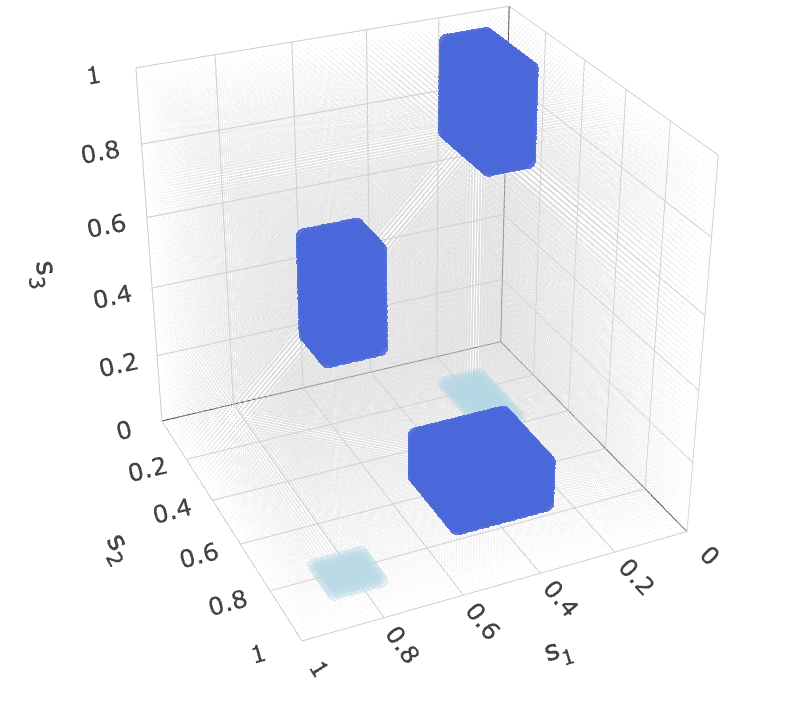}
		\caption{Design of the true IV effect $\alpha_\ell^\ast$ for $\ell\in \{1,\ldots,\lfloor p/2+1 \rfloor\}$ in the 3D simulations.}
		\label{fig:patt_3d}
	\end{figure}

	For $\ell=1,\ldots,q$,
	\[
	\alpha_\ell^\ast(\bm{s})=1.5I(\bm{s}\in\mathcal R_{3\mathrm D}),
	\]
	where $\mathcal R_{3\mathrm D}$ is the union of three cuboids, specified as follows (shown in Figure~\ref{fig:patt_3d}):
	$\mathcal R_{3\mathrm D}=\mathcal R_1\cup\mathcal R_2\cup\mathcal R_3$,
	where
	\[
	\begin{aligned}
		\mathcal R_1
		&=[0.125,0.250]\times[0.125,0.375]\times[0.750,1.000],\\
		\mathcal R_2
		&=[0.750,0.875]\times[0.750,0.875]\times[0.625,0.875],\\
		\mathcal R_3
		&=[0.250,0.500]\times[0.500,0.750]\times[0,0.125].
	\end{aligned}
	\]
	For $\ell=q+1,\ldots,p$, $\alpha_\ell^\ast(\bm{s})=0$. The noise settings are
	\[
	(\sigma,\rho_1,\rho_2,\rho_3)=
	\begin{cases}
		(1.5,0.20,0.02,0.01),&\text{low noise},\\
		(3.0,0.50,0.05,0.02),&\text{high noise}.
	\end{cases}
	\]
	
	We set the triangulation scheme as $K_1=40$, $K_2=2560$. The spline degree and smoothness pairs are $(d_1,r_1)=(5,1)$ for $\beta$ and $(d_2,r_2)=(5,0)$ for each $\alpha_\ell$. The increasing number of $K_2$ in the 3D case is due to the dimensionality of the spatial domain. A finer mesh for invalid IV effects is needed to provide sufficient spatial resolution for capturing localized and spatially varying invalid-IV effects in three dimensions. 
	
	\subsection{Signal-to-noise ratios}
	\label{sec:supp-snr}
	
	We characterize the simulation settings using two signal-to-noise ratios. The first-stage signal-to-noise ratio,
	\begin{equation}
		\label{EQU:snr-1}
		\mathrm{SNR}_D=\frac{\operatorname{Var}(\mathbf Z^\top\bm\gamma^\ast)}{\operatorname{Var}(\xi)},
	\end{equation}
	quantifies the strength of the candidate instruments. The imaging signal-to-noise ratio is defined as
	\begin{equation}
		\label{EQU:snr-2}
		\mathrm{SNR}_Y=
		\frac{
			N^{-1}\sum_{j=1}^{N}
			\operatorname{Var}\left[
			D_i\beta^\ast(\mathbf s_j)+
			\sum_{\ell=1}^{p}Z_{i\ell}\alpha^\ast_\ell(\mathbf s_j)
			\right]
		}{
			N^{-1}\sum_{j=1}^{N}
			\operatorname{Var}\left\{\eta_i(\bm{s}_j)+\varepsilon_i(\mathbf s_j)\right\}
		},
	\end{equation}
	which measures the average signal strength of the imaging outcome relative to the observation noise. Throughout the simulations, we report both $\mathrm{SNR}_D$ and $\mathrm{SNR}_Y$ to distinguish the effects of instrument strength and imaging noise.
	The realized SNR values among all scenarios are reported in Table~\ref{tab:supp-snr}.
	
	\begin{table}[ht]
		\centering
		\caption{ Signal-to-noise ratios for the simulation settings.}
		\label{tab:supp-snr}
		\resizebox{0.55\textwidth}{!}{
			\begin{tabular}{ccccccc}
				\toprule
				Dimension & Design & $p$ & Noise & $\operatorname{SNR}_D$ &
				$\operatorname{SNR}_Y\;$\\
				\midrule
				\multirow{4}{*}{2D} & \multirow{4}{*}{Smooth} & \multirow{2}{*}{3} & Low & 0.22 &4.75  \\
				&  & & High & 0.22 & 1.51 \\
				\cline{3-6}
				&  & \multirow{2}{*}{10} & Low & 0.27 & 5.27 \\
				&  &  & High & 0.27 & 1.67 \\
				\hline
				\multirow{4}{*}{2D} & \multirow{4}{*}{Nonsmooth} & \multirow{2}{*}{3} & Low &0.22  & 8.03\\
				&   &  & High &0.22  &2.54  \\
				\cline{3-6}
				& & \multirow{2}{*}{10} & Low &0.27  &8.55  \\
				& &  & High &0.27  &2.70  \\
				\hline
				\multirow{4}{*}{3D} & \multirow{4}{*}{Smooth}    & \multirow{2}{*}{3}  & Low  &0.32  &6.90 \\
				&    &   & High &0.32  &1.70 \\
				\cline{3-6}
				&   & \multirow{2}{*}{10} & Low  &0.39  &7.60 \\
				&    &  & High &0.39  &2.05 \\
				\bottomrule
			\end{tabular}
		}
	\end{table}
	
	\subsection{Implementation and tuning}
	\label{sec:supp-implementation}
	
	Hi-FIVE represents $\beta(\cdot)$ and $\alpha_\ell(\cdot)$ using Bernstein--B\'ezier
	splines over nested triangulations $\triangle_1$ and $\triangle_2$, respectively, where
	$\triangle_2$ is obtained by uniform refinement of $\triangle_1$. In two dimensions,
	each parent triangle is divided into four subtriangles; in three dimensions, each parent
	tetrahedron is divided into eight subtetrahedra. Locations lying exactly on shared edges
	or faces are removed when constructing local basis matrices to avoid duplicate simplex
	assignments.
	
	For each candidate $\lambda^{\rm R}$, a solution path over $\lambda^{\rm S}$ is fitted and
	$\lambda^{\rm S}$ is selected by mBIC. The final $\lambda^{\rm R}$ is selected by
	subject-level $K$-fold cross-validation. The optimization uses local linear approximation
	for the nonconvex group penalty, coordinate updates for the instrument--simplex blocks,
	and consensus ADMM for the continuity constraints on $\beta(\cdot)$.
	
	Across implementations, the numerical implementation uses a maximum of 20 outer LLA iterations and 100 inner coordinate-descent passes. The LLA, coordinate-descent, and ADMM convergence tolerances are $10^{-4}$; the ADMM penalty is $2$ with a maximum of $10^3$ ADMM iterations and $10^3$ conjugate-gradient iterations. The sparsity path is set to 60 logarithmically spaced values, decreasing from $10^3$ to $10^{-6}$ in the 2D case and from $e^{-1.5}$ to $e^{-5}$ in the 3D case.
	
	
	\subsection{Evaluation criteria}
	\label{sec:supp-metrics}
	
	For Monte Carlo replication $r=1,\ldots,R$, define
	\[
	\operatorname{ISE}^{(r)}
	=\frac{1}{N}\sum_{j=1}^{N}
	\{\widehat\beta^{(r)}(\bm{s}_j)-\beta^\ast(\bm{s}_j)\}^2.
	\]
	We report
	\[
	\operatorname{MISE}=\frac{1}{R}\sum_{r=1}^R\operatorname{ISE}^{(r)},
	\qquad
	\operatorname{SD}_{\mathrm{ISE}}
	=\left[\frac{1}{R-1}\sum_{r=1}^R
	\{\operatorname{ISE}^{(r)}-\operatorname{MISE}\}^2\right]^{1/2}.
	\]
	
	For a nominal simultaneous confidence level $1-\tau$, the empirical coverage rate is
	\(
	\operatorname{ECR}
	=\frac{1}{R}\sum_{r=1}^R
	I\left\{\beta^\ast(\bm{s}_j)\in\widehat{\mathcal C}^{(r)}_{1-\tau}(\bm{s}_j),
	\ j=1,\ldots,N\right\}.
	\)
	
	\subsection{Additional estimation results}
	\label{sec:supp-estimation-results}
	
	\begin{table}[H]
		\centering
		\caption{2D smooth: Mean integrated squared error and standard deviation (MISE $\times 100$, $\text{SD}_{\text{ISE}} \times 100$) of the causal effect $\beta$.}
		\label{tab:2dcont-mise-ecr}
		\resizebox{\textwidth}{!}{
			\begin{tabular}{cccccccc}
				\toprule
				\multirow{2}{*}{$n$}  & \multirow{2}{*}{Noise}  & \multicolumn{3}{c}{$p$=3} & \multicolumn{3}{c}{$p$=10}\\
				\cmidrule(lr){3-5} \cmidrule(lr){6-8}
				&  & sisVIVE & sisVIVE+  & Hi-FIVE  & sisVIVE & sisVIVE+  & Hi-FIVE \\
				\hline
				\hline
				\multirow{2}{*}{100} & Low & 58.45 (39.16) & 27.52 (19.10) & 1.03 (1.22) & 40.66 (33.40) & 18.90 (16.28) & 0.74 (0.55) \\
				\cline{2-8}
				& High & 67.72 (45.52) & 28.60 (21.06) & 1.86 (2.30) & 48.69 (32.87) & 20.71 (15.65) & 1.47 (1.03) \\
				\cline{1-8}
				\multirow{2}{*}{300} & Low & 55.08 (13.48) & 26.60 (6.54) & 0.29 (0.22) & 47.01 (36.73) & 22.41 (18.14) & 0.23 (0.20) \\
				\cline{2-8}
				& High & 57.38 (14.07) & 26.73 (6.65) & 0.51 (0.43) & 50.57 (35.09) & 23.36 (17.34) & 0.41 (0.38) \\
				\bottomrule
			\end{tabular}
		}
	\end{table}

	\begin{table}[H]
		\centering
		\caption{2D nonsmooth: Mean integrated squared error and standard deviation (MISE $\times 100$, $\text{SD}_{\text{ISE}} \times 100$) of the causal effect $\beta$. }
		\label{tab:jump-mise}
		\resizebox{\textwidth}{!}{
			\begin{tabular}{cccccccc}
				\toprule
				\multirow{2}{*}{$n$} & \multirow{2}{*}{Noise} & \multicolumn{3}{c}{$p=3$} & \multicolumn{3}{c}{$p=10$}\\
				\cmidrule(lr){3-5} \cmidrule(lr){6-8}
				& & sisVIVE & sisVIVE+ & Hi-FIVE & sisVIVE & sisVIVE+ & Hi-FIVE \\
				\hline
				\hline
				\multirow{2}{*}{100} & Low & 34.46 (18.96) & 34.78 (16.57) & 21.57 (11.28) & 40.46 (23.50) & 41.15 (21.30) & 18.05 (9.24) \\
				\cline{2-8}
				& High & 43.24 (22.67) & 36.16 (17.29) & 29.45 (12.84) & 46.59 (24.85) & 42.39 (21.02) & 30.79 (15.43) \\
				\cline{1-8}
				\multirow{2}{*}{300} & Low & 32.44 (7.44) & 34.66 (6.80) & 14.20 (5.94) & 56.74 (15.44) & 57.37 (14.26) & 11.33 (1.24) \\
				\cline{2-8}
				& High & 35.03 (7.99) & 34.99 (6.93) & 24.83 (13.46) & 58.77 (15.59) & 57.64 (14.19) & 19.52 (14.84) \\
				\bottomrule
			\end{tabular}
		}
	\end{table}

	\begin{table}[H]
		\centering
		\caption{3D smooth: Mean integrated squared error and standard deviation (MISE $\times 100$, $\text{SD}_{\text{ISE}} \times 100$) of the causal effect $\beta$. }
		\label{tb:mise-3d}
		\resizebox{\textwidth}{!}{
			\begin{tabular}{cccccccc}
				\toprule
				\multirow{2}{*}{$n$}  & \multirow{2}{*}{Noise}  & \multicolumn{3}{c}{$p$=3} & \multicolumn{3}{c}{$p$=10}\\
				\cmidrule(lr){3-5} \cmidrule(lr){6-8}
				&  & sisVIVE & sisVIVE+  & Hi-FIVE  & sisVIVE & sisVIVE+  & Hi-FIVE \\
				\hline 
				\hline
				\multirow{2}{*}{100} & Low & 11.42 (5.04) & 2.94 (2.05) & 1.72 (1.92) & 9.39 (3.22) & 3.12 (1.38) & 1.23 (1.05) \\
				\cline{2-8}
				& High & 36.16 (15.79) & 5.95 (6.21) & 5.55 (6.74) & 26.71 (9.83) & 4.54 (2.65) & 3.01 (2.91) \\
				\cline{1-8}
				\multirow{2}{*}{300} & Low & 4.92 (1.07) & 1.93 (0.67) & 0.53 (0.52) & 6.50 (1.47) & 3.31 (0.90) & 0.47 (0.40) \\
				\cline{2-8}
				& High & 12.42 (2.78) & 2.50 (1.45) & 1.21 (1.34) & 12.76 (2.51) & 3.82 (1.25) & 1.01 (0.86) \\
				\bottomrule
			\end{tabular}
		}
	\end{table}
	
	\paragraph{Illustration: comparison of estimators for the 3D smooth case}\mbox{}
	\begin{figure}[H]
		\centering
		
		\begin{tabular}{c@{\hspace{0.01\textwidth}}cccc}
			& \makebox[0.21\textwidth][c]{True}
			& \makebox[0.21\textwidth][c]{Hi-FIVE}
			& \makebox[0.21\textwidth][c]{sisVIVE}
			& \makebox[0.21\textwidth][c]{sisVIVE+}
			\\[-0.1cm]
			
			\raisebox{4\height}{\makebox[0.08\textwidth][c]{$s_3=9$}}
			&
			\includegraphics[
			width=0.21\textwidth,
			trim={0cm 1cm 2.5cm 1.2cm},
			clip
			]{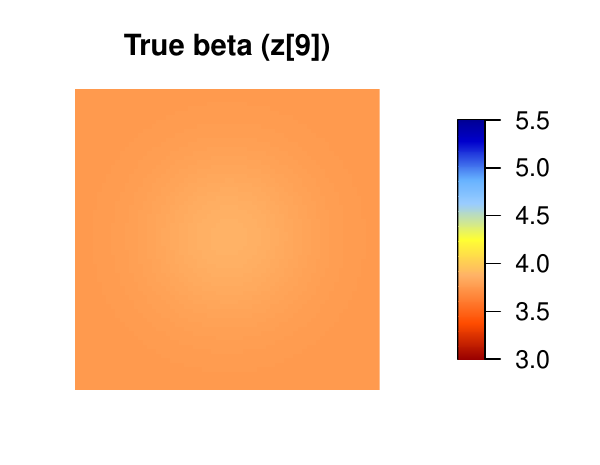}
			&
			\includegraphics[
			width=0.21\textwidth,
			trim={0cm 1cm 2.5cm 1.2cm},
			clip
			]{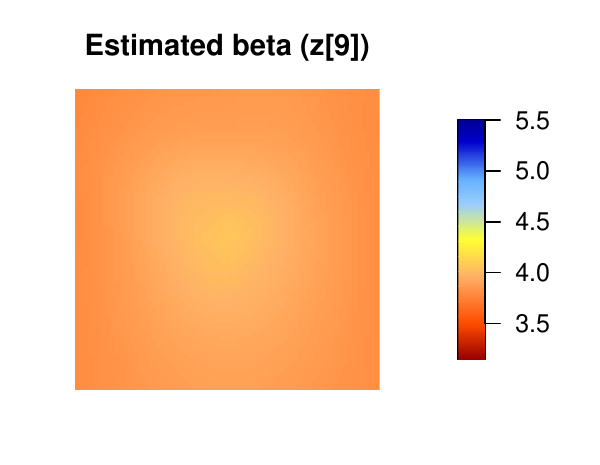}
			&
			\includegraphics[
			width=0.21\textwidth,
			trim={0cm 1cm 2.5cm 1.2cm},
			clip
			]{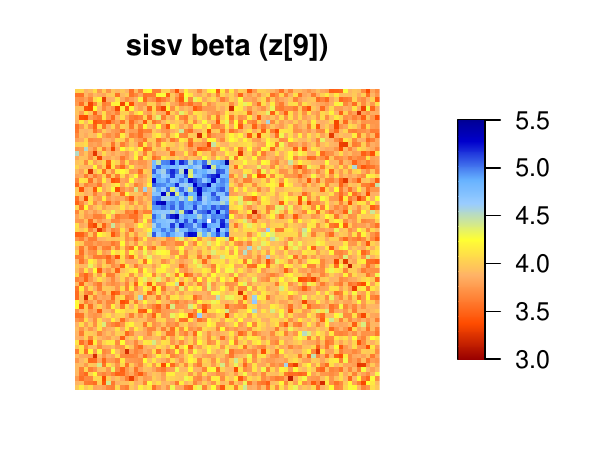}
			&
			\includegraphics[
			width=0.21\textwidth,
			trim={0cm 1cm 2.5cm 1.2cm},
			clip
			]{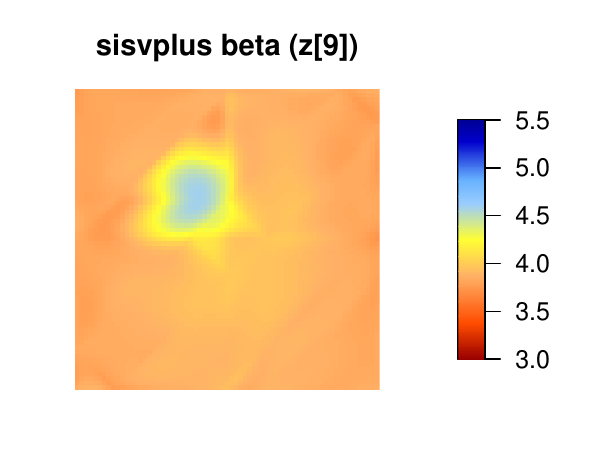}
			\\[-0.05cm]
			
			\raisebox{4\height}{\makebox[0.08\textwidth][c]{$s_3=23$}}
			&
			\includegraphics[
			width=0.21\textwidth,
			trim={0cm 1cm 2.5cm 1.2cm},
			clip
			]{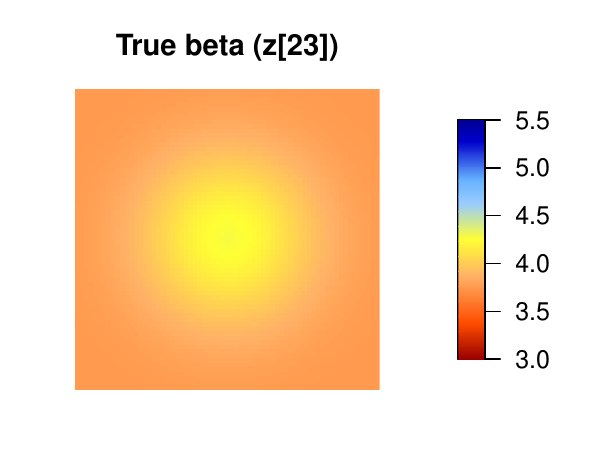}
			&
			\includegraphics[
			width=0.21\textwidth,
			trim={0cm 1cm 2.5cm 1.2cm},
			clip
			]{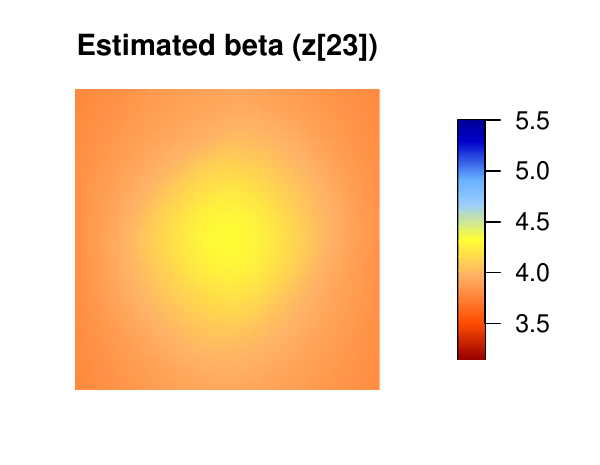}
			&
			\includegraphics[
			width=0.21\textwidth,
			trim={0cm 1cm 2.5cm 1.2cm},
			clip
			]{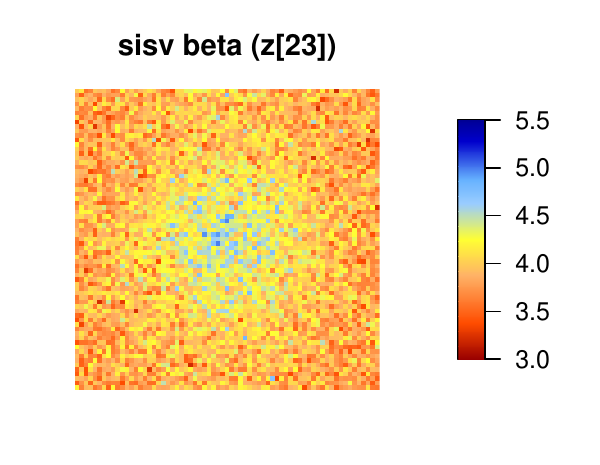}
			&
			\includegraphics[
			width=0.21\textwidth,
			trim={0cm 1cm 2.5cm 1.2cm},
			clip
			]{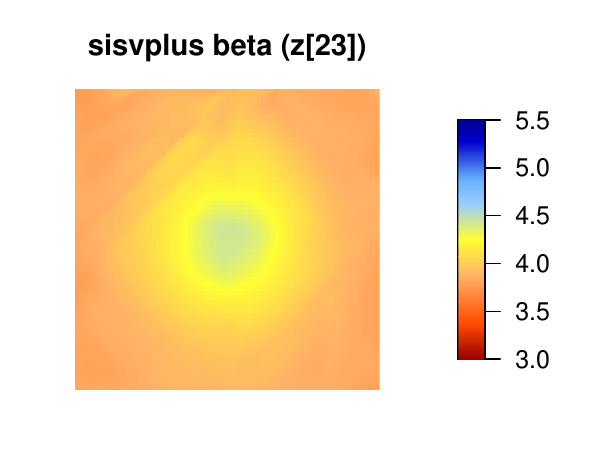}
			\\[-0.05cm]
			
			\raisebox{4\height}{\makebox[0.08\textwidth][c]{$s_3=51$}}
			&
			\includegraphics[
			width=0.21\textwidth,
			trim={0cm 1cm 2.5cm 1.2cm},
			clip
			]{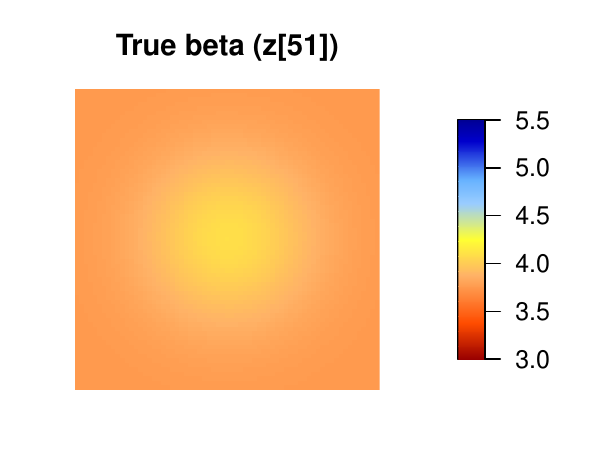}
			&
			\includegraphics[
			width=0.21\textwidth,
			trim={0cm 1cm 2.5cm 1.2cm},
			clip
			]{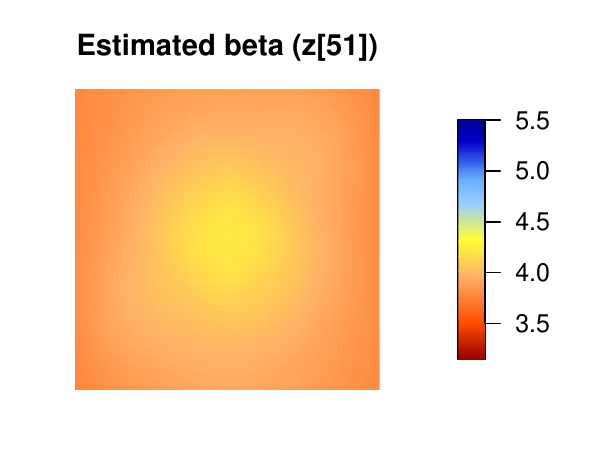}
			&
			\includegraphics[
			width=0.21\textwidth,
			trim={0cm 1cm 2.5cm 1.2cm},
			clip
			]{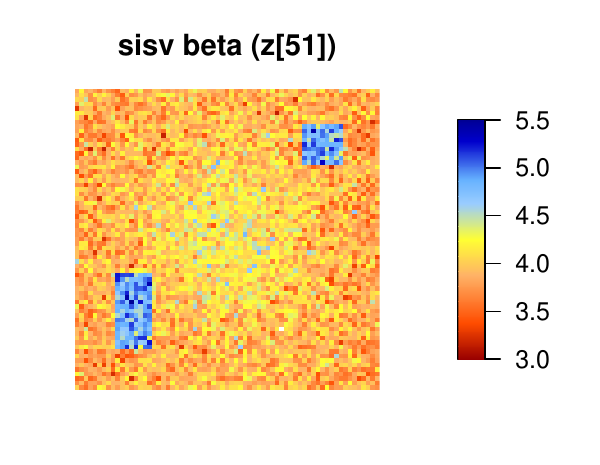}
			&
			\includegraphics[
			width=0.21\textwidth,
			trim={0cm 1cm 2.5cm 1.2cm},
			clip
			]{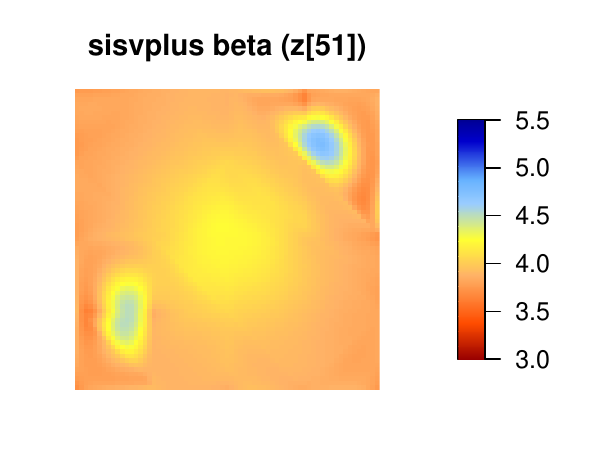}
		\end{tabular}
		
		\vspace{0.1cm}
		
		\includegraphics[
		width=0.4\textwidth
		]{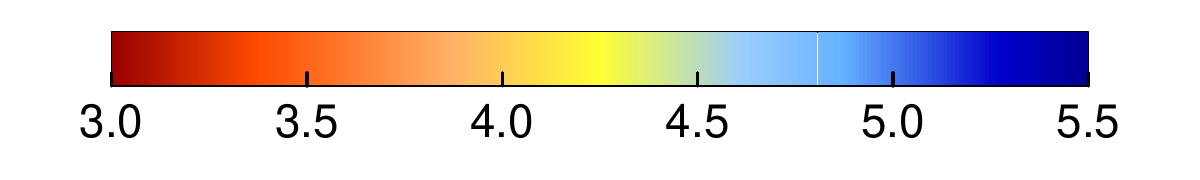}
		
		\caption{Estimation of $\beta$ in the 3D smooth example with $n=100$, 10 candidate IVs, and high noise. }
		\label{S-fig:bias_allmethod_cont}
	\end{figure}
    \subsection{Triangulation sensitivity}
	\label{sec:supp-triangulation}
	
	The smooth two-dimensional design with $n=300$ and $p=10$ is fitted using coarse,
	default, and fine triangulations. In addition, the interior vertices of the default
	triangulation are independently perturbed while preserving the domain boundary and mesh validity. All estimates are evaluated on the common pixel/voxel grid.
	
	\begin{table}[H]
		\centering
		\caption{MISE and standard deviation of the Hi-FIVE estimator under different mesh resolutions for $n=300$ over 200 replications.  Each entry is MISE (SD of ISE) multiplied by 100.}
		\label{tab:supp-triangulation}
		\begin{tabular*}{\textwidth}{@{\extracolsep{\fill}}lcccccc}
			\toprule
			& \multirow{2}{*}{$K_1$} & \multirow{2}{*}{$K_2$} & \multicolumn{2}{c}{$p=3$} & \multicolumn{2}{c}{$p=10$} \\
			\cmidrule(lr){4-5} \cmidrule(lr){6-7}
			& & & Low & High & Low & High \\
			\midrule
			Coarse &32  &128  & 0.30 (0.21) & 0.52 (0.42) & 0.24 (0.19) & 0.43 (0.38) \\
			Default &50  &200  &0.29 (0.22)  &0.51 (0.43)   &0.23 (0.20)  &0.41 (0.38)  \\
			Fine &72  &288  & 0.29 (0.21) & 0.63 (0.45) & 0.23 (0.20) & 0.59 (0.39) \\
			\bottomrule
		\end{tabular*}
	\end{table} 
	
	The triangulation sensitivity analysis shows that the Hi-FIVE estimator is generally stable across reasonable mesh resolutions. The coarse and default triangulations yield nearly identical MISE and variability for both $p=3$ and $p=10$, indicating that the estimation results are not sensitive to moderate changes in the number of simplices. Under low noise, the fine triangulation also produces comparable performance. Under high noise, the fine mesh results in a slight increase in MISE. Additionally, the corresponding standard deviations remain similar across the three resolutions. Thus, further refinement beyond the default mesh provides little additional benefit.
	Overall, these results indicate that Hi-FIVE is robust to the choice of triangulation within a practical range, while the default mesh resolution provides a favorable balance between approximation accuracy and estimation stability.

	\section*{Section D. Additional Details for Application}
	\paragraph{Covariate adjustment}
	
	Age, sex, and diagnosis are included as adjustment covariates. Age is standardized, sex is coded as female versus male, and diagnosis is coded as cognitively normal (CN) versus cognitively impaired (EMCI/MCI). To adjust for these covariates, we residualize the imaging outcome, p-tau exposure, and genetic instruments with respect to age, sex, and diagnosis prior to estimation. The subsequent first-stage regression and Hi-FIVE analysis are performed using the resulting residualized variables.
	
	\paragraph{Implementation details} For the real-data analysis, we use a triangulation consisting of $K_1=317$ tetrahedra for modeling the spatially varying causal effect. The same triangulation is used to define the subregions for the invalid-IV effects, yielding $K_2=317$. The spline degree and smoothness pairs are $(d_1,r_1)=(5,1)$ for $\beta$ and $(d_2,r_2)=(5,0)$ for each $\alpha_\ell$. The first-stage SNR is 0.176. We use a maximum of 20 outer LLA iterations and 100 inner coordinate-descent passes. The LLA, coordinate-descent, and ADMM convergence tolerances are all set to \(10^{-4}\), with an ADMM penalty parameter of 2, a maximum of $10^3$ ADMM iterations, and $10^3$ conjugate-gradient iterations. The smoothing parameter grid ranges from  $10^{1}$ to $10^{-7}$, while the sparsity parameter is selected over a path of 30 logarithmically spaced values decreasing from $e^{-1}$ to $e^{-7}$. 
	
\begin{center}
		\begin{figure}[!htbp]
			\centering      \includegraphics[width=1.1\linewidth]{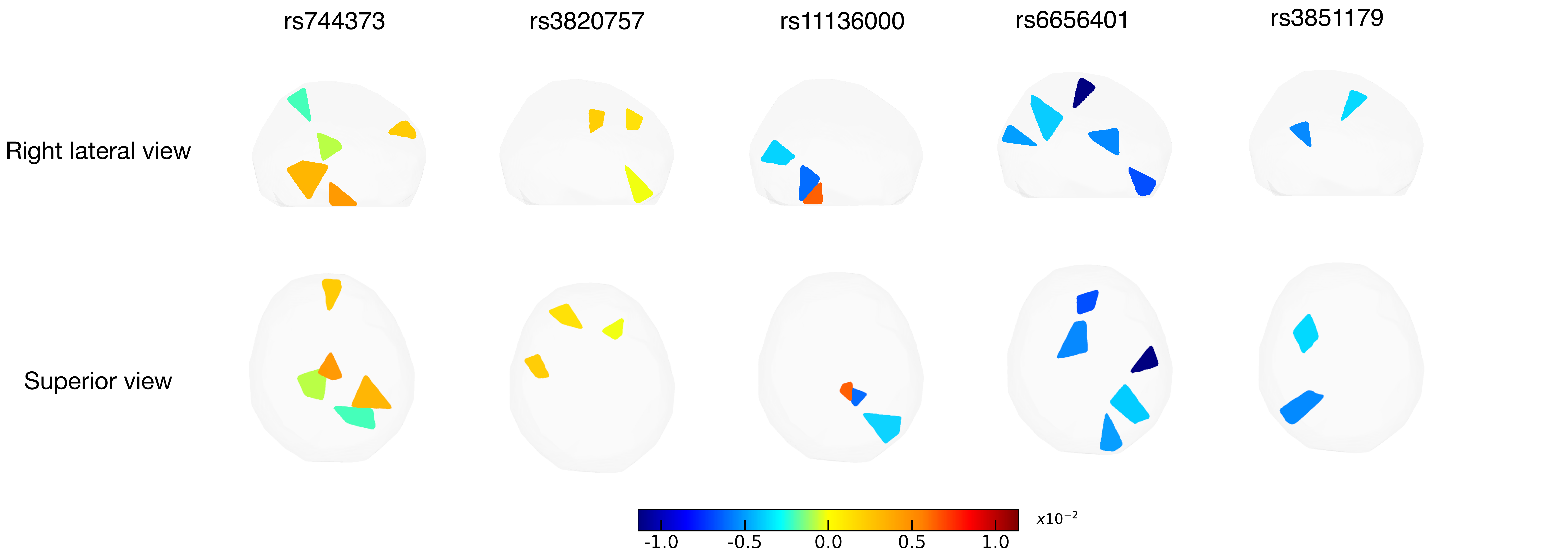}
			\vspace{-0.8cm}
			\caption{IV effects of five SNPs estimated by Hi-FIVE: Right lateral view (right = anterior); Superior view (top = anterior).}
			\label{fig:3drealdata-alpha_est}
		\end{figure}
	\end{center}

	\paragraph{Invalid IV illustration}
	The IV effects for five SNPs (rs744373, rs3820757, rs11136000, rs6656401 and rs3851179) are shown in Figure \ref{fig:3drealdata-alpha_est}. The colored regions represent the estimated nonzero invalid-IV effects, with blue indicating negative signals and orange to red indicating positive signals. Overall, the invalid-IV effects are sparse and spatially clustered, with clear SNP-specific patterns: rs744373 and rs3820757 show mainly positive clusters, whereas rs11136000 shows a smaller mixed pattern with both negative and positive regions. In contrast, rs6656401 and rs3851179 show predominantly negative clusters. Across the right lateral and superior views, these effects remain spatially localized, supporting the use of spatially varying invalid-IV estimation in the Hi-FIVE analysis.
	
\end{document}